\documentclass[english]{article}
\usepackage[T1]{fontenc}
\usepackage{geometry}
\usepackage{amsmath}
\usepackage{amssymb}
\usepackage{dsfont}
\usepackage{import}
\usepackage{mathtools}
\usepackage{url}
\usepackage{xcolor}
\usepackage{cite}
\usepackage{comment}
\usepackage{caption}
\usepackage{subcaption}
\usepackage{svg}
\usepackage{xcolor}
\usepackage{tikz}
\usepackage{tensor}
\usepackage{tikz}
\usepackage{geometry} 
\usepackage{float}
\usepackage{lipsum}

\definecolor{EdwardsLinkColor}{rgb}{0.0,0.0,0.3}

\usepackage[unicode=true,
 bookmarks=true,bookmarksnumbered=false,bookmarksopen=false,
 breaklinks=false,pdfborder={0 0 0},pdfborderstyle={},backref=false,colorlinks=true]
 {hyperref}
\hypersetup{pdftitle={Open-Closed-Open Triality},colorlinks=true,urlcolor=EdwardsLinkColor,anchorcolor=EdwardsLinkColor,citecolor=EdwardsLinkColor,filecolor=EdwardsLinkColor,linkcolor=EdwardsLinkColor,menucolor=EdwardsLinkColor,linktocpage=true}

\DeclareMathOperator{\Tr}{Tr}

\makeatletter
\usepackage{braket}

\newcommand\be{\begin{equation}}
\newcommand\ee{\end{equation}}

\newcommand{\inv}{\frac{1}}
\newcommand{\half}{\frac{1}{2}}

\newcommand{\bea}[1]{\begin{eqnarray}\label{#1} }
\newcommand{\eea}{\end{eqnarray}}

\newcommand{\w}{\omega}

\newcommand{\de}{\delta}

\newcommand{\dg}{\dagger}

\newcommand{\s}{\sigma}

\newcommand{\psid}{\psi^{\dg}}

\newcommand{\chid}{\chi^{\dg}}

\definecolor{darkgreen}{cmyk}{0.9,0,0.9,0}

\definecolor{darkblue}{cmyk}{0.9,0.9,0,0}

\makeatother

\usepackage{babel}

\begin{document}

\title{\bf{Strings from Feynman Diagrams}}

\author{ Rajesh Gopakumar${^\Phi}$,  Rishabh Kaushik${^S}$,  Shota Komatsu$^{\chi}$ ,\\  Edward A. Mazenc${^{\psi}}$ \&  Debmalya Sarkar$^{S^{\dagger}}$}
\date{}

\maketitle

\makeatletter{\renewcommand*{\@makefnmark}{}
\footnotetext{${}^\Phi$rajesh.gopakumar@icts.res.in}\makeatother}
\makeatletter{\renewcommand*{\@makefnmark}{}
\footnotetext{${}^S$rishabh.kaushik@icts.res.in}\makeatother}
\makeatletter{\renewcommand*{\@makefnmark}{}
\footnotetext{${}^\chi$shota.komatsu@cern.ch}\makeatother}
\makeatletter{\renewcommand*{\@makefnmark}{}
\footnotetext{${}^\psi$emazenc@ethz.ch}\makeatother}
\makeatletter{\renewcommand*{\@makefnmark}{}
\footnotetext{${}^{S^\dagger}$dsarkar@g.harvard.edu}\makeatother}

\begin{center}
    \small{${}^{\Phi, S, S^{\dagger}}$\it{International Centre for Theoretical Sciences-TIFR,\\ Shivakote, Hesaraghatta Hobli, Bengaluru North 560089, India.}}\\
    \vspace{0.1cm}
  \small{${}^\chi$\it{CERN, Theoretical Physics Department,\\ CH-1211 Geneva 23, Switzerland.}}\\
\vspace{0.1cm}
  \small{${}^\psi$\it{Institut für Theoretische Physik, ETH Zürich, \\ CH-8093 Zürich, Switzerland.}}\\
  \vspace{0.1cm}
  \small{${}^{S^{\dagger}}$\it{Jefferson Physical Laboratory, Harvard University, \\Cambridge, MA 02138, USA.}}
\end{center}

\vspace{2cm}

\begin{abstract}
\noindent For correlators in $\mathcal{N}=4$ Super Yang-Mills preserving half the supersymmetry, we manifestly recast the gauge theory Feynman diagram expansion as a sum over dual closed strings. Each individual Feynman diagram maps on to a Riemann surface with specific moduli. The Feynman diagrams thus correspond to discrete lattice points on string moduli space, rather than discretized worldsheets. This picture is valid to all orders in the $1/N$ expansion. Concretely, the mapping is carried out at the level of a two-matrix integral with its dual string description. It provides a microscopic picture of open/closed string duality for this topological subsector of the full AdS/CFT correspondence. At the same time, the concrete mechanism for how strings emerge from the matrix model Feynman diagrams predicts that multiple open string descriptions can exist for the same dual closed string theory. By considering the insertion of determinant operators in $\mathcal{N}=4$ SYM, we indeed find six equivalent open-string descriptions. Each of them generates Feynman diagrams related to one another via (partial) graph duality, and hence encodes the same information. The embedding of these Kontsevich-like duals into the 1/2 SUSY sector of AdS/CFT is achieved by open strings on giant graviton branes.

\end{abstract}

\pagebreak

\tableofcontents

\pagebreak

\section{Overview \& Summary of Results}

't Hooft's seminal insight into the correspondence between large $N$ gauge theories and string theories \cite{tHooftPlanar} relied on viewing the gauge theory Feynman diagrams (double-line or ribbon graphs) as some sort of closed string worldsheets. Making this picture precise has been difficult, and presents one of the primary obstacles to a deeper understanding of gauge-string duality. The goal of this paper is to concretely illustrate an explicit reconstruction algorithm of the string theory dual to a simple gauge theory, namely a matrix model. This can also be viewed as a topological subsector of the standard AdS/CFT correspondence \cite{Maldacena:1997re,Adsandholog,Gubser:1998bc}, in which the gauge theory description reduces to a two-matrix integral. We will show how to extract both the worldsheet and the target space embedding map of the dual closed strings entirely from the matrix model Feynman diagrams (for a class of correlators). Very schematically, the one-equation summary of this work takes the form shown in Fig. 1.

\begin{figure}[H]
    \centering
    \includegraphics[scale=0.7]{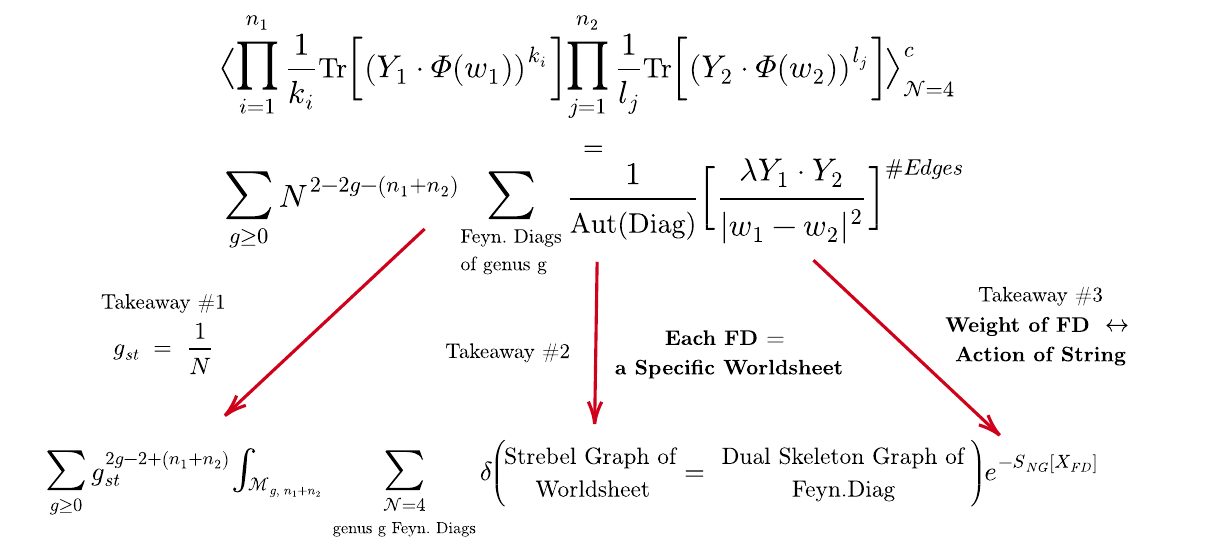}
    \caption{\textbf{Schematic Summary of Main Result } The purpose of this paper is to manifestly recast the Feynman diagram expansion of certain protected correlators in $\mathcal{N}=4$ SYM as a dual sum over closed string configurations.}
    \label{fig:introsummary}
\end{figure}

The top line is a connected correlation function of multiple 1/2 BPS operators in $SU(N)$ $\mathcal{N}=4$ Super Yang Mills, inserted at two points on the boundary. The middle line shows its expansion in terms of Feynman diagrams of a given (graph) genus. The bottom line is the dual sum over closed string worldsheets. The precise worldsheet theory which gives rise to such an integrand on moduli space will be presented elsewhere \cite{DSDIII} - a quick summary can be found in Sec. \ref{sec:Amodel}. For now, we highlight three main lessons to take away from this equality: 
 
 \begin{enumerate}
    \item{As expected, the $1/N$ expansion translates to the genus expansion of the dual closed string, with $g_{st}=1/N$. }
     \item Perhaps more surprising is the equivalence at the level of the summand. Each individual Feynman diagram of the gauge theory maps precisely onto one specific worldsheet: this is the delta-function in the integrand on moduli space $\mathcal{M}_{g,n}$.
     \item The weight of each Feynman diagram (here a simple product of position space propagators) corresponds to the exponential of the Nambu-Goto action. This is the area of the worldsheet, computed via the pull-back of the target space metric under the embedding map of the string $X_{FD}(z)$. The ribbon graph also encodes the map $X_{FD}(z)$. 
 \end{enumerate}

\begin{figure}[H]
    \centering
\includegraphics[scale=0.70]{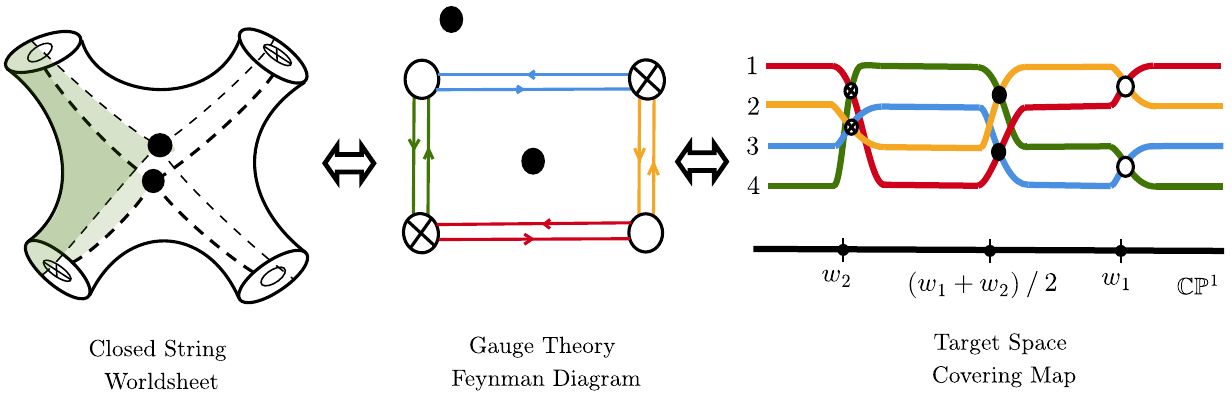} 

    \caption{\textbf{\textit{Each} Feynman Diagram as a Closed String Configuration } We map each matrix model Feynman diagram to a specific closed string worldsheet. Each edge of the graph corresponds to one strip building up the worldsheet. The same diagram also encodes the embedding map of the string into the target space. The embedding turns out to be a covering map of the Riemann sphere, branched over precisely three points. To each edge, we associate one sheet of the cover. The string picture is exact; in particular, it holds even when the Feynman diagrams are 'sparse'. }
    \label{fig:FDasBoth}
\end{figure}

 We have structured the paper around explaining how precisely these three main features arise, first focusing on the worldsheet perspective on the duality, and then shifting gears to the target space. These results build on previous work \cite{DSDI,gopakumar2011simplest} which outlined a derivation of the simplest gauge-string duality - the dual of a single Hermitian matrix model. One may thus view the present example as a derivation of the `next simplest gauge/string duality'. In the rest of this section, we will give an overview of the various ingredients that go into the reconstruction of the dual string theory in this case, as well as some of its features and the embedding into the AdS/CFT correspondence. These will then be fleshed out in the following sections. We will illustrate many of the ideas of this paper with a particularly simple, `sparse' Feynman diagram, shown in Fig. \ref{fig:FDasBoth}.

\subsection{The Worldsheet Perspective} \label{sec:introWS}

't Hooft famously showed how one might associate a genus $g$ to each double line Feynman diagram that perturbatively contributes to an $n$-point correlator\cite{tHooftPlanar}. 
A refinement of 't Hooft's picture was proposed in \cite{freefieldsadsI,freefieldsadsII,freefieldsIII}, where one would associate to \textit{each} Feynman diagram of a gauge theory a specific $n$-punctured closed string Riemann surface of genus $g$. The idea was that the sum over these Feynman diagrams (for fixed $g$) would thus transmute into a sum (or integral) over the closed string moduli space ${\cal M}_{g,n}$. 

\begin{figure}[H]
    \centering
\includegraphics[scale=0.65]{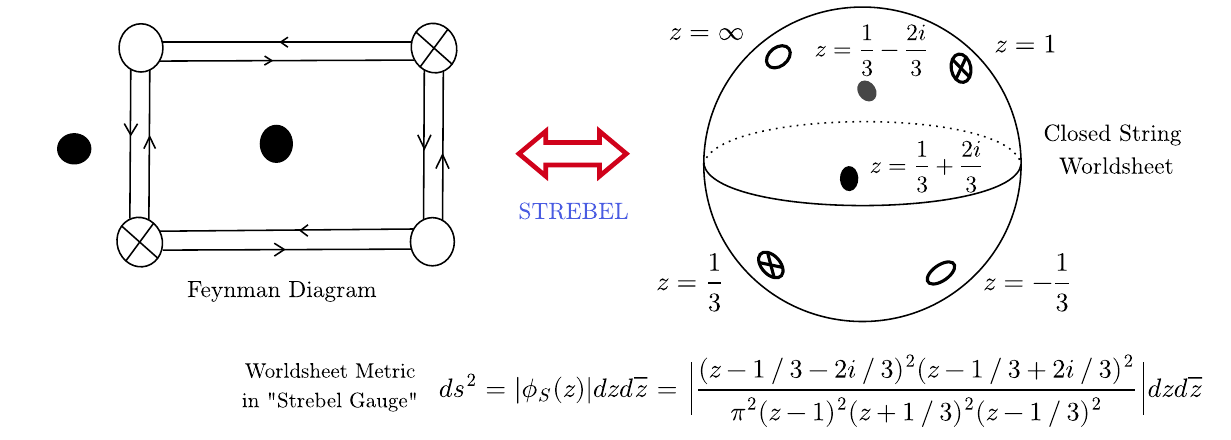} 

    \caption{\textbf{Reconstructing the Worldsheet } There exists a unique Strebel differential $\phi_{S}(z)dz^2$ on every Riemann surface. It is fully specified by a (metrized) ribbon graph. This allows us to precisely translate between gauge theory Feynman diagrams and specific closed string worldsheets, including an explicit metric on the surface.}
    \label{fig:FD=WSIntro}
\end{figure}

We will explicitly implement this idea in this paper, as illustrated in Fig. \ref{fig:FD=WSIntro}. In Sec. \ref{sec:StrebGraphsVsFD}, we demonstrate precisely how to assign a closed string worldsheet to each Feynman diagram of our matrix model. At first sight, this extremely fine-grained notion of open/closed string duality might appear surprising. An older picture, arising from the minimal model string dualities of the early '90s \cite{gross-migdal,brezin1990exactly,douglas1990strings}, suggested that the Feynman diagrams represented a mere discretization of the worldsheet. There, the dual minimal string theories appeared only after taking a further continuum limit, which required a double-scaling limit of the matrix model \cite{DavidDiscretizedWS,klebanov1991string, ginsparg1993lectures}. The original Feynman diagrams therefore did not individually have any significance as string worldsheets, being akin to a non-universal latticisation. However, most examples of higher-dimensional gauge/string dualities operate in the regular 't Hooft limit - without any double scaling \footnote{We are certainly not the first to emphasize this point. Dijkgraaf and Vafa gave beautiful topological string interpretations for certain non-double-scaled matrix integrals in  \cite{DijkgraafVafa02}, showing how the emergent target space was captured by the matrix model spectral curve. Our focus here is on the closed string worldsheet.}. If we want to understand how something like the AdS/CFT correspondence works at the most microscopic level, along the lines of 't Hooft's vision, we therefore need to step away from the double-scaling limit\footnote{Ooguri and Vafa \cite{OV2002worldsheet} also adopted this literal worldsheet perspective in deriving the conifold duality of \cite{GV98,GV99}. The double-lines of the ribbon graphs were understood as phase boundaries on the worldsheet. In the language of Open-Closed-Open triality, this duality between Chern-Simons and an A-model topological string is of `F-type', and hence works somewhat differently than the one advocated here. }. In particular, we need to rethink the dictionary between Feynman diagrams and the worldsheet, and this is what was advocated in \cite{freefieldsadsI,freefieldsadsII,freefieldsIII}.

A central role in this program is played by a natural quadratic differential on the worldsheet, the so-called Strebel differential \cite{freefieldsIII}. In Sec. \ref{sec:StrebelReview}, we will explain how the Strebel differential provides a powerful tool to translate between ribbon graphs of a gauge theory and Riemann surfaces. In a nutshell, it gives a precise dictionary between any genus $g$ ribbon graph with $n$ faces, along with a set of length assignments to its edges, and points on the (decorated) moduli space of genus $g$ Riemann surfaces $\mathcal{M}_{g,n} \times \mathbb{R}_{+}^{n}$ (summarized in Fig. \ref{fig:StrebelDict}) \cite{strebel1984, KontsevichAiry, harer1988cohomology, mukhireview}. Such a parametrization of moduli space has also found important applications in string field theory \cite{ZwiebachProof, WittenOSFT, Atakan}. The dictionary relies on a canonical graph on the corresponding Riemann surface (the `Strebel Graph') associated with the Strebel differential, as illustrated in Fig. \ref{fig:HorTraj}. 

In addition, the Strebel differential defines a metric on the worldsheet, discussed in Sec. \ref{sec:strebelmetric}. This metric is flat everywhere, except for conical singularities mapping on to the vertices and faces of the corresponding ribbon graph. This can be viewed as a special choice of gauge (`Strebel gauge') for the dual string theory. One way to frame our proposal is that this particular choice of gauge for the worldsheet metric makes manifest the relation between gauge theory Feynman diagrams and the dual strings. In Sec. \ref{sec:ExplicitWS}, we provide an example reconstructing the metric on the worldsheet dual to a particularly simple Feynman diagram (shown in Fig. \ref{fig:FD=WSIntro}).

 The Strebel differential is the key object in a mathematical construction tailor made for gauge-string duality. Sec. \ref{sec:PhysPic} shows that it paints a very physical picture, illustrated in Fig. \ref{fig:StrebelAsStrips}, as to how precisely each closed string worldsheet can be obtained by gluing together open string strips. These strips are in one-to-one correspondence with the (double line) edges of the gauge theory Feynman diagram. This will be the starting point of Sec. \ref{sec:howto}, where we provide a concrete algorithm to reconstruct a closed string from a Feynman diagram. Ultimately, these strips can be viewed as descending from the worldsheet of open strings between the D-branes giving rise to the gauge theory. The Strebel construction is thus literally instructing us how the open strings reassemble themselves into closed ones, directly at the level of the worldsheet.

 \begin{figure}[H]
    \centering
    \includegraphics[scale=0.65]{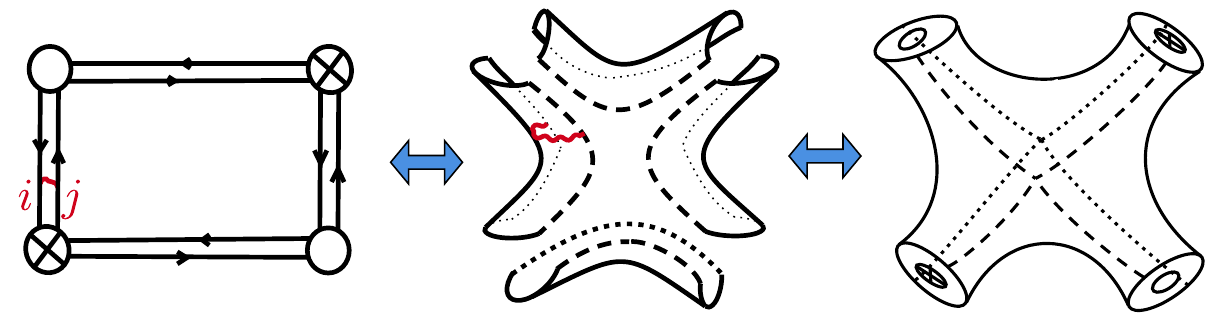}
    \caption{\textbf{Gluing Open String Strips } The Strebel differential decomposes each Riemann surface into a collection of strips. We identify the edges of the gauge theory Feynman diagram with these strips. Precise gluing rules, discussed in Sec. \ref{sec:howto}, allows one to assemble the closed string worldsheet from these open string strips.  }
    \label{fig:IntroGlueing}
\end{figure}

The Strebel differential furthermore provides a rigorous understanding of the appearance of {\it string bits} \cite{KogutBits,Klebanov-susskind,Thorn:1991fv}: each such strip making up the closed string surface can be thought of as the worldsheet of a single string bit. We will see how this construction makes manifest the known relation between single trace operators and asymptotic closed string states in the dual description \cite{Polyakovbook}. Each matrix (`letter') appearing in the gauge-invariant words corresponds to one bit making up the string, with the $U(N)$-invariance of the trace tying together the endpoints of the string bit chain. See Fig. \ref{fig:StringBits}.  

\begin{figure}[ht!]
    \centering

    \includegraphics[scale=0.68]{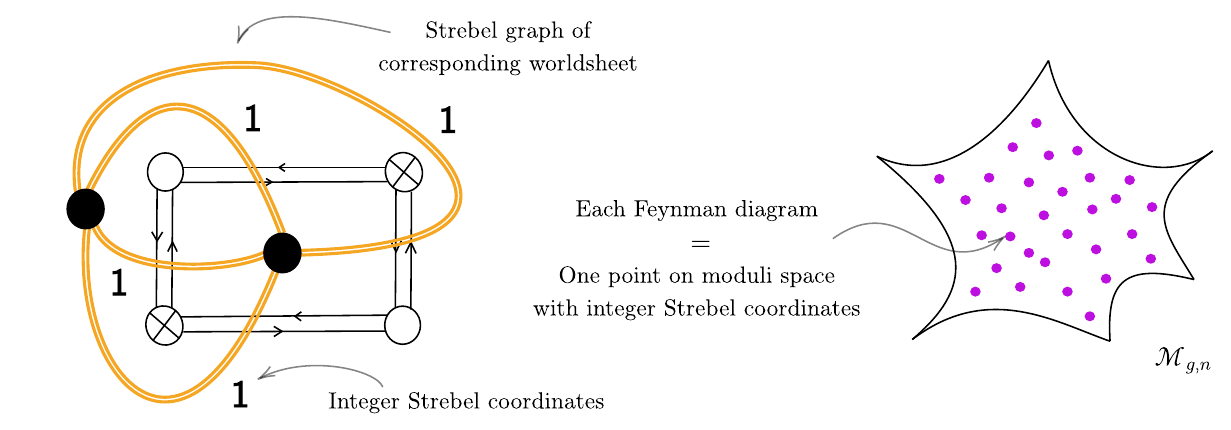}

    \caption{\textbf{Feynman Diagrams as Lattice Points on $\mathcal{M}_{g,n}$ } Each Feynman diagram, contributing to a particular $n$-point correlator, maps onto a particular point on moduli space, labeled by a set of integers. These integers are the edge lengths of the worldsheet's Strebel graph, related to the original Feynman diagram via graph duality.}
    \label{fig:IntroMgnLattice}
\end{figure}

 One of the striking consequences of applying this construction to our matrix model is that the dual closed string theory admits a presentation in which only a discrete set of worldsheet configurations contribute to the path integral! The Strebel parametrization of the moduli space makes manifest what is special about these points: as we will see, all the lengths associated to the edges of the (Strebel) graph are integers (as was first prefigured in \cite{razamatGauss}). Since these lengths essentially provide coordinates on $\mathcal{M}_{g,n} \times \mathbb{R}_{+}^{n}$, their integrality defines a lattice on moduli space. We thus reach the conclusion that the Feynman diagrams of the matrix model, instead of discretizing the worldsheet of the string, directly latticize the moduli space of Riemann surfaces - see Fig. \ref{fig:IntroMgnLattice}.  

The physical observables of the gauge theory, i.e. the matrix model correlators, will in fact count these discrete points (subject to some constraints determined by the precise form of the correlators). These objects have been studied before in the mathematics literature as computing `discrete volumes of moduli space'. They provide a natural discretisation of the continuum Kontsevich and Weil-Petersson volumes of moduli space \cite{ChekovNorb, NorbStringVols, OkuyamaDiscWP} which have been at the centre of much recent work in low dimensional gravity and string theory \cite{SSSJTmatrix,LowensteinVolsTR,VirMinString,MaloneyWP,WittenVolsReview,NorburySuperVols,BlommaertDSSYK,JohnsonVols}\footnote{Another way that these discrete points on moduli space can be seen to be special is in their characterization as arithmetic Riemann surfaces. These are special points on moduli space where the Riemann surface admits a realisation in terms of a polynomial equation defined over the algebraic closure of the rationals $\bar{\mathbb{Q}}$.}.
In an appropriate `BMN-like' limit, the discrete volumes do go over to the more familiar continuum volumes of Kontsevich \cite{KontsevichAiry} or Mirzakhani \cite{Mirzakhani,Eynard:2007kz}.

The claim that there is a surprising localization of the worldsheet theory to these special surfaces provides a stringent check of our derivation. This is a prediction that can be verified  in the dual worldsheet theory, as will be shown in \cite{DSDIII}. We should note that a similar localisation on moduli space was discovered in the case of the worldsheet theory for the tensionless limit of $AdS_3\times S^3\times T^4$ (i.e. with one unit of NS-NS flux) \cite{eberhardt2019worldsheet,eberhardt2020deriving}. In fact, in a particular limit of correlators with `large twist' operators, it was shown that this localisation occurs precisely onto the same special lattice points mentioned above \cite{gaberdiel2021symmetric}. There too, the Feynman diagrams associated with the dual symmetric orbifold CFT could be associated with each string worldsheet configuration contributing to the localised path integral.

\subsection{The Target Space Perspective}

\begin{figure}[H]
    \centering
\includegraphics[scale=0.70]{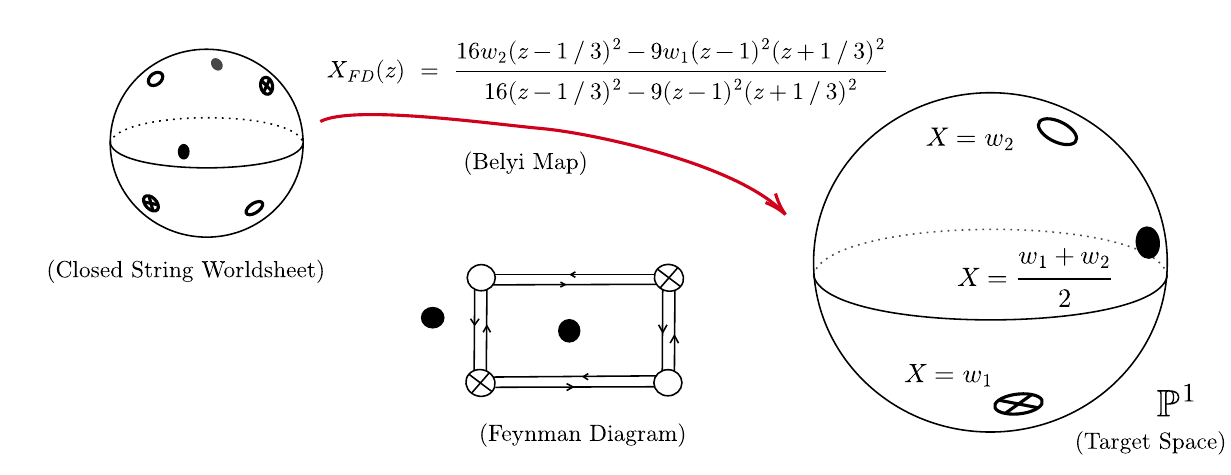} 

    \caption{\textbf{Reconstructing the Map into Target Space } Each Feynman diagram encodes a specific holomorphic covering map of the Riemann sphere, $X_{FD}(z)$, branched over exactly three points. The vertices and faces of the diagram specify the ramification profiles over these three points. In simple cases, such as here, $X_{FD}(z)$ can be written down explicitly. This so-called Belyi map is the embedding of the worldsheet we reconstructed into a target space $\mathbb{CP}^{1}$. From the point of view of $AdS/CFT$, we would like interpret this $\mathbb{CP}^{1}$ as the compactified plane on the $AdS_{5}$ boundary where the $\mathcal{N}=4$ SYM gauge theory operators were inserted.}
    \label{fig:IntroFDBelyi}
\end{figure}

Reconstructing the worldsheet is, however, only half the story. 
A full derivation of the closed string theory from the Feynman diagrams requires understanding how these worldsheets are embedded in the emergent target space (the `bulk'). If we want to literally identify the sum over Feynman diagrams with a dual path integral over string configurations, we need to able to reproduce the \textit{weight} of each individual Feynman diagram. On the closed string side, each configuration is weighted by the exponential of the string's action, $e^{-S_{WS}[X]}$, evaluated on the embedding map $X$. We will show how each Feynman diagram precisely encodes the map $X$ into target space, see Fig. \ref{fig:IntroFDBelyi}.

Specifically, we compute correlators of single traces of multiples of either one of the two matrices, and consider their perturbative expansion in terms of Feynman diagrams. It will turn out that we can specify certain holomorphic maps into a target space sphere, purely from the combinatorial data associated to any given diagram \cite{gopakumar2011simplest,deMelloKoch:2014khl, DiFrancescoItzykson}. The weight of each diagram precisely maps onto the Nambu-Goto action of the string. From a topological string point of view, this action is the pullback of the target space Kähler form to the worldsheet \cite{WittenTopSigma,MarinoAmodelreview}. We can thus manifestly re-express the sum over gauge theory Feynman diagrams in terms of a sum over embedding maps of Riemann surfaces into an emergent target space. Furthermore, the scaling with $N$ of the ribbon graph determines the genus of the given Riemann surface in the topological expansion of this string theory. The size of the matrices, $N$, thus plays the role of the inverse string coupling constant, as in the usual 't Hooft counting, but distinct from the double scaling limit.

The maps that arise are special branched coverings of the Riemann sphere known as \textit{Belyi maps}\cite{zbMATH03668723,Lando2003GraphsOS}. They are branched over exactly three points of the target Riemann sphere. The dictionary between these maps and the Feynman diagrams goes roughly as follows. The total number of Wick contractions (i.e. the number of edges of the graph) specifies the degree of the cover. The branching profile over two of the three points can be read off from the two types of vertices of the Feynman diagram (corresponding to the two types of matrices). The faces, in turn, parametrize the branching over the third and final point. The fact that these maps can be encoded as graphs was popularized by Grothendieck \cite{grothendieck1984}. He called them `Childrens' Drawings', or \textit{dessins d'enfants}. For us, these drawings are none other than the gauge theory Feynman diagrams. The surprising equivalent description of these drawings in terms of Riemann surfaces covering a sphere can then be explained as one of the simplest instances of gauge-string duality \cite{deMelloKoch:2010hav}.

This construction, starting from the Feynman diagrams, predicts that the dual closed string theory should somehow localize to these Belyi maps. This is a highly non-trivial requirement of the worldsheet theory. As mentioned above, in the context of deriving the $AdS_3/CFT_2$ correspondence at the symmetric orbifold point, the tensionless string theory \cite{Gaberdiel:2018rqv} was indeed shown to localize to certain branched coverings of a sphere \cite{eberhardt2019worldsheet,eberhardt2020deriving, gaberdiel2021symmetric}. Among the key ingredients of this localisation were the special properties of the $SL(2,\mathbb{R})$ Wess-Zumino-Witten (WZW) model, at the  (supersymmetric) level $k=1$.

The dual to our two-matrix model is a topologically-twisted coset of this $AdS_3$ string at exactly this level. How this worldsheet theory precisely reproduces the Belyi maps will be discussed at length in \cite{DSDIII}. This string theory also turns out to be equivalent to (a subsector of) the $c=1$ string at self-dual radius \cite{MukhiVafa,ashok2006topological}, which had already been shown to be dual to the 2-matrix model via open-closed-open triality in \cite{DSDI}. As a sanity check, we include in this paper certain genus 0 and genus 1 computations, which match the $c=1$ string theory predictions made over thirty years ago in \cite{klebanov1991string}. Once again, we emphasize that the matrix models considered here are not double-scaled, unlike the Matrix Quantum Mechanics which is used to describe the $c=1$ string theory.\\ 

\subsection{Tying the Two Perspectives Together}

One might wonder how these two parts of the story relate to each other, namely the reconstruction of the worldsheet itself versus that of the string's embedding into the target space. Recall that the Strebel differential gives a precise reconstruction of the worldsheet by gluing various infinite strips together. More precisely, it specifies a set of transition functions between local coordinate charts on each strip \cite{mulase1998ribbon,mulaseStrebel}. In this paper, we will be able to understand the emergent closed string embedding map as a the composition of two operations. First, we find a simple unramified map from each strip (making up the worldsheet) onto the target space $\mathbb{CP}^{1}$. Geometrically, each strip wraps the bulk Riemann sphere precisely once. If we then compose this simple map for each strip with the Strebel gluing procedure, we derive the global map for the full closed string worldsheet onto the target. The branching of this total covering map is thus dictated by the way the various strips meet to build up the string worldsheet. This construction allows us to identify the metric on the worldsheet defined by the Strebel differential with the pullback, under the Belyi map, of a natural Kähler metric on the target space sphere. This explains microscopically how the gauge theory Feynman diagrams contain all the necessary information to simultaneously reconstruct both the worldsheet and the bulk embedding. 

From the worldsheet perspective, we had stated in Sec. \ref{sec:introWS} that we have contributions from a discrete set of special points on moduli space. How is this compatible with the above picture of the embedding into a target sphere? A theorem proved by mathematician G.V. Belyi in 1979 provides the answer \cite{zbMATH03668723}. He showed that not all Riemann surfaces admit Belyi maps, i.e. not all worldsheets can cover the sphere in the precise way we derived from the Feynman diagrams. Instead, these maps only exists for precisely those points on moduli space which correspond to arithmetic Riemann surfaces. As mentioned above, those are precisely the points which have integer length Strebel differentials. Hence the localization of the dual string path integral to Belyi maps gives rise to an integrand on moduli space which is a sum of delta-functions with support exclusively on the discrete set of points we derived from the matrix model Feynman diagrams via the Strebel construction.

\subsection{Embedding in $\mathcal{N}=4$ SYM \& AdS/CFT}
One obvious question is how the lessons we learn from deriving the closed string dual to a two-matrix integral generalize to more complicated examples of gauge/string duality. Since the Strebel parametrization of the string worldsheet in terms of ribbon graphs is completely general, we could hope this particular instance might be embedded in a full-fledged holographic duality. This is indeed the case. Concretely, we show how the generating function of the correlators in $\mathcal{N}=4$ SYM which preserve half the supersymmetry (1/2 SUSY) can be reduced to the two-matrix integral studied in this paper. These correlators consists of multiple single trace $1/2$-BPS operators inserted at two points on the boundary of AdS. The two matrices thus correspond to two different linear combinations of the six scalar fields of $\mathcal{N}=4$ evaluated at these two points respectively. This further provides a physical interpretation of the two $N \times N$ matrices: they describe transverse oscillations of open strings on the $N$ D3-branes giving rise to $\mathcal{N}=4$ at low energies. The Feynman diagrams of the matrix model are exactly those used to compute the 1/2 SUSY correlators. Mapping these diagrams to closed string configurations thus realizes a microscopic picture of open/closed duality for this highly supersymmetric sector of the full AdS/CFT correspondence. 

\begin{figure}[H]
    \centering

    \includegraphics[scale=0.6]{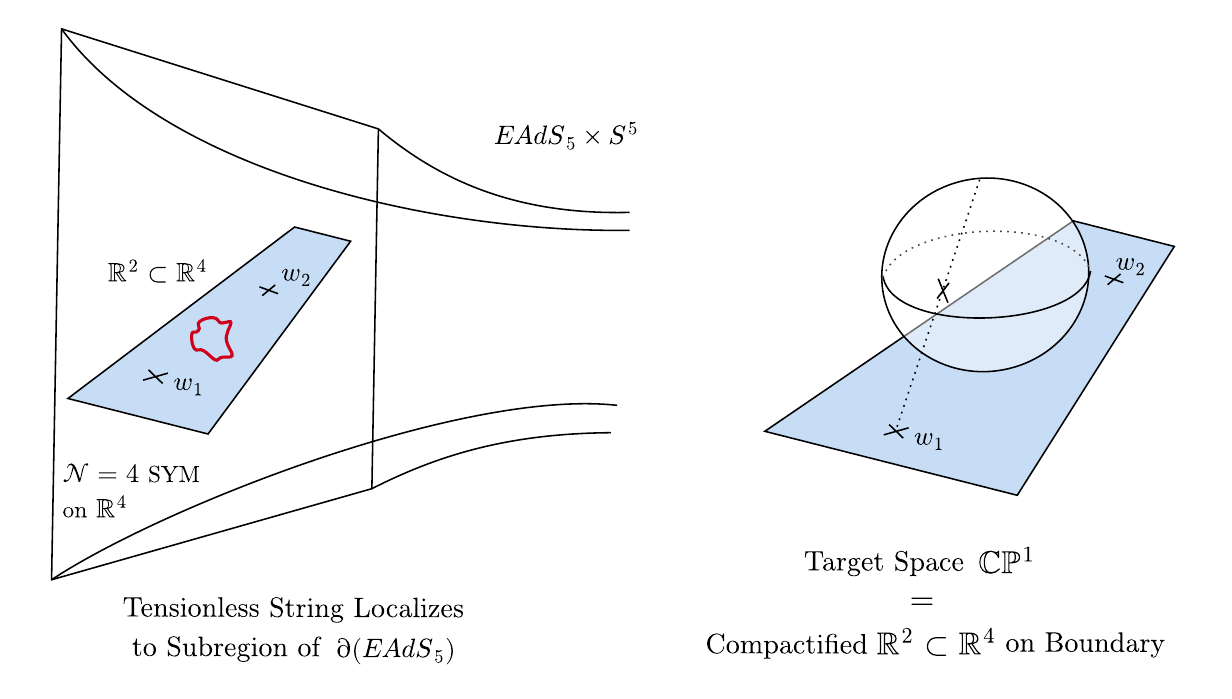}

    \caption{\textbf{Proposed Embedding in AdS/CFT }  On the open string side, 1/2 BPS determinant correlators inserted at two points $w_1$ and $w_2$ in $\mathcal{N}=4$ SYM reduce to the two-matrix integral studied in this paper. (\textit{Left}) On the closed string side, the picture we suggest is that of a tensionless string localizing to the boundary of AdS. (\textit{Right}) The string wraps the plane on which the gauge theory operators were inserted, compactified to a $\mathbb{CP}^{1}$. This is the Riemann sphere target space discussed in the context of Belyi Maps. }
    \label{fig:IntroEmbedding}
\end{figure}

 Embedding the Belyi maps into Type IIB string theory on $AdS_{5} \times S^5$ is less straightforward. However, since these gauge theory correlators we study are essentially independent of the 't Hooft coupling, we can appeal to the proposed worldsheet dual of free $\mathcal{N}=4$ SYM \cite{gaberdiel2021string, gaberdiel2021worldsheet}. A recent computation of such simple two-point functions of $1/2$ BPS operators showed that, under certain assumptions, the closed twistor string localizes precisely to covering maps of a $\mathbb{P}^{1}$ \cite{Bhat:2021dez}. The north and south poles of this sphere correspond to the two insertion points in the gauge theory correlator. The branching over the poles was confirmed to be given by the $R$-charge of the operators. This matches our finding here that single trace operators built of $l$-matrix elements creates an order $l$ branchpoint on the closed string worldsheet. So far, only the spectrum of free $\mathcal{N}=4$ SYM has been matched to the that of this proposed twistor dual (again, under a certain assumption regarding the physical state condition on the worldsheet). The fact that these protected correlators could also be expressed in terms of covering maps of $\mathbb{P}^1$, just as we find here, buttresses the picture articulated in \cite{Bhat:2021dez} and a direct embedding of the Belyi maps into string theory on $AdS_{5} \times S^{5}$.

\subsection{Multiple Open String Descriptions for the Same Closed String Dual}
When we view open-closed string duality through the lens of the Strebel construction, a surprising possibility arises. Both the Strebel graph and its graph dual encode the same information. As we will see, the (skeleton graphs of) matrix model Feynman diagrams are {\it not} the Strebel graphs of the dual closed strings, but rather their graph dual. One can then ask, is there another gauge theory whose Feynman diagrams are directly the Strebel graphs? Its diagrammatic expansion would then reproduce the same sum over dual closed string worldsheets. If we think of these two gauge theories as arising from open strings, this would suggest the existence of multiple open string descriptions for the same closed string dual \cite{Joburg}. In \cite{DSDI}, this was dubbed as an {\it Open-Closed-Open Triality}. The existence of multiple open string description might seem rather far-fetched. However, even one of the simplest string theories, pure topological gravity on the worldsheet (i.e. the $(2,1)$-minimal string), admits two open string duals: the double-scaled one-matrix model and the cubic Kontsevich model \cite{gross-migdal,Witten2dintersec,KontsevichAiry}. In the double-scaled matrix model, vertices of the Feynman diagrams go over to vertex operator insertions in the minimal string. In the Kontsevich model, it is the faces of the diagrams which correspond to the marked points on the worldsheet. Graph duality, exchanging vertices and faces, thus indeed relates their Feynman diagrams, as was first pointed out in \cite{Joburg}. As shown by Maldacena, Moore, Seiberg and Shih \cite{Maldacena_2004}, these two matrix integrals should be understood as capturing the open strings on two different set of branes, the $ZZ$ and $FZZT$ branes of the minimal string. Such a picture was generalized to the $(p,1)$ minimal string in \cite{HashimotoShih}.

We will find an even richer picture of open-closed-open triality for the two-matrix integral, where the two types of vertex operator insertions can be realized in terms of either faces\footnote{Matrix models with nontrivial weights for faces are studied previously in the context of the so-called dually-weighted matrix models \cite{Kazakov:1995ae,Das:1989fq}. Our open-closed-open triality encompasses generalization of such models.} or vertices of ribbon graphs.  This gives a total of six equivalent but different matrix models. Their Feynman diagrams are related to one another via (partial) graph duality, see Figs. \ref{fig:OCOfromBranes}. All of them generate the same sum over closed string worldsheets. In fact, we are able to identify the edges of their respective diagrams as various special curves traced out on the closed string Riemann surface by the Strebel differential (shown in Fig. \ref{fig:OCOfromWS}).  
Since our starting matrix model could be neatly embedded in $\mathcal{N}=4$ SYM, one can ask what these other open string descriptions represent, in that context. The punchline will be that the equivalent of the Kontsevich-like models will arise from open strings on $1/2$ BPS giant graviton branes, as anticipated in \cite{Joburg,brown2011complex}. This idea has also been explored recently in the context of $\mathcal{N}=4$ SYM under the guise of the "$\rho$-matrix model" \cite{komatsuOCO, Caron-Huot:2023wdh}, which allowed for a large $N$ saddle point evaluation of determinant correlators. 

\subsection{Plan of the Paper}

Section \ref{sec:WSmain} focuses on the worldsheet. We introduce the matrix model and its Feynman diagrams in Sec. \ref{sec:MatrixFDs}. How this matrix model arises from $\mathcal{N}=4$ SYM is postponed to later in the paper (Sec. \ref{sec:embedding}). We then review the Strebel parametrization of the closed string moduli space $\mathcal{M}_{g,n}$ in Secs. \ref{sec:StrebelReview} and \ref{sec:strebelmetric}. We show how it allows us to translate between individual Feynman diagrams and specific Riemann surfaces in \ref{sec:ExplicitWS} and \ref{sec:howto}, based on the physical picture painted in \ref{sec:PhysPic}.

Section \ref{sec:targetpersp} derives the target space picture of this simple gauge/string duality. In Sec. \ref{subsec:deriveBelyi} Each Feynman diagram is mapped to a particular holomorphic covering map of a target $\mathbb{CP}^{1}$. We explicitly reconstruct such a map for a particular Feynman diagram in Sec. \ref{sec:ExplicitBelyi} before justifying the construction in Sec. \ref{subsec: Wicktoperm}.

The purpose of Sec. \ref{sec:tyingpersps} is to tie together the two perspectives. It relates the localization on the moduli space of Riemann surfaces to the path integral localization on Belyi maps into target space. We then give an overview of the topological A-model worldsheet theory, which describes the closed string dual to the two-matrix integral. This motivates the derivation in Sec. \ref{sec:StrebelAsPullback} showing how the Strebel differential arises from the pull back to the worldsheet of a Kähler form on target space. This suffices to show, in Sec. \ref{sec:weight=action}, that the Nambu-Goto action of the string indeed reproduces the weighting factor of each dual Feynman diagram . 

Section \ref{sec:embedding} spells out how the matrix model is embedded within $\mathcal{N}=4$ SYM, and briefly outlines how we believe the Belyi maps, on the closed string side, fit into a recently proposed twistor string description dual to the free gauge theory. 

We explore the ideas of Open-Closed-Open triality in Sec. \ref{sec:OCOgen}. We find six different matrix integral representations for the generating function of $\mathcal{N}=4$ SYM correlators preserving half the supersymmetry in Secs. \ref{sec:intinout}  \& \ref{sec:OCOpartials}. We show, in Sec. \ref{sec:OCOGiants}, how strings on giant graviton branes provide the equivalent of the Kontsevich matrix model for this 1/2 SUSY subsector of the AdS/CFT correspondence. The closed string worldsheet and target space perspectives on this Open-Closed-Open triality are discussed in Secs. \ref{sec:OCOonWS} and \ref{sec:OCOinTS}, respectively. 

For a lightning summary of this work, we refer the reader to the short bullet list of the most important ideas in Sec. \ref{sec:Takeaways}. We then conclude in Sec. \ref{sec:lookingforward} by outlining some of the most immediate next steps suggested by this derivation of this simple model of gauge/string duality. 
\pagebreak

\section{Worldsheets from Feynman Diagrams} \label{sec:WSmain}

The purpose of this section to explain how each Feynman diagram in a perturbative expansion of matrix model correlators maps onto a specific point on the moduli space of Riemann surfaces, which we identify as the worldsheet of the dual closed string. We first define the two matrix model, which we will be focusing on, and describe the $1/N$ expansion of its correlators. We then introduce the Strebel parametrization of the moduli space and explain how the ribbon graph data suffices to precisely reconstruct the Riemann surface of the string. We emphasize that the Feynman diagram should not be thought of as a mere discretization of the worldsheet. In particular, we never appeal to any double-scaling limit.  Instead, for the purpose of demonstration, we choose a particularly simple (sparse) Feynman diagram, and map it onto a particular point on the stringy moduli space $\mathcal{M}_{h,n}$ \footnote{In order to avoid any confusion with the coupling $g$ of the matrix model, we have decided to refer to the genus as $h$ in this section.}. Finally, we provide a constructive algorithm to assemble the closed string worldsheet by gluing together open string strips corresponding to the edges of the gauge theory Feynman diagrams. In the next section we will connect this worldsheet perspective to the target space one.

\subsection{The Matrix-Model \& Its Feynman Diagrams } \label{sec:MatrixFDs}

The two-matrix model we will consider in this paper is also the simplest two-matrix model: 
\begin{equation}
   Z_{N}= \int dK dM_{N \times N} \ e^{-\frac{N}{g} \Tr(KM)}.
\end{equation}
We will take $K,M$ to be Hermitian $N\times N$ matrices \footnote{As we will see in Sec. \ref{sec:embedding}, $K$ and $M$ arise as complex linear combinations of the six scalars of $\mathcal{N}=4$ SYM evaluated at a point: $K = Y_{1}^{I} \Phi^{I}(w_1)$ and $M = Y_{2}^{I} \Phi^{I}(w_2)$. While the scalars are hermitian matrices, the coefficients of the six-vectors $Y_{j}^{I}$ are necessarily complex since the vectors square to 0, i.e. $Y_{j}^{I}Y_{j}^{I}=0$. The Wick combinatorics for Hermitian matrices suffices to reproduce the $\mathcal{N}=4$ SYM result. } 

We will be studying correlators built out of traces of either matrix\footnote{The $i,j$ indices in the formula below should not be confused with matrix indices, but rather label the various powers of the matrices appearing in the traces.}
\begin{align}\label{eq:2MM}
    \Braket{\prod_{i=1}^{V_{K}} \frac{1}{l_{i}}\Tr\left( K^{l_i}\right) \prod_{j=1}^{V_{M}} \frac{1}{n_{j}}\Tr\left( M^{n_j} \right)} & \equiv  \frac{1}{Z_{N}}\int dK dM_{N \times N} \ e^{-\frac{N}{g} \Tr(KM)}\prod_{i=1}^{V_K} \frac{1}{l_{i}}\Tr\left( K^{l_i}\right) \prod_{j=1}^{V_M} \frac{1}{n_{j}}\Tr\left( M^{n_j} \right).
\end{align}
In fact, we will primarily be interested in the connected contribution to these correlators, which we denote by a subscript $c$. We will ultimately map these connected correlators onto expectation values of vertex operators in the dual closed string theory. There will be one insertion of a vertex operator for each insertion of a trace on the matrix model side, as is familiar from the usual AdS/CFT dictionary. We will find two natural sets of vertex operators, corresponding to insertions of traces built from either $M$ or $K$\footnote{The factors of $\frac{1}{l_i}$ and $\frac{1}{n_j}$ are chosen so that the final answer for the correlator is precisely equal to that of a counting problem in the bulk.}. The disconnected parts of the matrix model correlator, in turn, will be dual to multi-string states. This reflects the fact that the gauge-theory provides a second-quantized description, while the worldsheet formalism is inherently first-quantized. We will explain later why, from the point of view of the higher-dimensional gauge theory, single trace "words" built from both $K$ and $M$ are not the appropriate operators to consider - even though they make perfect sense to study from a purely matrix-model perspective. 

Following 't Hooft, the connected correlators admit a $1/N$ expansion
\begin{equation}
     \Braket{\prod_{i=1}^{V_{K}} \frac{1}{l_{i}}\Tr\left( K^{l_i}\right) \prod_{j=1}^{V_{M}} \frac{1}{n_{j}}\Tr\left( M^{n_j} \right)}_c = \sum_{h\geq 0} N^{2-2h-(V_{K}+V_{M})} \Braket{\prod_{i=1}^{V_K} \frac{1}{l_{i}}\Tr\left( K^{l_i}\right) \prod_{j=1}^{V_M} \frac{1}{n_{j}}\Tr\left( M^{n_j} \right)}_{c, h} .
\end{equation}
One should view the RHS of the above equation as defining what we mean by the "genus $h$" correlator in the matrix model. This expansion will translate to the usual genus expansion for the dual worldsheet computation. It is worth noting that the genus expansion of such correlators truncates at a finite order in $1/N$. This is another highly non-trivial requirement of the dual closed string. We will ultimately understand it as the restricted existence of non-trivial covering maps of the Riemann sphere, to which the worldhseet theory localizes (see Sec. \ref{sec:finitegenusexp}).

Let us now turn our attention to the computation of such quantities, and their diagrammatic representation. The $n$-point function of this two-matrix model can be computed via Wick's theorem since the action is quadratic. Each Wick contractions pairs an $M$ and $K$ matrix element, since the two-point function reads
\begin{equation}\label{eq:prop}
    \braket{K_{ij}M_{lm}} = \frac{g}{N} \delta_{jl}\delta_{im} \,\,\,; \quad \quad \quad  \braket{K_{ij}K_{lm}}= \braket{M_{ij}M_{lm}} = 0. 
\end{equation}
The truncation of the genus expansion discussed above can be restated simply as the fact there exists only a finite number of Wick contractions for any correlator, and among these, there will be one with the smallest power of $N$. 

\begin{figure}[ht!]
    \centering
    \includegraphics[scale=0.8]{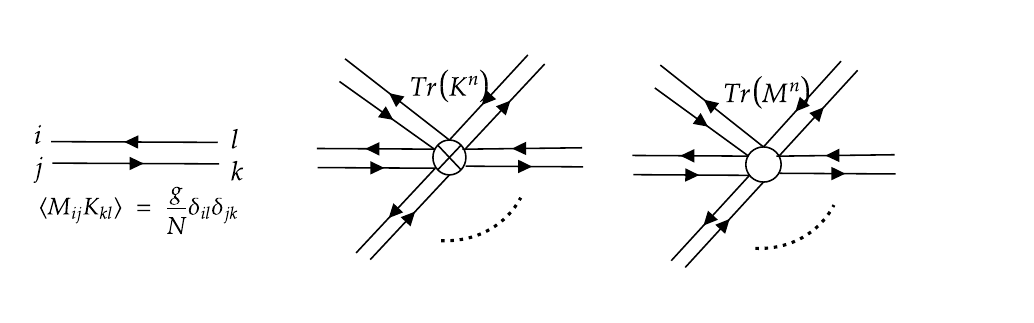}

    \caption{\textbf{Feynman Rules for the $K,M$ Matrix Model } The propagator, and the two kinds of (external) vertices, corresponding to insertions of $\Tr(K^n)$ and $\Tr(M^n)$. }
    \label{fig:KMbasic}
\end{figure}
 
We can keep track of these various Wick contractions contributing to a correlators by using Feynman diagrams. These diagrams will be built starting with two types of vertices. These correspond to the insertions of traces of either the $M$ or $K$ matrices (but not both) as defined by the correlator. Since each Wick contraction pairs a $K$ and $M$ matrix element, with no $KK$ or $MM$-propagator, the edges of the graph can only connect vertices of different type. Graphs with this property are called (fully) bi-partite graphs. As usual, the valency of the vertices is determined by the power of the matrix in the trace. See Figure. \ref{fig:KMbasic}. The Feynman diagram expansion of the genus $h$ correlator,
\begin{equation}
     \Braket{\prod_{i=1}^{V_{K}} \frac{1}{l_{i}}\Tr\left( K^{l_i}\right) \prod_{j=1}^{V_{M}} \frac{1}{n_{j}}\Tr\left( M^{n_j} \right)}_{c,h}, \label{eq:GenCorrelator}
 \end{equation}
will therefore consist of a sum over connected ribbon graphs of genus $h$, with $V_{K}$ "crossed" vertices with valency $l_i$, $V_{M}$ "un-crossed" vertices with valency $n_j$, and $E=\sum_{i=1}^{V_{K}} l_{i} = \sum_{j=1}^{V_{M}} n_{j}$ edges. We can quickly confirm its scaling with $N$: each edge comes with a factor of $N^{-1}$ while the sum over the color index in each face (loop) give a positive power of $N$. Each edge also gives a factor of $g$.  The vertices do not contribute any further factors, so that such a diagram scales as $g^E N^{(V_{K}+V_{M})-E+F-(V_{K}+V_{M})}=g^E N^{2-2h-(V_{K}+V_{M})}$.   

One of the simplest examples we can consider is the following four-point function:
\begin{equation}
    \Braket{\Tr\left( K^2\right)\Tr\left( K^2\right)\Tr\left( M^2 \right)\Tr\left( M^2\right)} = 2 \Braket{\Tr\left( K^2\right)\Tr\left( M^2\right)}^2 +  \Braket{\Tr\left( K^2\right)\Tr\left( K^2\right)\Tr\left( M^2 \right)\Tr\left( M^2\right)}_c.
\end{equation}
\begin{figure}[h!]
    \centering
    \includegraphics[scale=0.65]{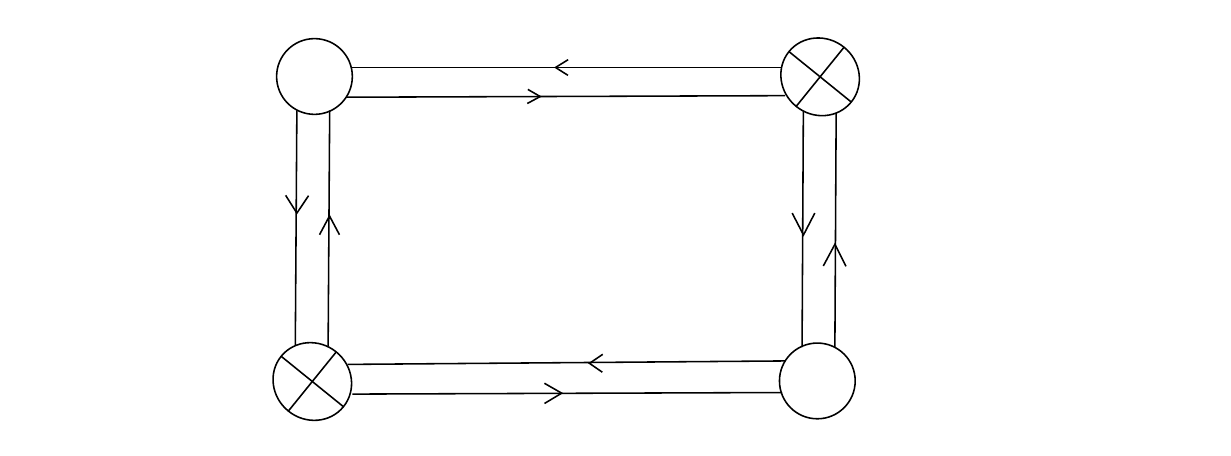}

    \caption{\textbf{The only Feynman diagram contributing to the computation of $\Braket{\left(\frac{1}{2}\Tr(K^2)\right)^2\left(\frac{1}{2}\Tr(M^2)\right)^2}_{c}$ } The two crossed vertices correspond to the (external) insertions of $\Tr K^2$, while the two uncrossed ones map onto the insertions of $\Tr M^2$. There are no internal vertices since the theory is free. It is clear there are no non-planar, connected (bipartite) graphs built out of this set of vertices, which is why there are no subleading in $N$ contributions to the correlator. }
    \label{fig:MainFD}
\end{figure}

We will follow this example throughout the entire paper, illustrating along the way how the relevant Feynman diagram gives rise to a specific worldsheet and Belyi map. Focusing our attention on the connected component, we find
\begin{equation}
    \Braket{\frac{1}{2}\Tr\left( K^2\right)\frac{1}{2}\Tr\left( K^2\right)\frac{1}{2}\Tr\left( M^2 \right)\frac{1}{2}\Tr\left( M^2\right)}_{c} = g^4\times N^{-2}.
\end{equation}
In this particularly simple example, there are no higher genus contributions beyond the leading order genus zero term. In fact, there is only one Feynman diagram we need to consider. This can be confirmed diagrammatically, see Fig.\ref{fig:MainFD}.

Note that the (single) coefficient in the genus expansion is an integer - up to an overall $g$ dependence, which is always set by the powers of $K$ (or $M$) appearing in the correlator . This is no coincidence. We will later discuss the combinatorial problem which the computation of these correlators is solving. The punchline will be that these expectation values are counting certain discrete points on the moduli space of Riemann surfaces $\mathcal{M}_{h,n}$. We will need the Strebel parametrization of moduli space to make that connection clear. 

%\pagebreak

\subsection{The Strebel reconstruction of Riemann surfaces from ribbons graphs}  \label{sec:StrebelReview}

The Strebel parametrization of the (decorated) moduli space of punctured Riemann surfaces, $\mathcal{M}_{g,n} \times \mathbb{R}_{+}^{n}$, provides an invaluable tool to elucidate open/closed string duality\cite{Rastelli2004}\footnote{In this section, $g$ will denote a genus, not the matrix model coupling.}. It provided the backbone for the construction of cubic open string field theory \cite{WittenOSFT}. Kontsevich famously applied Strebel's ideas \cite{KontsevichAiry} to prove Witten's conjecture on the integrable hierarchy governing $\psi$-class intersection theory on the moduli space of curves. We will use it as a way to precisely translate between the gauge theory Feynman diagrams and the dual closed string worldsheet.
%In this section, 
We will present it here as an algorithm to reconstruct the worldsheet, and focus on those features most salient to deriving gauge/string duality. First, we briefly review the Strebel parametrization of moduli space, focusing on the way a certain quadratic differential leads to a foliation of every Riemann surface by a particular family of curves. These curves are known as horizontal trajectories (together with an associated family of vertical trajectories) and provide the bridge between worldsheet moduli and graph data.
\begin{figure}[ht!]
    \centering

    \includegraphics[scale=0.75]{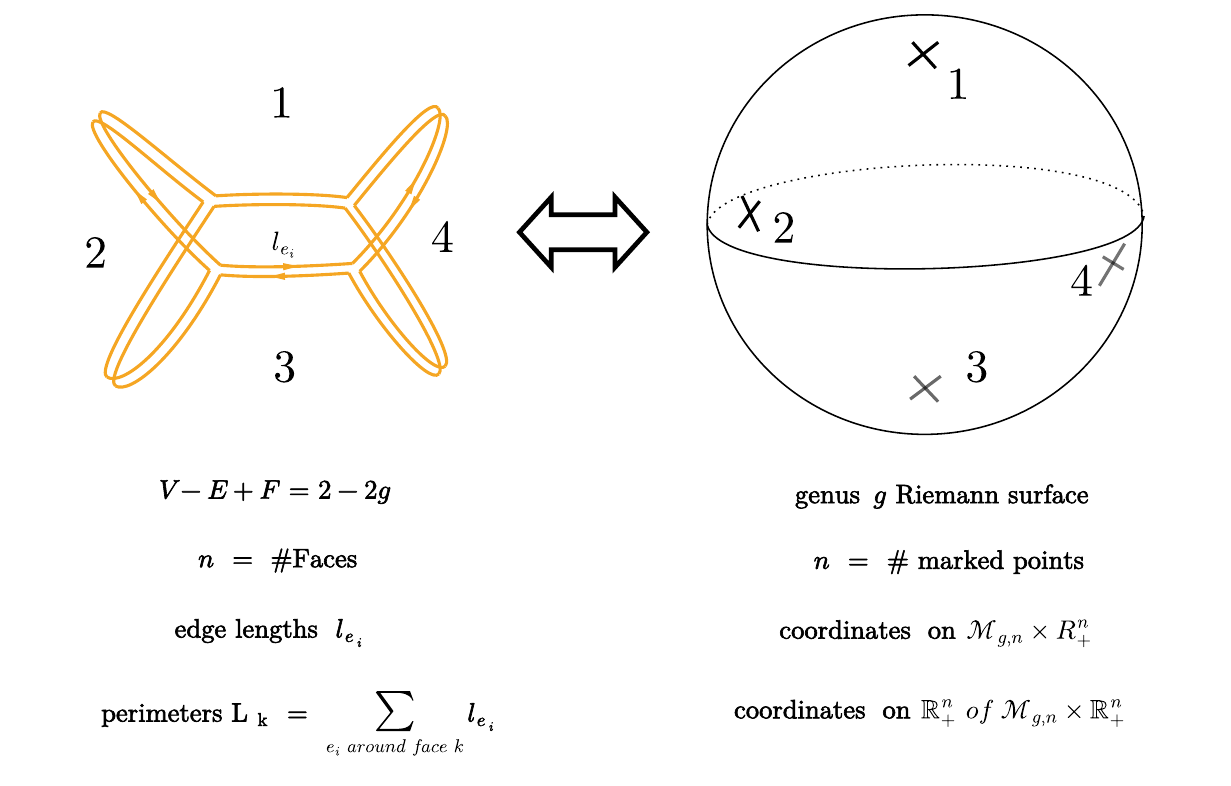}
   
    \caption{\textbf{A crash-course on the Strebel parametrization of $\mathcal{M}_{g,n}\times \mathbb{R}_{+}^{n}$ } Every metrized ribbon graph maps onto a particular point on moduli space. The genus of the graph, defined by $V-E+F=2-2g$, determines the genus of the Riemann surface it encodes. The number of faces matches the number of marked points, which will ultimately be where vertex operators are inserted on the worldsheet of the dual closed string. The lengths assigned to the edges of the graph serve as coordinates on (a patch of) $\mathcal{M}_{g,n}\times \mathbb{R}_{+}^{n}$. Along with the discrete graph data, they single out a particular point on $\mathcal{M}_{g,n} \times \mathbb{R}_{+}^{n}$. Different length assignments to the individual edges, keeping the perimeters around each face fixed, distinguish different closed string worldsheets (i.e. points on $\mathcal{M}_{g,n}\times \mathbb{R}_{+}^{n}$ with the same coordinates in the $\mathbb{R}_{+}^{n}$ fiber). We will see that, in the case of our matrix models, we land on worldsheets parametrized by integer Strebel lengths. The perimeters will be fixed by the power of the matrices in the trace insertions for the correlators at hand. }
    \label{fig:StrebelDict}
\end{figure}

At its heart, the Strebel parametrization establishes an \textit{exact} equivalence between the set of genus $g$ metrized ribbon graphs with $n$ faces, and the set of points on the decorated moduli space,  $\mathcal{M}_{g,n} \times \mathbb{R}_{+}^{n}$ \footnote{To be precise, this is an orbifold isomorphism. This means that for any point on $\mathcal{M}_{g,n}$ with a non-trivial automorphism group, the corresponding ribbon graph has the same automorphism group. See Thm. 6.3.2 of \cite{CountingSurfaces}.}. The genus of the diagram is defined using Euler's formula relating the number of faces, edges and vertices: $V-E+F=2-2g$. The vertices of the so-called Strebel graphs are generically trivalent. Metrized implies that to each edge of the graph, we assign a non-negative real number, called the {\it length}. These lengths serve as coordinates on the (cells of the) decorated moduli space. 

As a quick sanity check, let us compare the number of edges of a trivalent genus $g$ ribbon graph with $n$ faces to the (real) dimension of (the top-dimensional cell of) $\mathcal{M}_{g,n} \times \mathbb{R}_{+}^{n}$, which is $(6g-6+2n)+n$. Since three edges emanate from every vertex, and each edge is shared by two vertices, we have the relation $E=3V/2$. Plugging this into $V-E+F=2-2g$, we find the total number of edges to be $E=6g-6+3n$. This at least shows that such a metrized ribbon graph indeed carries enough information to encode a particular point on  $\mathcal{M}_{g,n} \times \mathbb{R}_{+}^{n}$.

One of the interesting outcomes of our construction for the 2-matrix model will be that all these edge lengths turn out to be integers. The Riemann surfaces specified by integer Strebel lengths are very special: these are the so-called arithmetic Riemann surfaces. That is, they are those (complex) curves which can be defined as the solution to polynomial equations defined over the algebraic numbers. 

How should we think of the extra $\mathbb{R}_{+}^{n}$ fiber of the decorated moduli space from a closed string perspective? The short answer is that the Strebel construction in fact defines a metric on each Riemann surface. We will see the $n$ positive real numbers corresponds to the proper lengths of the $n$ asymptotic closed strings, computed using this Strebel metric. They will thus characterise the external states (i.e. play the role of momenta). 

\begin{figure}[ht!]
    \centering
    \includegraphics[scale= 0.6]{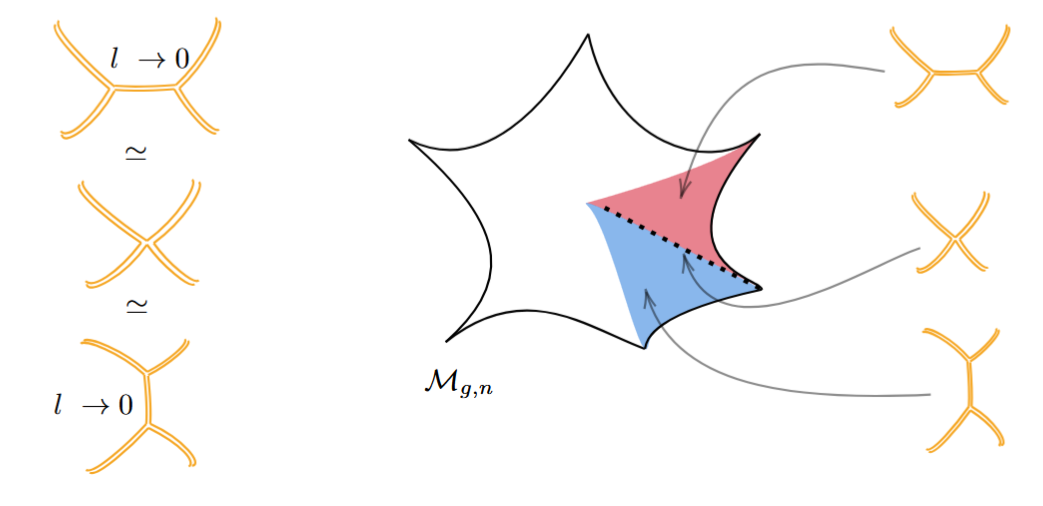}
    \caption{\textbf{Higher Valency Vertices in the Strebel Parametrization of Moduli Space } The set of genus $h$ metrized ribbon graphs with $n$ faces provide a simplicial decomposition of $\mathcal{M}_{g,n} \times \mathbb{R}_{+}^{n}$. Varying the lengths of the edges of a trivalent graph (keeping the $n$ perimeters fixed) sweeps out a top-dimensional cell in $\mathcal{M}_{g,n}$. Graphs with vertices of higher valency can be thought of as the result of one or more Strebel lengths of the trivalent graphs going to zero. For example, a four-valent vertex can be thought of as two trivalent vertices connected by a zero length edge. These graphs map onto points lying on higher co-dimension slices of moduli space, at the boundary of neighboring cells.}
    \label{fig:CollapseEdge}
\end{figure}

There are of course multiple genus $g$ ribbon graphs with $n$ faces. How do these different diagrams relate to the moduli space? A given diagram with all possible length assignments covers only a portion of $\mathcal{M}_{g,n} \times \mathbb{R}_{+}^{n}$. In fact, as shown in Fig. \ref{fig:CollapseEdge}, the finitely many such graphs provide a simplicial decomposition of the moduli space: varying the lengths on the edges of each inequivalent diagram sweeps out one such cell. As the length of any one edge goes to zero, we lie on a (real) co-dimension one slice of the moduli space, a boundary between two top cells. We can think of the Strebel graphs with vertices of valency greater than three as the result of collapsing the edges between various trivalent vertices, i.e. that we have sent one of the lengths to zero. We can then move to a different  adjoining cell by expanding in a different channel, so to say, to get an inequivalent graph. These are the so-called Whitehead moves of graph theory. See Fig. \ref{fig:CollapseEdge}. Thus higher-valent graphs encode worldsheets living on these higher co-dimension slices of moduli space. We mention this fact, since the Feynman diagrams of the $K,M$-matrix model will always give rise to Strebel graphs with valency at least four. 

Note there are as many \textit{faces} of the Strebel graph as there are marked points on the Riemann surface, namely $n$. The sum of the edge lengths around a given face is called the {\it perimeter} of that face. These parametrize the $\mathbb{R}_{+}^{n}$ fiber of the decorated moduli space. In turn, if we vary the edge lengths of the ribbon graphs keeping the perimeters fixed (as we will later do in a Feynman diagram expansion of our matrix model correlators), we will move purely along directions within $\mathcal{M}_{g,n}$. 

How is such a relation between Riemann surfaces and ribbon graphs established? Underlying this exact equivalence is the existence and uniqueness of a particular meromorphic quadratic differential $\phi_{S}$ on the worldsheet, known as the Strebel differential. We have provided a pedagogical introduction to the mathematics of these differentials in \cite{DSDI}, see Section 5, so we will be brief and revisit only the most relevant features. 

In local coordinates $z$ on the Riemann surface in question, such a differential always takes the form 
\begin{equation}
     \phi_{S} = \left( \phi_{S} \right)_{zz} dz \otimes dz = \phi(z) dz^2 . 
\end{equation}
The Strebel differential has the unique property that its only singularities are double poles at the location of the marked points. Near the $k$-th marked point, in local coordinates $u_k$, it takes the form
\begin{equation}
    \phi_{S}  \approx - \frac{L_{k}^2}{(2\pi)^2} \frac{du_{k}^2}{u_{k}^2} .\label{eq:StrebelNearPole}
\end{equation}

\begin{figure}[ht!]
    \centering

    \includegraphics[scale=0.7]{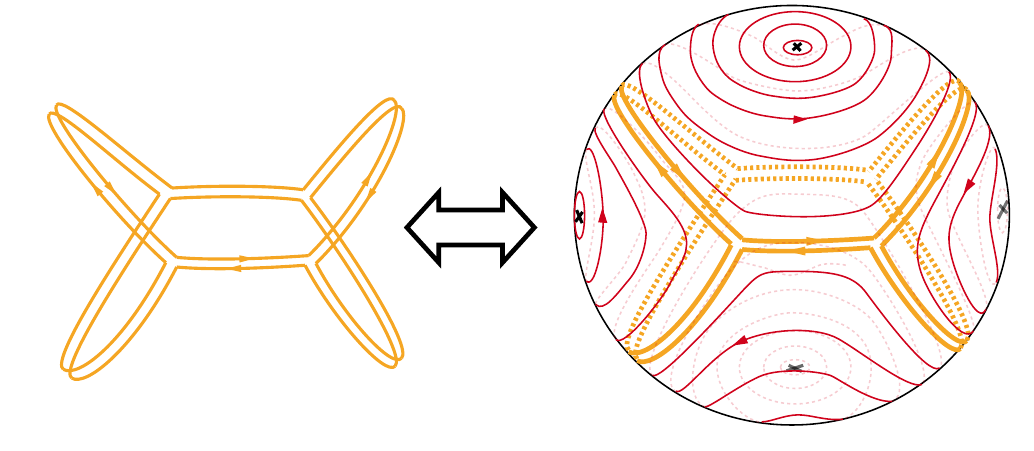}
   
    \caption{\textbf{Strebel Graph from Horizontal Trajectories } Each Riemann surface can be foliated by a set of curves known as Strebel's horizontal trajectories (in red). Along these curves, the square root of the Strebel differential is real and defines a positive line element. They can therefore be assigned a notion of length. It is a theorem of Strebel that for the most part, the horizontal trajectories of a Strebel differential form closed curves around each of the marked points. The exceptions are the so-called "critical" horizontal trajectories (in orange). These are, instead, open segments connecting the zeroes of the Strebel differential. It is the union of these segments, along with their lengths, that assigns to each Riemann surface a unique Strebel graph. Its vertices therefore correspond to the zeroes of the differential. $m$ edges of the Strebel graph meet at an $(m-2)$-th order zero of the Strebel differential. The faces, on the other hand, are in one-to-one correspondence with the poles of the Strebel differential, which coincide with the marked points on the Riemann surface.}
    \label{fig:HorTraj}
\end{figure}

For now, all that will really be needed to understand the connection to ribbon graphs is that the Strebel differential defines a special set of trajectories on any particular punctured Riemann surface.
One family of such curves are called horizontal trajectories. Parametrizing them by some $t$, they satisfy the relation\footnote{In other words, on horizontal trajectories, ${\rm Im}\left(\sqrt{\phi(z)}dz\right)=0$.} 
\begin{equation}
    \phi\left(z_{H}(t)\right) \left( \frac{dz_{H}(t)}{dt} \right)^2 > 0. \label{eq:horizontal}
\end{equation}
The square-root of the Strebel differential defines a line-element $d\tau = \sqrt{\phi(z)} dz$. It is real along such curves. We can therefore use it to define a notion of length. For the most part, the horizontal trajectories form closed curves, which surround each of the marked points. See Fig. \ref{fig:HorTraj}. 

 For instance, in the local coordinates $u_k$ near the $k$-th marked point (as in Eq.(\ref{eq:StrebelNearPole})). The family of curves $u_{k}^{\text{H}}(t) = r e^{i t}$, for $r\in \mathbb{R}_{+}$ and $t \in [0,2\pi)$, can be readily checked to satisfy Eq.(\ref{eq:horizontal}). Integrating the line-element $d\tau$ along such a closed trajectory\footnote{The sign ambiguity in choosing a branch for the square root of the Strebel differential corresponds to a choice of orientation of the face of the Strebel graph.},
\begin{equation}
    \oint_{C_k} d\tau = L_{k} \label{eq:PerimeterInt}
\end{equation}
(where $C_k$ is any horizontal trajectory around the k-th marked point),
we find that its total length is $L_{k}$, independent of $r$. These $L_{k}$ are the perimeters discussed above, i.e. the coordinates along the $\mathbb{R}_{+}^{n}$-fiber of the decorated moduli space. Once we specify these residues, $L_{k}$, for the double-poles, there  exists a unique Strebel differential for each (inequivalent) Riemann surface. Note that this set of closed horizontal trajectories foliates an open disc region around each marked point. We can view this disc as conformally equivalent to a semi-infinite cylinder, with its asymptotic infinity mapped to the marked point. 

There is a special set of horizontal trajectories for every Strebel differential, called the critical horizontal trajectories, which are not closed, in the strict sense of the word. They form a set of measure zero on the surface. These trajectories trace out a connected graph on the worldsheet, formed of various edges. The edges begin and end at the zeroes of the Strebel differential. To obtain a metrized ribbon graph, we integrate the line element $d\tau$ along each segment connecting the zeroes of the Strebel differential. Since these are horizontal trajectories, the line element is again real and the lengths of each edge are positive, real numbers. By thickening the lines of this graph into ribbons (i.e. by keeping track of the orientation assigned to each face), we thereby obtain the Strebel graph associated to the surface, as illustrated in Fig. \ref{fig:HorTraj}.

Let us mention a few more important properties of the Strebel graph. The valency of the vertex is dictated by the order of the zero of the Strebel differential. In other words, near a point where $m$ critical horizontal trajectories meet, there exists local coordinates $w$, such that the Strebel differential takes the form, in the vicinity of the zero,\footnote{The overall constant is simply chosen for later convenience.} 
\begin{equation}\label{eq:strebelzero}
     \phi_{S}  \approx \frac{m^2}{4} w^{m-2}dw^2
\end{equation}
Note that a simple zero, i.e. $m=3$, corresponds to a trivalent vertex of the Strebel graph. 
This will be important later when we present an algorithm to assemble the closed string worldsheet from the matrix model Feynman diagrams. Our reconstruction rules will need to guarantee that we reproduce the correct local behaviour of the Strebel differential on the dual Riemann surface.

\subsection{Strebel Graphs versus Matrix Model Feynman Diagrams} \label{sec:StrebGraphsVsFD}

How do the graphs used in the Strebel parametrization of the decorated moduli space relate to the Feynman diagrams of the gauge theory? There are two issues we need to address. Firstly, as graphs, are they equivalent? Secondly, how do we assign a notion of length to the various edges of the Feynman diagram?

We argued previously \cite{DSDI} that there are two natural answers to the first question: either we identify the Feynman diagram directly with the Strebel graph, or we think of it as its graph dual. These in fact provide two different notions of open/closed string duality, which were dubbed V(ertex)-type and F(ace)-Type. We will, in fact, see an interesting refinement of this mechanism in Sec. \ref{sec:OCOgen}.

\begin{figure}[ht!]
    \centering

    \includegraphics[scale=0.7]{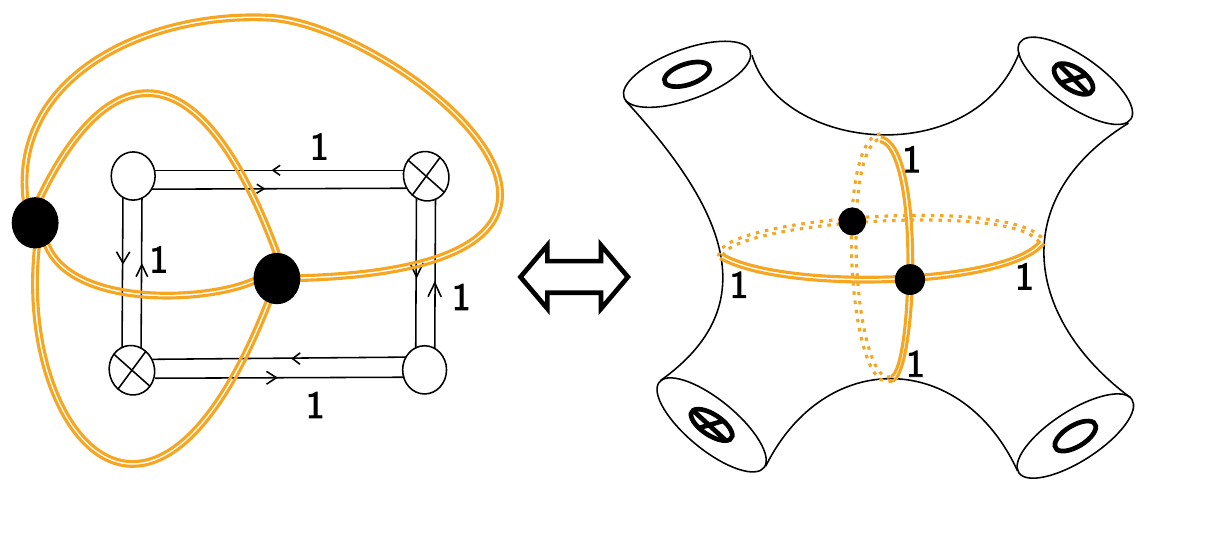}

\caption{\textbf{The Closed String Worldsheet Dual to Our Feynman Diagram } An $n$-point correlator in the $K,M$-matrix model should map onto an $n$-point function of vertex operators in the dual closed string. The worldsheet dual to our Feynman diagram with four vertices should therefore asymptotically looks like four closed strings, i.e. it is a Riemann surface with four marked points. However, the Strebel graph of such a surface, drawn in orange, has four faces, not four vertices. The Strebel graph is therefore not the original Feynman diagram itself, but rather its graph dual. Once we assign lengths to each edge of the Strebel graph, in this case length one to every edge, we have singled out a precise point on $\mathcal{M}_{g,n}\times \mathbb{R}_{+}^{n}$. The $\mathbb{R}_{+}^{n}$ fiber coordinates, which are the sum of lengths around each face of the Strebel graph, are fixed by the powers of the matrices in the traces. In this case, each perimeter has length $2$, since we considered the operators $\Tr(K^2)$ and $\Tr(M^2)$. Since Strebel graphs are generically trivalent, we should view the quartic vertices here as as the collapse of two trivalent vertices, i.e. two of the Strebel lengths have been set to zero.}

    \label{fig:dualtomainFD}
\end{figure}

For now, let us first focus on the concrete Feynman diagrams capturing the various Wick contractions used to compute the correlators of section \ref{sec:MatrixFDs}. We ultimately want to argue this matrix model captures a topological subsector of the standard AdS/CFT correspondence (see section \ref{sec:embedding}). In that context, we know that single trace operators map onto vertex operators in the dual string theory. From a Feynman diagram perspective, each trace inserted in the correlator corresponds to an external \textit{vertex}. We therefore want our Feynman diagram to capture a worldsheet with as many marked points as vertices. The Strebel graph, on the other hand, has as many \textit{faces} as marked points of the Riemann surfaces it parametrizes. We therefore come to the important conclusion that the correspondence of our matrix model Feynman diagrams can only be to the graph {\it dual} to the Strebel graph! See Fig. \ref{fig:dualtomainFD}. 

\begin{figure}[ht!]
    \centering

    \includegraphics[scale=0.7]{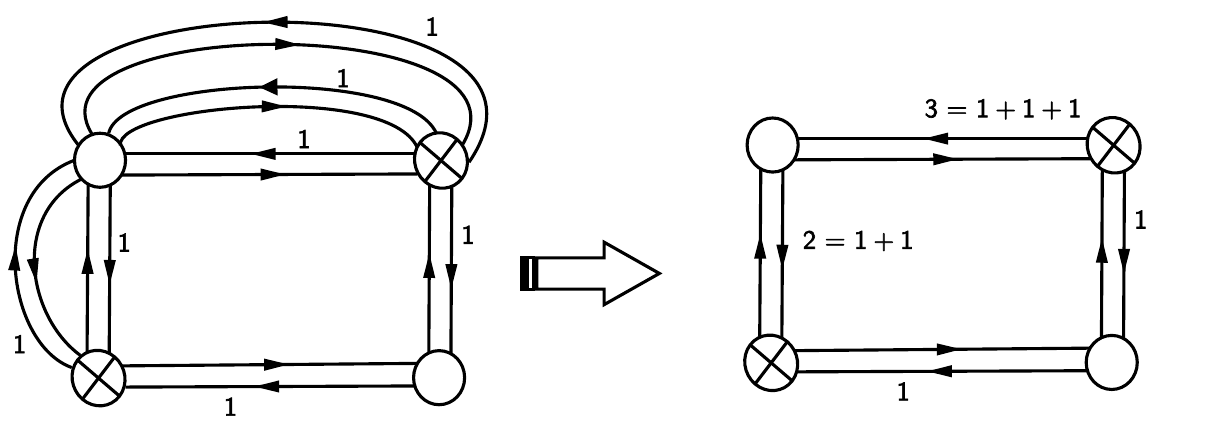}

    \caption{\textbf{A Feynman Diagram Contributing to the Computation of $\Braket{\frac{1}{4}\Tr(K^4) \frac{1}{2}\Tr(K^2)\frac{1}{5}\Tr(M^5) \frac{1}{2}\Tr(M^2)}_{c}$ and its Skeleton Diagram } Edges connecting the same two vertices, bordering a two-sided face, are called homotopic. The skeleton graph is formed by bundling together all homotopic edges of the Feynman diagram. Assigning unit length to each edge of the original Feynman diagram, the process of bundling has the effect of assigning an effective length given by the number of edges collapsed together. This means that ultimately all Strebel graphs will have integer length assignments. In this particular case, we obtain a skeleton graph identical to our favorite Feynman diagram, but with different length assignments to its edges. We will see in Sec. \ref{sec:howto} why this bundling procedure translates to simply adding together the individual (unit) lengths of each original edge.}
    \label{fig:SkeletonGraph}
\end{figure}

One important comment is in order. Strebel graphs do not have vertices of order 2. This can be traced back to the fact that the valency $m$ of a vertex corresponds to an $(m-2)$-th order zero (see Eq.(\ref{eq:strebelzero})). Edges of the dual Feynman graph which border faces with only two sides must therefore be bunched together. Identifying such homotopic edges of the graph results in what is termed as the skeleton of the graph\footnote{Homotopic edges are edges connecting the same two vertices which can be deformed into one another without crossing any other edge.}. We have illustrated this in Fig.\ref{fig:SkeletonGraph}. Technically speaking, it is this skeleton graph which is graph dual to the Strebel graph. We will see why this is a natural thing to do in \ref{sec:howto}. See \cite{freefieldsadsII,freefieldsIII} for more a detailed discussion of this construction in the higher-dimensional case. 

This bunching of edges is closely related to the second question we still need to address. A Feynman diagram does not, {\it per se},  inherently carry a notion of length assigned to its edges. In some sense, we need to choose a certain prescription, which we can then {\it a posteriori} determine to be a consistent choice. In \cite{freefieldsadsI,freefieldsadsII,freefieldsIII}, one of the authors crafted a proposal for gauge theories with all fields transforming in the adjoint representation of the gauge group. Since to every edge of a Feynman diagram there is an associated propagator, the idea was to first re-express this propagator in the Schwinger parametrization, and then interpret the Schwinger time (more precisely, the inverse time) as the Strebel length associated to that edge. In the case of the zero-dimensional matrix integrals we are considering in this paper, there are no real propagators to Schwinger-parametrize. This was already highlighted by Razamat \cite{razamat2010matrices} in his search for a string dual to the Gaussian matrix model. He put forward the idea to simply assign unit length to any given edge connecting two vertices of the Feynman diagram. In the process of bunching together homotopic edges, we simply add the lengths together. In other words, the length associated to the effective edge is simply the number of edges that were collapsed together in the process of reducing the Feynman diagram to its skeleton graph. This is the length assignment which we will use for the skeleton Feynman graph, but also for its graph dual. Again, we will see from Section \ref{sec:howto}, why this is quite a natural thing to do. Fig. \ref{fig:SkeletonGraph} illustrates an example where the skeleton graph of a particular Feynman diagram reduces to our favorite Feynman diagram, but with different length assignments.

This simple length assignment to the edges has a striking consequence: the worldsheets dual to the matrix model Feynman diagrams are all parameterized by Strebel graphs with integer lengths\footnote{Integer Strebel differentials were also considered in the context of Seiberg-Witten theory in \cite{IntegerStrebs}.}. Since these lengths serve as coordinates on the decorated moduli space, we find that the closed string theory path integral must admit a presentation which only receives contributions from a discrete set of worldsheets, labeled by integers. One of the most important take-away lessons of this construction is that instead of viewing the Feynman diagrams as discretizing the worldsheet, they have instead provided {\it a latticization of the moduli space}! 

\begin{figure}[H]
    \centering

    \includegraphics[scale=0.9]{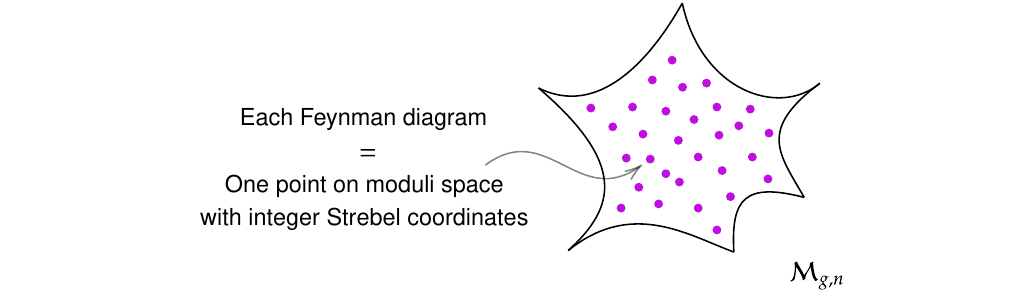}

    \caption{\textbf{Latticization of $\mathcal{M}_{g,n}$ instead of Discretization of the Worldsheet } Each Feynman diagram, contributing to a particular $n$-point correlator, maps onto a particular point on moduli space. Since all the lengths of the corresponding Strebel graph are integers, and these serve as coordinates on $\mathcal{M}_{g,n}$, all dual worldsheets reside on a lattice in moduli space. These special points are the so-called arithmetic Riemann surfaces. In Sec. \ref{sec:StringLocalization}, we will see how the string theory path integral localizes so as to only receive contributions from this discrete set of surfaces.}
    \label{fig:latticeMgn}
\end{figure}

The points on moduli space parametrized by these integer length Strebel graphs are in fact very special worldsheets: they are so the so-called arithmetic Riemann surfaces. This means they can be viewed as the zero locus of a polynomial defined over the algebraic closure of the rational numbers, $\bar{\mathbb{Q}}$. At this point, it should be quite surprising that any string theory path integral would somehow select out these particular points on moduli space. We will later find a rather satisfying answer through a theorem due to Belyi.

\begin{figure}[ht!]
    \centering

    \includegraphics[scale=0.6]{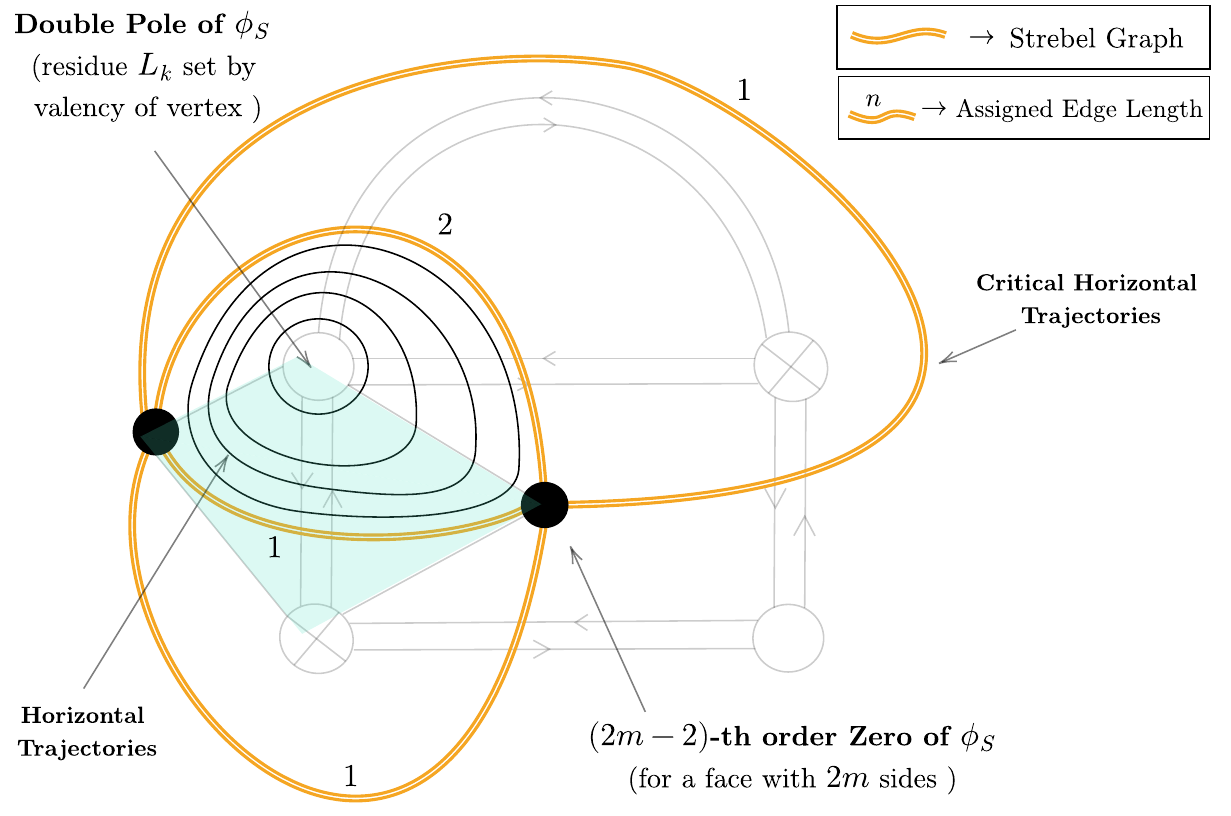}

    \caption{\textbf{Feynman Diagrams vs. Strebel Differentials } To summarize, each vertex of the Feynman diagram corresponds to a double-pole of the Strebel differential $\phi_{S}$. These reside at the marked point on the closed string worldsheet, where the associated vertex operators would be inserted. The valency of the vertex determines the residue $L_{k}$ (see Eq.(\ref{eq:StrebelNearPole})). They therefore fix the $\mathbb{R}_{+}^{n}$ fiber coordinates in $\mathcal{M}_{g,n} \times \mathbb{R}_{+}^{n}$ appearing in Strebel's parametrization of the moduli space. Each face of the diagram bordered by $2m$ faces ($m>1$) corresponds to a zero of the differential of order $2m-2$. The light blue shaded region will correspond to one of the strips making up the dual Riemann surface, discussed in Secs. \ref{sec:PhysPic} and \ref{sec:howto} and shown in Fig. \ref{fig:StrebelAsStrips}. }
    \label{fig:FDvsStrebelDiff}
\end{figure}

So far, we have described how to precisely assign to each matrix model Feynman diagram a point on the decorated moduli space $\mathcal{M}_{g,n} \times \mathbb{R}_{+}^{n}$. To properly implement open-closed string duality, we need to understand how the Feynman diagram expansion of each correlator transmutes to the usual sum (or integral) over the closed string moduli space $\mathcal{M}_{g,n}$. In other words, we need to elucidate the role of the extra $\mathbb{R}_{+}^{n}$ coordinates. We begin by noting that these $\mathbb{R}_{+}^{n}$  coordinates are, from the point of view of the Strebel graph, the sum of the lengths associated to the edges bordering each of the $n$ faces. As graph duals, each face of the Strebel graph maps onto a vertex of the original Feynman diagram. As such, the perimeter associated to each face of the Strebel graph is equal to the valency of the vertex of the original Feynman diagram. Moreover, for the free theory we are considering here, the valency of the original vertices is equal to the power of the matrices appearing on the traces of the correlators (they are all "external" vertices). \footnote{All Strebel graphs dual to the skeleton graphs of Feynman diagrams contributing to a given correlators have the same set of perimeters. What does depend on the skeleton graph construction is the length assignment to the \textit{individual edges} bordering a given face. The perimeter, i.e. the sum of these lengths around a face, is however always the same.} Hence, when we consider the Feynman diagram expansion of an arbitrary genus $g$ correlator as in Section \ref{sec:MatrixFDs},    
\begin{equation}
    \Braket{\prod_{i=1}^{V_K} \frac{1}{l_{i}}\Tr\left( K^{l_i}\right) \prod_{j=1}^{V_M} \frac{1}{n_{j}}\Tr\left( M^{n_j} \right)} _{c,g},
\end{equation}
each diagram will be mapped onto a point of the decorated moduli space ${\mathcal{M}}_{g, V_{K}+V_{M}} \times {\mathbb{R}}_{+}^{V_{K} +V_{M}}$, with coordinates $(l_1,...,l_{V_{K}},n_1,...,n_{V_{M}})$ on the ${\mathbb{R}}_{+}^{V_{K} +V_{M}}$ fiber. Since the fiber coordinates are fixed for all diagrams, the sum over Feynman diagrams indeed becomes a sum over discrete points of the closed string moduli space ${\mathcal{M}}_{g, V_{K}+V_{M}}$\footnote{This is in fact slightly different from the prescription advocated for higher-dimensional gauge theories in \cite{freefieldsadsI,freefieldsadsII,freefieldsIII}. Since the Strebel lengths there arose from the Schwinger-parametrization of the propagators, the perimeters were free variables that were integrated over, leaving behind a moduli space integral over just $\mathcal{M}_{g,n}$. Here, the perimeters are instead fixed by the precise form of the correlator. We thank Z. Komargodski for discussions on this point.}. We have summarized the relationship between the Feynman diagram and the Strebel differential of the dual (closed string) worldsheet in Fig. \ref{fig:FDvsStrebelDiff}.

Let us make one final comment on the prescription outlined above to assign lengths to the edges of the Feynman diagram. As we mentioned previously, there was a certain arbitrariness to this choice. The best we can do is to check its internal consistency {\it a posteriori}. In the case at hand, we will be lucky that the dual closed string theory has multiple presentations in terms of three known equivalent worldsheet theories. The Feynman diagram approach we are following here, including the integer length assignment presented above, will naturally land us on one of these three presentations (an A-model topological string). The fact that this A-model string agrees with the other two presentations - a B-model Landau-Ginzburg theory and the $c=1$ string at self-dual radius - which we can derive as closed string duals by other (non Feynman diagram) means, confirms the validity of this prescription for the matrix model case at hand. While this prescription is adequate and natural for matrix models, it may arise more generally in higher dimensional theories as well, in appropriate limits. Indeed, as we describe, our matrix models describe a sector of ${\cal N}=4$ Super Yang-Mills theory. Further, these integer length Strebel differentials also arise in the large twist limit (Gross-Mende-like limit) of correlators in the free symmetric orbifold case. This suggests a certain universal nature to these special points in the moduli space which would be very good to understand better.

\subsection{The Worldsheet Metric in Strebel Gauge} \label{sec:strebelmetric}

In the Polyakov formulation of string theory, we integrate over the allowed possible metrics on the worldsheet. The action is invariant under local diffeomorphisms and Weyl rescalings (we are assuming the cancellation of the conformal anomaly). We treat these reparametrizations as redundant and gauge them. The remaining physical degrees of freedom parametrization of the worldsheet are its moduli, i.e. its coordinates on $\mathcal{M}_{g,n}$. For each such point on $\mathcal{M}_{g,n}$, we can choose a representative metric. This is a choice of gauge.

One of the important properties of the Strebel differential is that it defines a local metric on the worldsheet given by 
\begin{equation}
    ds^2 = |\phi(z)| dz d\bar{z} \label{eq:StrebelMetric}
\end{equation}
 
We call this the metric in "Strebel gauge". Much like the conformal gauge, all the non-trivial information of the metric resides in the conformal factor. What does the metric geometry of the worldsheet look like in this gauge?

While the total integrated curvature on a Riemann surface is related to its genus via the Gauss-Bonnet theorem,
\begin{equation}
    \frac{1}{4\pi} \int \sqrt{g} R \ d^2z = 2-2h,
\end{equation}
how this curvature is distributed depends upon a choice of representative metric. In Strebel gauge, the metric is flat everywhere, except for conical singularities at the zeroes and poles of the Strebel differential. 

To see this, note that we can define (complex) coordinates $\zeta$, away from the zeroes and poles of the differential,

\begin{equation}
    \zeta(z) = \int_{z_{0}}^{z} \sqrt{\phi(z')} dz'
\end{equation}
where we have chosen an arbitrary zero of the Strebel differential $z_{0}$ as the basepoint of integration. In $\zeta$ coordinates, again, valid away from the poles and other zeroes, the Strebel differential takes the simple form $d\zeta^2$. The metric thus indeed becomes the flat Euclidean metric
 \begin{equation}
     ds^2\big|_{\overset{\text{away from}}{\text{zeroes and poles}}} = d\zeta d\bar{\zeta}
 \end{equation}
 showing that the curvature vanishes almost everywhere on the worldsheet.

 At the zeroes of the Strebel differential, the metric is singular. In fact, the metric vanishes there. How quickly it vanishes determines the precise angular deficit. Near an $m$-th order zero, the metric looks like (cf. Eq.(\ref{eq:strebelzero}))

 \begin{equation}
     ds^2 \big|_{\overset{\text{near a}}{\text{zero}}}=\frac{(m+2)^2}{4}|\w^{m}|d\w d\bar{\w}
 \end{equation}
 We can use, for example, the transformation of the Ricci scalar under Weyl rescalings to compute the local curvature\footnote{Under a Weyl rescaling of a two-dimensional metric, $\tilde{g} = e^{2 \Omega}g$, the Ricci curvature transforms as $\tilde{R} = e^{-2 \Omega} \left( R - 2 \nabla^{2} \Omega \right)$.}
 \begin{equation}
      \frac{1}{4\pi} \sqrt{g} R\big |_{\overset{\text{at} \ (m)-\text{th}}{\text{order zero}} } = - 2 \frac{\partial}{\partial \w} \frac{\partial}{\partial \bar{\w}} \log\left( |\w|^{m} \right) =  -\frac{m}{2} \delta(\w) \delta(\bar{\w}) 
 \end{equation}
We thus find a conical singularity at $w=0$, with an angular excess $\alpha = m\pi$. Similarly, in coordinates $u_{k}$ near the $k$-th double-pole of the Strebel differential, the metric takes the form 
\begin{equation}
     \left.ds^2\right|_{\substack{\text{near} \ k\text{-th}\\\text{pole}}} = \frac{L_{k}^2}{(2\pi)^2 |u_{k}|^2}du_k d\bar{u}_k
\end{equation}
Hence, at each marked point on the Riemann surface, there is a positive curvature singularity (independent of the residue at that double-pole)

\begin{equation}
      \left.\frac{1}{4\pi} \sqrt{g} R \right|_{\substack{\text{at} \ k\text{-th}\\\text{marked point}} } =  \delta(u_{k})\delta(\bar{u}_{k})
\end{equation}

If the Strebel differential on a Riemann surface of genus $g$ has $s_{m}$ zeroes of order $m$ and $n$ double poles, then the Gauss-Bonnet theorem enforces

\begin{equation}
   \frac{1}{4 \pi} \int_{WS} \sqrt{g_{S}} R_{S} d^2 z = \sum_{m \geq 1} s_{m} \left(-\frac{m}{2}\right) + n = 2-2g \label{eq:GBforStrebel}
\end{equation}
This same formula has nice interpretation from the point of view of the Strebel graph. Recall that each vertex of valency $m+2$ corresponds to an $m$-th order zero, while each double pole maps onto a face. The Strebel graph for the above differential would therefore have $s_{m}$ vertices of valency $m+2$, $n$ faces and $E= (1/2) \sum_{m \geq 1} (m+2) s_{m} $ edges. Euler's formula then neatly reproduces the Gauss-Bonnet constraint in Strebel gauge  (Eq.(\eqref{eq:GBforStrebel})), as it must:  

\begin{equation}
     V-E+F = \sum_{m \geq 1} s_{m} -    \sum_{m \geq 1} s_{m} \frac{(m+2)}{2} + n = 2-2g .
\end{equation}

\subsection{The Explicit Worldsheet \& Metric Dual to Our Feynman Diagram} \label{sec:ExplicitWS}

\begin{figure}[ht!]
    \centering
\includegraphics[scale=0.7]{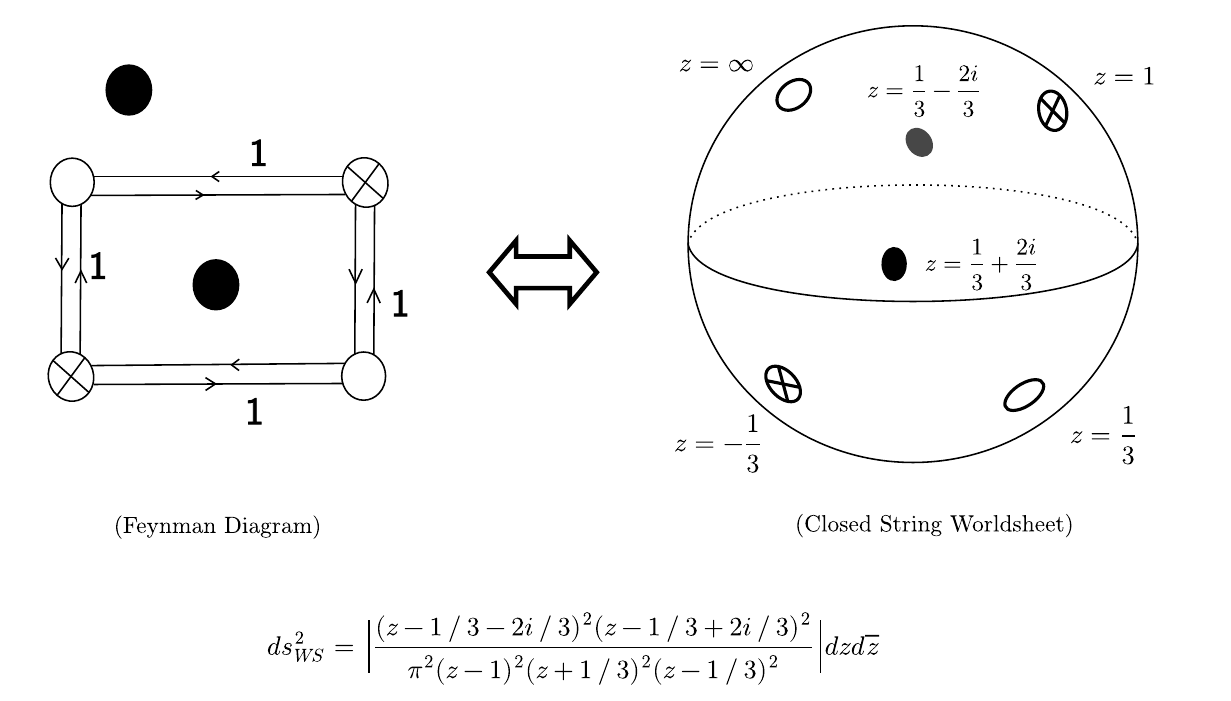}

    \caption{\textbf{The Explicit Worldsheet \& Metric Dual to our Feynman Diagram } We can explicitly reconstruct the full Strebel differential $\phi_{WS}(z) dz^2$ from the metrized graph data on the left. This determines the exact point on $\mathcal{M}_{0,4}$ dual to our Feynman diagram. It is a sphere with punctures at $z=1,\infty, \pm \inv{3} $ (up to a global $SL(2,\mathbb{C})$ action), where the vertex operators dual to $\Tr(K^2)$ and $\Tr(M^2)$ are inserted. The faces of the diagram map onto the zeroes of the differential, at $z =\inv{3} \pm \frac{2i}{3}$. The metric on the worldsheet, in Strebel gauge, is then given by $ds^2_{WS} = |\phi_{WS}(z)|dz d\bar{z}$. }
    \label{fig:FDexplicitMetric}
\end{figure}

A (metrized) Strebel graph suffices, in principle, to uniquely pinpoint a particular point on moduli space. We can illustrate this by explicitly constructing the Strebel differential for our Feynman diagram in Fig. \ref{fig:MainFD}. 
This is the unique Strebel differential $\phi(z)dz^2$ whose critical graph is the dual to the Feynman diagram, as explained in the previous section (see Fig. \ref{fig:dualtomainFD}). 
Recalling that the metric in the Strebel gauge is then given by $ds_{WS}^2=|\phi(z)|dzd\bar{z}$,  we can also explicitly reconstruct the metric on the worldsheet. We summarize our findings in Fig. \ref{fig:FDexplicitMetric}.

 The four vertices of the Feynman diagram in Fig. \ref{fig:MainFD} will translate to four double poles of the Strebel differential, which lie at the marked points. The planarity of the diagram means we are working on a worldsheet of genus 0. As usual, by the residual $SL(2,\mathbb{C})$ invariance on the sphere, we can fix the location of the vertex operator insertions to lie at $z=1,\pm t,\infty$. $t$ parametrizes the modulus of the four-punctured sphere. We will specify a unique point on $\mathcal{M}_{0,4}$ by determining $t$. A version of this computation appeared in the context of Seiberg-Witten theory in \cite{IntegerStrebs}.

Since the matrix model correlator is built out of only $\Tr(K^2)$ and $\Tr(M^{2})$, the residues at these double poles are all equal\footnote{The sign of the residue depends on the appropriate orientation of the face of the Strebel graph.}, $L_{1}=L_{t}=L_{-t}=L_{\infty}=2$. The two faces of the Feynman diagram require the Strebel differential to have two zeroes. We will denote the position of these zeroes on the worldsheet by $z_a$ and $z_b$. These too will be determined explicitly. Since each face is bordered by four edges (i.e. the dual (Strebel) graph vertices are quartic), the order of these zeroes is fixed to be $4-2=2$.

The most general meromorphic differential satisfying these requirements takes the form 
\begin{equation}
    \phi(z)dz^2 = -\frac{1}{\pi^2} \frac{(z-z_{a})^2(z-z_{b})^2}{(z-1)^2(z^2-t^2)^2}dz^2,
\end{equation}
while the condition on the residues all being equal to $2$ for the double poles at $z=1$ and $z=\pm t$, respectively, further imposes
\be 
\begin{split}
    \frac{(1-z_{a})^2(1-z_{b})^2}{(1-t)^2(1+t)^2}&=1,\\
    \frac{(t-z_{a})^2(t-z_{b})^2}{4t^2(t-1)^2} &= 1, \\
    \text{ and }\ \frac{(t+z_{a})^2(t+z_{b})^2}{4t^2(t+1)^2} &= 1.
\end{split}
\ee

One can solve these polynomial equations for $z_{a}$ and $z_{b}$, the zeroes of the Strebel differential. We obtain for now two (non-trivial) families of solutions, \footnote{ One also finds $z^{(3)}_{a, b} = -t \pm \sqrt{2t(1+t)}$, which however is simply related to $z^{(2)}_{a, b}$ via $t \rightarrow -t$.}
\begin{equation} \label{eq:solszeroes}
    \begin{split}
    z^{(1)}_{a,b} &= 1\pm \sqrt{1-t^2},\\
    z^{(2)}_{a, b} &= t\pm \sqrt{2t(t-1)}.\\
    \end{split}
\end{equation}
We will find that only one of the two is admissible. What remains is to determine $t$ itself. To do so, we use the metric information of the Strebel graph. In other words, we impose that each edge of the Strebel graph has length $1$, corresponding to the single contraction between each vertex of the Feynman diagram. 
\be \label{reallength}
\int_{z_{a}}^{z_{b}} \sqrt{\phi(z)}dz = \pm \frac{i}{\pi} \int_{z_{a}}^{z_{b}} \frac{(z-z_{a})(z-z_{b})}{(z-1)(z^2-t^2)}dz = 1 + i0
\ee 
The sign ambiguity is due to the symmetric choice between $z_{a}$ and $z_{b}$. Due to the special case of quadratic zeros, the integral is rather straightforward as it involves neither square roots nor branch cuts. Note that this simplification, in fact, always occurs for the Strebel differentials dual to our matrix model Feynman diagrams. This is due to the fact that all faces have an even number of bordering edges (since each edge can only connect vertices of different types). This means that all Strebel zeroes are of even order. The line-element defined by the square root of the differential is thus itself a meromorphic function on the worldsheet.

Once we plug in the $\{z^{(i)}_{a, b}\}$ of Eq.(\ref{eq:solszeroes}), which are functions of $t$, into Eq.(\ref{reallength}), we can then finally determine $t$. For $z^{(1)}_{a,b}$, no solution exists. For the second family $z^{(2)}_{a,b}$, we find $t=1/3$. Hence, the zeroes dual to the faces of the diagram, lie at 
\be 
z_{a,b} = \inv{3} \pm \frac{2i}{3}.
\ee 
The Strebel differential on the Riemann surface dual to the Feynman diagram is therefore completely fixed
\be 
\phi_{WS}(z) dz^2 = -\inv{\pi^2} \frac{(z- 1/3 - 2i/3)^2(z-1/3 +2i/3)^2}{(z-1)^2(z + 1/3)^2(z-1/3)^2}dz^2 \label{eq:ExplicitStrebel}
\ee 

Summarizing, we have determined the exact point on $\mathcal{M}_{0,4}$ dual to our Feynman diagram to be a sphere worldsheet with punctures at $z=1,\infty, \pm \inv{3} $. The vertex operators dual to $\Tr(K^2)$ get inserted at $z=-\inv{3},1$, while those dual to $\Tr(M^2)$ sit at $z=\inv{3},\infty$. Of course, only the relative location of the marked points is physical, since we can always act by an overall $SL(2,\mathbb{C})$ transformation on this set. A representative metric for this point on moduli space, i.e. the metric on the worldsheet in Strebel gauge, is then given by $ds^2_{WS} = |\phi_{WS}(z)|dz d\bar{z}$. It can be checked that this metric is flat everywhere, except for conical singularities at the marked points and at $z =\inv{3} \pm \frac{2i}{3}$ (the zeroes of the Strebel differential).

\subsection{A Physical Picture of the Strebel Parametrization: Gluing Open String Strips} \label{sec:PhysPic}

Thus far, the Strebel parametrization as described above might seem like a specialised mathematical construct. The purpose of this section is to argue that this construction is actually very physical and concrete. It gives a realization of several of the pieces of physical intuitions that have accumulated over the years about gauge/string duality. In fact, it will provide a constructive algorithm to build up the closed string worldsheet from the ribbon graph as we will describe explicitly in Sec. \ref{sec:howto}. 

\begin{figure}[ht!]
    \centering

    \includegraphics[scale=0.7]{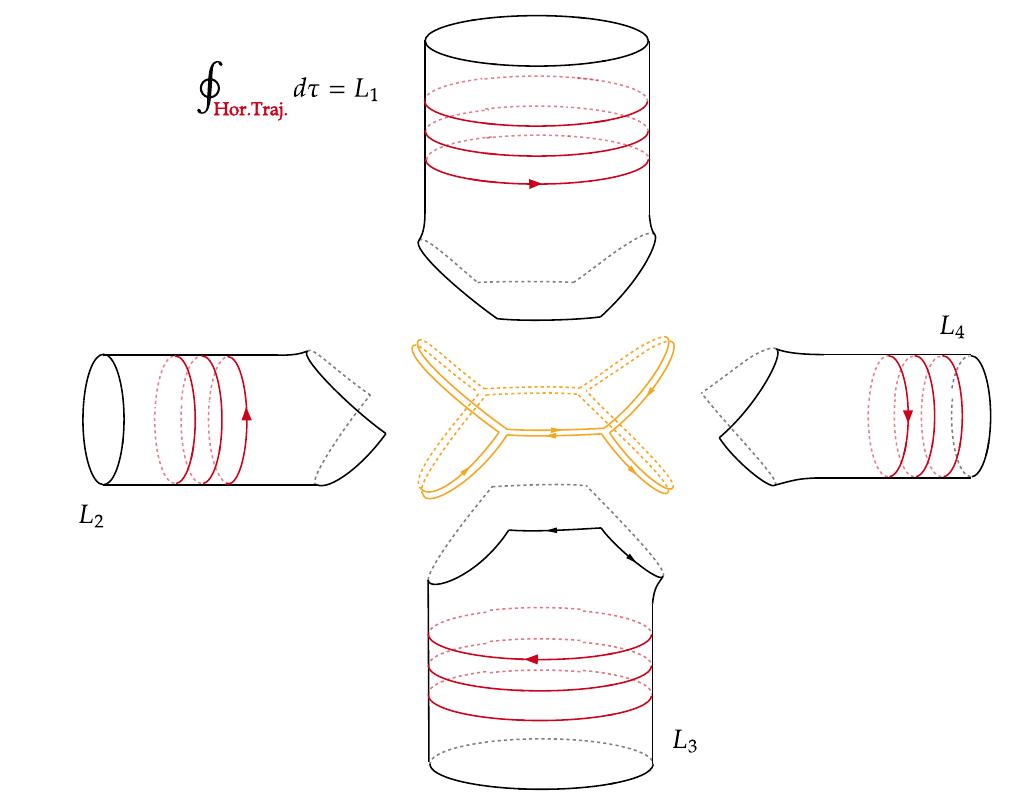}
    \caption{\textbf{Strings Interact at the Vertices of the Strebel Graph } Similar to light-cone gauge, the Strebel construction decomposes every Riemann surface into a collection of semi-infinite cylinders glued together along the Strebel graph. Each semi-infinite cylinder corresponds to one family of closed horizontal trajectories surrounding the marked points. Physically, these represent the free propagation of closed strings. All string interactions are therefore completely localized to the vertices of the Strebel graph, where they split or join. In the Strebel gauge, the integral of the square root of the Strebel differential along the horizontal trajectories defines the proper length of each string. It is constant along each semi-infinite cylinder and equal to the perimeter $L_k$ (see Eq.(\ref{eq:PerimeterInt})). This is the physical interpretation of the $\mathbb{R}_{+}^{n}$ coordinates from the closed string perspective.}
    \label{fig:StrebelAsCylinders}
\end{figure}

In this subsection, we will focus on three aspects of the Strebel parametrization which directly connect the open and closed string pictures. Firstly, we will see how the Strebel picture expresses every Riemann surface as a collection of half-infinite cylinders glued together along the Strebel graph. These cylinders rare the dual closed strings inserted at the punctures. Secondly, each such cylinder will be further decomposed as a collection of strips - one strip for every edge of the Strebel graph. This strip decomposition will be crucial in understanding how to reconstruct the world sheet of a closed string from the gauge theory Feynman diagrams. Thirdly, we will see how these strips provide, in some sense, a rigorous and covariant description of the string bit picture \cite{KogutBits,Thorn:1991fv, Klebanov-susskind}, which has long been advocated as a geometrization of gauge-string duality.

The first point is rather straightforward to see. We already encountered all the ingredients in Sec. \ref{sec:StrebelReview}, though, perhaps, it is still helpful to spell out the details. We saw that the (compact) horizontal trajectories form a set of closed curves around each marked point. This decomposes the Riemann surface into a union of open discs, whose closure is given by the Strebel graph. Each of these discs should be viewed as the conformal mapping of a semi-infinite cylinder. The asymptotic infinity of each cylinder maps onto one of the marked points. Physically, this corresponds to one asymptotic string state created by a vertex operator inserted at the marked point. The foliation of each disc by the horizontal trajectories can then be viewed as Euclidean time propagation of this string, as familiar from radial quantization. 

Recall that the Strebel differential defines a metric on the worldsheet via Eq.(\ref{eq:StrebelMetric}). In this Strebel gauge, the proper length of each asymptotic string is given by Eq.(\ref{eq:PerimeterInt}) as the residue of the corresponding double pole, $L_{k}$. The circumference of the cylinder is thus fixed throughout. These cylinders are then glued together along the Strebel graph. In that sense, the picture that arises is that of $n$ strings coming in from infinity, and all interactions taking place along the Strebel graph. In particular, this is quite similar to light-cone gauge string theory, which localizes the string vertex to a point, allowing strings to join and split. See Fig.\ref{fig:StrebelAsCylinders}.
In some sense the interactions occur at the vertices of the Strebel graph. 
\begin{figure}[ht!]
    \centering
    \includegraphics[scale =0.85]{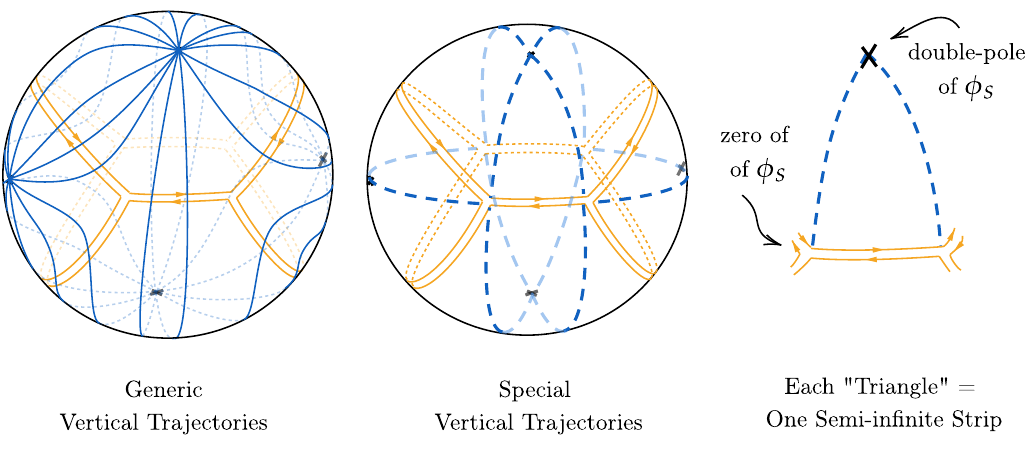}

    \caption{\textbf{From Vertical Trajectories to Strips } There exists a second family of curves on every Riemann surface, determined by its Strebel differential, known as the vertical trajectories. \textit{(Left)} For the most part, they run from marked point to marked point (in solid blue lines), crossing each edge of the Strebel graph exactly once.  \textit{(Middle)} Special vertical trajectories (dashed blue lines) connect marked points to vertices of the Strebel graph (in dashed lines). \textit{(Right)} Each edge of the Strebel graph along with the two curves connecting a marked point (double-pole of $\psi_{S}$) to its vertices (zeroes of $\phi_{S}$) border a triangular shape on the worldsheet, isomorphic to a semi-infinite strip. The (homotopic) edges of the gauge theory Feynman diagrams are in one-to-one correspondence with these strips.}
    \label{fig:VertTraj}
\end{figure}

The second takeaway is that the Strebel differential further dissects each cylinder into a collection of strips. This is a crucial additional step: it connects closed strings to open strings. We identify each of these strips with the worldsheet of an open string. From the point of view of gauge/string duality, these are the same open strings stretched between the D-branes, which, at low energies (or in a topological sector, as here), gave rise to the gauge theory. 

From the point of view of the Strebel differential, we are exploiting a different type of curve traced out on the worldsheet, known as a vertical trajectory. See the left panel of Fig. \ref{fig:VertTraj}. What distinguishes these curves is that the square root of the Strebel differential, $d\tau$, is purely imaginary along them\footnote{In other words, on vertical trajectories, we have ${\rm Re}\left(\sqrt{\phi}dz\right)=0$.}:
\begin{equation}
    \phi\left(z_{V(t)}\right) \left( \frac{dz_{V}(t)}{dt} \right)^2 < 0 \label{eq:vertical}
\end{equation}

There are special vertical trajectories that connect the zeroes and poles of the Strebel differential. We may think of them as the vertical analog of the critical horizontal trajectories (which gave rise to the Strebel graph). These special vertical trajectories form a set of measure zero as well. Their union also traces out a graph on every Riemann surface, see the dashed lines in the middle panel of Fig. \ref{fig:VertTraj}. 
\begin{figure}[ht!]
    \centering

\includegraphics[scale=0.65]{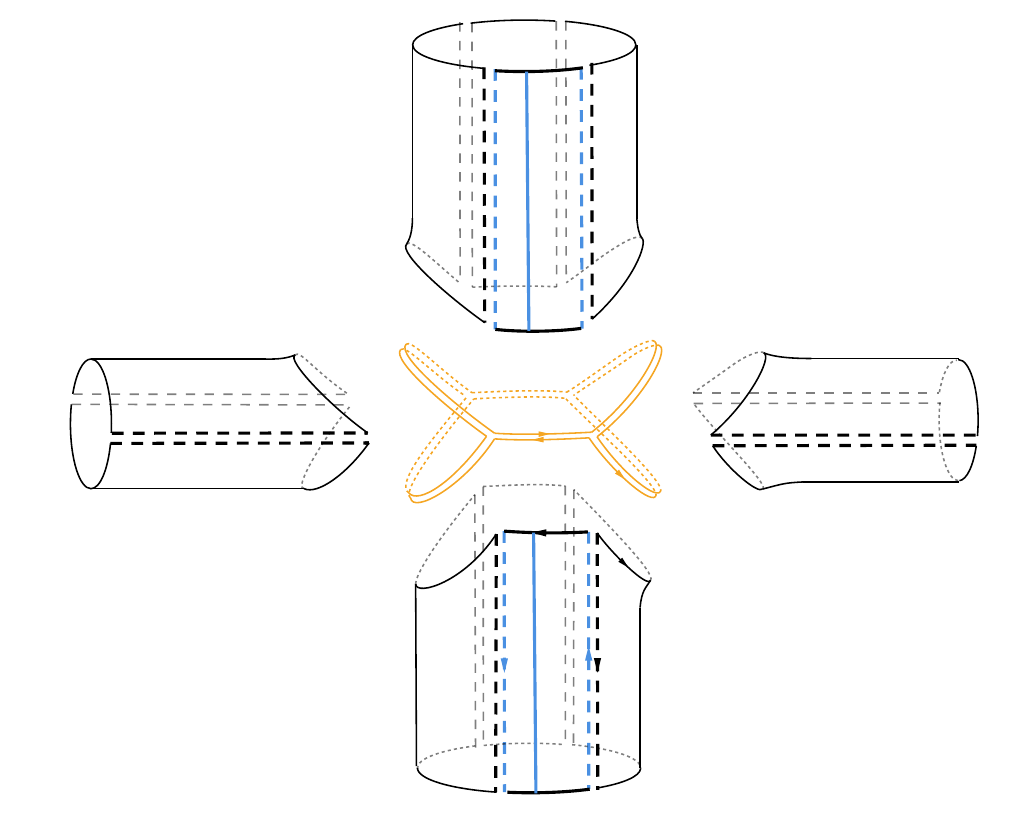}

    \caption{\textbf{Every Riemann Surface as a Collection of Strips } The solid blue line is a vertical trajectory connecting two poles of the Strebel differential. The same line was also drawn in Fig. \ref{fig:VertTraj}. By exploiting the special vertical trajectories (in dashed lines), each cylinder is further decomposed into a set of strips. Each (unit length) strip corresponds to exactly one edge of the gauge theory Feynman diagrams (for V-type dualities). From a D-brane perspective, each strip is the worldsheet of an open string. The Strebel construction thus makes explicit how open strings reassemble themselves into closed ones, directly at the level of the worldsheet.}
    \label{fig:StrebelAsStrips}
\end{figure}

From the point of view of the Strebel graph, these are lines connecting its vertices to the center of its faces, where the poles are located. As shown in the right panel of Fig. \ref{fig:VertTraj}, these lines provide a sort of triangulation of the Riemann surface. The base of each triangle is formed by one of the (horizontal) edges of the Strebel graph, the two other sides are special vertical trajectories connecting the endpoints of the edge to a marked point. Each of these triangles is conformally equivalent to a half-infinite strip. In particular, on the $j$-th strip, the Strebel differential can always be brought to the local form
\begin{equation}
     q = dz_{j}^2.
     \label{eq:StrebelOnStrip}
\end{equation}
This relies on the fact that we can always define local coordinates $z_j$ on the strip by 
\begin{equation}
    z_{j} \equiv \int^{z_{j}}_{0} \sqrt{\phi(s)}ds.
\end{equation}

The physical picture that emerges is therefore that the entire Riemann surface can be decomposed into a collection of strips. We have already seen how every face corresponds to a semi-infinite cylinder. The refinement of that statement is now that every edge bordering a given face of the Strebel graph corresponds to one semi-infinite strip making up that cylinder. The width of the panel is dictated by the Strebel length associated with that edge. A face with $k$ sides should therefore be thought of as a cylinder built out of $k$ strips, sewn together along the special vertical trajectories - see Fig. \ref{fig:StrebelAsStrips}.

\begin{figure}[ht!]
    \centering
    
    \includegraphics[scale=0.7]{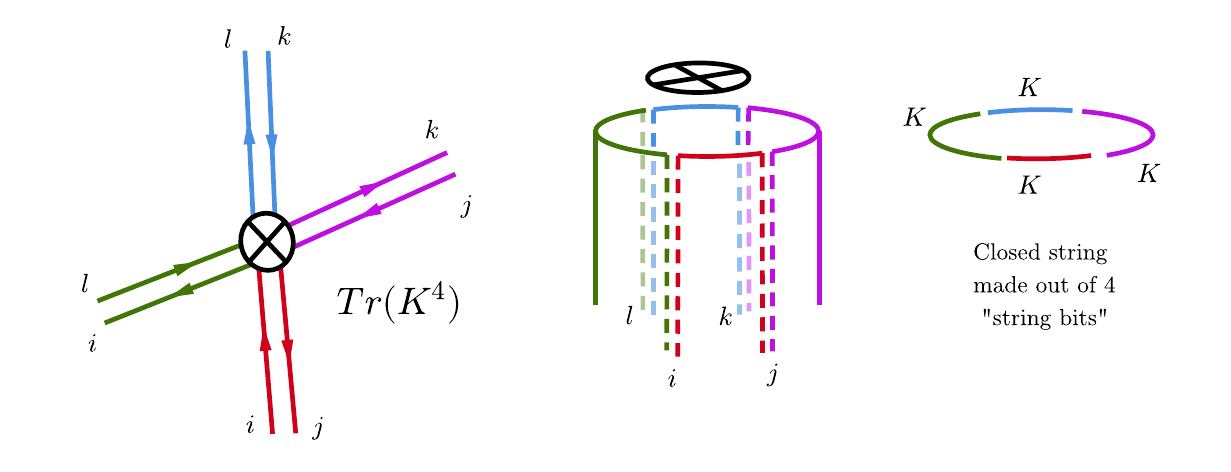}
    \caption{\textbf{From Strips to String Bits } Each strip in the Strebel construction can be viewed as the worldsheet of a "string bit". A single trace operator built of $k$-matrices (in this example, $k=4$) will give rise to a closed string built out of $k$ string bits: one bit per matrix element. Matrix multiplication adjoins string bits together while the $U(N)$-invariance of the trace sows the final endpoints of the string together to form a closed string. Each bit will carry one unit of (vertex operator) momentum in the dual closed string worldsheet.}
    \label{fig:StringBits}
\end{figure}

The third and final point we would like to make regards the natural appearance of string bits \cite{Thorn:1991fv,Klebanov-susskind}. Each of the strips making up the cylinder can be viewed as the worldsheet of a single string bit. Indeed, recall that each such cylinder corresponds to a vertex of the original Feynman diagram (or, equivalently, a face of the Strebel graph). A vertex of valency $k$ arises via the insertion of a single trace operator built out of $k$ matrices. Each edge emanating from the vertex corresponds to a single matrix element in the trace. We will spell out shortly how each edge of the Feynman diagram becomes one of the strips in the Strebel picture. Hence, what we will find is that each asymptotic closed string state dual to a single trace operator built out of $k$ matrix elements consists of $k$ string bits: one bit per matrix insertion. Matrix multiplication enforces two neighboring matrix elements in the trace to share a common index. This maps onto two neighboring string bits being joined together. The $U(N)$-invariance of the trace finally joins the chain of string bits into a closed string. See Fig. \ref{fig:StringBits}. The cyclicity of the trace gives the residual $\mathbb{Z}_{k}$ symmetry of the string (broken by the string bits down from a full $U(1)$-reparametrization along the string). 

Moreover, on the dual closed string worldsheet theory, $k$ will correspond to a "momentum" label for the vertex operator. We thus see how, very much in the spirit of the BMN-limit of AdS/CFT \cite{BMN}, each bit carries one unit of momentum\footnote{The momentum discussed above in the BMN picture was the lightcone momentum $p_{+}$. We already noted the parallel between the localization of string interactions in the light-cone gauge and the Strebel gauge. This is yet another similarity.}. Since, by graph duality, $k$ will also become the perimeter of the face of the Strebel graph, this gives another closed string picture of the $\mathbb{R}_{+}^{n}$ component of the decorated moduli space (which will only take on integer values).

\subsection{Building a closed string worldsheet from a Feynman Diagram} \label{sec:howto}

In this subsection, we spell out an explicit reconstruction algorithm of the dual closed string worldsheet starting from the matrix model Feynman diagrams. 
To do so in full generality, the basic idea will be to define the worldsheet in terms of a collection of local coordinate charts and various transition functions amongst them. We use the strips from the previous section as the basic building blocks and identify them with the edges of the Feynman diagram. We then define a set of gluing rules, so as to reproduce the known properties of the Strebel differential, such as the exact behavior near its poles and zeroes. The full Strebel differential on the worldsheet, along with the corresponding metric it defines, will therefore be defined patch by patch. We will follow closely the discussion in \cite{mulaseStrebel}, adapting it to more general bi-partite ribbon graphs. 

\begin{figure}[ht!]
    \centering
    \includegraphics[scale=0.83]{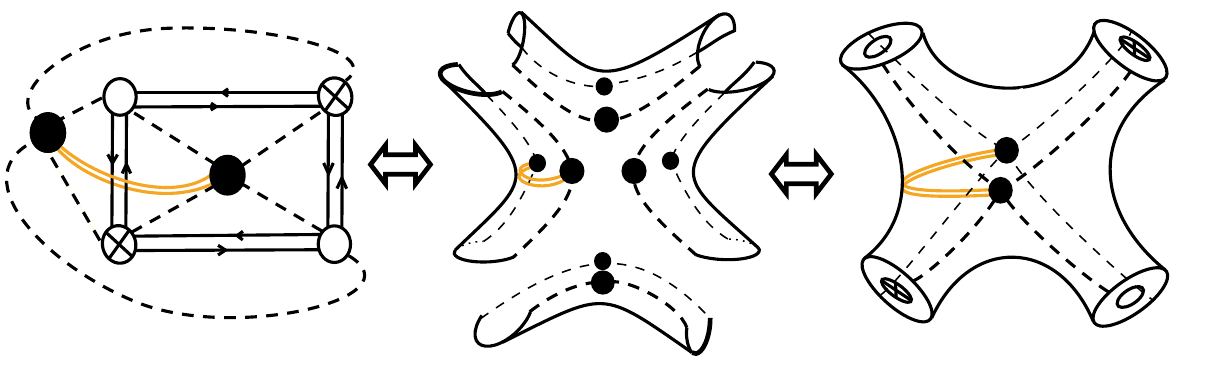}

    \caption{\textbf{Assembling the Closed String Worldsheet from Feynman Diagram Edges } Each edge of the matrix model Feynman diagram corresponds to one strip. This closed string worldsheet will be built out of four strips since our diagram here has four edges. The dashed lines show where the strips are sewn together. Multiple strips meet at the vertices of the Feynman diagram (the crossed and uncrossed white vertices), as well as the center of the faces (the black dots). The original vertices of the Feynman diagram are conformally mapped to the marked points on the Riemann surface. In the language of the Strebel differential residing on the closed string worldsheet, the (white) vertices lie at the poles, while the center of the faces (black dots) will reside at the zeroes. The dashed lines then map onto the special vertical trajectories. }
    \label{fig:FDtoStripstoWS}
\end{figure}

Let us first summarize the reconstruction algorithm. Roughly speaking, it goes as follows:

\begin{enumerate}
    \item \textbf{Identify the Unit Strips:} Associate to each edge of the Feynman diagram one of the strips making up the Riemann surface. Define a local coordinate chart $z_j$ covering each strip, with the corresponding Strebel differential $\phi(z_j) dz_{j}^2 = dz_{j}^2$. Each such strip is assigned unit length.
    \item \textbf{Glue along the (homotopic) edges:} In order to construct the Strebel graph with its metrized lengths, we can first form the skeleton graph of the Feynman diagram. This means we need to bunch together homotopic edges. Since to each edge, we have assigned a strip, we need to glue these strips together. The integer Strebel lengths will come from the fact that by juxtaposing various strips side by side, their widths simply add. \footnote{The various glueing procedures need not to be done in any particular order. In the rest of this subsection, we are therefore referring to the individual strips of the Feynman diagram, not the effective strips of its skeleton graph.}  
    \item \textbf{Glue at the vertices:} The vertices of the Feynman diagram map onto the marked points of the worldsheet. We need to join all the strips meeting at a vertex in such a way that we reproduce the double-pole behavior of the Strebel differential near the marked point. 
    \item \textbf{Glue at the faces:} The faces of the matrix model Feynman diagram correspond to the zeroes of the Strebel differential. We glue all the strips belonging to the edges bordering a face by defining a local coordinate system, valid on an overlapping patch. The transition function needs to guarantee that the Strebel differential vanishes as a simple power, dictated by the number of edges bordering the face.  
\end{enumerate}

We now spell out some of the details involved in each step. To illustrate this general procedure, we will implement this algorithm for our favorite Feynman diagram.
\\

\noindent\textbf{1. Identifying the Unit Strips}

\begin{figure}[ht!]
    \centering
    \includegraphics[scale=0.77]{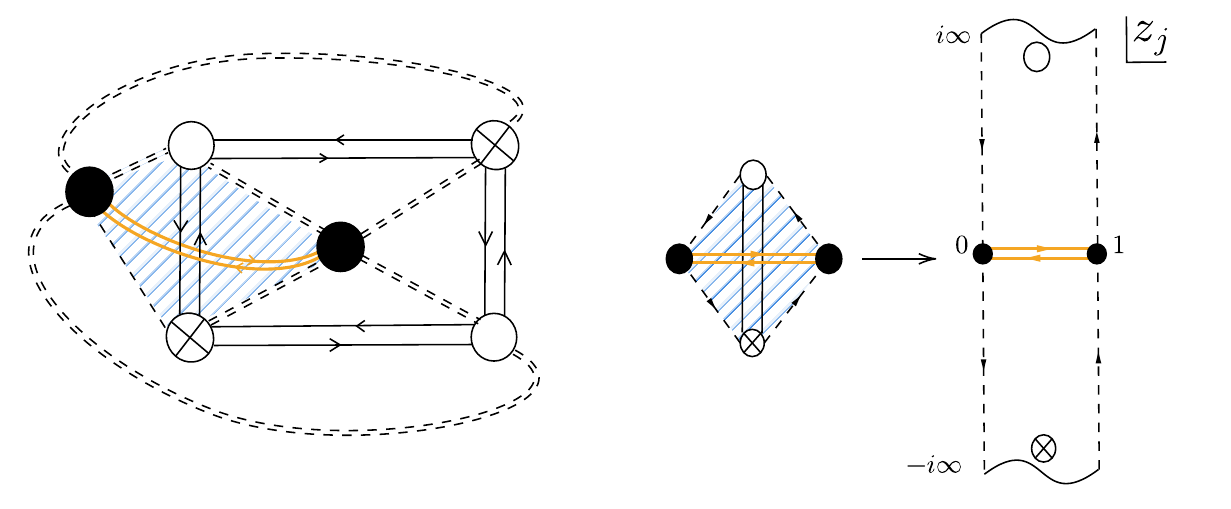}
   \caption{\textbf{1. Each Edge of the Feynman Diagram as an Open String Strip } To the $j$-th edge of the original Feynman diagram, we associate a strip with local coordinates $z_j$. Such an edge connects connects an $M$-vertex (uncrossed) and a $K$-vertex (crossed). The dual edge (in orange) connects the center of the two neighboring faces, which we have labeled by black dots. The $K, M$-vertices and the center of the faces trace out a diamond-shaped region associated with each edge, here blue-white striped. Strictly speaking, it is this diamond-shaped region that we map onto a strip. In the local coordinates $z_j$ on the strip, the black dots lie on the real axis, at $z_j=0$ and $z_j=1$ respectively, while the $M$-vertex gets sent to $z_j=+i \infty$ and the $K$-vertex to $z_j=-i\infty$. Note that each edge of the gauge theory Feynman diagram carries two color indices, which are none other than the Chan-Paton indices of the open strings between D-branes which gave rise, at low energies or via localization, to the gauge theory. These strips are thus literally the string worldsheets of the open string theory.}
    \label{fig:DiamondAsStrip}. 

 \end{figure}

Sec. \ref{sec:PhysPic} showed how the Strebel differential dissects every Riemann surface into a collection of strips. We want to identify each such strip with one of the edges of the Feynman diagram. Fig. \ref{fig:DiamondAsStrip} illustrates how we do this, which we also describe below.
 Recall that, from the point of view of the Strebel differential, the imaginary axis of those strips runs parallel to the vertical trajectories connecting marked points. The marked points correspond to the external vertices of the matrix model Feynman diagram. Hence, we want to align each strip with an edge connecting two vertices. We further pick an orientation for each strip. In local coordinates $z_{j}$ on the $j$-th strip, we choosing the $M$-vertices (uncrossed white vertex) to lie at $z_{j}=+i \infty$ and the $K$-vertices (crossed white vertex) at $z_{j}=-i\infty$.

The real axis of this strip is given by the edge of the Strebel graph (orange), a (critical) horizontal trajectory. It is indeed perpendicular to the edge of the original Feynman diagram, since the Strebel graph is graph dual to the Feynman diagram. The vertices of the Strebel graph lie at zeroes of the Strebel differential, which translates to the faces of the matrix model Feynman diagrams (again, by graph duality). We will denote the center of the faces with solid black dots. Finally, there are lines connecting the black dots to the external (white) vertices of the Feynman diagrams: these correspond to the special vertical trajectories connecting zeroes and poles of the Strebel differential. We thus identify the diamond-shaped region on the Feynman diagram, whose corners are given by a $K$- and $M$-vertex and two face centers, as one of the strips in the Strebel construction (cf. Fig. \ref{fig:StrebelAsStrips}).  

On the $j$-th strip, we define local coordinates $z_j$, as shown in Fig. \ref{fig:DiamondAsStrip}. We choose the origin to be one of the black dots, i.e. endpoints of the Strebel edge, consistent with the orientation of the strip. On this strip, the Strebel differential takes the simple form
\begin{equation}
    \phi(z_j) dz_{j}^{2}= dz_{j}^2 .
\end{equation}
We choose to assign unit width to each strip, so that $\Re(z_{j}) \in [0,1]$. These provide the basic building blocks of our construction. Note that since each edge of the Feynman diagram is assigned two color indices (on either side of the ribbon), one should identify these strips as the worldsheet of the open strings of the gauge theory. The edge's color indices are the Chan-Paton factors of this open string. 
\\
\usetikzlibrary{patterns}

\begin{figure}[ht!]
    \centering

    \includegraphics[scale=0.65]{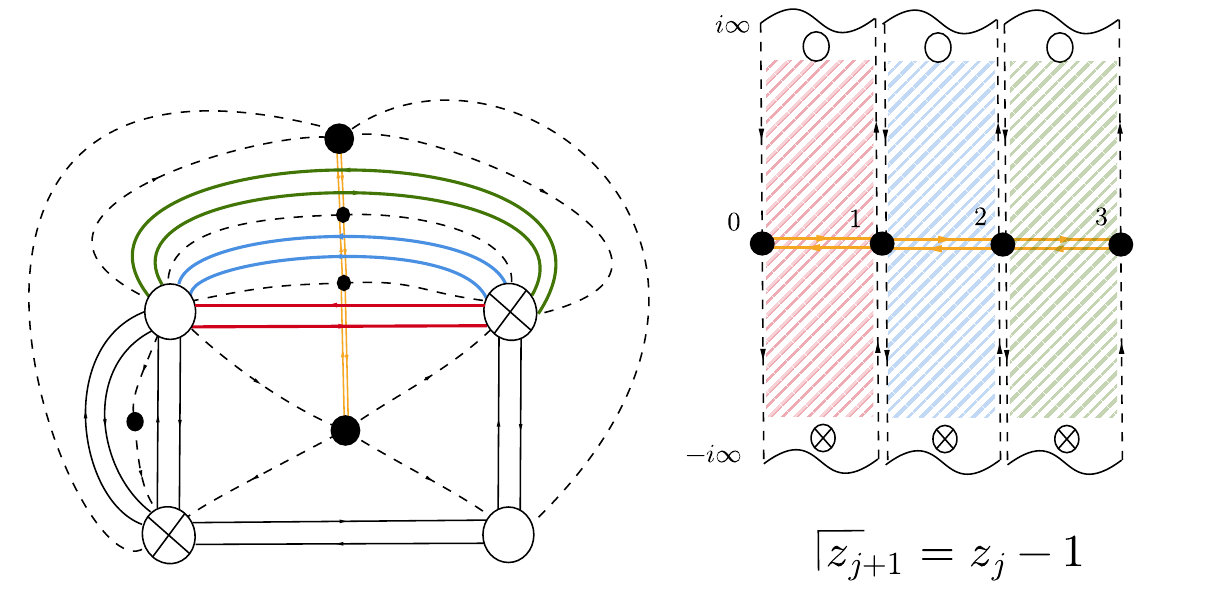}

    \caption{\textbf{2. Gluing Along (Homotopic) Edges } Returning to the example of Fig. \ref{fig:SkeletonGraph}, we see how bundling homotopic edges in the skeleton construction corresponds to simply adjoining their associated strips one after another. We have highlighted in color three homotopic edges (in red, blue, and green), as well as their associated strips (in red, blue, and green respectively). The transition function between the coordinates on each strip is a simple shift. This gives an effective strip with a width equal to the number of homotopic edges bundled together (in this case, $1+1+1=3$). This further explains why all the Strebel lengths (e.g. the length of the orange segment) are (positive) integers in our construction. }
    \label{fig:GluingAtEdges}
\end{figure}

\noindent\textbf{2. Gluing along (Homotopic) Edges}

We discussed at the end of Sec. \ref{sec:StrebGraphsVsFD} that it was really the skeleton graph of the matrix model Feynman diagram that was graph dual to the Strebel graph of the corresponding closed string worldsheet. This was due to the fact that a two-sided face in the Feynman diagram would give rise, via graph duality, to a vertex of valency two for the Strebel graph. However, Strebel graphs only have vertices of valency three and greater. We thus need to collapse such 2-sided faces by bundling together homotopic edges. While this part of the construction is not required for our chosen Feynman diagram of Fig. \ref{fig:MainFD}, it will generically be useful to obtain the various lengths of the Strebel graph. 

Since each edge is associated with an infinite strip, we want to glue these strips together. All these edges connect the same vertices, we thus want to attach the strips along the vertical trajectories of the Strebel differential, which correspond to the lines running parallel to the imaginary axis on each strip. In other words, we simply juxtapose the strips one next to another, using the simple transition function relating neighboring coordinates $z_j$ and $z_{j+1}$
\begin{equation}
    z_{j+1}=z_{j}-1 . \label{eq:EdgeLenghtsAdd}
\end{equation}
The shift by one traces back to the unit length assignment to each individual strip. The net result of this gluing procedure is clear: we obtain another infinite strip with a flat Strebel differential, but with a total width set by the number of collapsed edges. This is the origin of the integrality of the Strebel lengths\footnote{Note that if we had assigned to each strip a length $l_{0}\in \mathbb{R}$, all Strebel lengths would be given in terms of integer multiples of $l_{0}$. We thus see it is truly the addition of the Strebel lengths originating with Eq.(\ref{eq:EdgeLenghtsAdd}) that gives the integer structure to our construction. Choosing $l_{0} \neq 1$ would still pick out the same lattice points on ${\cal M}_{g,n}$, whose location depends only on the ratio of Strebel lengths. The only effect would be to simply rescale the ${\mathbb R}_+^n$ factor which will not change any of the conclusions.}. 
Since there are no homotopic edges in our favorite Feynman diagram, in Fig. \ref{fig:GluingAtEdges}, we have returned to the example of the skeleton graph construction we had already considered in  Fig. \ref{fig:SkeletonGraph}. We see clearly how the bundling of the three colored edges gives an effective Strebel length of $3$ (in orange). \\

\noindent\textbf{3. Gluing at the Vertices}

The vertices of the $K, M$-matrix model Feynman diagram should map onto the marked points of the worldsheet. This must be the case if we want single trace operators to correspond to the insertion of vertex operators on the closed string side. We will need to distinguish the marked points, depending on whether the vertex operators are dual to traces of $K$ or $M$. This will lead to ever so slightly different gluing rules for the $K$- versus the $M$-vertices.  

We will use local coordinates $u_{k} \ (v_{k})$ near the $k$-th marked point dual to an $M \ (K)$-vertex, with the origin chosen to coincide with the marked point. These will be coordinates on the cylinder, when conformally mapped to the plane.
 From Eq.(\ref{eq:StrebelNearPole}), we know the Strebel differential needs to have a second order pole with residue $L_{k}$ at $u_k=0 \ (v_{k}=0)$. From Sec. \ref{sec:StrebelReview}, the perimeter $L_{k}$ should equal the sum of the Strebel lengths around the face. In other words, it is the sum of the widths of all strips making up the cylinder, so for a marked point corresponding to the operator $\Tr(M^l)$ or $\Tr(K^l)$, the residue would simply be $L_k = l$.  
\begin{figure}[ht!]
    \centering

    \includegraphics[scale=0.77]{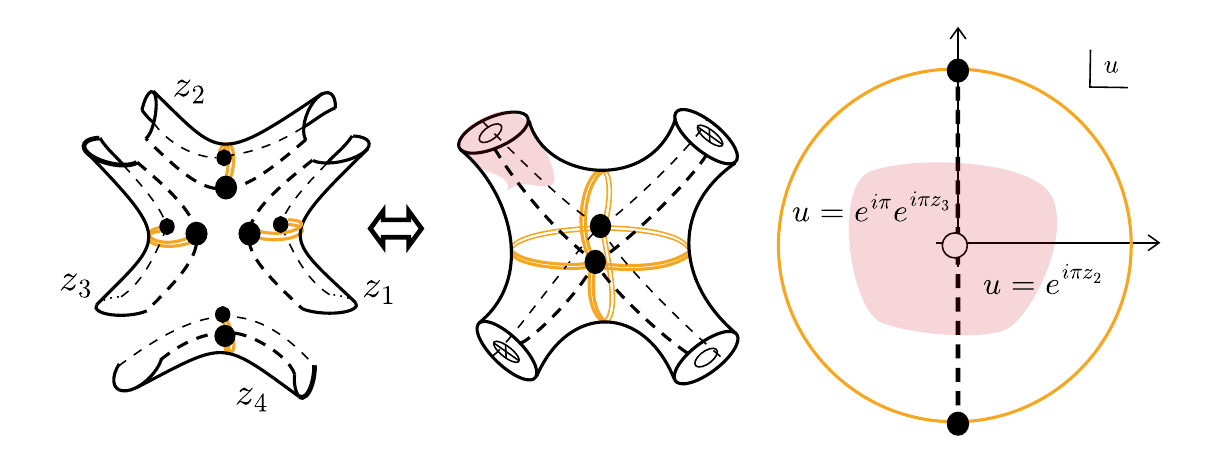}

    \caption{\textbf{3. Gluing Strips at Vertices } Consider the edges emanating from a common vertex of the (skeleton) Feynman diagram. Recall vertices of the $K, M$-model map onto the marked points of the dual worldsheet. These marked points are the conformal mapping of a semi-infinite cylinder. We glue strips associated to these edges together near such a vertex (here, the white uncrossed dot) by defining a local coordinate system $u$ valid in an open neighborhood shared by all strips, along with the appropriate transition function. For the $k$ th $M$-vertex corresponding to, say $\Tr(M^l)$, $l$ strips with local coordinates $z_j$ come together. We define $u_{k}(z_{j})= \exp(2 \pi i \frac{j+z_j}{l})$. For $K$-vertices, an almost identical transition function is used. This will give rise to a Strebel differential with a second-order pole at the origin of the $u$-coordinate system. Here, the cylinder is made out of $l=2$ strips, and $j$ takes values $\{0,1\}$.}
    \label{fig:GlueingAtVertices}
\end{figure}

Concretely, for the $k$-th marked point dual to, say the operator $\Tr(M^l)$, $l$ strips come together to form the (conformally mapped) cylinder near the marked point. $z_j$ is the local coordinate on the $j$-th strip ($j = 0, 1, \dots, l-1$). We glue these strips together by defining the transition function to the $u_k$ coordinates via
\begin{equation} \label{eq:zTOu}
    u_{k}(z_j)= \exp({2 \pi i \frac{j+ z_j}{l}}) , 
\end{equation} 

This indeed maps the Strebel differential $dz_{j}^2$ on each strip to 
\begin{equation}
   dz_{j}^{2} = - \frac{l^2}{(2\pi)^2} \frac{du_{k}^2}{u_{k}^2},
\end{equation}
giving the required double-pole behaviour. 
We have illustrated this procedure for one of the vertices of our beloved Feynman diagram in Fig. \ref{fig:GlueingAtVertices}, showing in orange the edges of the Strebel graph.

For a $K$-vertex, say $\Tr(K^l)$, Eq.(\eqref{eq:zTOu}) needs to be altered by changing the sign in the exponent:
\begin{equation} \label{eq:zTOv}
    v_{k}(z_j)= \exp({-2 \pi i \frac{j+z_j}{l}}),
\end{equation}
This can ultimately be traced back to the fact that the $K$- and $M$-vertices will always have an opposite orientation. The orientation of the vertices is inherited from the flow of the color indices in the faces of the fully bipartite ribbon graphs we are studying. \\

\noindent\textbf{4. Gluing at the Faces}

The final step of our reconstruction algorithm focuses on reproducing the behavior of the Strebel differential near its zeroes. Consider the edges bordering a common face of the (skeleton) Feynman diagram. We need to specify how all the strips associated to these edges should be joined together at the center of the face.

The first thing to notice is that there is always an even number of strips joining around a face center. This goes back to the fact the graph is fully bi-partite. Different strips join at $z_{j}=0$ or $z_{j}=1$, in an alternating fashion. Say we have $2m$ strips joining at a face center. Half of them will be glued near $z_{2j}=0$ and the other half at $z_{2j-1}=1, j \in \{1, \dots, m\}$. The gluing map is given by 
\be \label{faceglueing}
\begin{split}
    \w(z_{2j}) &= e^{\frac{(2j-1)i\pi}{m}}\big(z_{2j}\big)^{1/m}\\
    \w(z_{2j-1}) &= e^{\frac{(2j-3)i\pi}{m}}\big(z_{2j-1}-1\big)^{1/m}
\end{split}
\ee 
\begin{figure}[ht!]
    \centering

    \includegraphics[scale=0.78]{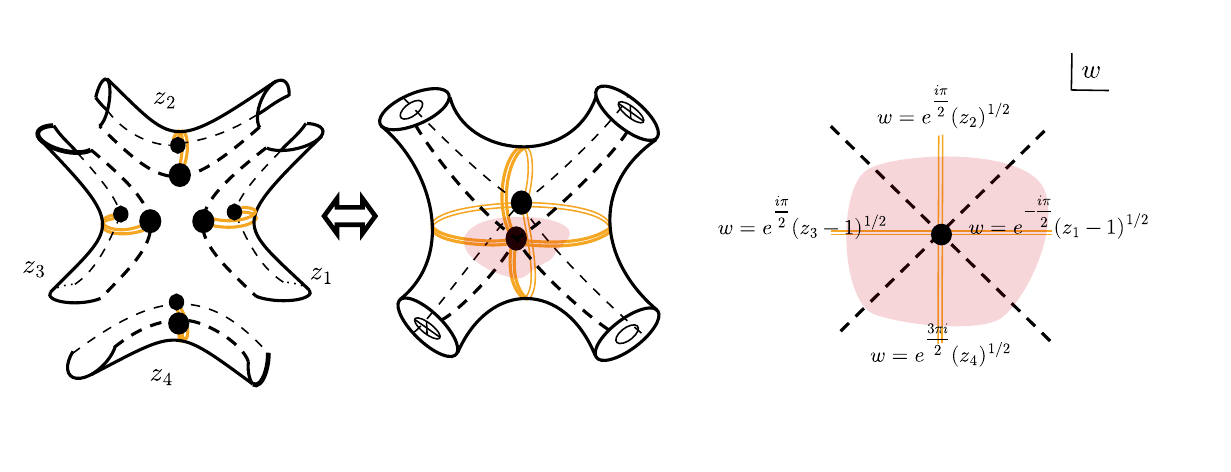}

    \caption{\textbf{4. Gluing Strips at Faces } Consider the edges bordering a common face of the (skeleton) Feynman diagram. We "glue" their associated strips together near the center  of the face (the black dot) by defining a local coordinate system $\omega$ valid in an open neighborhood shared by all strips, along with the appropriate transition function. There are always an even number of strips meeting at a point. When $2m$ strips with local coordinates $z_{k}$ come together, half of them are joining at $z_{2j}=1$ and the other half at $z_{2j-1}=1$. We define the local "face coordinates" $\omega$ such that $\omega(z_{2j})= e^{ \frac{(2j-1)i\pi}{m}} z_{2j}^{1/m}$ and  $\w(z_{2j-1}) = e^{\frac{(2j-3)i\pi}{m}}(z_{2j-1}-1)^{1/m}$. This will give rise to a Strebel differential with a $2m-2$-th order zero at the origin of the $w$-coordinate system (the black dot), as expected for an $2m$-valent vertex of the Strebel graph (in orange). }
    \label{fig:GlueingAtFaces}
\end{figure}

The faces of the Feynman diagram become, by graph duality, the vertices of the Strebel graph. These correspond to the zeroes of the Strebel differential. Recall that the valency of the vertex dictates the order of the zero. 
If the $p$-th face of the matrix model Feynman diagram is bordered by $2m$-edges (it has to be even due to the nature of our matrix model propagator), we should, through this gluing map,  land on a $(2m-2)$-th order zero of the Strebel differential in the region where all the associated strips meet. We can check that 
\be 
\big(dz_{2j}\big)^2 = m^2\w^{2m-2} d\w^2, \ \ \text{and,} \ \ \big(dz_{2j-1}\big)^2 = m^2\w^{2m-2} d\w^2,
\ee 
So under this gluing rule, the face center is indeed mapped to a $(2m-2)$-th order zero of the Strebel differential.

As can be seen in the example of Fig. \ref{fig:GlueingAtFaces} (where $m=2$), this coordinate transformation essentially sends each strip into a wedge, whose opening angle $\frac{\pi}{m}$ is set by the number of strips $2m$ meeting at that face. The real axis along each strip (which, again by construction, lies on the Strebel graph (in orange)) is mapped to lines emanating from the origin of the $w$-coordinate system. They are offset from the dashed lines (the special vertical trajectories of the Strebel differential) by an angle $\frac{\pi}{2m}$.

The system of coordinates $u_{k}$, ($k \in 1,..,V_{M}$), $v_{l}$, ($l \in 1,..,V_{K}$), $\w_p$, ($p \in 1,...F$) and $z_{j}$, ($j \in 1,..,E$), along with the respective transition functions, completely specify the closed string worldsheet built from the genus $g$ Feynman diagram (with $F$ faces and $E$ edges) used in the computation of the single trace correlators studied in Sec \ref{sec:MatrixFDs}. We have therefore achieved the goal we set out for this section, namely explicitly mapping every Feynman diagram to a closed string configuration. However, so far, we have only discussed which points on the moduli $\mathcal{M}_{g,n}$ contribute to the dual string path integral computation of said correlator. To understand the weight associated to each string configuration, we need to go further and construct the embedding of the string into the target space. This is the purpose of the following section.

\pagebreak

\section{The Closed String Target Space from Feynman Diagrams}\label{sec:targetpersp}

So far, we have recast the Feynman diagram expansion of a matrix model correlator as a sum over particular worldsheets. However, each Feynman diagram comes with a particular weight. From the point of view of the dual string theory, the weight associated to each worldsheet should come from the action functional of the string evaluated on the embedding map, $e^{-S[X]}$. In order to reproduce the weighting factor associated to each Feynman diagram we therefore need to reconstruct the action and the embedding map. For the particular A-model topological string dual we have found, the action is known \cite{WittenTopSigma,MarinoAmodelreview}. In Sec. \ref{sec:weight=action}, we will find it reduces to the Nambu-Goto action.

In this section, we want to reconstruct the embedding map of the worldsheet into the target space, directly from the Feynman diagrams. This might sound like an impossible task. However, we will show that the embedding maps are quite simple: they can be fully characterized by discrete graph data. The string embedding will be given in terms of holomorphic covering maps of a target space $\mathbb{CP}^{1}$, branched over exactly three points of the sphere. These are known as Belyi maps. What is the significance of there being three branch points? We will see they are in one-to-one correspondence with the two types of vertices and the faces of the Feynman diagrams. In Sec. \ref{sec:ExplicitBelyi}, we provide an explicit, analytic expression for the map dual to our favorite Feynman diagram. 
\begin{figure}[ht!]
    \centering
\includegraphics[scale=0.7]{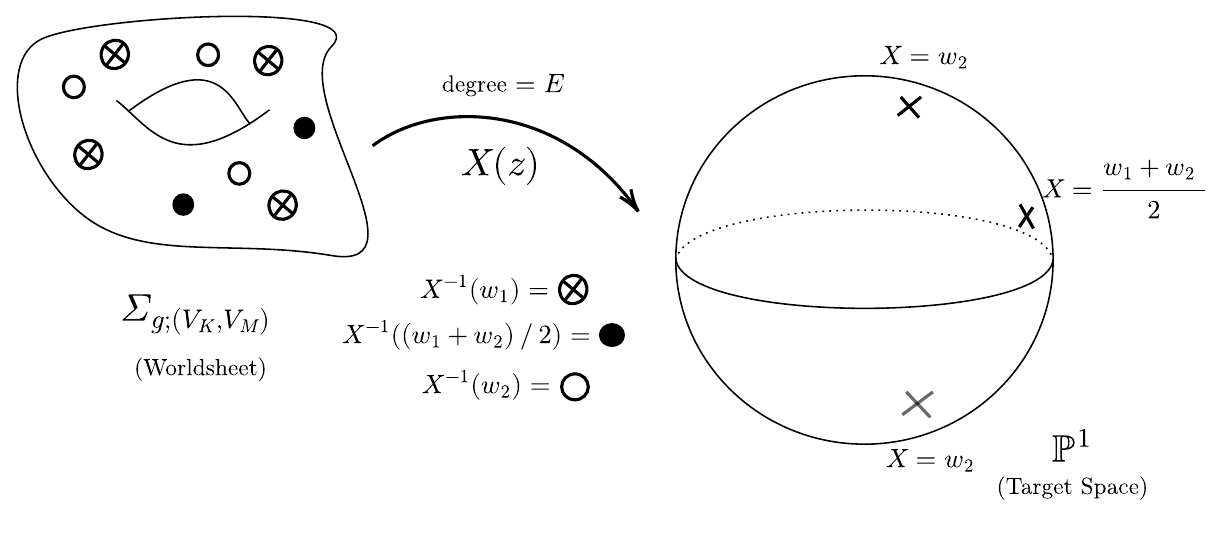}

    \caption{\textbf{Deriving the String Embedding Map } We already reconstructed the worldsheet of the dual closed string from the Feynman diagram. Here, we derive its embedding map $X(z)$ into the target space, also purely from the Feynman diagram. The target turns out to be the Riemann sphere, $\mathbb{CP}^{1}$. In this string theory, the string worldsheet wraps the $\mathbb{P}^1$ as many times as there are edges of the ribbon graph, $E$. The $K$-vertices, which are marked points on the worldsheet $\Sigma_{g;(V_{K},V_{M})}$ turn out the be the pre-images of $X=w_1\in \mathbb{P}^1$, while the set $X^{-1}(w_2)$ give the $M$-vertices. The pre-images of $X=(w_1+w_2)/2$ correspond to the faces of the Feynman diagram, or alternatively, the zeroes of the Strebel differential on the worldsheet. From the point of view of $AdS_{5}/CFT_{4} $, we will interpret this $\mathbb{P}^1$ as a particular subspace on the boundary of $AdS_5$ to which the string localizes. }
    \label{fig:EmbeddingMap}
    
\end{figure}

We later discuss, in Sec. \ref{sec:embedding}, how this target space can viewed from the perspective of the $AdS_{5}/CFT_{4} $ correspondence. We would like to interpret the $\mathbb{CP}^{1}$ as a two-dimensional subspace of the boundary of $AdS_{5}$. Roughly speaking, for the protected subsector captured by the two-matrix integral, we can take the zero coupling limit of $\mathcal{N}=4$ SYM. We believe the string then localizes to the boundary, much as in the tensionless limit studied extensively in the $AdS_{3}/CFT_{2}$ correspondence. It wraps the compactified plane on which certain gauge theory operators were inserted, as shown in Fig. \ref{fig:IntroEmbedding}. Exactly this picture had been recently suggested by a computation, in \cite{Bhat:2021dez}, of simple $1/2$ BPS correlators using the twistor string proposed as the string dual to free $\mathcal{N}=4$ SYM \cite{gaberdiel2021string,gaberdiel2021worldsheet}. That said, we will mostly focus on the Riemann sphere alone as the string's target space for the remainder of this section.

Belyi maps are completely specified by their branching structure above the three points, which can be encoded in terms of three permutations. Our basic strategy is therefore to proceed in two steps. We first associate each Feynman diagram in the expansion of a correlator to a set of three permutations. We then translate these permutations into a Belyi map. This mirrors the strategy of Gross and Taylor in obtaining the string dual to 2d Yang-Mills theory. They used Schur-Weyl duality to recast the gauge theory observables in terms of permutation data. They then similarly interpreted this data in terms of branched covering maps \cite{Gross:1993cw, Gross:1993hu, Cordes:1994sd}. One particularly nice feature of our construction is that the Feynman diagram itself has a simple interpretation from the point of view of the target space. Its edges are the pre-image of a line (segment) on the target space sphere under the covering map. For the $K,M$-matrix model, this line in target space connects the two points $X = w_1$ and $X = w_2 \in \mathbb{P}^{1}$, as illustrated in Fig. \ref{fig:FDasPreim}.

The above construction is not new. It is perhaps better known under the name of \textit{dessin d'enfants}, or children's drawing, as popularized by Grothendieck \cite{grothendieck1984} and beautifully reviewed in \cite{Lando2003GraphsOS}. To the best of our knowledge, this perspective was brought to matrix models in \cite{DiFrancescoItzykson}, see also \cite{KostovMAPS}. It is also very nicely articulated in the context of gauge-string duality by de Mello Koch and Ramgoolam in \cite{deMelloKoch:2010hav}. What is particularly satisfying here is that we now have a worldsheet theory which indeed localizes to these maps (see Sec. \ref{sec:Amodel}); though the proper discussion of that WZW model is postponed to \cite{DSDIII}. This worldsheet theory had also been found via other means in \cite{DSDI}, giving a further consistency check on the duality. Overall, the takeaway will therefore be that each Feynman diagram encodes both a particular worldsheet, via the Strebel construction, \textit{as well as} the embedding of the worldsheet into the closed string geometry (here a $\mathbb{P}^1$).

\subsection{How to Derive a Belyi Map from a Feynman Diagram.}\label{subsec:deriveBelyi}

In Sec. \ref{sec:howto}, we presented an algorithm to reconstruct a worldsheet from a matrix model Feynman diagram. Now we will provide an algorithm to derive the embedding map of that string worldsheet from that same Feynman diagram. We will later explain why this algorithm works. As mentioned previously, the algorithm consists of two steps: first extracting the permutation data from the diagram, and then translating the said permutation data into a covering map. We will denote by $X$ the coordinates on the target Riemann sphere, while $z$ refer to (local) coordinates on the worldsheet.

Consider now a Feynman diagram of genus $h$, with $V_{K}$ crossed vertices, $V_{M}$ uncrossed vertices, $E$ edges and $F$ faces\footnote{Note these are not all independent, since $V-E+F=2-2h$. For example, once we specify the number of single trace insertions, $V_{K}, V_{M}$, along with the powers of matrices appearing, the $\{l_{i},n_{j}\}$ and the power of $N$ associated the diagram, i.e. the genus $h$, then its number of faces $F$ is determined.} contributing to the correlator
\begin{equation}
    \Braket{\prod_{i=1}^{V_K} \frac{1}{l_{i}}\Tr\left( K^{l_i}\right) \prod_{j=1}^{V_M} \frac{1}{n_{j}}\Tr\left( M^{n_j} \right)} _{c,h},
\end{equation}
In particular, as already noted below Eq.(\ref{eq:GenCorrelator}), the form of the propagator forces the number of edges to satisfy $E= \sum_{i=1}^{V_{K}}l_{i}=\sum_{j=1}^{V_{M}}n_{j}$. We will consider the Feynman diagram directly, not its skeleton graph.

\begin{figure}[H]
    \centering
\includegraphics[scale=0.5]{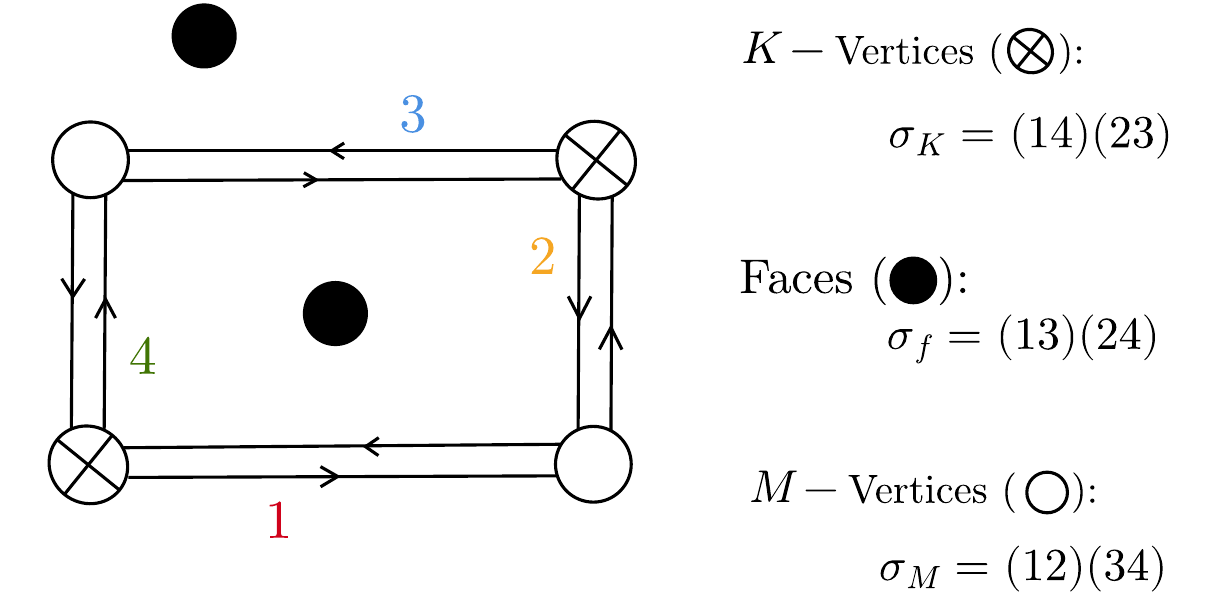}

    \caption{\textbf{From Feynman Diagrams to Permutations } We label all edges $1,2,..,E$, working our way around the diagram, starting from any vertex. We then read off three permutations $(\sigma_{K}$, $\sigma_{f}$, $\sigma_{M})  \in S_{E}$, associated to the $K$-vertices, the faces and the $M$-vertices, respectively. There are as many cycles in each permutation as there are $M$-vertices, faces and $K$-vertices. These three permutations will encode the branching structure of the string covering map above three points of the target space, $X=w_1$, $X=(w_1 + w_2)/2$ and $X=w_2 \in \mathbb{P}^{1}$.}
    \label{fig:FDtoPerms}
    
\end{figure}

To extract the map $X(z)$, we first determine three permutations from the diagram, associated to the two types of vertices and one type of face. We have done so for our favorite Feynman diagram in Fig.\ref{fig:FDtoPerms}. This provides a simple example of the more general steps outlined below.

\noindent\textbf{From Feynman Diagrams to Permutations}
\begin{enumerate}
    \item Pick an arbitrary vertex of the graph. Assign an overall orientation to each edge (this is on top of the orientation given to the faces, induced by the arrows in each color line of the ribbon graph), say the edge points from crossed to uncrossed vertices. To each edge of the graph, assign an integer in $\{1,2,...,E\}$, which we draw to the right of the edge (according to the aforementioned orientation). We proceed from the starting vertex, by going around the faces in an order consistent with the orientation of the faces. These are the colored labels $1$ through $4$ in Fig. \ref{fig:FDtoPerms}.
    \item We now assign an element $\sigma_{K}$ of the permutation group $S_{E}$ to the crossed vertices. $\sigma_{K}$ will consist of as many cycles as there are crossed vertices in the graph, $V_{K}$. Each such vertex inherits an orientation, clock-wise or anti-clock-wise, from the orientation of the faces it corners. We assign to a particular vertex a single cycle by writing down all the labels of the edges that emanate from that vertex, according to the vertex's orientation. For example, in Fig. \ref{fig:FDtoPerms}, all vertices have a clockwise orientation, so the two $K$-vertices gives the cycles $(14)$ and $(23)$.  The total permutation is simply the union of these cycles. $\sigma_{K}$, given in this case by $(14)(23)$, will later dictate the branching structure of the Belyi map above $X=w_1\in \mathbb{P}^{1}$.
    \item We similarly assign a permutation $\sigma_{M}$ to the uncrossed vertices. $\sigma_{M}$ will therefore consist of $V_{M}$ cycles. From the two $M$-vertices in Fig. \ref{fig:FDtoPerms}, we can read off the cycles $(12)$ and $(34)$, so that $\sigma_{M}= (1\thinspace2)(3\thinspace4)$. 
    \item We now assign a permutation $\sigma_{f}$ to the faces of the diagram \footnote{As mentioned before, we are working directly with the Feynman diagram, not its skeleton graph. We therefore include faces bordered only by two edges in the reconstruction of the Belyi map. They will give rise to length-1 cycles in $\sigma_{f}$}. To each face, we assign a cycle, so that $\sigma_{f}$ is the product of $F$ cycles. We build each cycle from all the labels of the edges bordering a given face, in an ordering dictated by the orientation of its color-index line. The diagram in Fig. \ref{fig:FDtoPerms} has two faces (the black dots), giving rise to the two cycles $(24)$ (for the inner face) and $(13)$ (for the outer face), so that the total permutation $\sigma_{f}$ reads $\sigma_{f}=(13)(24)$. 
\end{enumerate}

\noindent \textbf{From Permutations to Maps}

This algorithm assigns to each Feynman diagram of the matrix model, three permutations $\sigma_{K}$, $\sigma_{M}$ and $\sigma_{f}$. We now need to translate this combinatorial data into the language of covering maps. The fact that (finite degree) covering maps of  $\mathbb{P}^{1}$ by generic worldsheets are completely classified by this permutation data is guaranteed by the Riemann existence theorem\cite{Lando2003GraphsOS}. 

The number of edges, $E$, dictates the degree of the cover. In other words, the genus $h$ worldsheet, with $V_{K}+V_{M}$ marked points, wraps the target space sphere $E$ times. More precisely, since the cover is branched at (exactly) three points, a generic point on the target sphere will have $E$ preimages on the worldsheet. Each edge of the Feynman diagram therefore corresponds to one sheet of the cover, a fact we will expand upon shortly.
Above the three branchpoints, there will be fewer pre-images. The number of cycles in $\sigma_{K}$,$\sigma_{f}$ and $\sigma_{M}$ determines the number of pre-images. There are therefore $V_{K}$ points on the worldsheet which all get sent to  $X=w_1$, while the other $V_{M}$ marked points are mapped to $X=w_2$. The $F$ zeroes of the Strebel differential on the worldsheet, corresponding to the faces, are sent to $X=(w_1+w_2)/2$. See Fig. \ref{fig:EmbeddingMap}. 

\begin{figure}[ht!]
    \centering
\includegraphics[scale=0.7]{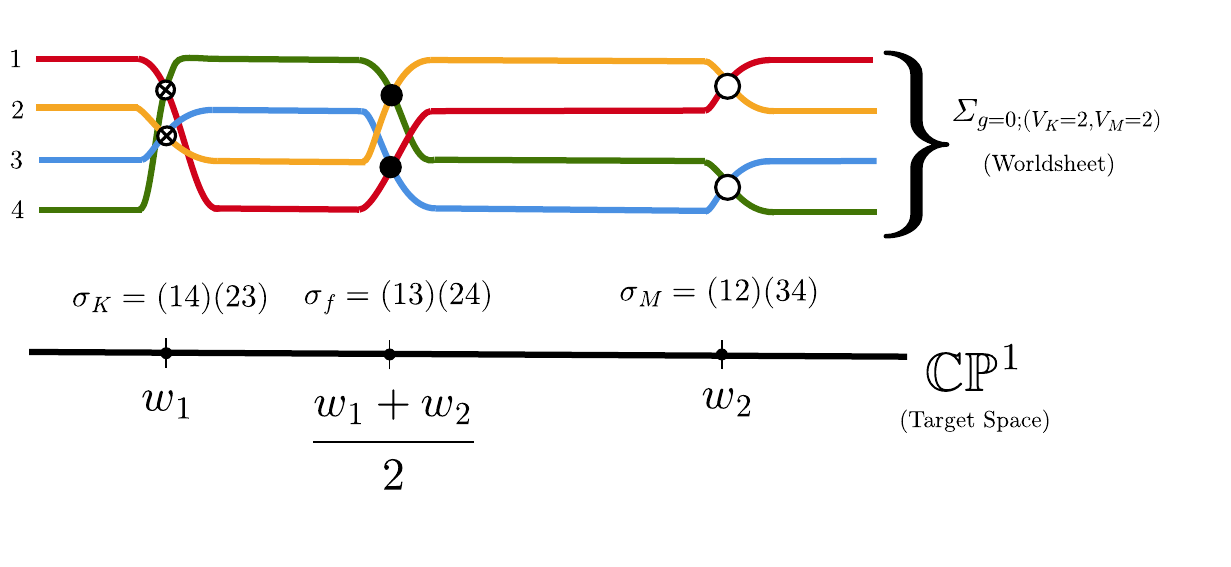}

    \caption{\textbf{From Permutations to Embedding Maps } In this string theory, the worldsheet $\Sigma_{g,n}$ (here the four colored lines) wraps the entire target space $\mathbb{CP}^{1}$ (in black) multiple times. Such holomorphic covering maps are fully characterized by their branching structure. Each edge of the Feynman diagram corresponds to one sheet of the cover (the colors of each sheet match those of the edges in Fig. \ref{fig:FDtoPerms}). The present example is thus a degree four cover. The permutation $\sigma_{K}$, read off from the $K$-vertices, shows there are two pre-images of $w_1\in\mathbb{P}^{1}$ on the worldsheet. These correspond to the two cycles of $\sigma_{K}$. Graphically, these are the two $K$-vertices of the Feynman diagram. Sheets 1 and 4, and sheets 2 and 3, meet and get permuted above $w_1\in\mathbb{P}^{1}$. The permutation only involves simple two-cycles, because we only considered traces of $K^2$. We can similarly read off the branching over $(w_1+w_2)/2$ and $w_2 \in \mathbb{P}^{1}$ from the other two permutations, coming from the faces and $M$-vertices respectively. }
    \label{fig:PermstoMaps}
\end{figure}

So far, we have only discussed the rough cycle structure of the three permutations. What does each individual cycle mean? Since $V_{K}$, $V_{M}$ and $F$ are all smaller than $E$, this means that various sheets must meet at the pre-images of the branchpoints. Each cycle of $\sigma_{K}$,$\sigma_{f}$ and $\sigma_{M}$ tells us which sheets are meeting at a given pre-image of $(w_1,(w_1 + w_2)/2,w_2) \in \mathbb{P}^1$, respectively. To be more precise, there is a non-trivial monodromy around each of the pre-images of the three branchpoints in the target space. \footnote{It is perhaps helpful to glance at Fig. \ref{fig:PermstoMaps} while reading the next few sentences.} This means that as we go around one of the branchpoints on the Riemann sphere, we end up on a different sheet of the cover compared to the one we originally started out on. Which sheet is determined by the exact structure of the corresponding cycle, with neighboring elements in the cycle being related by the monodromy of going around the branchpoint once. Finally, the statement that there must be trivial monodromy if we encircle all three branchpoints on the sphere (we can always contract such loop) translates to the important fact relating the product of the three permutations
\begin{equation}
    \sigma_{K} \sigma_{M} \sigma_{f} = \mathbb{Id}
\end{equation}
where $\mathbb{Id}$ is the identity element in $S_{E}$.

Fig. \ref{fig:PermstoMaps} exemplifies how the three permutations associated to our Feynman diagram encode the corresponding covering map. The four edges of the Feynman diagram give rise to a degree four cover. The two pre-images of $X=w_1$ correspond to the two cycles of $\sigma_{K}$, or, more diagrammatically, to the two crossed vertices of the Feynman diagram. On the worldsheet $\Sigma_{h=0;(V_{K}=2,V_{M}=2)}$, these are the marked points where the vertex operators dual to $\Tr(K^2)$ are inserted. We determined those to lie at $z=1,-1/3 \in \Sigma_{h=0,(2,2)}$ in Sec. \ref{sec:ExplicitWS}. The individual cycles of $\sigma_{K}$,$(14)$ and $(23)$, mean that if we start with a point near $X= w_1$ and its preimage on sheet $1$ of the worldsheet, and we make a $2\pi$ rotation in $\mathbb{P}^1$ around $X = w_1$ while following its preimage, we end up on the preimage of the same point, but on sheet $4$ in the worldsheet. A similar thing happens for preimages on sheets $2$ and $3$. In the same way, we translate the permutations $\sigma_{M}$ in terms of the monodromy around the other two marked points $z=1/3,\infty$. The pre-images of $X=(w_1+w_2)/2$ lie at the zeroes of the Strebel differential on the worldsheet, here at $z=1/3 \pm 2i/3$.

\subsection{Feynman Diagrams From Target Space} \label{sec:FDaspreimage}

There exists a nice target space picture as to how the Feynman diagram encodes a Belyi map, $X(z)$. We know the two types of vertices of the Feynman diagram correspond to the two types of marked points on the worldsheet of the string. The $K$-marked points all get mapped onto $X=w_1$ while the $M$-marked points are sent to $X=w_2$. 

\begin{figure}[ht!]
    \centering
\includegraphics[scale=0.6]{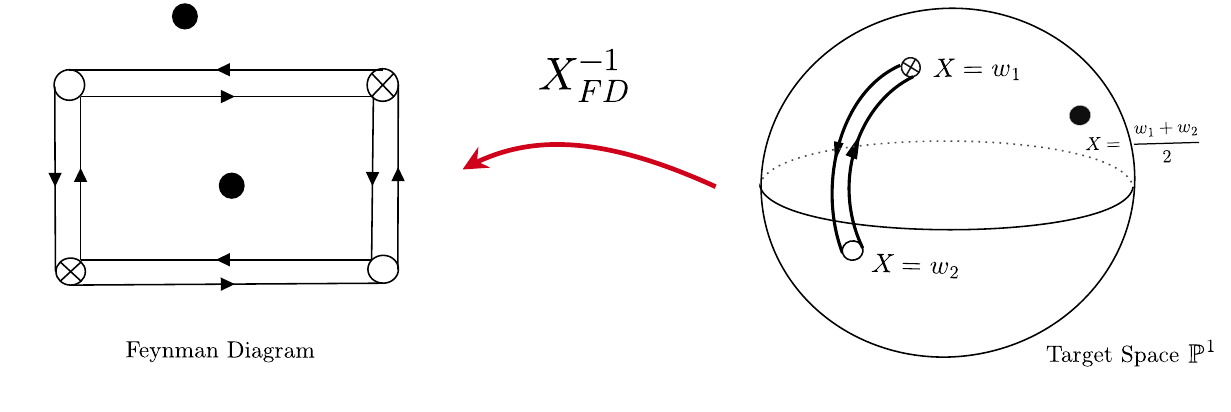}

    \caption{\textbf{Feynman Diagram from Target Space Interval } An equivalent way to understand the encoding of the Belyi map $X_{FD}$ via a Feynman diagram is to view the ribbon graph as the pre-image of a target space interval. The endpoints of the interval lie at the two branchpoints $X=w_1$ and $X=w_2$. In particular, this makes manifest why the degree of the covers equals the number of edges. }
    \label{fig:FDasPreim}
\end{figure}

Each edge of the Feynman diagram connects one type of vertex to another. It can therefore be mapped into a line in target space connecting $w_1$ to $w_2$. The entire Feynman diagram can then be viewed as the pre-image of such a target space interval, see Fig. \ref{fig:FDasPreim}. Schematically, 

\begin{equation}
    \text{Feynman Diagram} = X^{-1}\left( [w_1,w_2] \right) 
\end{equation}
A generic point away from $X=w_1,w_2$ has as many pre-images as the degree of the map. This perspective makes clear why there are as many edges to the diagram as the degree of the cover. There are of course fewer vertices of any one type as there are edges, which simply reflects the fact that at the branchpoints, there are fewer pre-images than the degree of $X$. We will have more to say about this picture in Sec. \ref{sec:backtostrips}, and will use this perspective on the Feynman extensively in Sec. \ref{sec:OCOinTS}.  

This observation about the Feynman diagram as the preimage of an interval under the covering map also makes clear the relation with Grothendieck's "\textit{dessin d'enfants}", which are defined as graphs with alternate crossed and uncrossed vertices, see \cite{Lando2003GraphsOS}. Associated to such a "\textit{dessin}", there exists a map to $\mathbb{P}^1$, under which preimage of the interval $[0,1]$ is the graph. We can now draw the parallel easily, and see that our Feynman diagram is precisely the "\textit{dessin}", and the map is (up to an $SL(2, \mathbb{C})$ transformation) our Belyi map.

\subsection{Reconstructing the Explicit Belyi Map Dual to our Feynman Diagram}\label{sec:ExplicitBelyi}

While the discrete permutation data suffices to uniquely specify a particular Belyi map, it would be satisfying to explicitly write down the embedding map $X(z)$. In parallel to our discussion regarding the worldsheet, we now do so for the case of our Feynman diagram from Fig. \ref{fig:MainFD}. We will temporarily ignore our findings from Sec. \ref{sec:ExplicitWS}, though our final result will need to show that the various pre-images, on the worldsheet, of $X=w_1,w_2$ and $X=(w_1+w_2)/2$ indeed agree with the location of poles and zeroes of the Strebel differential we found there. 

Since our Feynman diagram is planar, we can use the fact that covering maps of the sphere by genus 0 Riemann surface admit a rational function parametrization. To construct the explicit Belyi map for this specific, we take the following strategy
\begin{itemize}
    \item First we find a map $\zeta(z): \Sigma_{0, 2+2} \to \mathbb{P}^1$, that maps the $M$ vertices to $\zeta =\infty$, $K$ vertices to $\zeta=0$, and the Strebel zeroes to $\zeta =1\in \mathbb{P}^1$, with the desired branching property at each of these points.
    \item Then we compose it with the following function
    \be \label{compose}
    X(z) = \frac{w_2\zeta(z) + w_1}{\zeta(z) +1}
    \ee 
    Which takes $\zeta=0\to X=w_1, \zeta=\infty \to X=w_2, \zeta = 1 \to X= (w_1+w_2)/2$, without disturbing the branching properties at these points. $X(z)$ will be our desired covering map.
\end{itemize}

Let us determine $\zeta(z)$ for the case at hand. Since the $K$-vertices get mapped to $\zeta=0$, and both vertices have valency two, the numerator can be written as the product of two squares. The two $M$-vertices will instead get mapped to $\zeta=\infty$. The order of the poles is again fixed by the valency of the uncrossed vertices (here 2), so the denominator can also be written as the product of two squares.

\begin{equation}
\zeta(z) =\alpha \frac{(z-a)^2 (z-b)^2}{(z-c)^2(z-d)^2} \label{eq:VertexConstraint}
\end{equation}
where $a,b$ are the locations on the worldsheet of the $K$-marked points and $c,d$ those of the $M$-marked points. $\alpha$ is some overall constant. We can see that a generic point $\zeta_{0} \in \mathbb{P}^1$ will indeed have four pre-images. This is the statement that the map is of degree four, since our Feynman diagram has four edges.

From the fact that the two faces get sent to $\zeta=1$, and that each face has two edge labels ($(13)$ and $(24)$), we know that the numerator of $\zeta(z)-1$ also takes the form of the product of two squares:

\begin{equation}
    \zeta(z)-1 = \beta \frac{(z-e)^2 (z-f)^2}{(z-c)^2(z-d)^2} \label{eq:FaceConstraint}
\end{equation}
$e,f$ will ultimately need to be the zeroes of the Strebel differential on the worldsheet, providing a further check of our result in Sec. \ref{sec:ExplicitWS}. $\beta$ is again a simple overall constant.

By plugging in Eq.(\ref{eq:VertexConstraint}) into the left-hand side (LHS) of $\zeta(z)$, and equating the same powers of $z$ in the two numerators, we obtain a system of five (i.e. degree of the map +1) algebraic equations for our eight unknown coefficients. This leaves three coefficients remaining, which we can fix by the residual $SL(2,\mathbb{C})$ invariance on the four-punctured sphere\cite{zvonkine:hal-00347348}.

Following the choice made in Sec. \ref{sec:ExplicitWS}, we set $a=1$, $b=t$, $c=-t$. Choosing the remaining marked point to lie at $z=\infty$ translates to keeping only $(z-c)^2$ in the denominator of $\zeta(z)$. The remaining equations to solve then read:
\begin{align}
 &(\alpha -1) t^2=\beta  e^2 f^2, \qquad
 t (\alpha +\alpha  t+1)=\beta  e f (e+f), \qquad
 \alpha =\beta ,  \nonumber\\
 &\alpha +\alpha  t (t+4)=\beta  \left(e^2+4 e f+f^2\right)+1,\qquad 
 \alpha  (t+1)=\beta  (e+f).
\end{align}

The unique solution (up to the trivial relabeling $e \leftrightarrow f$) fixes the normalization of $\zeta(z)$ and the position of the other marked points 
\begin{equation}
    \alpha = \beta = -\frac{9}{16} \qquad  t = - \frac{1}{3}\,,
\end{equation} 
along with the pre-images of $\zeta=1$ on the worldsheet
\begin{equation}
    e = \frac{1}{3} -\frac{2i}{3} \qquad f = \frac{1}{3} +\frac{2i}{3} \,.
\end{equation}

The values for $e$ and $f$ indeed agree with our computation of the Strebel zeroes in Sec. \ref{sec:ExplicitWS}. The intermediate map $\zeta(z)$ thus takes the explicit form

\be
\zeta(z) = -\frac{9}{16} \frac{(z-1)^2 (z+1/3)^2}{(z-1/3)^2}\, .
\ee 

We now simply need to compose this answer with the map (\ref{compose}) to get the precise covering map dual to our Feynman diagram
\begin{equation} \label{eq:explicitBelyi}
    X(z)_{FD} = \frac{16w_2(z-\inv{3})^2 - 9w_1(z-1)^2(z+\inv{3})^2}{16(z-\inv{3})^2 - 9(z-1)^2(z+\inv{3})^2} 
\end{equation}

Of course, at higher genus and more marked points, the above manipulations quickly become complicated. That said, it is anyhow rare in string theory that we can solve so explicitly for the string embedding into the target space\footnote{A notable counterexample are the many solutions of the Gross-Mende saddle point equations found in \cite{Gross:1987kza, Gross:1987ar}}. This simple example however illustrates the power of a single Feynman diagram to reconstruct the entire string configuration uniquely.

\begin{figure}[H]
    \centering
\includegraphics[scale=0.75]{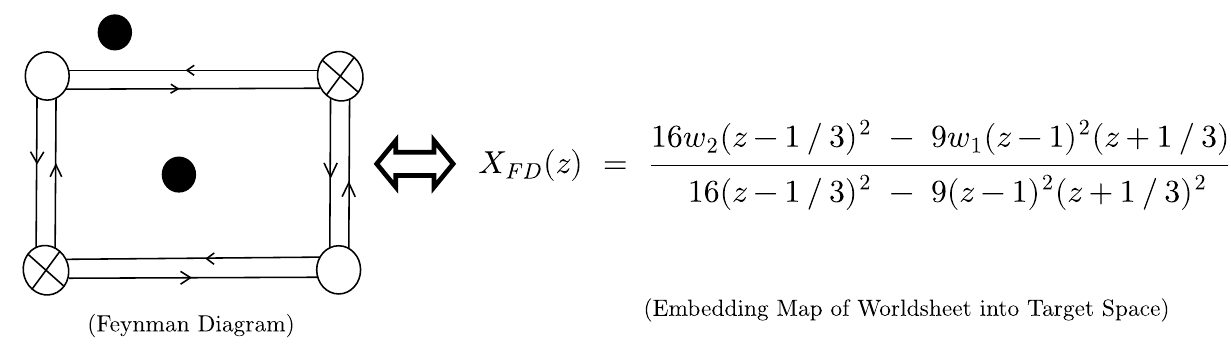}

    \caption{\textbf{The String Embedding Dual to our Feynman Diagram } We can explicitly reconstruct the Belyi map, from the worldsheet into the target space, purely from the Feynman diagram data. This function is unique up to an overall $SL(2,\mathbb{C})$ action, which corresponds to picking the location of three of the marked points on the worldsheet. $(X(z)-w_2)$ has two double-roots at $z=1$ and $z=-1/3$ because the $M$-vertices have valency two. Similarly, $(X(z)-w_1)$ has two double roots at $z=1/3$ and $z=\infty$, corresponding to the two $K$-vertices of valency two. These values agree with the location of the marked points on the worldsheet we reconstructed in Sec. \ref{sec:ExplicitWS}. One can similarly check the the solutions of $X=(w_1+w_2)/2$ give the zeroes of that explicit Strebel differential (Eq.(\ref{eq:ExplicitStrebel})).}
    \label{fig:FDexplicitMap}
\end{figure}

\subsection{Why The Algorithm Works: from Wick Contractions to Permutations}\label{subsec: Wicktoperm}

We would like to briefly explain why the algorithm of Sec. \ref{subsec:deriveBelyi} works. In other words, we try to answer why we should expect a direct map between Feynman diagrams and permutation triples. The basic idea will be to manifestly reformulate the sum over Wick contractions (which graphically becomes the Feynman diagram expansion) as a sum over permutations\cite{deMelloKoch:2010hav}. For those readers primarily interested in the application of the two-matrix model to gauge/string duality, this subsection can certainly be skipped on a first reading.

In the algorithm presented in Sec. \ref{subsec:deriveBelyi}, the first two permutations $\sigma_{K}$ and $\sigma_{M}$ were read off from the external vertices of the diagram. This means they are already completely specified by the choice of correlator. To warm up, let us consider the simplest example: the computation of $\langle \Tr (M^2) \Tr (K^2) \rangle$. In this case, we can write (repeated indices are summed over):

\begin{equation}
    \Tr(M^2) \Tr(K^2) =  M^{a_{2}}_{a_{1}}M_{a_{2}}^{a_{1}} \ K^{b_{2}}_{b_{1}} K_{b_{2}}^{b_{1}} =  M^{a_{\sigma_{M}(1)}}_{a_{1}}M_{a_{2}}^{a_{\sigma_{M}(2)}} K^{b_{\sigma_{K}(1)}}_{b_{1}} K_{b_{2}}^{b_{\sigma_{K}(2)}}
\end{equation}

where $\sigma_{M}(1)=2$, $\sigma_{M}(2)=1$ and similarly for $\sigma_{K}$. Due the cyclicity of the trace, only the conjugacy class of $\sigma_{K}$ and $\sigma_{M}$ are truly determined. Let us now turn to the more general case. For the correlator 

\begin{equation}
    \Braket{\prod_{j=1}^{V_M} \Tr\left( M^{n_j} \right) \prod_{i=1}^{V_K} \Tr\left( K^{l_i}\right) }_{c,h}
\end{equation}

not to vanish, there must be a many $M-$ and $K$-matrix elements appearing: $\sum_{j=1}^{V_{M}}n_{j} = \sum_{i=1}^{V_{K}}l_{i} =E$. This is also the number of edges $E$ in any Feynman diagram contributing to this observable. We can now define two permutations $\sigma_{K},\sigma_{M} \in S_{E}$ encoding which single trace operators we are looking at. We can write 

\begin{equation}
   \prod_{j=1}^{V_M} \Tr\left( M^{n_j} \right) =  M^{i_1}_{i_{\sigma_{M}(1)}}...M^{i_E}_{i_{\sigma_{M}(E)}}
\end{equation}

where $\sigma_{M}$ has the cycle structure $(n_1)(n_2)\dots (n_{V_M})$, and similarly for $\sigma_{K}$. To understand the appearance of the third permutation associated to the face, we first compute the correlator using Wick's theorem: 

\begin{align}\label{eq_perm_1}
	 \Braket{\prod_{j=1}^{V_M} \Tr\left( M^{n_j} \right) \prod_{i=1}^{V_K} \Tr\left( K^{l_i}\right) } &=\langle M^{i_1}_{i_{\sigma_{M}(1)}}...M^{i_E}_{i_{\sigma_{M}(E)}}K^{j_1}_{j_{\sigma_K(1)}}...K^{j_E}_{j_{\sigma_K(E)}}\rangle\\
 &= \langle M^{i_1}_{k_1}...M^{i_E}_{k_E}K^{j_1}_{l_1}...K^{j_E}_{l_E}\rangle \delta^{k_1}_{i_{\sigma_{M}(1)}}...\delta^{k_E}_{i_{\sigma_{M}(E)}}\delta^{l_1}_{j_{\sigma_K(1)}}...\delta^{l_E}_{j_{\sigma_K(E)}}\\
 &= \left( \frac{g}{N} \right)^{E} \left(\sum_{\gamma\in S_E}\delta^{i_1}_{l_{\gamma(1)}}...\delta^{i_E}_{l_{\gamma(E)}}\delta^{j_{\gamma(1)}}_{k_1}...\delta^{j_{\gamma(E)}}_{k_E} \right) \delta^{k_1}_{i_{\sigma_{M}(1)}}...\delta^{k_E}_{i_{\sigma_{M}(E)}}\delta^{l_1}_{j_{\sigma_K(1)}}...\delta^{l_E}_{j_{\sigma_K(E)}}\nonumber\\
&=\left( \frac{g}{N} \right)^{E}\sum_{\gamma\in S_E}\delta^{i_1}_{l_{\gamma(1)}}...\delta^{i_E}_{l_{\gamma(E)}}\delta^{j_{\gamma(1)}}_{i_{\sigma_{M}(1)}}...\delta^{j_{\gamma(E)}}_{i_{\sigma_{M}(E)}}\delta^{l_1}_{j_{\sigma_K(1)}}...\delta^{l_E}_{j_{\sigma_K(E)}}
 \end{align}

 The sum over $\gamma \in S_{E}$ is simply the sum over all Wick pairings between the $M$'s and $K$'s. It is not yet $\sigma_{f}$.

 Using the Wick's theorem \eqref{eq:prop} now, we get the following by doing some simple manipulations
 \begin{align}
	\left\langle \prod_{i=1}^{V_K}\Tr(K^{n_i})\prod_{j=1}^{V_M}\Tr(M^{l_j})\right\rangle=& \left( \frac{g}{N} \right)^{E}\sum_{I,J,K,L}\sum_{\gamma\in S_E}\delta^{i_1}_{l_{\gamma(1)}}...\delta^{i_E}_{l_{\gamma(E)}}\delta^{j_{\gamma(1)}}_{k_1}...\delta^{j_{\gamma(E)}}_{k_E}\delta^{k_1}_{i_{\sigma_{M}(1)}}...\delta^{k_E}_{i_{\sigma_{M}(E)}}\delta^{l_1}_{j_{\sigma_K(1)}}...\delta^{l_E}_{j_{\sigma_K(E)}}\nonumber\\
 =& \left( \frac{g}{N} \right)^{E}\sum_{I,J,L}\sum_{\gamma\in S_E}\delta^{i_1}_{l_{\gamma(1)}}...\delta^{i_E}_{l_{\gamma(E)}}\delta^{j_{\gamma(1)}}_{i_{\sigma_{M}(1)}}...\delta^{j_{\gamma(E)}}_{i_{\sigma_{M}(E)}}\delta^{l_1}_{j_{\sigma_K(1)}}...\delta^{l_E}_{j_{\sigma_K(E)}}\nonumber\\
    =& \left( \frac{g}{N} \right)^{E}\sum_{I,J,L}\sum_{\gamma\in S_E}\delta^{i_1}_{l_{\gamma(1)}}...\delta^{i_E}_{l_{\gamma(E)}}\delta^{l_{\gamma(1)}}_{j_{\sigma_K \gamma(1)}}...\delta^{l_{\gamma(E)}}_{j_{\sigma_K\gamma(E)}}\delta^{j_{\gamma(1)}}_{i_{\sigma_{M}(1)}}...\delta^{j_{\gamma(E)}}_{i_{\sigma_{M}(E)}}\nonumber\\
    =& \left( \frac{g}{N} \right)^{E}\sum_{I,J}\sum_{\gamma\in S_E} \delta^{i_1}_{j_{\s_K\gamma(1)}}...\delta^{i_E}_{j_{\s_K\gamma(E)}}\delta^{j_{\gamma(1)}}_{i_{\sigma_{M}(1)}}...\delta^{j_{\gamma(E)}}_{i_{\sigma_{M}(E)}}\nonumber\\
    =& \left( \frac{g}{N} \right)^{E}\sum_{I,J}\sum_{\gamma\in S_E}\delta^{i_1}_{j_{\s_K\gamma(1)}}...\delta^{i_E}_{j_{\s_K\gamma(E)}}\delta^{j_{\s_K\gamma(1)}}_{i_{\sigma_{M}(\gamma^{-1}\s_K\gamma)(1)}}...\delta^{j_{\s_K\gamma(E)}}_{i_{\sigma_{M}(\gamma^{-1}\s_K\gamma)(E)}}\nonumber\\
    =& \left( \frac{g}{N} \right)^{E} \sum_{I} \sum_{\gamma\in S_E} \de^{i_1}_{i_{\sigma_{M}\gamma^{-1}\s_K\gamma(1)}} \dots \de^{i_{E}}_{i_{\sigma_{M}\gamma^{-1}\s_K\gamma(E)}}\nonumber\\
	=&\left( \frac{g}{N} \right)^{E}\sum_{\gamma\in S_E}\text{tr}_E(\sigma_{M}\gamma^{-1}\sigma_K\gamma)\label{eq: sum_gamma}
\end{align}
where the trace in the last line is defined as the product of delta functions that appear in the previous line.

We can now sum over the conjugacy class $[\sigma_K]$ of $\sigma_K$ on both sides of \eqref{eq: sum_gamma}. Since the left-hand side depends only on the cycle structure i.e. the conjugacy class $[\sigma_K]$ and not on the particular representative, we will just get the size of the conjugacy class $\sigma_K$, $|[\sigma_K]|$ times the summand upon summation,
\begin{align}\label{eq: perm_main}
	\left\langle \prod_{i=1}^{V_K}\Tr(K^{n_i})\prod_{j=1}^{V_M}\Tr(M^{l_j})\right\rangle=&\frac{g^EN^{-E}}{|[\sigma_K]|}\sum_{\sigma_K\in [\sigma_K]}\sum_{\gamma\in S_E}\text{tr}_E(\sigma_{M}\gamma^{-1}\sigma_K\gamma)=\frac{g^EE!}{|[\sigma_K]|}\sum_{\sigma_K\in [\sigma_K]}tr_E(\sigma_K\sigma_{M})\nonumber\\
	=&\frac{g^EN^{-E}E!}{|[\sigma_K]|}\sum_{\sigma_K\in [\sigma_K]}N^{C_{\sigma_K\sigma_{M}}}
\end{align}
where we have used $\text{tr}_E(\sigma)=N^{C_{\sigma}}$ with $C_\sigma$ being the number of cycles in permutation $\sigma \in S_E$, this is clear from the definition of $\text{tr}_E$ in terms of product of delta functions. We now introduce a third permutation $\sigma_f\in S_E$ and write it in a form similar to that we see in the Gaussian one matrix model,
\begin{align}\label{eq:permresult_main}
     \left\langle \prod_{j=1}^{V_M}\text{Tr}(M^{l_j})\prod_{i=1}^{V_K}\text{Tr}(K^{k_i})\right\rangle&=\frac{g^EE!}{|[\sigma_K]|}\sum_{\substack{\sigma_K\in [\sigma_K],\\
     \sigma_f\in S_E}}N^{C_{\sigma_f}-E}\delta(\sigma_M\sigma_f\sigma_{K})
\end{align}
where we have used $C_{\sigma_K\sigma_{M}}=C_{\sigma_f^{-1}}=C_{\sigma_f}$(permutations and their inverses lie in same conjugacy class and hence, have same cycle structure). The delta function,  $\delta(\sigma_M\sigma_f\sigma_{K})$, contributes $1$ to the sum only whenever $\sigma_M\sigma_f\sigma_{K}=\mathbb{I}$, and is otherwise zero. It is the permutation $\s_f$ which turns out to be the permutation corresponding to the faces of the Feynman diagram. We have included a simple but nontrivial example in Appendix \ref{sec:FDperm}. This tells that three permutations satisfying the $\sigma_M\sigma_f\sigma_K=\mathbb{I}$ are indeed enough to capture the information of correlators for our two matrix models. The form of the quadratic term in the matrix model action
 ($\sim\Tr(KM)$) we chose to work with is crucial for this description in terms of these three permutations.
\pagebreak
\section{Tying Together the Worldsheet \& Target Space Perspectives} \label{sec:tyingpersps}

Up until now, we have succeeded in assigning to each diagram in the Feynman expansion of an arbitrary correlator both a particular worldsheet along with its embedding into the target space. This is the sense in which a single diagram encodes a full "string configuration". However, the language used in reconstructing the Riemann surface via Strebel differentials seems a priori quite distinct from the language of branched covering maps, so that the two perspectives might appear quite disjoint. The purpose of this section is to reconcile these two pictures.

\subsection{Localization to Discrete Points on $\mathcal{M}_{g,n}$ via Belyi's Theorem} \label{sec:StringLocalization}
First, we want to understand why the path integral of the dual closed string theory should only receive contributions from arithmetic Riemann surfaces. Indeed, in reconstructing the worldsheet from the Feynman diagram, we landed on Strebel graphs with integer length assignments to each edge. The points on $\mathcal{M}_{g,n}$ labeled by integer Strebel graphs are exactly the arithmetic Riemann surfaces. Our construction starting from the Feynman diagrams therefore predicts that the closed string integrand on moduli space should localize to a sum of delta-functions supported on a subset of these discrete points. This is how the usual continuous integral over moduli space of the string would then replicate the discrete (and finite) sum over Feynman diagrams. This is a very unusual thing for a string theory to do. In \cite{razamat2010matrices,razamatGauss}, where the integer length assignment was first proposed, this localization was \textit{assumed} for the dual string. A concrete mechanism was then laid out by one of the authors in \cite{gopakumar2011simplest}.

The argument relies on a mathematical theorem, proved by G.V. Belyi in 1979\cite{zbMATH03668723}. He showed that not all Riemann surfaces admit Belyi maps. In other words, not all worldsheets can wrap the target space sphere in the precise way laid out in Sec. \ref{sec:targetpersp}. The only worldsheets which do are precisely the arithmetic Riemann surfaces! This means that one way for the integrand on moduli space to pick out these special discrete points is for the string theory to localize to holomorphic covering maps of the sphere, branched over exactly three points (which we have taken to lie at $X=w_1,w_2 \  \& \thinspace (w_1+w_2)/2$. It is then very satisfying that all the embedding maps we derive from the Feynman diagrams do satisfy these properties. 

Of course, the question remains whether we can find an explicit worldsheet theory which localizes to Belyi maps. For example, in the case of $2d$ Yang-Mills, Gross and Taylor had long ago recast its correlators in terms of covering maps. However, it remains until today an open problem to find a string path integral which reduces to such a sum over branched covers. For exciting recent progress in that direction, see \cite{Aharony:2023tam,KomatsuWIP,GaberdielWIP}. Below, we briefly discuss the concrete closed string dual to our two-matrix integral. The details of this theory will appear in \cite{DSDIII}.

\subsection{Overview of the Proposed Closed String Dual}\label{sec:Amodel}

The A-model topological string with target $\mathbb{P}^1$ has long been known to localize to holomorphic covering maps of the sphere. As such, it was explored as a potential candidate dual the Gaussian matrix model in \cite{gopakumar2013correlators}. However, the string path integral generically admit maps with arbitrarily complicated branching, which are not fixed by the precise form of the vertex operator correlators. Fortunately, the derivation of the $AdS_{3}/CFT_{2}$ correspondence at the symmetric orbifold point recently uncovered a string theory where the branching structure could also be specified. The authors of \cite{eberhardt2019worldsheet} showed that correlation functions of certain operators in a particular $SL(2,\mathbb{R})$ WZW-model were computed by a sum over branched covering maps of the sphere. In that case, the target sphere was understood as the conformal boundary of $AdS_{3}$. This relied crucially on the level of this supersymmetric WZW being minimal, $\hat{k}=1$. This translates to the tensionless limit of the closed string. For each covering map appearing in the computation of the correlators, the branch points on the worldsheet lied exactly at the insertions of the vertex operators. The order of the branching was determined by the choice of operators. To be precise, the vertex operators $\mathcal{V}_{j=1/2}^{\omega}(X,z)$ in the $\omega$-spectrally flowed $j=1/2$ continuous representations of $SL(2,\mathbb{R})$ created a branching of order $\omega$ at $z$ on the worldsheet. $z$ gets mapped to $X$ on the target space sphere under the covering map. Such maps are of course more general than Belyi maps. They can be branched over more than three points. 

The string theory dual to our two-matrix model is a topological cousin of this $AdS_{3}$ string. It consists of a topologically twisted (A-twist) $\mathcal{N}=2$ supersymmetric Kazama-Suzuki coset $SL(2,\mathbb{R})/U(1)$, coupled to topological gravity on the worldsheet. The (susy) level of the coset takes on its critical value, $\hat{k}=1$, so that the matter central charge $\hat{c}=9$ is the same as that of a Calabi-Yau. Mukhi and Vafa \cite{MukhiVafa} had found this precise theory as a topological string reformulation of the $c=1$ string at self-dual radius. This $c=1$ string, in turn, is defined in terms of a $c=1$ compact boson (compactified at the self-dual radius, hence the name), a $c=25$ Liouville theory and the standard $c=-26$ bc-ghost system. In \cite{DSDI}, we had already found an operator dictionary between single trace operators $\Tr(K^n)$ ($\Tr(M^n)$) and the standard $c=1$ tachyon momentum operators $T_{+n}$ ($T_{-n}$). This was achieved via an exact mapping of the matrix integral to the generating function of $c=1$ correlators (which we encounter again in Sec. \ref{sec:IMmodel}). Unfortunately, the $c=1$ theory does not immediately present itself in terms of a sum over covering maps.\footnote{There does seem to be an intriguing direct connection between, on one side, the Feynman diagram expansion of matrix model correlators, and, on the other hand, the screening charge expansion of $c=1$ correlators. Work is in progress to make this precise.} Instead, we can use the duality relating the $c=1$ string to the coset and translate this dictionary there. Following the work of \cite{ashok2006topological}, updating the construction of Mukhi \& Vafa to include the effects of $SL(2,\mathbb{R})$ spectral flow, we find that traces built out of $K$'s or $M$'s can be written in terms of simple vertex operators on the worldsheet.
 
In \cite{DSDIII}, it will be shown explicitly, how the string theory path integral computation of correlators built out of products of these vertex operators localizes to a sum over Belyi maps. The insertions of $\Tr(K^{n})$ create the order $n$ branching over $X=w_1$, and similarly for $M$, just as as we found in Sec. \ref{sec:targetpersp}. The branching over the third point, dual to the faces of the Feynman diagrams, is more subtle, and we postpone that analysis to \cite{DSDIII}. 

At any rate, the main takeaway is that the physical correlators of the twisted $SL(2,\mathbb{R})/U(1)$ coset can be recast as a sum over Belyi maps, as predicted from the target space perspective on the Feynman diagrams. Via Belyi's theorem, these maps only exists for the points on moduli space parametrized by integer Strebel graphs, so that only those worldsheets which are dual to the Feynman diagrams, contribute to the path integral.

\subsection{Back to the Strips: Building Up a Belyi Map} \label{sec:backtostrips}
The second main point we need to explain is why the same Feynman diagram can be viewed both as (graph dual to) the Strebel graph of the worldsheet, but also as a graphical encoding of a Belyi map. We will do so by returning to the strips we used as basic building blocks to assemble the closed string worldsheet. Here, we will first define a simple embedding map for each strip, $X_{\text{s}}(z)$, where each strip wraps the target space $\mathbb{P}^1$ exactly once. We then show how the composition of the gluing procedures of Sec \ref{sec:howto} along with the simple map $X_{\text{s}}(z)$ rise to the full branched covering map. In some sense, the embedding map of each strip is the "bit" of the Belyi map, while the strip itself is the worldsheet of a "string bit" making up each closed string. This explains why the Strebel graph itself, which encodes the complex structure of the worldsheet, is at the same time the  "dessin d'enfant" parametrizing the covering map. 

We begin by going back to each edge of the Feynman diagrams, which corresponds to one of the strips making up the worldsheet. Each infinite strip is of unit width, i.e. in the local coordinates $z_{j}$ we defined on each strip, $\Re(z_{j})\in [0,1]$. Recall that the $M$- and $K$-vertices at the endpoint of each edge lie at $z_{j}=i \infty$ and $z_{j}=-i\infty$ respectively. The two faces of any edge borders correspond to $z=0$ and $z=1$. We know from the Belyi maps that the $M$-vertices are pre-images of $X=w_2$, the $K$-vertices preimages of $X=w_1$ and the faces are mapped to $X=(w_1+w_2)/2$. The map $X_{\text{strip}}(z_j)$ therefore needs to satisfy

\begin{equation}
    X_{\text{s}}(i\infty) = w_2 \qquad   X_{\text{s}}(-i\infty)=w_1 \qquad   X_{\text{s}}(0)=X_{\text{s}}(1)=\frac{w_1+w_2}{2}
\end{equation}

A simple map which maps each point $z_{j}$ (with $\Re(z_{j}) \in [0,1)$ and $|\Im(z_{j})|<\infty$) on the strip to the Riemann sphere exactly once is

\begin{equation} \label{eq:Xstrip}
     X_{\text{s}}(z_{j})=\frac{w_{2} e^{-2\pi i z_{j}}+w_1}{e^{-2\pi i z_{j}}+1}
\end{equation}

\begin{figure}[ht!]
    \centering
\includegraphics[scale=0.65]{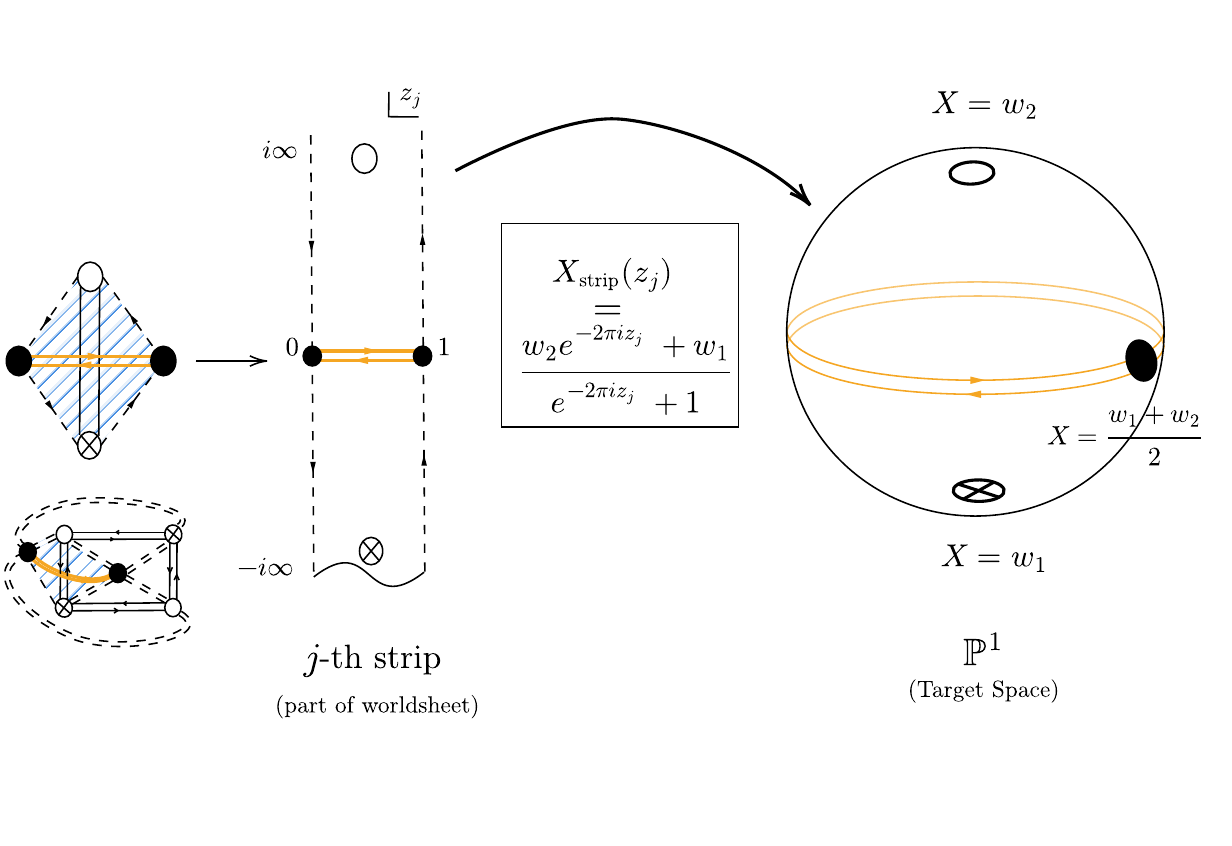}

    \caption{\textbf{"Belyi Map Bits" } To understand how the reconstruction of the worldsheet and the covering map are related, we study a single edge of a Feynman diagram. We have seen how each edge is associated to one strip making up the worldsheet, which we interpreted as the worldsheet of one string bit. From the target space perspective, each edge gives one sheet of the covering map. The basic idea is to build up the full Belyi map from the simple map $X_{\text{strip}}(z)$, embedding each strip into the target $\mathbb{P}^1$. On the $j$-th strip, $X_{\text{strip}}(z_{j})$ sends the $M$-vertex, at $z_{j}=i\infty$, to the insertion point of one of the gauge theory operators at $X=w_2$ and the $K$-vertex, at $z_{j}=-i\infty $, to $X=w_1$. The two black dots (which would lie on the faces of the Feynman diagram) get mapped to the midpoint $X=(w_1+w_2)/2$. Each strip therefore wraps the target space exactly once.}
    \label{fig:MapStrip}
\end{figure}

In other words, $X_{\text{s}}(z_{j})$ wraps the strips around the Riemann sphere exactly once, while sending its asymptotic ends to the points $w_2$ and $w_1$, respectively - see Fig. \ref{fig:MapStrip}. This explains why the reconstruction algorithm assigned to each edge of the Feynman diagram, one sheet of the covering map.

We now want to explicitly show how composing the three strip gluing rules, presented  Sec. \ref{sec:howto}, with the embedding map for each strip, results in the full Belyi map.

\subsubsection*{Gluing along (Homotopic) Edges $\rightarrow$ $n$-width strips wrap the $\mathbb{P}^1$ $n$ times}
Recall that in the Strebel construction of the worldsheet, the (integer) Strebel lengths emerged from bundling together homotopic edges. This gave an effective edge with a length assignment equal to the number of homotopic edges collapsed together. This reduced the original Feynman diagram to its skeleton graph. In deriving the Belyi map in the Sec. \ref{subsec:deriveBelyi}, we did not need to resort to the skeleton graph. 

As summarized in Fig. \ref{fig:GluingAtEdges}, the gluing rules for the local coordinates on homotopic edges are quite simple: $z_{j+1}=z_{j}-1$. In other words, we simply adjoin the $n$ strips side by side, resulting in an effective strip of width $n$. The map $X(z)$ for the effective strip with $\Re(z) \in [0,n]$ is a degree $n$ cover of the Riemann sphere. There are $n+1$ pre-images of $X=(w_1+w_2)/2$ at $z=0,1,2,...,n-1,n$ for $\Re(z) \in [0,n]$. The $n-1$ preimages in the middle ($z= 1, 2, \dots, n-1$) correspond to the face centers of two-sided faces bordered by the $n$ homotopic edges. There is no branching around these points, which is why they give rise to single cycles in the Belyi map for the permutation $\sigma_{f}$. To properly understand the behavior of the map near $z_j= \pm i\infty$, i.e. where the vertices live on the strip, we will need to go back to the gluing rules near vertices.

\subsubsection*{Glueing at Vertices $\rightarrow$ Branching above $X=w_1$  and $X=w_2 \in \mathbb{P}^1$}

The reconstruction algorithm assigns a cycle of length $l$ to the permutation $\sigma_{K}$ for each $K$-vertex of valency $l$. This means that the vertex operator dual to $\Tr(K^l)$ must somehow create a branching of order $l$ at the insertion point. An analogous statement holds for $\Tr(M^l)$. Let us see how this comes about. 

We know that the $l$ edges emanating from a vertex correspond to a particular gluing of $l$ strips on the worldsheet. Near the $k$-th marked point dual to a $M$-vertex, we constructed in Sec. \ref{sec:howto} local coordinates $u_{k}$. The transition function to the coordinates at each strip meeting at the vertex read

\begin{equation}
    u_{k}(z_{j}) = e^{2\pi i \frac{j}{l}}e^{2 \pi i \frac{z_{j}}{l}}, \quad j = 0, \dots, l-1
\end{equation}

Using the expression for $X_{\text{s}}(z_{j})$ in Eq.(\ref{eq:Xstrip}), we can find a local expression for the strip embedding map $X_s$ in terms of the $u_{k}$ coordinates:

\begin{equation}
   X_{\text{s}}(u_{k})= \frac{w_{2}u_{k}^{-l}+w_{1}}{u_{k}^{-l}+1}
\end{equation}

As we approach the $k$-th marked point where the vertex operator dual to $\Tr(M^{l})$ is inserted, i.e. as $u_{k} \rightarrow 0$, we see that the map simplifies to

\begin{equation}
    X_{\text{s}} = w_{2} + (w_1-w_2)u_{k}^{l} + \mathcal{O}(u_{k}^{2l})
\end{equation}

We therefore see that the marked point at $u_{k}=0$ indeed corresponds to $X=w_2$. Furthermore, a point in the neighborhood of $X=w_2$ has exactly $l$ pre-images near $u_k=0$, matching the desired local branching structure of the Belyi map. The $l$-elements in the particular cycle of $\sigma_{M}$ associated to this vertex are the $l$-strips meeting at the marked point. Since each strip covers the target space sphere exactly once, this explains why we can read off the cycles from the edges meeting a given vertex of the Feynman diagram. 

Recall that in local coordinates $v_{k}$ near a marked point on the worldsheet dual to a $K$-vertex, the transition function gets slightly modified
\begin{equation}
    v_{k}(z_{j}) = e^{-2\pi i \frac{j}{l}}e^{-2 \pi i \frac{z_{j}}{l}}
\end{equation}
so that, by using the explicit form of $X_{\text{s}}(z_{j})$, we instead arrive at 
\begin{equation}
   X_{\text{s}}(v_{k})= \frac{w_{2}v_{k}^{l}+w_{1}}{v_{k}^{l}+1}
\end{equation}.
This means that near the $k$-th marked point where a vertex operator dual to $\Tr(K^{l})$ is inserted, the map locally looks like
\begin{equation}
    X_{\text{s}}(v_{k}) = w_{1} + (w_2-w_1)v_{k}^{l} + \mathcal{O}(v_{k}^{2l})
\end{equation}

This shows that all the $K$-vertices which lie at $v_{k}=0$ on the worldsheet get mapped to the other gauge theory insertion point at $X=w_1$, with a similar $l$-fold branching. This reconciles why the vertex operator $\Tr(K^{l})$ both creates an asymptotic closed string made out of $l$- string bits (whose worldsheets are the $l$ strips), as well as having the effect of creating an $l$-fold branching above $X=w_1$ from a Belyi map perspective.

\subsubsection*{Gluing at Faces $\rightarrow$ Branching above $X=\frac{w_1+w_2}{2}$}

The final gluing rule will dictate the local branching near the midpoint $X=(w_1+w_2)/2$. Recall that the preimages of $X=(w_1+w_2)/2$ are the zeroes of the Strebel differential on the worldsheet, which correspond to the faces of the Feynman diagram.

Note that we have two sets of expressions for the gluing map near the face center (\ref{faceglueing}), and we should make sure that the covering map $X$ is well-behaved, in particular is independent of the different patches for smoothness. Say we have a face with $2m$ edges surrounding it. Then the gluing rules for the individual strips are given by (\ref{faceglueing}). Now let's see what the ramification property of the covering map looks like near this face center. \\ 
Near the face centres, the covering map takes the following form in terms of the local coordinate $\w$
\be 
\begin{split}
    X(\w) = X_{\text{s}}(z_{2j}(\w)) &= \frac{w_2\exp({-2\pi i z_{2j}})+w_1}{\exp{(-2\pi i z_{2j})} + 1} =  \frac{w_2\exp({-2\pi i (e^{(2j-1)i\pi}\w^m)})+w_1}{\exp({-2\pi i (e^{(2j-1)i\pi}\w^m)})+1}\\
    & \sim \frac{w_1 + w_2}{2} - ({w_1-w_2})\pi i  \w^m,
\end{split}
\ee 
and,
\be 
\begin{split}
    X(\w) = X_{\text{s}}(z_{2j-1}(\w)) &=\frac{w_2\exp({-2\pi i z_{2j-1}})+w_1}{\exp({-2\pi i z_{2j-1}}) + 1}= \frac{w_2\exp(-2\pi i (1+ e^{(2j-3)i\pi}\w^m))+w_1}{\exp(-2\pi i (1+ e^{(2j-3)i\pi}\w^m)) + 1}\\
    &\sim \frac{w_1 + w_2}{2} - ({w_1-w_2})\pi i  \w^m.
\end{split}
\ee 
As we can see, the covering map $X$ maps the face center to the target space point $X=(w_1 + w_2)/2$, and near a $2m$ sided face center, it is branched with order $m$, i.e. a point near $X=(w_1 + w_2)/2$ has $m$ pre-images near that face-center of the worldsheet. Looking at all the faces, we can see that the permutation governing the ramification property of the covering map near $X=(w_1 + w_2)/2$ will be given by,  
\be 
\s_f = (m_1)\dots(m_f)
\ee 
where $f$ is the number of faces in the Feynman graph, and the numbers $m_1, \dots, m_f$ signify that the $i$th face has $2m_i$ edges surrounding it.\\
We can verify that the covering map $X(z)$ has no other branch point, and since the target space is $\mathbb{P}^1$, the permutations will by construction satisfy 
\be
\s_M \s_f \s_K = 1
\ee 
which is the desired property of the Belyi map.

\subsection{Strebel Metric as Pull-back of Target Space Kähler Form} \label{sec:StrebelAsPullback}

There is another way to make manifest the relation between the Strebel differential on the worldsheet and the embedding map into the target space. This will further clarify why the same Feynman diagram can encode both these objects. The starting point is to notice that we can use the embedding map $X_{\text{s}}(z)=\frac{w_2e^{-2\pi i z}+w_1}{e^{-2\pi i z}+1}$ to relate the Strebel differential on each strip, $dz^2$, to a quadratic differential on the target Riemann sphere

\begin{equation}
     \begin{split}
         &\frac{d X_{\text{s}}(z)}{dz} = -2\pi i \frac{(X_{\text{s}}-w_1)(X_{\text{s}}-w_2)}{w_1-w_2}
         \implies \phi_{S}(z) dz^2 |_{\text{strip}} =dz^2 = -\frac{1}{4 \pi^2} \frac{(w_1-w_2)^2}{(X_{\text{s}}-w_1)^2 (X_{\text{s}}-w_2)^2 }dX_{\text{s}}^2
     \end{split}
\end{equation}

Since the embedding map is holomorphic, both the left and right hand sides of the second equality indeed transform as quadratic differentials under a (holomorphic) change of coordinates $z \rightarrow w(z)$. In Sec. \ref{sec:howto}, we explicitly constructed a complete set of coordinate charts, for all worldsheets, with holomorphic transition functions to the local coordinate on each strip. This means that we can in fact promote the above relation to an equality over the entirety of the Riemann surface, not just on an individual strip \footnote{This is a familiar proof strategy in say general relativity, where, to establish the equality between two covariant objects, one can go to a local coordinate frame in which the two expressions simplify greatly, and the equality becomes manifest. Their covariance guarantees that the equality holds in all coordinate systems. }. This establishes a general relation between the target space embedding map and the Strebel differential on the worldsheet

\begin{equation} \label{eq:StrebelAsPullback}
    \phi_{\text{S}}(z) dz^2 =  -\frac{1}{4 \pi^2} \frac{(w_1-w_2)^2}{(X-w_1)^2 (X-w_2)^2 }dX^2
 \end{equation}

Reassuringly, one can check this equality directly for the explicit Strebel differential and Belyi map dual to our prototypical Feynman diagram, computed in Secs. \ref{sec:ExplicitWS} and \ref{sec:ExplicitBelyi} (cf. Eq.(\ref{eq:ExplicitStrebel}) and Eq.(\ref{eq:explicitBelyi})).  

Eq.(\ref{eq:StrebelAsPullback}) implies that the volume form on the worldsheet, in Strebel gauge, is simply the pull-back of the following Kähler form $\Omega$ on the target $\mathbb{CP}^{1}$:

\begin{equation} \label{eq:Kahlerformtarget}
    |\phi_{\text{S}}(z)| dz \wedge d\bar{z} = \frac{1}{4\pi^2}\frac{(|w_1-w_2|^2)}{|X-w_1|^2 |X-w_2|^2 } dX \wedge d\bar{X} = X^{*} \left( \Omega_{\text{Target}}
 \right) \end{equation}

The precise form of $\Omega$ defines a target space Kähler potential 

\begin{equation}
    \mathcal{K}_{\text{Target}}(X,\bar{X}) = \frac{1}{4\pi^2 } \log\left( \frac{X-w_1}{X-w_2}\right) \log\left( \frac{\bar{X}-\bar{w_1}}{\bar{X}-\bar{w_2}}\right)
\end{equation}.

\subsection{Action of String reproduces Weight of Dual Feynman Diagram} \label{sec:weight=action}

The purpose of this subsection is to demonstrate that the (regulated) action of the closed string dual to a Feynman diagram reproduces the weight of that Feynman diagram. For the simple $\mathcal{N}=4$ correlators we are considering, the weight of each Feynman diagram is simply given by the product of free field propagators, one for each edge\footnote{We have, so far, only successfully matched the $w$-dependence of the propagator. There might be an interesting fermionic origin to the $(Y_1-Y_{2})^2 = 2 Y_{1} \cdot Y_{2}$ prefactor, perhaps by considering maps into $\mathbb{CP}^{1|2}$. We are also ignoring some overall factors of $4 \pi^2$ in this discussion, relative to the $\mathcal{N}=4$ SYM propagator of Sec. \ref{sec:embedding}. \label{fn:y1y2}}:

\begin{equation}
    \text{weight of} \ \mathcal{N}=4 \ \text{Feyn. Diags. for our correlators} = \left( \frac{Y_1 \cdot Y_{2}}{|w_1-w_2|^2} \right) ^{E}
    \end{equation}    
  
On the dual side, each closed string configuration appears in the path integral, weighted by $e^{-S}$. For the A-model topological string dual to our matrix integral, we expect the action of each string configuration to be the integral of the target space Kähler form pulled back to the worldsheet 

\begin{equation} \label{eq:TopstringAction}
    S_{\text{A-model}}[X] = 2\pi \int_{WS} X^{*}(\Omega_{\text{Target}}) = 2\pi \int_{WS} |\phi_{\text{S}}(z)| dz \wedge d\bar{z}
\end{equation}

The final integral over the worldsheet in Eq.(\ref{eq:TopstringAction}) is however just the area of the worldsheet in Strebel gauge! We have thus landed on the equivalent of the Nambu-Goto action, which defines the worldsheet metric as the pullback of the target space metric under the embedding of the string.

Unfortunately, this area is divergent. Physically, we can trace this back to the fact that, in the Strebel gauge, the worldsheet is a collection of half-infinite cylinders of fixed circumference, glued to the Strebel graph (cf. Fig. \ref{fig:StrebelAsCylinders}). The divergence simply reflects that the area of each cylinder is infinite, because it is infinitely long. One way to regulate the area of the worldsheet is to imagine imposing a \textit{IR} cutoff on the length of any one cylinder, which we denote by $L_{c}$. Locally mapping the cylinder to the plane, this is equivalent to excising a small hole around each marked point on the worldsheet. In other words, we introduce a worldsheet \textit{UV} cutoff, $\epsilon_{WS}$.

To precisely relate these various cutoffs and compute the regularized area of the worldsheet, we can exploit the fact that each cylinder consists of strips glued together (cf. Fig. \ref{fig:StrebelAsStrips}). We truncate the length of each semi-infinite strip directly, by restricting the range of the local coordinates $z_{j}$ on the $j$-th strip, $\mathfrak{Im}(z_{j}) \in [ - L_{c},+L_{c} ]$. This is in fact a covariant statement, since it coincides with the proper length of the strip computed using the Strebel metric. We can then use the transition function between the local coordinates $u_{k}$ near the marked point and those on each strip, see Eq.(\ref{eq:epsilonWS}) to relate $L_{c}$ to $\epsilon_{WS}$. If there are $m$ strips meeting at a vertex dual to the $k$-th marked point, i.e. if the $k$-th cylinder has circumference $m$, then the IR cutoff $L_{c}$ translates to UV cutoff in $u_{k}$ 

\begin{equation} \label{eq:epsilonWS}
    \epsilon_{WS} = e^{-2 \pi \frac{L_{c}}{m}}
\end{equation}

While explicitly finding the Strebel metric on each worldsheet for every gauge theory Feynman diagram would be impossible, we can nevertheless compute its regularized area. Since the Strebel construction dissects every Riemann surface into a collection of strips, we can simply add up the area of each strip, computed in the Strebel gauge,  to compute the total area of the worldsheet. 

In reconstructing the worldsheet from the gauge theory Feynman diagrams, we identified every edge of the Feynman diagram with one unit-width strip. Since the Strebel metric is flat on each strip, the area of an individual strip is simply

\begin{equation}
    \text{(Regulated) Area of Single Strip} = \text{height} \times \text{width}= 2 L_{c} \times 1
\end{equation}

The integral of the Strebel area element over the entire surface reconstructed from a Feynman diagram with $E$ edges is therefore 

\begin{equation}
 A_{WS} = \int_{WS} |\phi_{\text{S}}(z)| dz \wedge d\bar{z} = \# \ \text{Strips} \times   \text{(Regulated) Area of Single Strip} = E \times 2L_{c}
\end{equation}

We now want to show how this regularized total area, which is also the Nambu-Goto action evaluated on the embedding map $X(z)$, will reproduce the weight of the Feynman diagram. 

\begin{figure}[ht!]
    \centering
\includegraphics[scale=0.75]{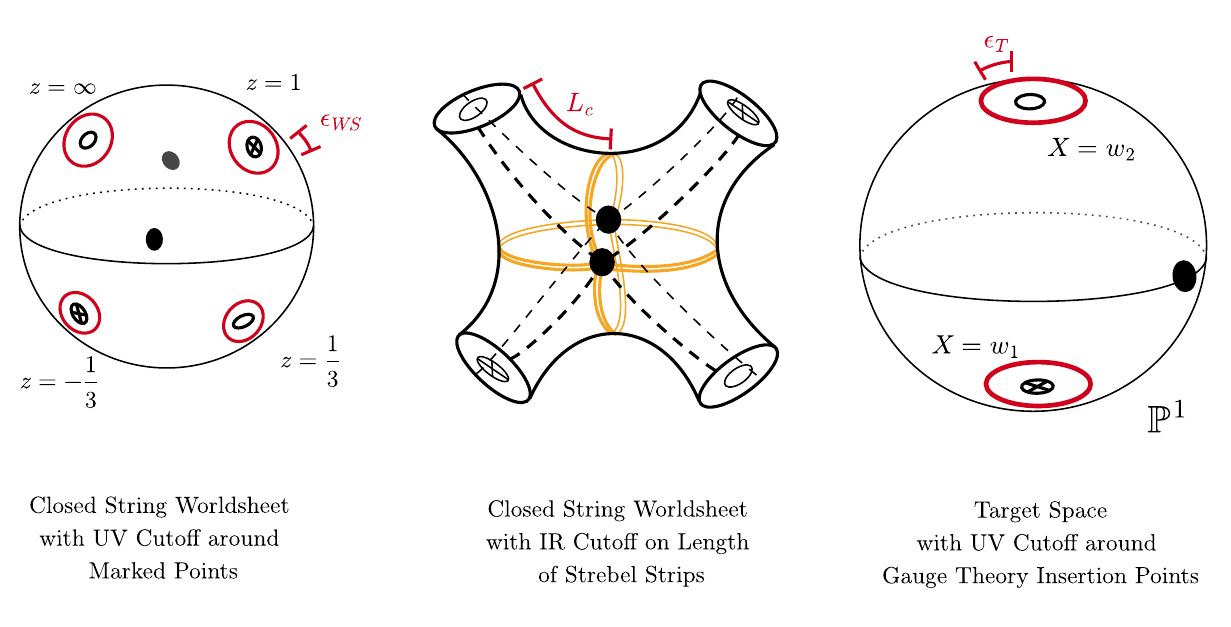} 

    \caption{\textbf{Regulating the Worldsheet Action } The action of the dual closed string reduces to the area of the worldsheet, computed using the Strebel metric. This area is a priori divergent, since the worldsheet geometry is that of flat semi-infinite cylinders glued to the Strebel graph. It can thus be regulated by setting an \textit{IR} cutoff, $L_{c}$, on the length of each strip making up said cylinders. Conformally mapping each cylinder to the disc, this translates to excising a small region around each marked point on the Riemann surface, tantamount to a worldsheet \textit{UV} cutoff $\epsilon_{WS}$. Finally, since each marked point on the worldsheet maps onto one of the singular points of the Kähler form in target space (at $X=w_1$ and $X=w_2$ on the $\mathbb{P}^1$), this procedure effectively introduces a target space \textit{UV} cutoff, $\epsilon_{T}$, around the two insertion points of gauge theory operators. }\label{fig:VariousCutoffs}
\end{figure}

The embedding map of the string into target space, $X(z)$, relates the worldsheet cutoffs to a cutoff in target space, $\epsilon_{T}$. To see how that is the case, recall that each of the marked points on the worldsheet map onto $X=w_1$ or $X=w_2$, the positions where the original gauge theory operators were inserted. The discs cut off around each marked point on the worldsheet map onto similar excisions around  $X=w_1$ and $X=w_{2}$, as illustrated in Fig. \ref{fig:VariousCutoffs}. From Eq.(\ref{eq:Kahlerformtarget}), we in fact see the target space Kähler form diverges at these two points. It is thus natural to expect we also need to introduce the cutoff $\epsilon_{T}$ around $X=w_1$ and $X=w_{2}$.

To find the precise relation between $\epsilon_{WS}, L_{c}$ and $\epsilon_{T}$, we can simply evaluate the embedding map on each strip at $z=iL_{c}$ and $z=-iL_{c}$, 

\begin{equation}
    |X(iL_{c})-w_2|  \equiv \epsilon_{T} \qquad  |X(-iL_{c})-w_1|  \equiv \epsilon_{T}
\end{equation}, 

which can then be solved for $L_{c}$ in terms of $\epsilon_{T}$ 

\begin{equation}
    L_{c} = \frac{1}{4 \pi} \log \left( \frac{|w_2-w_1 -\epsilon_{T}|^2}{\epsilon_{T}^2} \right)  =   \frac{1}{4 \pi} \log\left( \frac{|w_2-w_1|^2}{\epsilon_{T}^2} \right) + \mathcal{O}(\epsilon_{T}) 
\end{equation}.
\begin{figure}[ht!]
    \centering
\includegraphics[scale=0.70]{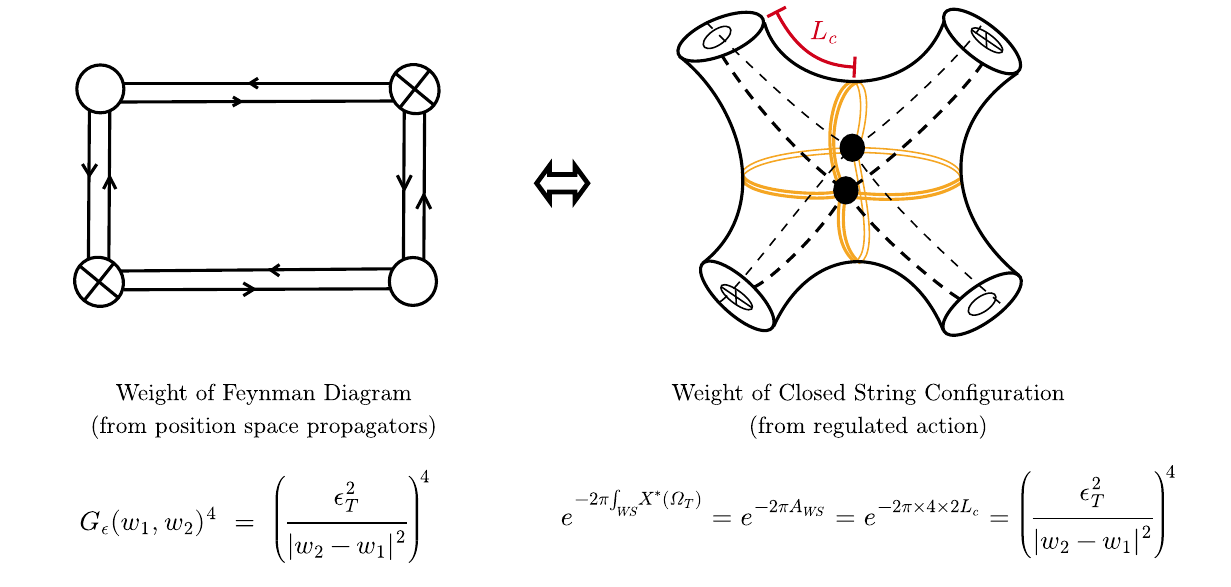} 

    \caption{\textbf{String Worldsheet Action = Weight of Feynman Diagram } The closed string action, evaluated on the embedding map $X(z)$, reduces to the Nambu-Goto action: $e^{-2\pi S[X] = e^{-2\pi A_{WS}}}$. While finding the explicit Strebel volume form on the worldsheet is complicated, the total (regulated) area of the worldsheet can be easily computed: it is the total number of strips making up the Riemann surface times the area of each strip. There are as many strips as there are edges in the dual Feynman diagram (in the above example, four strips for the four edges). The exponential of the (regulated) area of each strip gives exactly the position space propagator of the gauge theory. The weighting of each string configuration by the exponential of the action thus reproduces the weight of each Feynman diagram, up to the overall factor of $Y_{1} \cdot Y_{2}$ (see footnote \ref{fn:y1y2}).}
    \label{fig:WeightFD}
\end{figure}

This establishes an interesting connection between the UV on the worldsheet and the UV in the target space, since we can use Eq.(\ref{eq:epsilonWS}) to equate $\epsilon_{WS}$ and $\epsilon_{T}$ (up to a constant of proportionality). In the limit of $L_{c} \rightarrow \infty $ (i.e. $\epsilon_{T} \rightarrow 0$), the regulated Nambu-Goto action of the string, which again computes the total area of the worldsheet $A_{WS}$ consisted of $E$ strips, becomes 

\begin{equation}
    e^{-2\pi S_{NG}[X]} = e^{-2\pi A_{WS}} =  e^{-2\pi A_{WS}} = e^{-4 \pi E L_{c}} = \left(  \frac{\epsilon_{T}^2}{|w_2-w_1|^2}\right)^{E}\left( 1+ \mathcal{O}(\epsilon_{T})\right)
\end{equation}

The last term is simply the product of position space propagators, one for each edge of the underlying Feynman diagram. The spacetime $UV$ cutoff $\epsilon_{T}$ appears just as expected. We therefore see that the action of each string configuration dual to a particular diagram exactly reproduces the weight of said diagram!

A few comments are in order. Firstly, this is a remarkable simplification of the action. The embedding maps derived from each Feynman diagram can be both complicated and take very different forms, diagram to diagram. This means that the explicit Strebel volume form is a rather daunting object. However, when integrated over the worldsheet, we get a very simple expression. While we explained the factor of $E$ as the total number of strips making up the worldsheet (since each strip is identified with one of the edges of the Feynman diagrams), it also appears naturally from the point of view of target space: it is the degree of the covering map. The degree appears precisely in this manner in the A-model topological string, where it is identified with a particular instanton sector. Secondly, one might wonder why the action reproduces that of Nambu-Goto. This can be traced back to the fact that the semi-classical approximation to the topological string path integral is exact. Each string configuration is thus weighted by an on-shell action. The Nambu-Goto action is indeed known to be equivalent on-shell to its Polyakov-like sigma model cousins. This goes some way in explaining why the weight of each string configuration takes such a simple form.

\subsection{Finite Genus Expansion: Euler vs. Riemann-Hurwitz} \label{sec:finitegenusexp}

We can compute any correlator in the matrix model using a finite number of Wick contractions. This implies that the $1/N$ expansion always truncates at some finite order. In other words, there is a maximal genus above which no worldsheet of that topology contributes to the correlation function of vertex operators. This is a surprising statement, since most string theories have infinitely many terms in $g_{s}$ genus expansion. In fact, it is usually an asymptotic expansion, which led to the celebrated discovery of non-perturbative $e^{-1/g_{s}}$ effects \cite{cargeseSteve}. A similar observation was made by Itzhaki and McGreevy in the context of the proposed string dual to the matrix harmonic oscillator \cite{McGreevy_2004}. Following the "Strings from Feynman Diagrams" program, such a truncation will in fact be universal to any string theory dual to a free gauge theory (since there will always be a finite number of Wick contractions contributing to any correlator). It is therefore worth understanding how this puzzle gets resolved in this particular example of gauge/string duality.

How do we see this from the Feynman diagrams? Consider the expansion of the correlator in Eq.(\ref{eq:GenCorrelator}). The genus of the dual worldsheet is defined by the Euler characteristic of the graph:

\begin{equation}
    (V_{K}+V_{M})- \frac{1}{2} \left( \sum_{i=1}^{V_{K}} l_{i} + \sum_{j=1}^{V_{M}} n_{j} \right) + F = 2-2g,
\end{equation}

where we have again used the fact that any non-zero correlator will have $E=\sum_{i=1}^{V_{K}} l_{i} + \sum_{j=1}^{V_{M}} n_{j} $ edges, since each $K$ and $M$ must be paired together. Such a diagram comes with a power of $N^{2-2g}$. The number of vertices and edges is dictated purely by the form of the correlator. The $1/N$ expansion is thus an expansion in the number of faces of the diagrams. Since any diagram has at least one face, the inequality $F \geq 1$, leads to the upper bound on $g$:

\begin{equation}
    2g \leq  \frac{1}{2} \left( \sum_{i=1}^{V_{K}} l_{i} + \sum_{j=1}^{V_{M}} n_{j} \right) - (V_{K}+V_{M})+1 \label{eq:maxgen}
\end{equation}

From a target space perspective, we will derive this bound from the Riemann-Hurwitz formula. Consider the Belyi map dual to one of the Feynman diagrams appearing in the expansion of the correlator. We characterize its branching profile above the three target space points using the three permutations $\sigma_{K},\sigma_{f}$, and $\sigma_{M}$

\begin{equation}
    \sigma_{K} = (l_1)...(l_{V_{K}}) \qquad  \sigma_{f} = (m_1)...(m_{F}) \qquad  \sigma_{M}= (n_1)...(n_{V_{M}}) \label{eq:branchingstructure}
\end{equation}

The degree of the covering map $d$ is given by the order of the symmetric group to which these permutations belong, in other words:

\begin{equation}
    d = \sum_{i=1}^{V_{K}} l_{i} = \sum_{r=1}^{F} m_{r} = \sum_{j=1}^{V_{M}} n_{j} = E
\end{equation}

The Riemann-Hurwitz theorem relates the genus of worldsheet, the genus of the target space, the degree of the map, and the local branching structure: 

\begin{equation}
    2-2g_{WS}= d(2-2g_{\mathbb{CP}^1}) - \sum_{p\in WS}(e_{p}-1)
\end{equation}

where $e_p$ denotes the number of sheets that meet at a pre-image of a given ramification point of the target space. Using the specified branching data from Eq.(\ref{eq:branchingstructure}), the Riemann-Hurwitz formula reduces to the Euler characteristic of the Feynman diagram: 

\begin{align}
2-2g_{WS}  & = 2d - \left( \sum_{i=1}^{V_{K}} (l_{i} - 1) + \sum_{r=1}^{F} (m_{r}-1) + \sum_{j=1}^{V_{M}} (n_{j}-1) \right) \\
  & =    2d - \left( (d - V_{K}) + (d-F) + (d-V_{M}) \right) \\
  & =  -E + (V_{K}+V_{M})+F
\end{align} 

This is yet another consistency check on the identification of the covering surface with the worldsheet constructed from the Feynman diagram via the Strebel differential. The condition that $F \geq 1$, which is the requirement that the covering map is branched over exactly three (and not two) points of the target Riemann sphere, then gives us the same bound on the genus of the worldsheet found in Eq.(\ref{eq:maxgen}).

In other words, we can now explain why the dual string correlator receives no contributions from worldsheets above some maximal genus $g_{max}$: the Riemann-Hurwitz theorem simply forbids the existence of such covering maps!

\pagebreak

\section{Embedding in $AdS_{5}/CFT_{4}$} \label{sec:embedding}

Does the simple instance of gauge/string duality discussed here teach us anything about the inner workings of more general examples of holography, such as the celebrated duality between $\mathcal{N}=4$ SYM and Type IIB string theory on $AdS_{5} \times S^{5}$? We believe it does. The general program initiated in \cite{freefieldsadsI,freefieldsadsII,freefieldsIII} to reconstruct strings from Feynman diagrams in fact directly targeted higher dimensional (free) gauge theories with all fields in the adjoint, as in $\mathcal{N}=4$ SYM. Here, we show that our more humble matrix model/topological string duality can be directly embedded in a highly supersymmetric subsector of the full $AdS_{5}/CFT_{4}$ correspondence. 

The correlators of the matrix model will compute correlation functions in  $\mathcal{N}=4$ SYM annihilated by half the supercharges, to all orders in $1/N$.   As shown \cite{Drukker:2008pi,Drukker:2009sf}, these are expectation values of multi-trace 1/2-BPS operators built out of two linear combinations of the six scalar $N \times N$ matrices $\Phi^{I}_{ij}$ ($I \in {1,2..,6}$), inserted at two points on the boundary. From a bulk perspective, they capture supergravity Kaluza-Klein modes. We will refer to such observables as 1/2 SUSY, as opposed to 1/2 BPS which usually refers directly to the operator\footnote{Individually, 1/2 BPS operators preserve $24$ super(conformal) charges. (Any superconformal primary operators, even if they are non-BPS, are annihilated by 16 superconformal charges. Additionally, $1/2$ BPS operators are annihilated also by 8 supercharges.) However, the two-point functions of 1/2 BPS operators preserve only $16$ super(conformal) charges, which is half of the total super(conformal) charges of $\mathcal{N}=4$ SYM.}. Concretely, we will study correlators of the form
\begin{equation}
   \Big\langle \prod_{i=1}^{V_{K}} \frac{1}{l_{i}} \Tr \left( (Y_1 \cdot \Phi(w_{1}))^{l_{i}} \right) \prod_{j=1}^{V_{M}} \frac{1}{n_{j}}\Tr \left( (Y_2 \cdot \Phi(w_{2})^{n_{j}} \right) \Big\rangle_{\mathcal{N}=4}  \label{eq:N=4correlator}
\end{equation}
where $w_1,w_2$ are two points on the boundary (here taken to be $\mathbb{R}^{4}$), while $Y_{1}^I$ and $Y_{2}^{I}$ ($I \in {1,2..,6}$) are two (complex) vectors parametrizing the R-charge of the operator under the three Cartans of $SU(4)_{R} \simeq SO(6)$. The $\cdot \ $-multiplication is simply shorthand for an inner-product taken with the metric $\delta_{IJ}$, i.e. $Y_{1} \cdot \Phi = Y_{1}^{I} \Phi^{J} \delta_{IJ}$, and similarly for $Y_{2} \cdot \Phi$. The single trace operators being 1/2 BPS requires $Y_{1} \cdot Y_{1} = Y_{2} \cdot Y_{2} = 0$. The correlator will vanish unless $Y_{1} \cdot Y_{2} \neq 0$, which we henceforth assume\footnote{If $Y_{1} \cdot Y_{2} = 0$, then the two operators share no $R$-charge quantum numbers, and charge conservation requires the correlator to be identically zero.}. 

We can succinctly write down a generating function of correlators by considering the insertion of determinant operators in the gauge theory. We will drop the subscript ${\cal N}=4$ henceforth since this should be clear from the context. First, define two diagonal matrices $X_{Q \times Q} = \text{diag}\{x_{1},...,x_{Q}\}$ and $V_{R\times R} = \text{diag}\{v_{1},...,v_{R}\}$. Using $\det(...) = e^{\Tr \log(...)}$, we can formally rewrite correlators of determinant operators in terms of a exponential of all single trace operators we might wish to consider:
\begin{equation}
    \frac{ \langle \prod_{\mu=1}^{R} \det \left( v_{\mu}-Y_1 \cdot \Phi(w_{1}) \right) \prod_{a=1}^{Q} \det\left(x_{a}-Y_2 \cdot \Phi(w_{2}) \right) \rangle}{\det_{Q \times Q}(X)^{N} \det_{R \times R}(V)^{N}} =  \langle e^{-\sum_{i=1}^{\infty} \frac{\tilde{t}_{i}}{l_{i}} \Tr \left( (Y_1 \cdot \Phi(w_{1}))^{l_{i}} \right)}  e^{-\sum_{j=1}^{\infty} \frac{{t}_{j}}{n_{j}}\Tr \left( (Y_2 \cdot \Phi(w_{2})^{n_{j}} \right)} \rangle \label{eq:genfunctN=4}
\end{equation}
where the {\it times}, sometimes also called 
{\it Miwa variables}\footnote{This terminology is borrowed from the integrability approach to matrix models.}, $t_{k}$ and $\tilde{t}_{k}$ are defined in terms of $X$ and $V$ via 
\begin{equation} \label{eq:tMiwa}
    t_{k}= \Tr_{Q}(X^{-k}) \qquad  \tilde{t}_{k}= \Tr_{R}(V^{-k})
\end{equation} 

At finite $Q$ and $R$, trace relations imply that only the first $Q$ ($R$) times are independent. We will imagine taking $Q$ and $R\rightarrow \infty $ (\textit{after} taking the $N \rightarrow \infty$ limit of the gauge theory), so as to neglect these effects\footnote{It would be very interesting to study the effect of trace relations along the lines of \cite{JiHoon} in the context of open-closed-open triality.}. The correlators considered in Eq.(\ref{eq:N=4correlator}) can then formally be obtained from Eq.(\ref{eq:genfunctN=4}) by differentiation with respect to the times, subsequently setting the $t_{k}, \tilde{t}_{k}$ to zero. 

From the string theory point of view, the insertion of these determinant operators corresponds to adding giant graviton branes in the bulk. These D3-branes wrap an $S^{3} \subset S^{5}$ and are point-like in the $AdS_5$ factor of the target space - see Fig. \ref{fig:GiantGravitons}. They are supported by angular momentum. The vectors $Y_{i}$ determine how the $S^{3}$ sits inside the $S^{5}$, while the $x_{a}$ and $v_{\mu}$ play the role of chemical potentials for their angular momentum.

Giant graviton branes are heavy semi-classical objects, since the dimension of each determinant operator equals $N$. However, with $Q,R <<< N$, their backreaction is not sufficient to alter the $AdS$ asymptotics of the background. In other words, it is not sufficient to shift the $\mathcal{O}(N^2)$ saddle of the gauge theory, or from a bulk perspective, an $AdS$-vacuum. Rewriting the determinant in terms of an exponential of single trace operators translates to replacing the giant graviton branes with a coherent state of closed strings in $AdS$. We may then regard Eq.(\ref{eq:tMiwa}) as a dictionary between the open string moduli of the D-brane and the precise deformation parameters of the closed string background.

\subsection{From $\mathcal{N}=4$ Super Yang-Mills to the 2-Matrix Model} \label{sec:N=4toMatrixModel}

The first step in embedding our simple gauge/string duality into the AdS/CFT correspondence is to reduce the full $\mathcal{N}=4$ SYM path integral computation of such correlators to that of a two-matrix integral. We will start directly with the generating function of Eq.(\ref{eq:genfunctN=4}). 
\begin{equation}
    \frac{1}{Z_{\mathcal{N}=4}} \int D(\text{fields}) e^{-\frac{N}{\lambda} S_{\mathcal{N}=4}} \prod_{a=1}^{Q} \det \left( x_{a}-Y_2 \cdot \Phi(w_{2}) \right) \prod_{\mu=1}^{R} \det\left(v_{\mu}-Y_1 \cdot \Phi(w_{1}) \right) \label{eq:N=4Start}
\end{equation}

Since this correlator of determinant operators is protected (i.e. $1/2$ BPS), we can take the zero 't Hooft coupling limit of the gauge theory, $\lambda \rightarrow 0 $, to simplify the computation. By rescaling the $\mathcal{N}=4$ fields by a factor of $\sqrt{\lambda}$, we observe that the six scalars decouple from the fermions and the gauge field in this limit. The path integral therefore factorizes into a product of path integrals over, on one hand, the gauge and fermionic fields, and on the other hand, the scalars\footnote{There is an overall Gauss law constraint but that will not play a role in the following.}. Since the determinant operators only contain the scalars, the fermionic and gauge field factors cancel between the numerator and denominator of Eq.(\ref{eq:N=4Start}). A further simplification arises from the fact that the quartic (commutator squared) interaction of the scalars is also suppressed in the zero coupling limit. We are left with a much simpler computation in a theory of six free Klein-Gordon fields:
\begin{equation}
    \frac{1}{Z_{\text{scalar}}} \int D\Phi^{I} e^{+\frac{N}{2\lambda} \int d^{4}w \Tr \left( \Phi^{I} \Box \Phi^{I} \right) } \prod_{a=1}^{Q} \det \left( x_{a}-Y_2 \cdot \Phi(w_{2}) \right) \prod_{\mu=1}^{R} \det\left(v_{\mu}-Y_1 \cdot \Phi(w_{1}) \right)
\end{equation}

We can then define a projector onto the complement of the shared R-charge sector of the two operators 
\begin{equation}
    P^{I J} = \delta^{IJ}-\frac{1}{ Y_1 \cdot Y_{2}} \left( Y_{1}^{I} Y_{2}^{J}+ Y_{2}^{I} Y_{1}^{J}\right)
\end{equation}
This allows us to further factor the remaining path integral. At the level of the action, we can now write
\begin{equation}
    \frac{1}{2} \int d^{4}w \Tr \left( \Phi^{I} \Box \Phi^{I} \right) =  \frac{1}{2} \int d^{4}w \Tr \left( (P \Phi)^{I} \Box (P \Phi)^{I} \right) + \frac{1}{Y_{1}\cdot Y_{2}}\int d^{4}w \Tr \left( Y_{1}\cdot \Phi \Box Y_{2} \cdot \Phi \right)
\end{equation}
We can decompose the path integral measure into a piece in the Kernel of $P^{IJ}$ and part in its image:
\begin{equation}
 \prod_{I=1}^{6} D\Phi^{I} = D(P\Phi)^{I} \times D (Y_{1} \cdot \Phi) D (Y_{2} \cdot \Phi)
\end{equation}
Again, since the determinants only involve $Y_1 \cdot \Phi$ and $Y_2 \cdot \Phi$, all contributions to the path integral involving $P \Phi^{I} $ will cancel between the numerator and denominator, leaving behind
\begin{equation}
    \frac{1}{Z_{Y\cdot \Phi}} \int D (Y_{1} \cdot \Phi) D (Y_{2} \cdot \Phi) e^{\frac{N}{\lambda Y_{1} \cdot Y_{2} } \int d^{4}w \Tr \left( Y_{1}\cdot \Phi \Box Y_{2} \cdot \Phi \right)} \prod_{a=1}^{Q} \det \left( x_{a}-Y_2 \cdot \Phi(w_{2}) \right) \prod_{\mu=1}^{R} \det\left(v_{\mu}-Y_1 \cdot \Phi(w_{1}) \right)
\end{equation}

The final step in the reduction to the matrix integral comes from parametrizing the matrix-valued fields as 
\begin{align}
    Y_1 \cdot \Phi(w)   \equiv  \frac{|w_1 -w_{2}|^2}{|w-w_{2}|^2}K +  F_1(w) \\
      Y_2 \cdot \Phi(w)   \equiv \frac{|w_1 -w_{2}|^2}{|w-w_{1}|^2} M + F_2(w)
\end{align}
where we choose $F_{1}(w_1)= F_{2}(w_2)=0$.  We have also integrated by parts. This ensures the two matrices $K$ and $M$ are simply the value of the (combination of) scalar fields at the two insertion points: $ Y_1 \cdot \Phi(w_1)=K$ and $ Y_2 \cdot \Phi(w_2)=M$. 
Using the identity 
\begin{equation}
    \Box \frac{1}{|w|^2} = -4\pi^2 \delta(w)
\end{equation}, 
the action becomes
\begin{equation}
    \int d^{4}w  \Tr \left( Y_1 \cdot \Phi(w)  \Box Y_2 \cdot \Phi(w) \right) = - 4\pi^2 |w_{1} -w_{2}|^2 \Tr \left( KM \right) +  \int d^{4}w \Tr (F_{1} (w) \Box F_{2}(w)) 
\end{equation}

Given the fact that the path integral measure will also factorize (in the loose sense in which it is even defined)
\begin{align}
    D \left( Y_{1}\cdot\Phi \right) = \prod_{w \in \mathbb{R}^{4}} d(Y_{1}\cdot\Phi(w)) = dK \prod_{w \neq w_1} dF_{1}(w)  \\
      D \left( Y_{2}\cdot\Phi \right) = \prod_{w \in \mathbb{R}^{4}} d(Y_{2}\cdot\Phi(w)) = dM \prod_{w \neq w_2}  dF_{2}(w), 
\end{align} 
we conclude that the contribution from the fields away from $w=w_1$ and $w=w_2$, i.e. from $F_1$ and $F_{2}$, again completely cancel between the numerator and denominator. We thus reach the desired equality between the path integral of $\mathcal{N}=4$ SYM with insertion of operators duel to $1/2$-BPS giant graviton branes and the two-matrix integral with determinant insertions for each matrix:
\begin{align} \label{eq:generatingfunc}
    &\langle \prod_{a=1}^{Q} \det \left( x_{a}-Y_2 \cdot \Phi(w_{2}) \right) \prod_{\mu=1}^{R} \det\left(v_{\mu}-Y_1 \cdot \Phi(w_{1}) \right) \rangle_{\mathcal{N}=4} \\ \nonumber
    &\hspace{30pt}=  \frac{1}{Z_{K,M}}\int dK dM_{N \times N} e^{-\frac{N}{\lambda} \frac{4 \pi^2 |w_1 -w_{2}|^2}{Y_{1}\cdot Y_{2}} \Tr \left( K M\right)} \prod_{a=1}^{Q} \det \left( x_{a}-M \right) \prod_{\mu=1}^{R} \det\left(v_{\mu}-K \right)
\end{align}

Comparing with our initial definition of the two matrix model, Eq.\ref{eq:2MM}, we see that the dependence of position and the charges of the $1/2$ SUSY operators is captured in the matrix model parameter \footnote{To be clear, this is not the coupling "$g_{int}$" often used in the $\mathcal{N}=4$ SYM integrability literature related to the 't Hooft coupling via $16 \pi^2 g_{int}^2  = \lambda $} 
\be \label{eq:gVSprops}
g \leftrightarrow \frac{\lambda Y_1\cdot Y_2}{4\pi^2|w_1 - w_2|^2}
\ee 

This derivation endows the $N \times N$ complex matrices $K$ and $M$ with a clear physical meaning. They are, respectively, the transverse oscillations of open strings in the $Y_{1}^{I}$ ($Y_{2}^{I}$) direction at $w=w_1$ ($w=w_2$) on the worldvolume of the $N$ D3 branes giving rise to $\mathcal{N}=4$ SYM at low energies. This also shows that the Feynman diagram expansion of the 1/2 SUSY observables of $\mathcal{N}=4$ SYM is identical to the Feynman diagram expansion of correlators of traces built out of $K$ or $M$.\footnote{From a purely matrix model perspective, it would be reasonable to consider traces built out of "words" involving \textit{both} the $K$ and $M$ matrices, such as $\Tr( M^2 K^3MK)$. However, the derivation above makes clear that $K$ and $M$ correspond to local fields at different points on the boundary, $K = Y_{1} \cdot \Phi (w_1)$ and  $M = Y_{2} \cdot \Phi (w_2)$. From the gauge theory perspective, it therefore does not make sense to build single trace operators involving both matrices.} The propagator of the matrix model is simply the position space propagator for the $\mathcal{N}=4$ fields, inserted at the two points $w_1$ and $w_2$. Mapping each Feynman diagram of the matrix model to a closed string worldsheet therefore explicitly implements open/closed duality for this highly supersymmetric sector of the full AdS/CFT correspondence. Moreover, it provides a clear interpretation of the determinant insertions in the matrix integral: they are 1/2 BPS giant graviton branes anchored on the boundary of $AdS_{5}$ at either $w_1$ or $w_{2}$. This will be important when we come to discuss the open-closed-open triality of the two-matrix model in Sec. \ref{sec:OCOgen}.

\subsection{Belyi Maps from Twistor Strings}

In this simple gauge/string duality, we recast the Feynman diagram expansion of 1/2 SUSY correlators into a sum over branched covering maps from 2d worldsheets onto a target Riemann sphere. We have already embedded the matrix model in $\mathcal{N}=4$ SYM. This explains how the matrix Feynman diagrams provide the open string description of this subsector of the AdS/CFT duality. One can now ask how the derived dual, which is a sum over covering maps, embeds itself on the closed string side, namely in Type IIB string theory on $AdS_{5} \times S^{5}$. Unfortunately, a well-defined closed-string worldsheet theory is still lacking due to difficulties in treating RR-flux. However, since the gauge theory correlators computed by the matrix model are protected, i.e. independent of the 't Hooft coupling, we can appeal to the recently proposed twistor string dual to free $\mathcal{N}=4$ SYM \cite{gaberdiel2021worldsheet}. 

This proposal gives a natural worldsheet prescription for obtaining the spectrum of the free theory. It has, however, not yet been derived from a first principles quantization. This is an obstacle to matching the correlators of the free gauge theory with those from the worldsheet in a rigorous way. Nevertheless, one may expect some features to be in parallel with the corresponding manifest match of correlators in the worldsheet description of the tensionless string theory on $AdS_3\times S^3 \times T^4$. There too, there is an underlying twistor string theory description \cite{dei2021free} which has recently been fleshed out \cite{McStay:2023thk, McStay:2024dtk}. 

Using this as a rough guide, 
a sketch of the dual string description of correlators was given in \cite{Bhat:2021dez}. 
The idea was that the string dual to correlators of BPS operators inserted at two points would get contributions from holomorphic maps into (a subspace of) the twistor space of $AdS_5$. In particular, this would be a ${\mathbb P}^1$ that would be parametrized by two of the twistor space coordinates (as homogeneous coordinates of the ${\mathbb P}^1$). Thus, at genus zero,
\begin{equation}
    X(z)=\frac{\mu(z)}{\lambda(z)}.
\end{equation}
Here $\mu(z), \lambda(z)$ are the two twistor field components which are non-zero. Each of these field configurations is a polynomial in $z$ whose degree depends on the total number of bits (letters) in the dual gauge theory operator. This gives rise to a branched covering of a target space $\mathbb{P}^1$ by the world sheet.  

In fact, for two-point functions of single trace operators, the discussion in   
\cite{Bhat:2021dez} (see Sec. 5.1 there) involved holomorphic maps 
corresponding to each Feynman diagram, using the Strebel construction. This reproduced the free Feynman propagator in a very natural way, when one assumes a Nambu-Goto weight given by the Strebel area. We can then expect a similar picture for our more general case (of multiple determinant operators inserted 
at two points) which has a more general branching behavior captured by the Belyi map spelled out in our Sec. 4. While there are many points here to be fleshed out, including the role of the third branch point, we are encouraged by the very similar ingredients that enter into the picture here.  
\pagebreak

\section{Open-Closed-Open Triality for the $1/2$ SUSY Sector of $\mathcal{N}=4$ SYM } \label{sec:OCOgen}

Open-Closed-Open triality is the idea that multiple open string descriptions can be dual to the \textit{same} closed string theory. This was first put forward in \cite{Joburg}, based on the two matrix descriptions of the $(2,1)$ minimal string. One of the main insights of \cite{Joburg} was to observe that the Feynman diagrams of the two theories were related by graph duality. While the closed string vertex operator insertions were dual to the \textit{vertices} of the Feynman diagrams in one model, they mapped on instead to the "faces" of the Kontsevich model ribbon graphs. As such, they were dubbed the V(ertex)- and F(ace)-Type open duals. The two matrix integrals were later understood to capture open strings on different sets of branes, the so-called ZZ or FZZT branes of Liouville theory.

Two of the authors argued in \cite{DSDI} that the existence of multiple open string descriptions for the same closed dual might in fact be quite generic. The gauge theory Feynman diagrams arising from these open string descriptions generate the same sum over closed string worldsheets via the Strebel parametrization of $\mathcal{M}_{g,n}$. For this to be true, their diagrams must be related via (partial) graph duality, so as to encode the same worldsheet moduli (with identical weighting factors for each diagram). For example, in one theory, such as the $K,M$ model we have studied so far, the gauge theory Feynman diagrams are graph dual to the Strebel graphs of the string worldsheets. A matrix integral that would directly generate those Strebel graphs as its own Feynman diagrams would therefore describe the same closed string theory.

In this section, we will show that six different matrix integrals can generate the same weighted sum over closed string worldsheets. This enriched notion of Open-Closed-Open triality is summarized in Fig. \ref{fig:OCOfromBranes}. We will see that the edges of their respective Feynman diagrams exploit different special trajectories traced out by the Strebel differential on the dual string worldsheet, as illustrated in Fig. \ref{fig:OCOfromWS}. 

\begin{figure}[ht!]
    \centering
\includegraphics[scale=0.75]{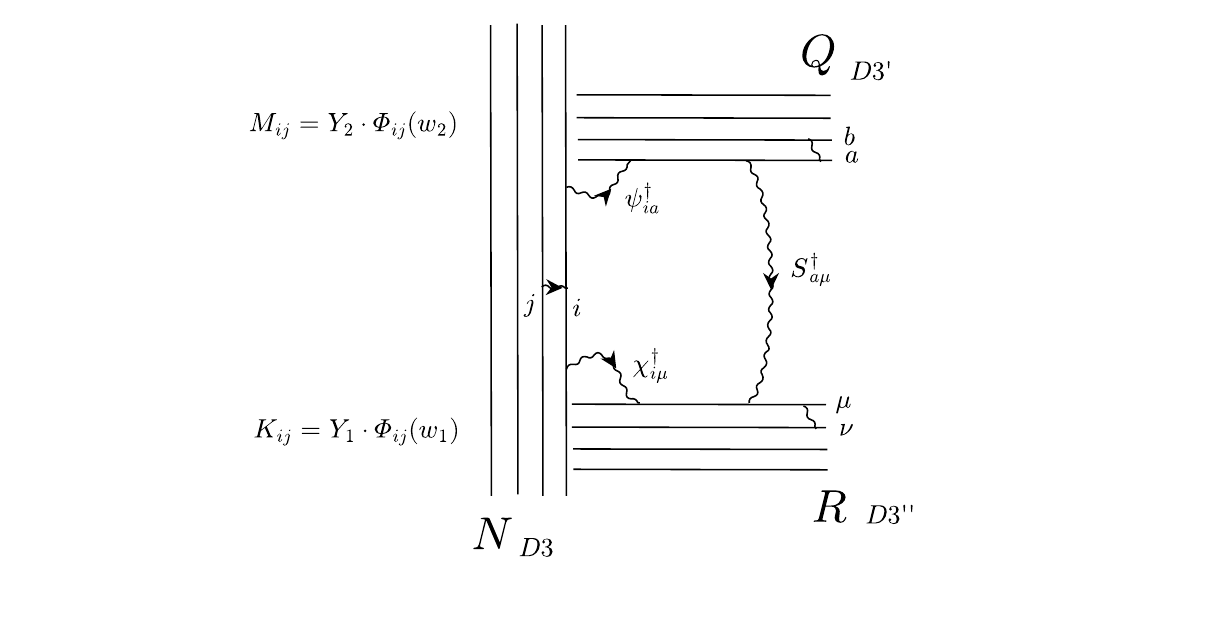} 

    \caption{\textbf{Three Stacks of D3 Branes and New Open Strings } From a brane perspective, the insertion of determinant operators corresponds to the addition of extra D3 branes. We can think of three separate stacks lying in ten-dimensional flat space: the "original" N D3s giving rise at low energy to  $\mathcal{N}=4$ SYM, along with Q D3' and R D3'' branes transverse to the D3s, in a manner dictated by the two six vectors $Y_{1}^{I}$ and $Y^{I}_{2}$. They intersect the worldvolume of the D3s at the two insertion points $w_1$ and $w_2$. There are fermionic open strings $\psi^{\dagger}_{ia}$ stretched between the D3-D3' and $\chi^{\dagger}_{i\mu}$ stretched between the D3-D3''. There are also D3'-D3'' bosonic strings, which will ultimately furnish the F-type description. }
    \label{fig:OtherOpenStrings}
\end{figure}

To demonstrate this phenomenon concretely, we first construct a generating function of correlators in the $K, M$ model via the addition of determinant insertions. This is the same generating function obtained from the reduction of determinant operator correlators in $\mathcal{N}=4$, see Eq.(\ref{eq:genfunctN=4}). The insertion of determinant operators in $\mathcal{N}=4$ is well-known to be holographically equivalent to inserting giant graviton branes in the bulk. We are thus studying a system of three stacks of branes: the "original" N D3 branes which gave rise to $\mathcal{N}=4$ SYM, along with two additional stacks of Q transverse D3' and R D3'' branes. The two six vectors $Y_{1}^{I}$ and $Y^{I}_{2}$ determine how these additional branes lie in the six dimensions transverse to the worldvolume of the original N D3s. They intersect this worldvolume at the two insertion points of the determinant operators, $w_1$ and $w_2$. The other open string theories we will encounter shortly will descend from open strings ending on these additional branes. See Fig. \ref{fig:OtherOpenStrings}.

We will be interested in how the Feynman diagrams of these various matrix models relate to one another. To do so, we introduce an additional matrix parameter in the action, so that the generating function\footnote{The tensor products are just convenient notation to denote the same product of determinants appearing in Eq.(\ref{eq:generatingfunc}).} now depends on three lists of parameters, $\{ t_{k},\bar{t}_{k},s_{k} \}$:

\begin{align} \label{eq:genfuncwithdets}
    &Z[t_{k},\bar{t}_{k},s_{k}]  \equiv   \left( \frac{\det_{N}(NY)}{\det_{Q}(X) \det_{R}(V)} \right)^{N}\inv{(2\pi g)^{N^2}} \int dK dM_{N \times N} e^{-\frac{N}{g} Tr_N(\sqrt{Y}K\sqrt{Y}M)} \\
    &\hspace{180pt}\det\left( X \otimes \mathbb{I}_{N}- \mathbb{I}_{Q} \otimes M \right) \det(V\otimes  \mathbb{I}_{N} - \mathbb{I}_{R} \otimes K)\nonumber
\end{align}

\begin{equation}  \label{eq:genfuncwithY}
     = \frac{\det_{N}(NY)^{N}}{(2\pi g)^{N^2}} \int dK dM_{N \times N} e^{-\frac{N}{g}Tr_N(\sqrt{Y}K\sqrt{Y}M) - \sum_{k \geq 1} \frac{t_{k}}{k} \Tr(K^k) - \sum_{k \geq 1} \frac{\bar{t}_{k}}{k} \Tr(M^k)}
\end{equation}

The parameters $\{ t_{k},\bar{t}_{k},s_{k} \}$ appearing as arguments on the LHS of Eq.(\ref{eq:genfuncwithY}) are related to three diagonal matrices $Y_{N\times N} = \text{diag}(y_1,...,y_N)$, $X_{Q \times Q}=\text{diag}(x_1,...,x_Q)$ and $V_{R \times R}=\text{diag}(v_1,...,v_R)$  via the Miwa transformation we already encountered in Eq.(\ref{eq:tMiwa}) 

\begin{equation}
    s_{k} = \Tr_{N}(Y^{-k}) \qquad    t_{k} = \Tr_{Q}(X^{-k}) \qquad    \bar{t}_{k} = \Tr_{R}(V^{-k})
\end{equation}

A particular (connected) correlator, in the free $K,M$ theory, can then be computed by taking derivatives with respect to the $t_{k}$  and $\bar{t}_{k}$:

\begin{equation} \label{eq:correlwithsk}
  \Braket{\prod_{i=1}^{V_K} \frac{1}{l_{i}}\Tr\left( K^{l_i}\right) \prod_{j=1}^{V_M} \frac{1}{n_{j}}\Tr\left( M^{n_j} \right)} _{c} = \prod_{i=1}^{V_K} \frac{\partial}{\partial t_{l_{i}}}  \prod_{j=1}^{V_M}  \frac{\partial}{\partial \bar{t}_{n_{j}}} \log \left(  Z[t_{k},\bar{t}_{k},s_{k}] \right) \big |_{t_{k}=\bar{t}_{k}=0}  
\end{equation}

The parameters $s_{k}$ instead give weights to the \textit{faces} of the Feynman diagrams contributing to any such correlator. Indeed, the propagator in Eq.(\ref{eq:genfuncwithY}) is modified to 

\begin{equation}
    \braket{K_{ij}M_{kl}} = \frac{g}{N} \sqrt{y_{i} y_{j}} \delta_{il}\delta_{jk}
\end{equation}

This means that a face bordered by $2k$ edges with an internal color index $i$ (to be summed over) flowing in the loop, will contribute an additional weighting factor of 

\begin{equation}
    \sum_{i=1}^{N} \left(\frac{1}{\sqrt{y_{i}}}\right)^{2k} =s_{k}
\end{equation}

to the Feynman diagram. We can thus read off the precise number of faces of a given type by looking at the powers of the $s_{k}$ appearing in Eq.(\ref{eq:correlwithsk}). With this notation in place, we can now proceed to derive the other equivalent matrix integrals. 

We will obtain these other open string descriptions, i.e. the other matrix integrals, by considering the open strings stretched between the various D-branes of the system. Operationally, this will be achieved by integrating in/out different open string degrees of freedom. We can then study the functional dependence of the other equivalent matrix integrals on the three lists of parameters. This allows us to track how the Feynman diagrams of the various models are related.

\subsection{Equivalent Open String Descriptions from Integrating In/Out} \label{sec:intinout}

The "F(ace)-Type" model is the matrix integral whose Feynman diagrams are directly the Strebel graphs of the worldsheets appearing in the closed string path integral we initially derived from the $K,M$ model. The marked points on those worldsheets corresponded to \textit{vertices} of the $K,M$ model Feynman diagram. They will now map on to the \textit{faces} of the Feynman diagrams of the "F-Type" model (hence the name). We will arrive at the F-type description in four steps, each involving integrating in or out certain other fields, corresponding to different open strings. These manipulations are exact. In particular, we do not resort to any of the numbers of branes $N, Q$ or $R$ being large. Many of the steps below closely mirror those of \cite{brown2011complex}, which had explored Open-Closed-Open triality in the same protected subsector of $\mathcal{N}=4$ SYM. There is also an interesting parallel to these manipulations in the quantum chaos literature \cite{Altland2022,AltlandSonner2020}. The one-matrix case gave rise to the source-determinant duality of \cite{Brezin:2016eax}\footnote{In \cite{BrezinReplicaCMP}, they showed the power of such manipulations. It allowed them to manifestly recast certain limits of the Gaussian matrix model correlators in terms of intersection numbers on the moduli space of $p$-spin curves \cite{Brezin:2020wqj,Hikami:2020dmc}. }. We refer to Appendix \ref{sec: diff_duals} for a more detailed presentation, including additional intermediate steps.

\textbf{STEP 1: Integrating in the Fermionic $D3/D3'$ and $D3/D3''$ Open Strings}

The first step is to re-express the determinant insertions in Eq.(\ref{eq:genfuncwithdets}). We can rewrite the $Q$ determinants in terms of $N \times Q$ complex fermions $\psi^{\dagger}_{ia}$. Physically, these are the open strings between the $N$ D3s and $Q$ D3's. Similarly, the $R$ determinants can be recast as an integral over $N \times R$ fermions $\chi^{\dagger}_{i\mu}$, which are now the open strings between the $N$ D3s and $R$ D3''s, as shown in Fig. \ref{fig:OtherOpenStrings}.

\begin{align} \label{eq:KMwithferms}
    Z=\frac{\mathcal{K}}{Z_N}\int dK dM d\psi d\psi^{\dagger}d\chi d\chi^{\dagger}\exp\bigg(&-\frac{N}{g}\text{Tr}_N(K\sqrt{Y}M\sqrt{Y})+\psi_{ia}^\dagger(X_{ab}\delta_{ij}-\delta_{ab}M_{ij})\psi_{jb}\nonumber\\
    &\hspace{100 pt}+\chi_{i\mu}^{\dagger}(V_{\mu\nu}\delta_{ij}-K_{ij}\delta_{\mu\nu})\chi_{j\nu}\bigg)
\end{align}

where we have abbreviated the overall normalization $\mathcal{K}=(\det_{N}(Y)/ (\det_{Q}(X)\det_{R}(V)))^{N}$, and $Z_N = (2\pi g/N)^{N^2}$ 

\textbf{STEP 2: Integrating out the Original $D3/D3$ Open Strings}

The original bosonic matrices $K$ and $M$, which captured the $D3/D3$ open strings, now appear linearly in the action. They can be explicitly integrated out \footnote{We adopt a purely imaginary contour for $K$, which results in a delta-function for $M$, which can then be trivially integrated. See Appendix \ref{sec:appFtype} for details.}. This gives the effective fermionic theory:   

\begin{align}
    Z=\bigg(\frac{1}{\det_{Q}(X)\det_{R}(V)}\bigg)^{N} \int d\psi d\psi^{\dagger}d\chi d\chi^{\dagger}\exp\bigg(\psi^{\dagger}_{ia}X_{ab}\psi_{ib}+\chi^{\dagger}_{i\mu}V_{\mu\nu}\chi_{i\nu}-\frac{g}{N}\psi^{\dagger}_{ia}(Y^{-\half})_{ik}\chi_{k\mu}\chi_{l\mu}^{\dagger}(Y^{-\half})_{lj}\psi_{ja}\bigg) \label{eq:fermionicint}
\end{align}

In this presentation, none of the Miwa times appear manifestly. We see that the sources $X$ and $Y$ give weights to the edges of the fermionic Feynman diagrams, since they now appear in the $\psi^{\dagger} \psi $- and $\chi^{\dagger} \chi$-propagators respectively. This will be important later when we aim to understand the relation between the Feynman diagrams of each matrix model.

\textbf{STEP 3: Integrating in the $D3'/D3''$ Open Strings}

We can decouple the quartic fermionic interaction term in Eq.(\ref{eq:fermionicint}) via the Hubbard-Stratanovich transformation. We introduce a complex bosonic rectangular matrix $S_{a \mu}$ that transforms in the bi-fundamental of $U(Q) \times U(R)$. It couples linearly to the fermions:

\begin{align} \label{eq:SSdagin}
 Z[t_{k},\bar{t}_{k},s_{k}] =&\frac{\mathcal{K}'}{Z_{QR}}\int d\psi d\psi^{\dagger}d\chi d\chi^{\dagger}dS dS^{\dagger}\exp\bigg( \frac{N}{g}S^{\dagger}_{\mu a}S_{a\mu}-S^{\dagger}_{\mu a}\chi^{\dagger}_{l\mu}(Y^{-\half})_{lj}\psi_{ja} - \psi_{ia}^{\dagger}(Y^{-\half})_{ik}\chi_{k\mu}\Phi_{a\mu} \nonumber\\ &\qquad\qquad\qquad\qquad\qquad\qquad\qquad\qquad\qquad\qquad\quad+\psi^{\dagger}_{ia}X_{ab}\psi_{ib}+\chi^{\dagger}_{i\mu}V_{\mu\nu}\chi_{i\nu}\bigg)
\end{align}

where $Z_{QR}= (2\pi g/N)^{QR}$.

\textbf{STEP 4: Integrating Back Out the Fermionic Strings}

The two fermionic integrals are now Gaussian. They can be carried out explicitly, resulting in a purely bosonic theory for the complex matrix $S$. This is the effective theory on the $D3'/D3''$ open strings. We have relegated the details to Appendix \ref{sec:appFtype}, simply quoting the final result here:

\begin{align} \label{eq:Fmodel}
  Z[t_{k},\bar{t}_{k},s_{k}] = &  \frac{(\text{det}_Q(X))^{R}(\text{det}_{R}(V))^{Q}}{\det_{N}(Y)^{N} Z_{QR}} \int dS dS^{\dagger}_{Q \times R} \exp(\frac{N}{g}\Tr(V S^{\dagger} X S) \prod_{i=1}^{N} \det(y_{i}-SS^{\dagger})\\
  = & \frac{(\text{det}_Q(X))^{R}(\text{det}_{R}(V))^{Q}}{Z_{QR}} \int dS dS^{\dagger}_{Q \times R} \exp(\frac{N}{g}\Tr(V S^{\dagger} X S) + \sum_{k\geq 1}  \frac{s_{k}}{k} \Tr((S S^{\dagger})^k)
\end{align}

In this presentation, the Miwa variables $t_{k}, \bar{t}_{k}$ are no longer manifest. This means that a single trace observable in the $K,M$ model does not correspond to any one single trace operator in this equivalent description. To understand the precise mapping, one would in principle need to invert the Miwa transformation to express derivatives with respect to $t_{k}, \bar{t}_{k}$ in terms of derivatives with respect to the matrix sources $X$ and $V$. Such derivatives would then insert various sums of matrix elements of $S$ and $S^{\dagger}$. More importantly though, note that the $s_{k}$, which gave weights to the faces of the $K,M$ model Feynman diagram, are now associated to $\textit{vertices}$ in the $S,S^{\dagger}$ matrix integral. This is the first diagnostic that the Feynman diagrams of the $S,S^{\dagger}$ matrix integral are exactly graph-dual to those of the $K,M$ model. 

We recover the (standard) generating function of correlators by setting the $y_{i}=1$ (and hence $s_{k}=N$), so that we arrive at the final equality

\begin{equation} \label{eq:finaleqgen}
\begin{split}
    \braket{e^{-\sum_{n \geq 1} \frac{t_{n}}{n}\Tr(K^n)} e^{-\sum_{n \geq 1} \frac{\tilde{t}_{n}}{n}\Tr(M^n) }}\hspace{40pt}&\\
    = \frac{(\text{det}_Q(X))^{R}(\text{det}_{R}(V))^{Q}}{Z_{QR}}& \int dS dS^{\dagger}_{Q \times R} \exp(\frac{N}{g}\Tr(V S^{\dagger} X S) + N \Tr \log \left( 1- SS^{\dagger}\right) )
\end{split}
\end{equation}
For the special case of $Q=R$ and $V=X$, the matrix integral above is related to the so-called {\it dually-weighted} matrix models studied in \cite{Kazakov:1995ae, Das:1989fq}. Thus the relationship to string theory discussed in this paper also provides a dual description of a special class of dually-weighted matrix models. This F-type description also appeared under the guise of the "$\rho$-matrix model" in \cite{komatsuOCO} and more recently using twistor string variables in \cite{Caron-Huot:2023wdh}. Specifically, after setting $g=1$, $X = \mathbb{I}_{Q\times Q}$ and $V = \mathbb{I}_{R\times R}$, we reproduce Eq.(5.37) of \cite{Caron-Huot:2023wdh} for $n=2$, $m_1=Q$ and $m_{2}=R$. 

\begin{figure}[ht!]
    \centering
\includegraphics[scale=0.7]{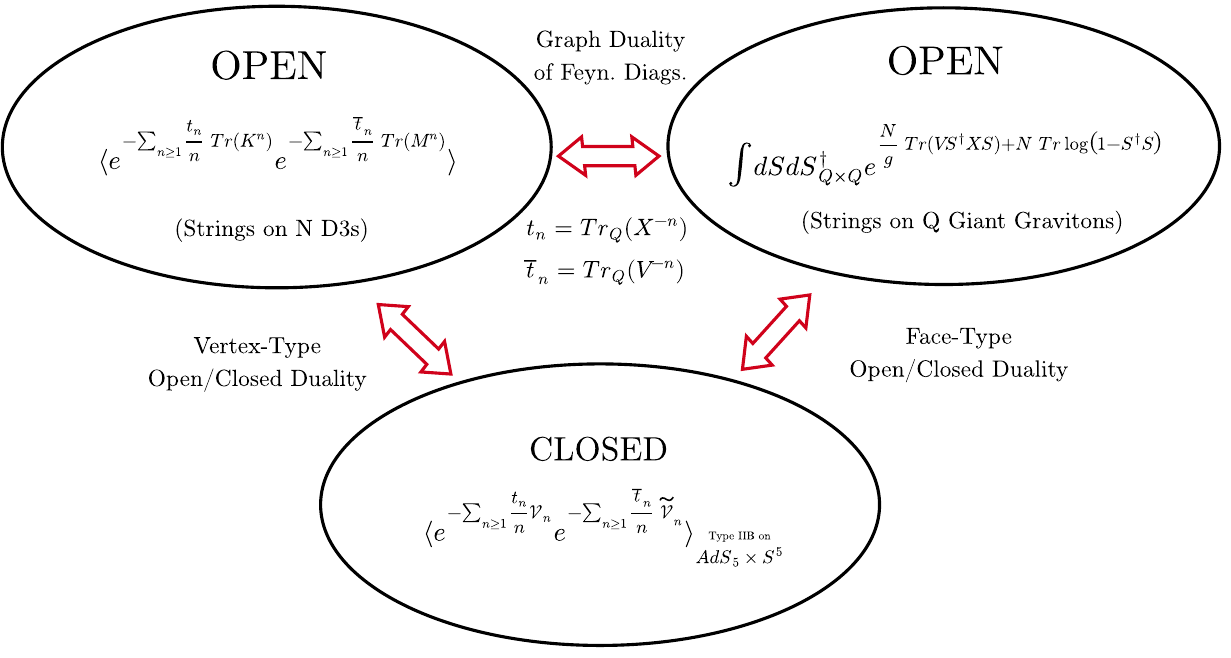} 

    \caption{\textbf{Open-Closed-Open Triality for the 1/2 SUSY Sector of $\mathcal{N}=4$ SYM } For the highly protected correlators considered in this paper, we find two open string descriptions in the form of two distinct matrix integrals. In the $K,M$ model, single trace operators map directly to closed string vertex insertions. From a Feynman diagram perspective, it is the vertices that go over to marked points on the dual worldsheet. We call this V(ertex)-type open/closed duality. The $S,S^{\dagger}$ model encodes the same generating function of closed string correlators, but in a very different way. This can be most clearly seen from the fact that the sources for vertex operators, the $t_{k}, \bar{t}_{k}$, appear non-linearly in that matrix integral. Its Feynman diagram expansion generates the same weighted sum over closed string worldsheets as the $K,M$ model, however its ribbon graphs are directly the Strebel graphs of the dual worldsheets. This means every face of the diagram goes over to a marked point. It plays the same role the Kontsevich model does for the $(2,1)$ minimal string. The complex matrices $S,S^{\dagger}$ can be interpreted as open strings on giant graviton branes, while the $K,M$ matrices capture open strings on the $N$ D3 branes giving rise to $\mathcal{N}=4$ SYM. }
    \label{fig:OCON=4}
\end{figure}

\subsection{The "Kontsevich"-like Model for the 1/2 SUSY Sector of $\mathcal{N}=4$ SYM from Giant Graviton Branes} \label{sec:OCOGiants}

By construction, both the $K,M$ and $S,S^\dagger$ matrix integrals encode the same generating function $Z[t_{k},\bar{t}_{k},s_{k}]$, see Eq.(\ref{eq:finaleqgen}). The $K,M$ model, which we deduced from a reduction of $\mathcal{N}=4$ SYSM, plays the role of the "V(ertex)"-type open string description, where single trace operators are dual to $1/2$ BPS vertex operators on the closed string worldsheet of Type IIB string on $AdS_{5} \times S^{5}$: $\Tr(K^n) \leftrightarrow \mathcal{V}^{1/2 BPS}_{n} $. There is no such obvious dictionary between closed string vertex operators and the complex matrices $S$. Instead, the physical picture is that the faces of the F-type model's ribbon graphs correspond to D-brane boundary conditions on the worldsheet, which can ultimately be replaced by a sum of local vertex operators. 

The $S, S^{\dagger}$ model plays the same role in the context of the 1/2 SUSY sector of Type IIB string on $AdS_{5} \times S^{5}$ as the Kontsevich model does for the $(2,1)$ minimal string. In the (2,1) minimal string, the two matrix descriptions descended from open strings on either ZZ branes or FZZT branes. In fact, Gaiotto and Rastelli even derived the  $Q \times Q $ Kontsevich model directly from cubic open string field theory on $Q$ FZZT branes in the  $(2,1)$ minimal string \cite{Gaiotto:2003yb}. The two types of branes underlying the V- and F-type descriptions are on a slightly different footing. In some sense, the ZZ branes discussed above "created" the $(2,1)$ background \footnote{Making this precise is difficult. It is not quite clear how to describe these branes as embedded in the equivalent of flat space, i.e. before their reaction. Identifying the matrix degrees of freedom as open strings on ZZ branes arises by treating a single eigenvalue of the matrix model as a probe. The probe dynamics of a single eigenvalue were matched to the expected properties of a ZZ instanton. If we treat the eigenvalues of the double-scaled matrix models in the free fermion picture, the emergence of the spacetime background is the statement of the existence of a Fermi sea of eigenvalues.\cite{Takayanagi:2004ge, Takayanagi:2005tq}}, while the FZZT branes can be treated in a probe limit, "living in" the $(2,1)$ background. The effect of the FZZT on the background can be equivalently described in terms of a complicated coherent state of closed string, explaining its equivalent to a generating function of closed string correlators.

\begin{figure}[ht!]
    \centering
\includegraphics[scale=0.70]{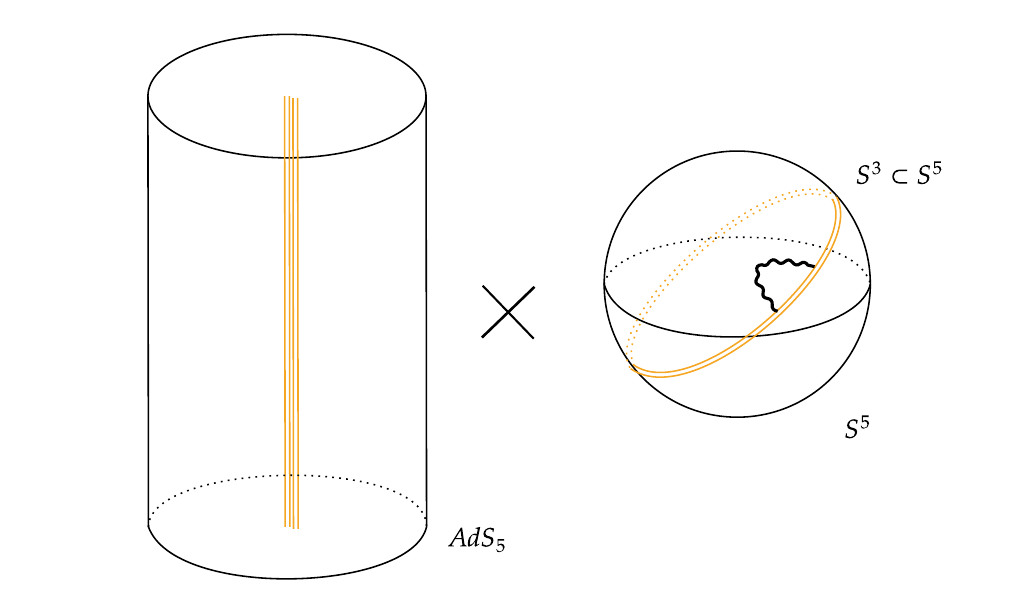} 

    \caption{\textbf{F-Type Model as Strings on Giant Graviton Branes } If we consider the back reaction of the original $N$ D3 branes on the ambient flat spacetime, we can view the additional branes dual to determinant insertions in the $K,M$-model as giant graviton branes in an $AdS_{5} \times S^{5}$ geometry. The complex matrices $S,S^{\dagger}$ appearing in the F-type model then descend from open strings on these giant graviton branes. These giant gravitons play a role similar to the FZZT branes in the $(2,1)$ minimal string, which also gave rise to the F-type open string dual, namely the cubic Kontsevich model. }
    \label{fig:GiantGravitons}
\end{figure}

In the present case, the V-type description arises from the original $D3/D3$ open strings, while the $D3'/D3''$ open strings furnish the F-type model. That said, we can consider the usual Maldacena large $N$ limit of the $D3$s. They famously backreact on the ten-dimensional flat space geometry and give rise to an $AdS_{5} \times S^{5}$ background. We can consider the other two stacks of branes as living directly in the emergent $AdS_{5} \times S^{5}$ geometry: these are precisely the giant graviton branes dual to determinant insertions in $\mathcal{N}=4$ SYM. Giant graviton branes wrap an $S^3$ inside the $S^{5}$ and are point-like in the $AdS_{5}$ factor. The $Y_{j}^{I}$ vectors determine how that $S^{3}$ sits inside the $S^{5}$, while the $w_{j}$ specify where their worldvolume intersects the AdS boundary. For simplicity, we can consider the case of $Q=R$ in Eq.(\ref{eq:Fmodel}). The physical picture is now, at leading order in $N$, that of $Q$ coincident giant graviton branes in the bulk. We can then interpret the matrices $S_{a \mu}$ as arising from open strings on the $Q$ giant gravitons, as illustrated in Fig. \ref{fig:GiantGravitons}. This realizes, in this very restricted subsector of the AdS/CFT correspondence, a conjecture of \cite{Joburg} sketching how open strings on giant graviton branes might furnish an F-type description of the AdS/CFT correspondence.

\subsection{Integrating In/Out as "Dynamical" Graph Duality} \label{sec:OCOgraphduality}

In this section, we would like to reinterpret the process of integrating in and out various open strings as implementing a dynamical graph duality of the Feynman diagrams of each theory. Viewing the gauge theory Feynman diagrams as open string worldsheets, this procedure can be understood as different holes opening or closing up. 

Since we have three stacks of branes, each open string worldsheet could in principle have three types of boundaries, one for each set of D-branes. When a particular type of boundary shrinks to zero size on the worldsheet, leaving behind a marked point, the associated Miwa-variable, $t_{k}, \bar{t}_{k}$ or $s_{k}$, appears manifestly in that description. 

To see this at play, we will first diagrammatically interpret each of the four steps of Sec. \ref{sec:intinout} at the level of the Feynman rules for each matrix integral. We will then bring all these ingredients together to track the transformation of our favorite Feynman diagram, and show how it goes over to its graph dual after the four steps.

\textbf{STEP 1:} In the $K,M$ model, we can compute the expansion of $\log Z[t_{k},\bar{t}_{k},s_{k}]$ in powers of $t_{k}$ and $\bar{t}_{k}$ using simple Feynman diagrams. Each factor of $t_{k}$ arises from a $K$-matrix vertex of valency $k$, and similarly for $\bar{t}_{k}$ and the $M$ vertices. Let us now try to identify this contribution from the point of view Eq.(\ref{eq:KMwithferms}), i.e. after we have integrated in the fermions $\chi$ and $\psi$. There are no longer any higher point vertices for $K$ and $K$, as in the original action of Eq.(\ref{eq:generatingfunc}). The only interaction vertices are the two Yukawa-like couplings shown below in Fig. \ref{fig:KMYukawas}.

\begin{figure}[ht!]
    \centering
\includegraphics[scale=0.70]{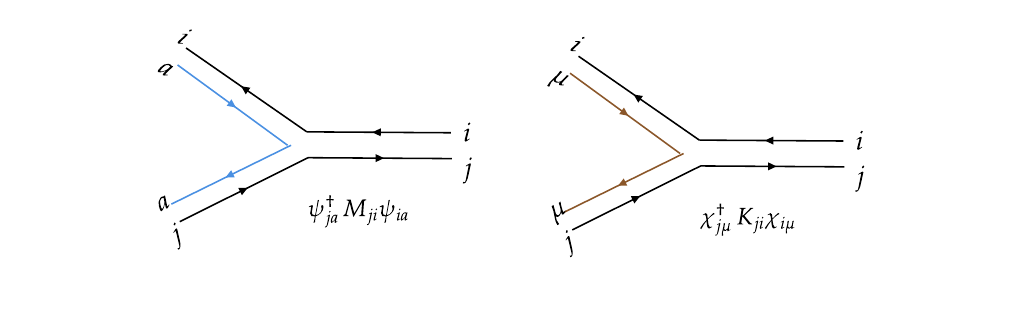} 

    \caption{\textbf{Yukawa Couplings in the $K,M + \psi,\chi$ Model} }
    \label{fig:KMYukawas}
\end{figure}

\begin{figure}[ht!]
    \centering
\includegraphics[scale=0.80]{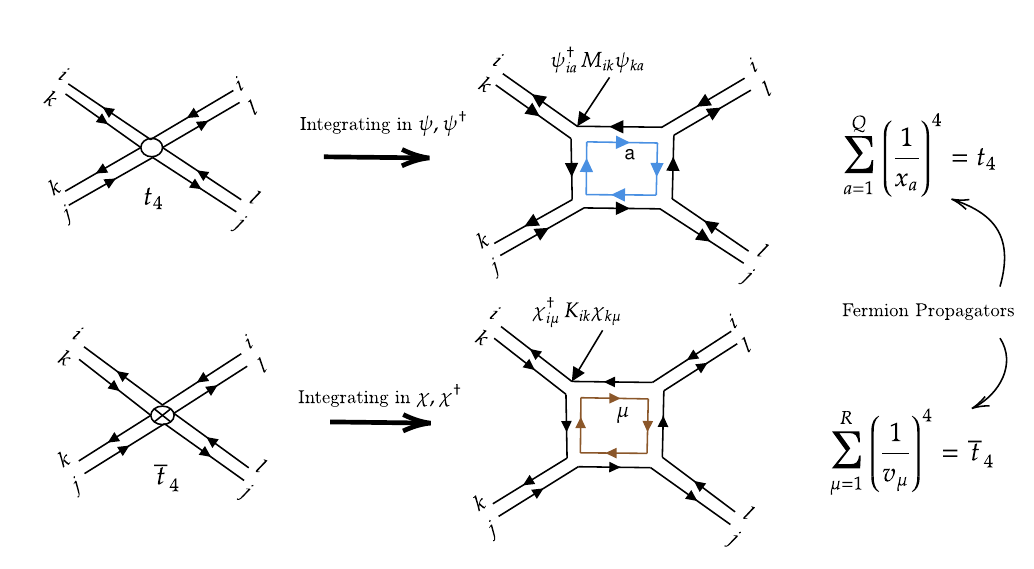} 

    \caption{\textbf{(STEP 1) Integrating in the Fermions } The two types of vertices of valency $k$ in the original $K,M$ model gets replaced by two new types of $k$-sided faces in the mixed bosonic/fermion theory. Each edge of these new faces is a fermionic propagator. The appearance of $X$ and $V$ in the "kinetic" term for the fermions shows that each edge receives a factor of $x_a^{-1}$ or $v_{\mu}^{-1}$. These propagators ensure that the new $k$-sides faces reproduce the weights $t_{k}$ and $\bar{t}_{k}$ of the original $K,M$ model vertices. }
    \label{fig:Vertexblowup}
\end{figure}

In the mixed fermion/boson theory, the two types of vertices of the $K,M$ model are replaced by a juxtaposition of cubic Yukawa couplings and fermionic propagator edges, see Fig. \ref{fig:Vertexblowup}. Effectively, step 1 of Sec. \ref{sec:intinout} replaces the two types of vertices with two new types of faces. The faces have either an $a \in {1,..,Q}$ or $\mu \in {1,..,R}$ color index flowing in it, depending on whether they originated as a $K$ or $M$ vertex, respectively. The appearance of these new types of faces reflects the fact that we have re-introduced the open strings ending on the other two stacks of branes: each type of face is one of the new possible boundary conditions for the open string giving rise to the Feynman diagram. 

How is the initial weighting of each original vertex of valency $k$, by $t_{k}$ or $\bar{t}_{k}$, reproduced in this fermionic picture? As shown in Fig. \ref{fig:Vertexblowup}, such a vertex is replaced by an emergent face with $k$ sides. Each side of this face corresponds to one fermionic propagator. As can be seen from the fermionic part of the action in Eq.(\ref{eq:KMwithferms}), each propagator contributes a factor of $x_{a}^{-1}$ (or $v_{\mu}^{-1}$). To obtain the weight of each new face, we multiply the contribution coming from each of its edges, then sum over the inner color index. This means that the two new types of faces, with $k$ sides, will receive an overall weighting factor of

\begin{equation}
     \sum_{a=1}^{Q} \left( \frac{1}{x_{a}}\right)^k = t_{k} \qquad  \sum_{\mu=1}^{R} \left( \frac{1}{v_{\mu}}\right)^k = \bar{t}_{k} ,
\end{equation},

thereby reproducing the original weighting factor of each vertex in the $K,M$ model.

\begin{figure}[H]
    \centering
\includegraphics[scale=0.80]{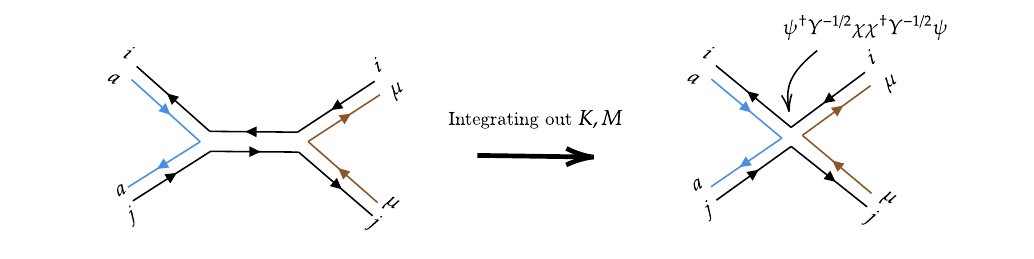} 

    \caption{\textbf{(STEP 2) Integrating out $K,M$ } Integrating out $K,M$ collapses the edge connecting two different Yukawa vertices. It generates a quartic vertex, representing the $\psi^{\dagger}\psi \chi^{\dagger} \chi$ term in the action of the purely fermionic matrix integral of Eq.(\eqref{eq:fermionicint}), }
    \label{fig:KMout}
\end{figure}

\textbf{STEP 2:} What does integrating out the bosonic matrices $K$ and $M$ look like at the level of the Feynman diagrams? We can see from the purely fermionic matrix integral in Eq.(\ref{eq:fermionicint}) that it gives rise to a quartic interaction of the form $\psi^{\dagger}\psi \chi^{\dagger} \chi$. This quartic interaction can be understood diagrammatically as the joining of the two different cubic Yukawa couplings. The $KM$-propagator edge connecting the two collapses, leaving behind an effective quartic vertex, as in Fig. \ref{fig:KMout}.

\begin{figure}[ht!]
    \centering
\includegraphics[scale=0.70]{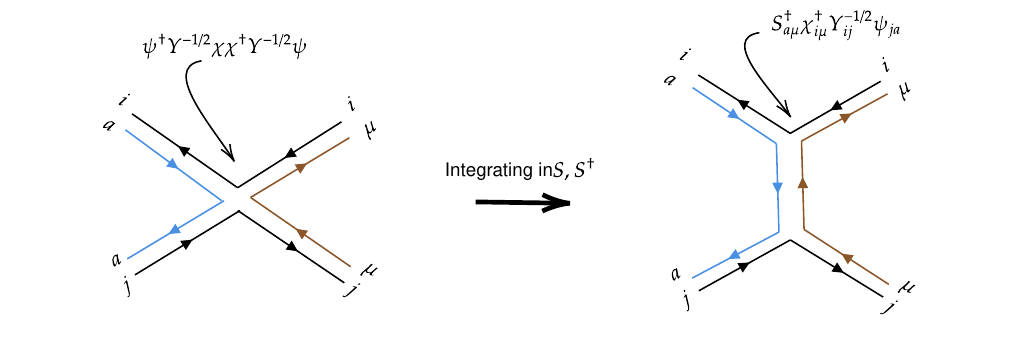} 

    \caption{\textbf{(STEP 3) Integrating in $S,S^{\dagger}$ } Integrating in the bosonic complex matrices $S,S^{dager}$ resolves the fermionic quartic vertex in a "different channel". It is now represented by two cubic interaction vertices, joined together by an $SS^{\dagger}$ propagator. Unlike $K$ and $M$, $S$ and $S^{\dagger}$ couple to both fermions simultaneously, $\psi^{\dagger}_{ai}S_{a \mu} \chi_{\mu i}$.}
    \label{fig:SSdagin}
\end{figure}

\textbf{STEP 3:} In some sense, the third step resolves the quartic fermion vertex in a "different channel", see Fig. \ref{fig:SSdagin}. By that, we mean the effect of the Hubbard-Stratanovich transformation is to rewrite the quartic vertex as two cubic vertices, joined together by a propagator for the complex matrices $S,S^{\dagger}$. These cubic interactions arise from a new set of Yukawa-like couplings between the complex matrix $S_{a\mu}$ and the fermions. As rectangular matrices, $S$ and $S^{\dagger}$ couple to both types of fermions simultaneously, as can be seen from the $\psi^{\dagger}_{ai}S_{a \mu} \chi_{\mu i}$ term in Eq.(\ref{eq:SSdagin}). Note that each such vertex comes with a factor of $y_{i}^{-1/2}$. At this point of the construction, all the original $k$-sided faces of the $K,M$ model, associated to color indices $i,j \in {1,..,N}$, have $2k$ cubic vertices at their corners. This guarantees that it remains weighted  by a factor of $s_{k}=\sum_{i=1}^{N} y_{i}^{-k}$.

\begin{figure}[H]
    \centering
\includegraphics[scale=0.75]{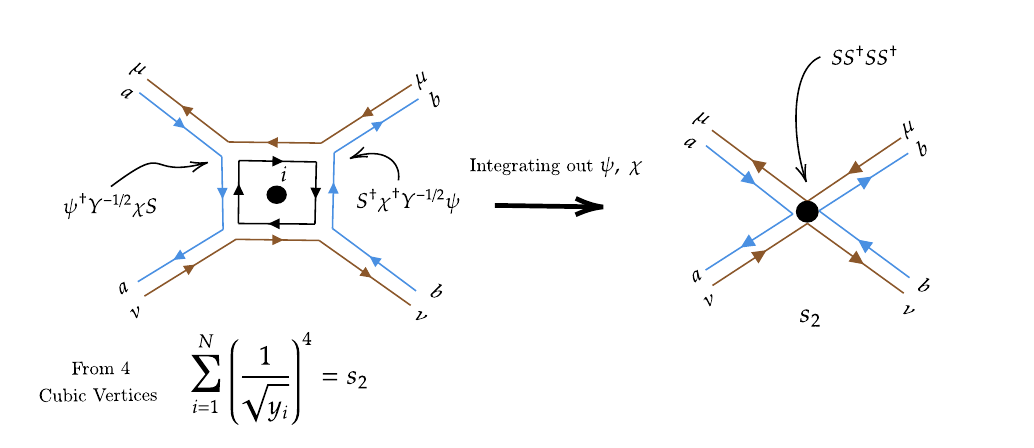} 

    \caption{\textbf{(STEP 4) Integrating the Fermions Out } The original faces of the $K,M$ model are bordered by fermion propagators. Upon integrating out the fermions, the original faces collapse, going over to vertices of the $S,S^{\dagger}$ model. From the point of view of the open string worldsheet, the boundaries associated to the original N D3-branes have shrunk to zero size. This final step implements the full graph duality, where vertices and faces of the V- and F-type open string descriptions are interchanged. }
    \label{fig:fermsout}
\end{figure}

\textbf{STEP 4:} The fourth and final step involves integrating back out the fermions. Recall that all the original faces of the $K,M$ model consist of edges corresponding to the fermionic propagators. Indeed, the $S$ and $S^{\dagger}$ edges only carry $a$ and $\mu$ indices. Integrating out the fermions collapses all such edges: the original faces of the $K,M$ shrink to a vertex of the $S,S^{\dagger}$ model, as shown in Fig. \ref{fig:fermsout}. This implements the total graph duality of the Feynman diagrams, where faces go over to vertices. As explained shortly above, this new vertex comes with a weighting factor $s_{k}$, if the original face of the $K,M$ model Feynman diagram had $2k$ sides.

Each vertex and face of the original $K,M$ model therefore obeys clear diagrammatic transformation rules upon integrating in/out of the various fields. We therefore reach the conclusion that each and every Feynman diagram appearing in the expansion of the generating function of correlators corresponds to exactly one Feynman diagram of the Kontsevich-like $S,S^{\dagger}$ model. In particular, it goes over to its graph dual, since the faces of the original Feynman diagram become vertices, and vice-versa. Recall that we associated the graph dual of the $K,M$ Feynman diagrams to the Strebel graphs of the worldsheet in Sec. \ref{sec:StrebGraphsVsFD}. We thus conclude that the $S,S^{\dagger}$ model generates the same weighted sum over dual closed strings, since its Feynman diagram expansion is literally the identical sum over Strebel graphs. 

Again, we stress this is not the same statement that single trace correlators of the $K,M$ model map onto single trace observables of the  $S,S^{\dagger}$ model. This, in turn, is not true. All Feynman diagrams contributing to one correlator in the V-type description will in general be mapped onto Feynman diagrams in the other model which individually are different contributions to several correlators. It is a not mapping between two correlators, but only between two Feynman diagrams.

\textbf{An Example: Back to our Main Feynman Diagram}

It is perhaps helpful to piece together all these various steps at once, and follow through the transformation undergone by a particular Feynman diagram. We will of course choose our favorite Feynman diagram, showing how it goes over to its graph dual. 

In the $K,M$ model, the contribution of order $t_{2}^{2}\bar{t}_{2}^2$ to $\log Z[t_{k},\bar{t}_{k},s_{k}]$ comes from the connected correlator 

\begin{equation}
    \left\langle \left( \frac{1}{2} \Tr K^2 \right)^2 \left( \frac{1}{2} \Tr M^2 \right)^2 \right\rangle_{c} 
\end{equation}

\begin{figure}[ht!]
    \centering
\includegraphics[scale=0.65]{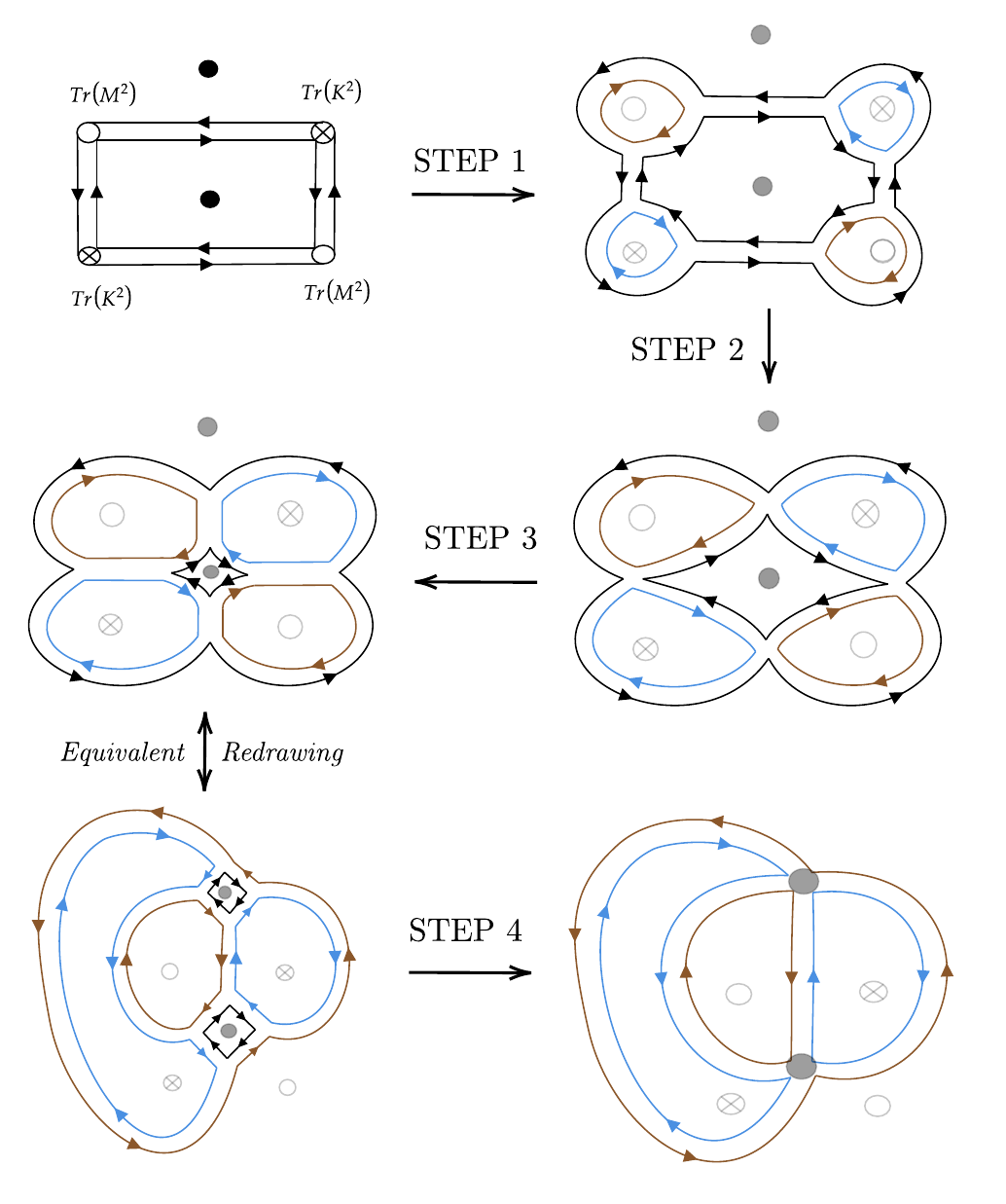} 

    \caption{\textbf{Steps 1 through 4 applied to our Feynman Diagram } The first diagram in the top left corner is the only Feynman diagram in the $K,M$ model contributing to $\log Z[t_{k},\bar{t}_{k},s_{k}]$ at order $t_{2}^{2}\bar{t}_{2}^2$. By integrating in the fermions, the vertices of valency two are replaced with new two-sided faces (in blue and brown). Integrating out $K,M$ gives rise to the quartic fermionic vertices. Integrating in the $S,S^{\dagger}$ matrices factorizes the quartic vertex into two cubic vertices once again. In the penultimate step, we have simply redrawn the same planar diagram for ease of visualization. Finally, integrating back out the fermions shrinks the original faces (black loops) to the vertices of the $S,S^{\dagger}$ Feynman diagram. We have included the original vertices (white crossed and uncrossed vertices) and faces (black dots) throughout the different diagrams, to highlight the process of dynamical graph duality. The first and last diagrams are (total) graph duals of one another. } 
    \label{fig:MainFDOCO}
\end{figure}

which receives contributions only from a single Feynman diagram, as we first encountered below Fig. \ref{fig:MainFD}. Fig. \ref{fig:MainFDOCO} shows how steps 1 through 4 transform the original Feynman diagram into its graph dual. More precisely, it identifies the Feynman diagrams of each matrix integral which evaluate to the same $t_{2}^{2}\bar{t}_{2}^2$ contribution to $\log Z[t_{k},\bar{t}_{k},s_{k}]$.

\subsection{Partial Graph Duals}\label{sec:OCOpartials}

One of the main differences between the open-closed-open triality of the one matrix model discussed in \cite{DSDI} and that of the two-matrix integral here is the existence of "partial" duals. These open descriptions generate Feynman diagrams which are only partially graph dual to the ribbons graphs of the $K,M$ model. More precisely, we will see that only one set of the vertices of the $K,M$ model goes over to faces of the ribbon graphs generated by these new matrix models. The $D3'/D3'$ and $D3''/D3''$ open strings will furnish these two partial dual descriptions, giving rise to bosonic two-matrix integrals, with matrices of size $Q \times Q$ or $R \times R$, respectively.

To obtain these bosonic partial duals, we first return to the purely fermionic integral in Eq.(\ref{eq:fermionicint}). We can in fact integrate out one fermion species at a time, since any one set of fermions appears at most quadratically. This lands us on the effective description purely in terms of either 
\begin{itemize}
    \item \textbf{the D3/D3' open strings }
\begin{align}\label{eq:justpsi}
    Z[t_{k},\bar{t}_{k},s_{k}] =\frac{\det_N(Y)^Q}{\det_Q(X)^N}\int d\psi d\psi^{\dagger}_{N \times Q}\exp\bigg(\psi^{\dagger}_{ia}X_{ab}Y_{ij}\psi_{jb} - \sum_{k=1}^{\infty}(\frac{g}{N})^{k}\frac{\bar{t}_{k}}{k}\Tr_N(\psi\psi^{\dagger})^k\bigg), 
\end{align}
or in terms of 
\item  \textbf{the D3/D3'' strings }
\begin{align}\label{eq:justchi}
    Z[t_{k},\bar{t}_{k},s_{k}] =\frac{\det_N(Y)^R}{\det_R(V)^N}\int d\chi d\chi^{\dagger}_{N \times R}\ \exp\bigg(\chi^{\dagger}_{i\mu}V_{\mu\nu}Y_{ij}\chi_{j\nu}- \sum_{k=1}^{\infty}(\frac{g}{N})^{k}\frac{t_{k}}{k} \Tr_N(\chi\chi^{\dagger})^k\bigg).
\end{align}
\end{itemize}

\begin{figure}[ht!]
    \centering
\includegraphics[scale=0.80]{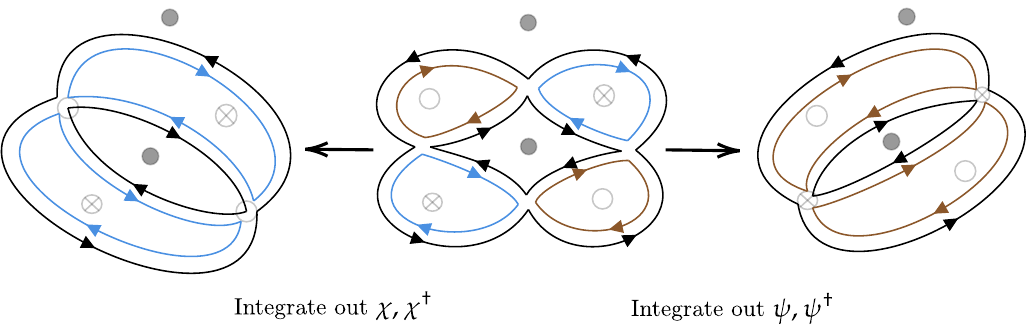} 

    \caption{\textbf{Feynman Diagrams of the One Fermion Description } The middle diagram is the transformation of our favorite Feynman diagram after step 2, as also shown in Fig. \ref{fig:MainFDOCO}. The quartic vertices of this middle graph arise from the $\psi^{\dagger}\psi \chi^{\dagger}\chi$ term in the two-fermion action. By integrating out one of the two fermion species, we arrive at the diagrams on either the left or right. These new diagrams share in common either the $K$ or $M$ vertices with the ribbon graphs of the $K,M$ model (in shadowed white vertices). This diagramatically reflects the fact that, as an intermediate step in the integrating in/out process, we find $M_{ij} \sim \psi^{\dagger}_{ia}\psi_{ja}$ and $K_{ij} \sim \chi^{\dagger}_{i\mu}\chi_{j\mu}$. The two remaining types of faces correspond to the original faces and the second type of vertex of the $K,M$ model. } 
    \label{fig:onefermout}
\end{figure}

The takeaway from the above two expressions for the generating function is that the original matrices $K_{ij}$ and $M_{ij}$ can equally be described in terms of mesonic operators $\psi^{\dagger}_{ia}\psi_{ja}$ and $\chi^{\dagger}_{i\mu}\chi_{j\mu}$. This can be seen by taking derivatives with respect to $\bar{t}_{k}$ ($t_{k}$) in Eq.(\ref{eq:justpsi}) (Eq.(\ref{eq:justchi})), which we know is dual to inserting $\Tr K^k$ ($\Tr M^k$) in the $K,M$ matrix integral. In either of these single fermion descriptions, one set of Miwa times appears directly at the level of the action. Geometrically, we have closed back up one of the holes on the open string worldsheet introduced in Step 1. These holes correspond to the faces of the fermionic degrees of freedom we have integrated back out. The Feynman diagrams of these single fermion theories share in common either the $K$ or $M$ vertices with the ribbon graphs of the $K,M$ model. In that sense, they are partial graph duals. The two remaining types of faces correspond to either the D3 \& D3' or D3 \& D3'' boundaries on the open string worldsheet. We have illustrated, in Fig. \ref{fig:onefermout}, the ribbon graphs generated by the single fermionic integrals which are equivalent to our main Feynman diagram.

There are two other types of strings we have neglected to mention so far: the D3'/D3' and D3''/D3'' open strings. There should exist an effective description of strings on either of these stacks of branes. We now find them. We expect these open strings to give rise to square matrices, of size $Q \times Q$ for the strings stretched solely between the D3's and size $R \times R$ for those on the D3''s. 

We can directly obtain those from the single species fermionic integrals in Eqs. \ref{eq:justpsi} and \ref{eq:justchi}. This was done by two of the authors previously in \cite{DSDI} for the fermionic integral in Eq.(\ref{eq:justpsi}). While we again refer the reader to the appendix for the intermediate steps, the basic idea is to introduce an "exact" collective field for either $\psi^{\dagger}_{ia}\psi_{ib}$ or $\chi^{\dagger}_{i\mu}\chi_{i\nu}$. Concretely, one can insert a representation of the identity in the fermionic integrals via a delta-function integral

\begin{align}
    1 = \int dB_{Q \times Q} \prod_{a,b=1}^{Q} \delta(A_{ba} + g\psi^{\dagger}_{ia}\psi_{ib}) = \inv{Z_Q}\int dA dB_{Q \times Q} e^{- \frac{N}{g}B_{ab} \left( A_{ba} + g\psi^{\dagger}_{ia}\psi_{ib} \right) } \\
    1 = \int dD_{R \times R} \prod_{\mu,\nu=1}^{R} \delta(D_{\nu \mu} + g\chi^{\dagger}_{i\mu}\chi_{i\nu}) = \inv{Z_R}\int dC dD_{R \times R} e^{-\frac{N}{g} C_{\mu \nu} \left( D_{\nu \mu} + g\chi^{\dagger}_{i\mu}\chi_{i\nu}\right) } 
\end{align}

One can then integrate out all fermions. We arrive at the following bosonic "partial duals":

\noindent $\bullet$ \textbf{The D3'/D3' Strings}
\begin{align}\label{eq:AB model}
    Z[t_{k}, \bar{t}_{k},s_{k}]=& \frac{\det_{Q}(X)^{Q}}{Z_Q}\int dA \ d B_{Q \times Q} \exp\bigg({-\frac{N}{g}\text{Tr}_Q(\sqrt{X} B\sqrt{X} A) - \sum_{k \geq 1} (-1)^k\frac{\bar{t}_{k}}{k} \Tr(A^{k}) -\sum_{k \geq 1} \frac{s_{k}}{k} \Tr(B^{k})  }\bigg) \\
    = &\bigg( \frac{\text{det}_{Q}(X) \text{det}_{R}(V)}{\text{det}_N(Y)}\bigg)^{Q}\frac{1}{Z_Q}\int dA \ d B e^{-\frac{N}{g}\text{Tr}_Q(\sqrt{X} B\sqrt{X} A)} \prod_{\mu=1}^R\text{det}_Q(v_{\mu} + A)^{-1}  \prod_{i=1}^N\text{det}_Q(y_i-B)\
\end{align}
\\
\noindent $\bullet$ \textbf{The D3''/D3'' Strings}
\begin{align} \label{eq:CD model}
Z[t_{k}, \bar{t}_{k},s_{k}] =& \frac{\det_R(V)^{R}}{Z_R}\int dC \ dD_{R \times R} \exp\left( -\frac{N}{g}\Tr(\sqrt{V}C\sqrt{V}D) - \sum_{k \geq 1} \frac{s_{k}}{k} \Tr(C^{k}) - \sum_{k \geq 1}(-1)^k \frac{t_{k}}{k} \Tr(D^{k})\right)\\
= & \bigg(\frac{\det(X)\det(V)}{\det(Y)}\bigg)^{R}\inv{Z_R}\int dC dD \exp(-\frac{N}{g}\Tr(\sqrt{V}C\sqrt{V}D)) \prod_{i=1}^N \det(y_i - C) \prod_{a=1}^{Q} \det(x_a + D)^{-1}
\end{align} 
where $Z_Q= (2\pi g/N)^{Q^2}, Z_R= (2\pi g/N)^{R^2}$. We now understand how the type of open-string dual we looked at in \cite{DSDI} is in fact one of these "partial dual" pictures. By that, we mean that if we compare their Feynman diagrams with those of the $K,M$-model, we see they share one vertex in common. This will translate to the fact that either traces of $K$ or traces of $M$ will become simple traces in the other matrix model. In fact, from the explicit Miwa times appearing in Eqs. \ref{eq:AB model} and \ref{eq:CD model}, we arrive at the following "operator" dictionary

\begin{equation}
    \Tr_{N}(M^{n}) \leftrightarrow (-1)^n \Tr_{Q}(A^{n}) \qquad  \Tr_{N}(K^{n}) \leftrightarrow (-1)^n \Tr_{R}(D^{n}) 
\end{equation}

This dictionary is, at face value, of limited use. The bosonic partial duals only share one set of Miwa variables in common with the $K,M$ model, so there is again no clean mapping of correlators between the two models.

\begin{figure}[H]
    \centering
    \includegraphics[scale=0.7]{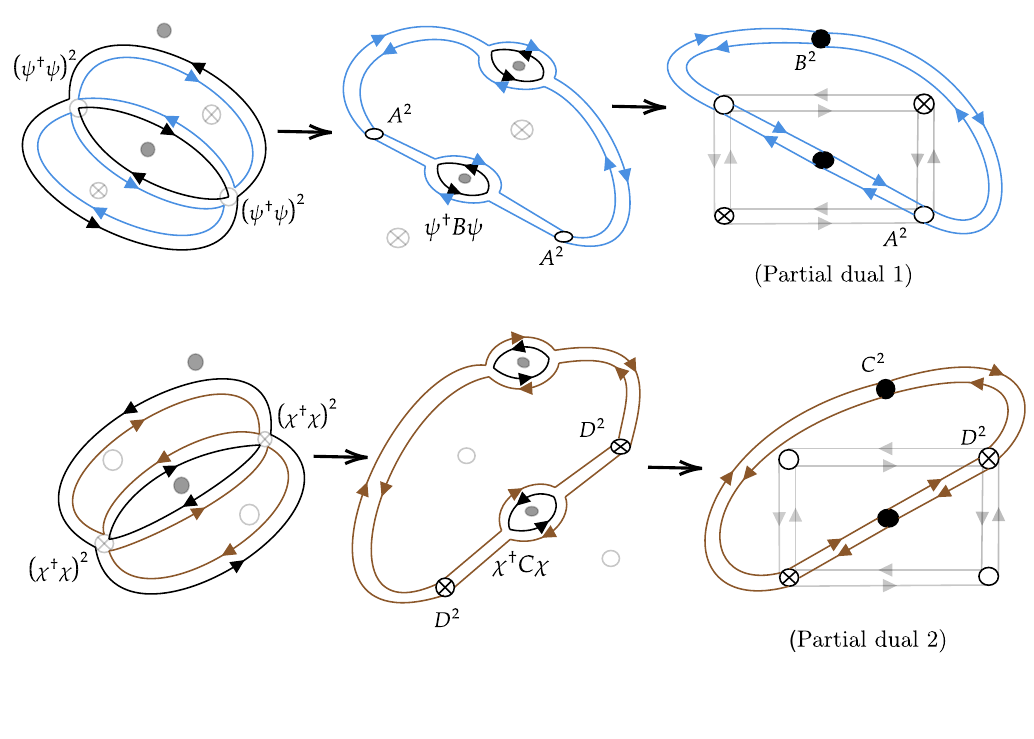}
    \caption{\textbf{Feynman Diagrams of the Bosonic Partial Duals } Partial graph duality is a novel feature of the open-closed-open triality of the two-matrix versus one-matrix models. The $A,B$ or $C,D$ models generate Feynman diagrams with two types of vertices and one type of face. Their ribbon graphs share a vertex in common with the diagrams of the $K,M$ model, determined by which of the two Miwa variables ($t_{k}$ or $\bar{t}_{k}$) appear manifestly in the action.  }
    \label{fig:partialdualFDs}
\end{figure}
\begin{figure}[H]
    \centering
\includegraphics[scale=0.7]{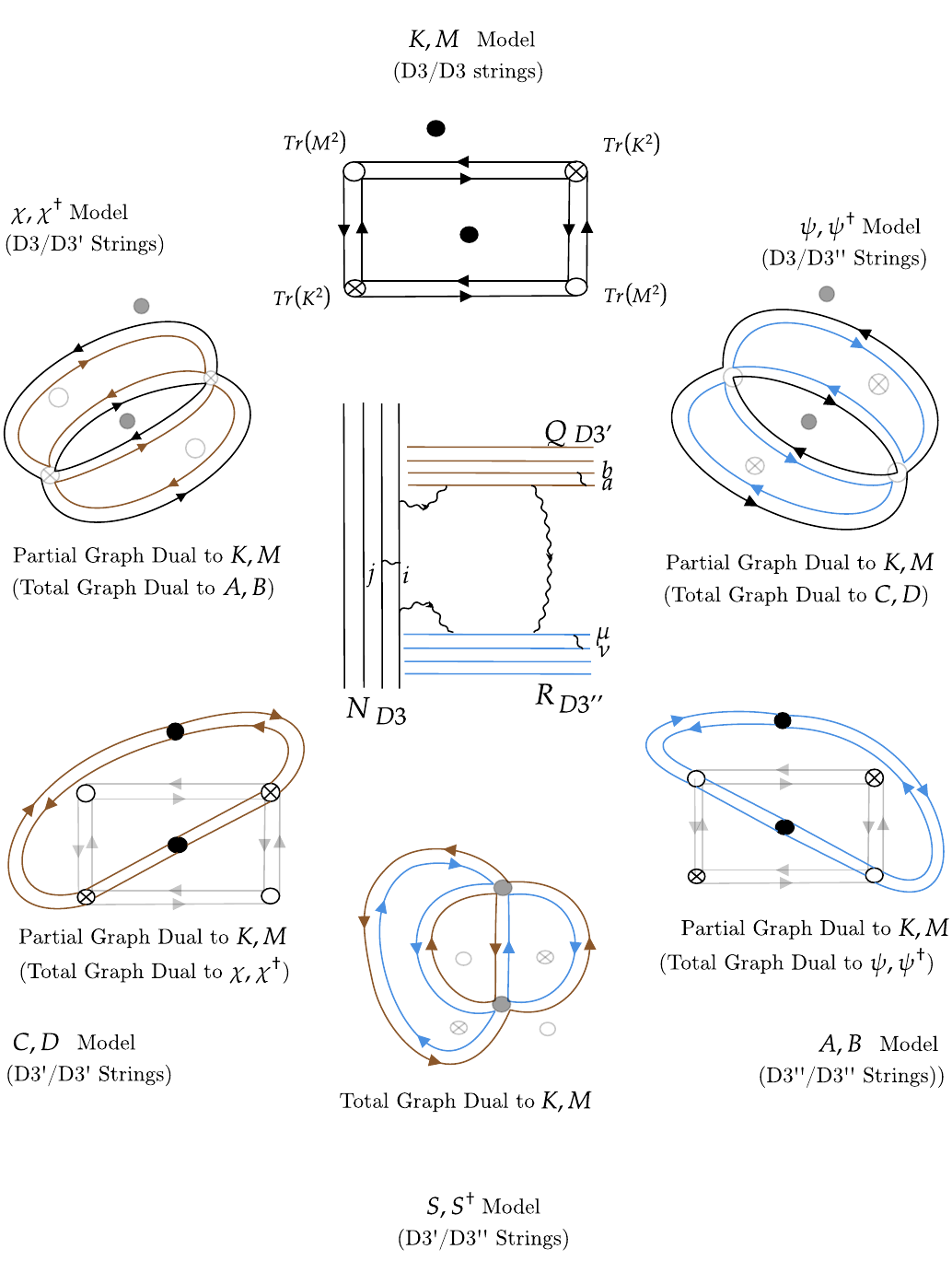} 

    \caption{\textbf{Summary of the Six Open String Descriptions \& their Feynman Graph Duality } By considering three stacks of branes, we find six equivalent open string descriptions in the guise of six different matrix integrals. There is a one-to-one correspondence between their Feynman diagrams: they are all related via (partial) graph duality. The $D3'/D3'$ and $D3''/D3''$ open strings furnish bosonic matrix models whose Feynman diagrams are partial duals to those of $K,M$: they share one set of vertices in common with the graphs of $K,M$. Strings stretched between two stacks of branes have two types of faces. These faces represent the two types of boundaries on the open string worldsheets. The two fermionic matrix integrals arise from the $D3/D3'$ and $D3/D3''$ strings. Their Feynman diagrams share one set of vertices and one set of faces common with the $K,M$ graphs. Finally, the $D3'/D3''$ open strings lead to the $S,S^{\dagger}$ model. Its Feynman diagrams are the total graph duals of those of the $K,M$ model. It thus plays the role of the F(ace)-type open description.} 
    \label{fig:OCOfromBranes}
\end{figure}

In summary, we have found a total of six different open string descriptions, whose Feynman diagrams are all related to each via (partial) graph duality. As such, they contain identical information, and all encode the same dual closed-string worldsheets. All open descriptions are manifestly equivalent, since they can all be obtained starting from a system of three stacks of branes and integrating out some subset of open strings. We summarize our findings in Fig. \ref{fig:OCOfromBranes}. 

\subsection{Extremal Correlators from the $c=1$ String} \label{sec:IMmodel}

One of the most surprising outcomes of open-closed-open triality in the 1/2 SUSY sector of $\mathcal{N}=4$ SYM is the appearance of the c=1 string. Concretely, a simple field redefinition takes the effective theory of the $D3'/D3'$ open strings in Eq.(\ref{eq:AB model}) into the so-called Imbimbo-Mukhi matrix model \cite{ImbimboMukhi}.

After setting $y_{i}=1$(which implies $s_k = \Tr(Y^{-k}) =N $), a simple  shift $B \rightarrow 1- B$ followed by a rescaling $B\rightarrow X^{-\half} B X^{\half}$ takes the $A,B$ model of Eq.(\ref{eq:AB model}) into 

\begin{align}
    Z[t_{k},\bar{t}_{k},s_{k}=N] &=(-1)^{Q^2}\frac{({\det(X) \det(V) })^Q}{Z_Q}\nonumber\\
    &\hspace{0pt}\times\int dA dB_{N \times N} \exp{\bigg(\frac{N}{g}\Tr(ABX)-\frac{N}{g}\Tr(AX)-\sum_{k\geq 1}(-1)^k \frac{\bar{t}_{k}}{k} \Tr(A)^k}\bigg)(\det(B) )^N
\end{align}

The Imbimbo-Mukhi model is however a one matrix integral. As shown in section 6.1 of \cite{DSDI}, one can factorize the above two-matrix integral into a product of one-matrix integrals. The only coupling between $A$ and $B$ comes from the term in the action $\Tr(ABX)$ involving their product. First we redefine $X \to -X$ (which flips the signs of the first two terms in the exponent, and takes care of the prefactor $(-1)^{Q^2}$). We define the matrix $\sigma=AXB$ and rewrite the integral in terms of $A$ and $\sigma$. 
\begin{align} \label{eq:appearanceIM}
    Z[t_{k},\bar{t}_{k},s_{k}=N]=& \frac{\det(X)^{Q} \det(V)^{Q}}{Z_Q}\int dA d\sigma \exp{\bigg(-\frac{N}{g}\Tr(\sigma)+\frac{N}{g}\Tr(AX)-\sum_{k \geq 1} (-1)^k\frac{\bar{t}_{k}}{k} \Tr(A)^{k}\bigg)}\nonumber\\
    & \hspace{200pt} \times(\det(\sigma) )^N (\det(A))^{-N-Q}(\det(X))^{-N-Q}\nonumber\\
    =& \frac{ (\det(V))^{Q} }{(\det(X))^{N}Z_Q}
    \underbrace{\int dA  \exp{\bigg(\frac{N}{g}\Tr(AX)+N\Tr(V(A))-(N+Q)\Tr\log{A}\bigg)}}_{\text{IM Model}}\nonumber\\
    &\hspace{210pt}\times\int d\sigma e^{-\frac{N}{g}\Tr(\sigma)}
    (\det(\sigma ) )^N 
\end{align}

where we have defined $N\Tr V(A) = -\sum_{k \geq 1} (-1)^{k}\frac{\bar{t}_{k}}{k} \Tr(A)^{k}$. After shifting $\sigma \rightarrow 1-\sigma$, the integral over $\sigma$ can then be recast as the Penner model \cite{penner1988perturbative}. The large $Q$ expansion of its free energy computes the Euler characteristic of $\mathcal{M}_{g,n}$. More importantly though, we witness the appearance of the Imbimbo-Mukhi model

\begin{equation}
   Z_{IM}[t_{k}, \bar{t}_{k}, \mu = -iN] = \inv{\det(X)^N}\int dA  \exp{\bigg(\frac{N}{g}\Tr(AX)+N\Tr(V(A))-(N+Q)\Tr\log{A}\bigg)}
\end{equation}

where we have identified the cosmological constant $\mu$ of the $c=1$ string theory with the the size of the original $K,M$ matrices: $\mu = -iN$. In Eq.(\ref{eq:appearanceIM}), all the dependence of the generating function $Z[t_{k},\bar{t}_{k},s_{k}=N]$ on the sources $t_{k}, \bar{t}_{k}$ resides solely in the Imbimbo-Mukhi factor. 

Building off previous work of Moore, Plesser and Rangoolam \cite{mooreRangPlessScattering} (including related work in \cite{mandal1991interactions, Dijkgraaf:1991qh}), Imbimbo and Mukhi wrote down their matrix integral in 1995 as a solution to the integrable hierarchy found to govern the generating function of tachyon momentum correlators in the $c=1$ string theory at self-dual radius (in our conventions, $R=1$). In other words, they established that 

\begin{equation}
    \log Z_{IM} [t_{k}, \bar{t}_{k}] = \langle e^{\sum_{k \geq 1} t_{k} T_{+k}}\ e^{\sum_{k \geq 1} \bar{t}_{k} T_{-k}} \rangle_{c=1, R=1} 
\end{equation}. 

From Eq.(\ref{eq:appearanceIM}), we have now rediscovered the Imbimbo-Mukhi model as the theory of open strings on the D3' branes! In particular, since $ Z[t_{k},\bar{t}_{k},s_{k}=1]$ is by construction the generating function of 1/2 SUSY correlators in $\mathcal{N}=4$ SYM, we arrive at the following dictionary: 

\begin{equation}
    \Big{\langle} \prod_{i=1}^{n_1} \frac{1}{k_{i}} \Tr \left[ \left(Y_1 \cdot \Phi(w_1)\right)^{k_i}\right] \prod_{j=1}^{n_2} \frac{1}{l_{j}} \Tr\left[ \left(Y_2 \cdot \Phi(w_2)\right)^{l_j} \right] \Big{\rangle}_{\mathcal{N}=4} =  \left( \frac{ \lambda Y_{1} \cdot Y_{2}}{4 \pi^2 |w_1-w_2|^2} \right)^{\sum_{i=1}^{n_1} k_{i}} \langle \prod_{i=1}^{n_1} T_{+k_{i}}\prod_{j=1}^{n_2} T_{- l_{j}} \rangle_{c=1}  \label{eq:N=4c=1}
\end{equation}

It equates certain extremal correlation functions in $\mathcal{N}=4$ SYM with correlators of tachyon momentum operators in the $c=1$ string theory at self-dual radius. The overall prefactor containing the spatial dependence is simply the free field propagator for the six scalars of $\mathcal{N}=4$ (encapsulated by the matrix model in the coupling $g$, see Eq.(\ref{eq:gVSprops})). The $c=1$ amplitude in turn computes the non-trivial matrix combinatorics. Since all steps of the integrating in/out procedure leading up to the Imbimbo-Mukhi model were exact, i.e. never required any large $N$ limit, the equality in Eq.(\ref{eq:N=4c=1}) is \textit{valid to all orders in the genus expansion}. The $1/N$ expansion in $\mathcal{N}=4$ SYM translates to the genus expansion of the $c=1$ string (in inverse powers of the cosmological constant $\mu$). From a purely matrix model perspective, this gives a dictionary between traces in the two-matrix integral and the tachyon momentum operators in the $c=1$ string, also found from an integrability perspective in \cite{BonoraXiong2matrix},

\begin{equation} \label{eq:tracesdict}
     \frac{1}{n}Tr(K^{n}) \leftrightarrow T_{+n} \qquad  \frac{1}{n}\Tr(M^{n}) \leftrightarrow T_{-n}
\end{equation} 

Jevicki and Yoneya \cite{Jevicki:2006tr} had in fact conjectured that precisely these correlators in $\mathcal{N}=4$ SYM could be computed in the $c=1$ string. Along with \cite{halfBPSc=1followup, Okuyamac=1}, they initially motivated the correspondence from the point of view of LLM geometries, equating the free fermion pictures on both sides. They provided several explicit checks of their conjecture. Here, we have used Open-Closed-Open triality to prove their conjecture for all correlators, to all orders in $1/N$.

As a sanity check for the equivalence between the matrix model correlators and $c=1$ string theory correlators, we have included some simple low genus results. \\

At genus zero, we can write down the following simple correlators by doing some explicit counting of Wick contractions
\be 
\begin{split}
\langle \Tr(M^l) \Tr(K^l) \rangle_{h=0} &= l g^lN^0\\
\langle  \Tr(K^l) \Tr(M^{l_1})\Tr(M^{l-l_1}) \rangle_{h=0} &= ll_1(l-l_1)g^lN^{-1}\\
\langle \Tr(K^{l})\Tr(M^{l_1})\Tr(M^{l_2})\Tr(M^{l_3})\rangle_{h=0} &= l(l-1)l_1 l_2 l_3 g^l N^{-2} \de_{l_1+l_2+l_3, l}  
\end{split}
\ee 
These can be explicitly matched with the planar correlators of $c=1$ Tachyon vertex operators, for example in \cite{klebanov1991string},
\be
\langle P(+l) P(-l) \rangle_{h=0} =  -(\Gamma(1-l))^2\frac{\mu^{l}}{l}
\ee 
with the dictionary for the operators
\begin{align}\label{eq: app_normalization_klebanov_1}
    \inv{l}\text{Tr}(M^l)\leftrightarrow\frac{-iP(l)}{\Gamma(1-|l|)}, \qquad \inv{l}\text{Tr}(K^l)\leftrightarrow\frac{-iP(-l)}{\Gamma(1-|l|)}
\end{align}
and dictionary for the parameters, 
\begin{align}\label{eq: app_normalization_klebanov_2}
    N \leftrightarrow -i \beta\mu \text{ and } g \leftrightarrow \mu.
\end{align}
At genus one, the two-point function of the matrix model can be computed (by expressing in terms of permutations \eqref{eq: perm_main} and solving some combinatorics problem) to be \cite{brown2011complex}, \cite{Brown:2010pb}
\begin{align}
    \langle\text{Tr}(M^l)\text{Tr}(K^l)\rangle_{h=1} = l {{l+1}\choose{4}}g^{l}N^{-2}
\end{align}
which again matches with the genus one, two point Tachyon correlator of \cite{klebanov1991string}
\be 
\langle P(+l) P(-l) \rangle_{h=1} = \Gamma((1-l)^2)\mu^{l}(\beta\mu)^{-2}\frac{(l+1)(l-1)(l-2)}{24}
\ee 
using the same dictionary Eq.(\ref{eq: app_normalization_klebanov_1}), (\ref{eq: app_normalization_klebanov_2}). Note that in all the matching above, we have the leg pole factors like $\Gamma(1-|l|)$ which diverge whenever $l$ is an integer\footnote{ For generic radius $R$, tachyon momenta $l$ is an integer multiple of $R^{-1}$. At self-dual radius $R\rightarrow 1$, $l$ becomes exactly an integer.}. We should really think of these divergent factors being absorbed in the tachyon vertex operator definition and the tachyon vertex operators getting renormalized as $R\rightarrow 1$. We have further done some higher genus checks for two point and three point functions using the permutation formalism of Eq.(\eqref{eq: perm_main}).

\subsection{Open-Closed-Open Triality on the Worldsheet}\label{sec:OCOonWS}

\begin{figure}[ht!]
    \centering
\includegraphics[scale=0.67]{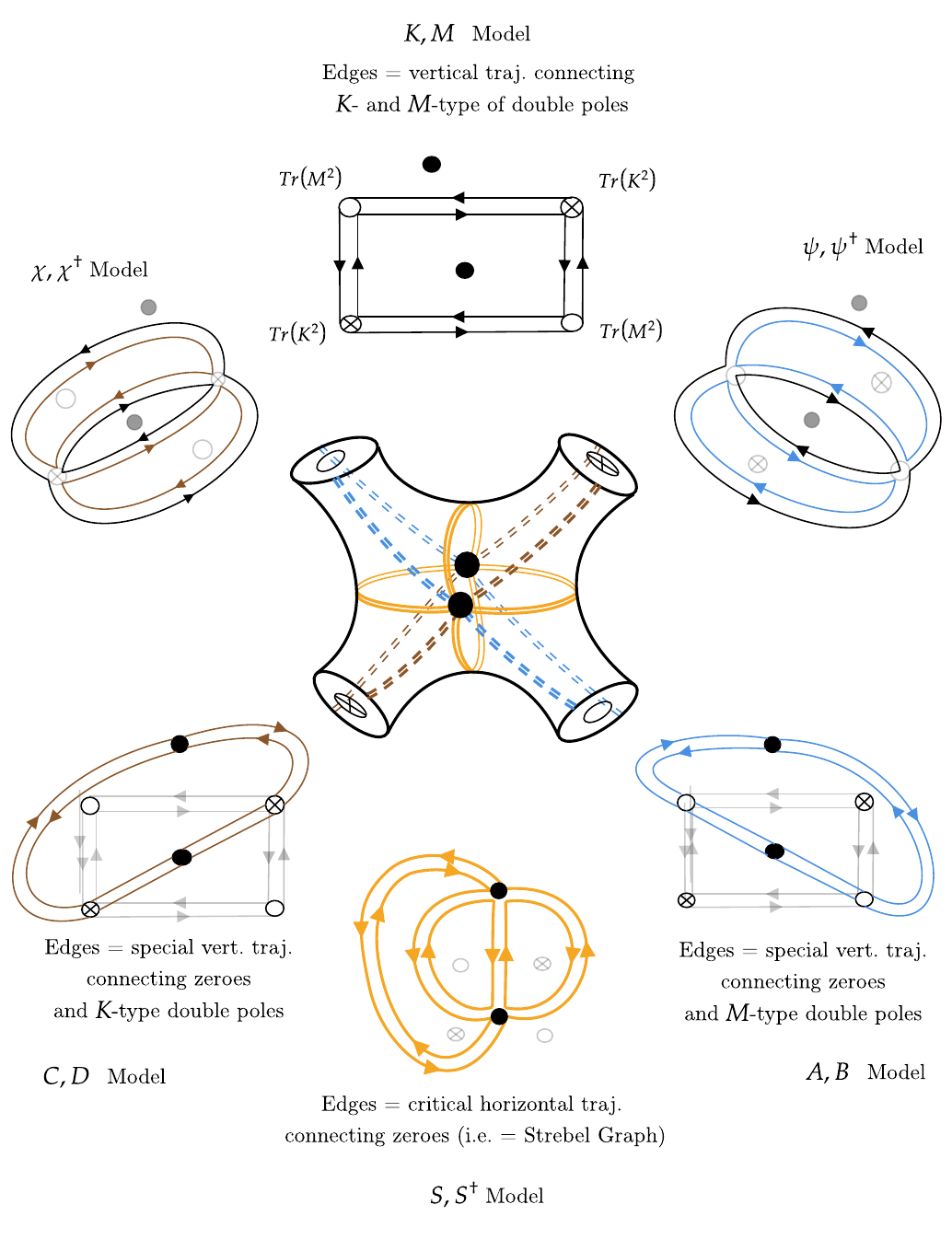} 

    \caption{\textbf{Open-Closed-Open Triality on the Worldsheet } The edges of the Feynman diagrams of the different (bosonic) open string descriptions map onto different trajectories of the Strebel differential on the worldsheet. They therefore all encode the same moduli of the dual Riemann surface. In the $K,M$ model, the edges correspond to (families of) vertical trajectories, connecting both types of marked points (i.e. double poles of the Strebel differential). The partial duals utilize the special vertical trajectories connecting zeroes and one type of double-pole, connecting a black dot to a white vertex. While the two-fermion diagrams (not depicted) seem to correspond to the two families of non-critical trajectories surrounding each type of marked point, we do not currently have a good understanding of the role of the fermionic partial duals. Finally, the edges of the $S,S^{\dagger}$ model Feynman diagrams lie directly on the critical horizontal trajectories, and as such can be directly identified with the Strebel graph of the closed string worldsheet.  } 
    \label{fig:OCOfromWS}
\end{figure}

We have shown that many open string descriptions, in the guise of matrix models, are dual to the same closed string theory in the very restricted setting of the 1/2 SUSY sector of $\mathcal{N}=4$ SYM. Up to six different matrix integrals generate the same generating function of correlators, which we showed admitted a closed string interpretation in terms of Riemann surfaces covering a target space $\mathbb{CP}^{1}$. One may wonder whether the closed string theory itself knows anything of this open-closed-open triality. We will argue it does. The punchline will be that the edges of each related Feynman diagram map onto the different trajectories traced out on the closed string worldsheet by the Strebel differential, as shown in Fig. \ref{fig:OCOfromWS}. For a quick review of the set of curves traced out on any Riemann surface via its unique Strebel differential, we refer the reader back to Secs. \ref{sec:StrebelReview} and \ref{sec:PhysPic}.

\textbf{$K,M$ Feynman Diagrams \&  Vertical Trajectories}\\
First, recall that the vertices of the $K,M$ model maps onto the marked points of the worldsheet. These marked points correspond to the location of the double poles of the differential. The edges of the $K,M$ model Feynman diagram thus run along the vertical trajectories connecting the two types of double-poles introduced in Sec. \ref{sec:PhysPic}. Such trajectories are not unique on the worldsheet. From the point of view of the Feynman diagrams, this reflects the fact that we needed to "bunch together" all homotopic edges connecting two vertices. The precise statement is that the \text{family} of all homotopic edges connecting two vertices map onto the \textit{family} of vertical trajectories connecting the two double poles. \\

\textbf{$S,S^{\dagger}$ Feynman Diagrams \&  Critical Horizontal Trajectories}\\
The zeroes of the Strebel differential correspond to the faces of the $K,M$ model, which we have denoted by black dots in Fig. \ref{fig:OCOfromWS}. These go over to the vertices of the $S,S^{\dagger}$ model Feynman diagrams. For that matrix integral, the edges of its ribbon graphs therefore lie on the critical horizontal trajectories connecting zeroes of the differential. The collection of critical horizontal trajectories forms the Strebel graph of the worldsheet, so we are simply back to the statement the $S,S^{\dagger}$ Feynman diagrams are directly identified with the Strebel graph of the worldsheet. \\

\textbf{Bosonic Partial Dual Feynman Diagrams \&  Special Vertical Trajectories}\\
The bosonic partial duals play an interesting in-between role. As shown in Fig. \ref{fig:OCOfromWS}, the edges of their Feynman diagrams connect one set of the original $K,M$ model vertices (white crossed or uncrossed) to the black dots. On the worldsheet, these represent the special vertical trajectories that connect a double pole to a zero of the differential. The $C,D$ model generates ribbon graphs that coincide with the special vertical trajectories connecting the $K$-type marked points to the zeroes, and similarly for the $A,B$ matrix integral. \\

\textbf{The Fermionic Feynman Diagrams \& Non-Critical Horizontal Trajectories}\\
As can be seen from in the first step of Secs. \ref{sec:intinout} and \ref{sec:OCOgraphduality}, the introduction of the fermions replace the vertices of the $K,M$ model by new faces. This parallels the non-critical trajectories surrounding the double-poles of the Strebel differential, illustrated in Fig. \ref{fig:HorTraj}. There are no canonical non-critical horizontal trajectories (just as there are no canonical vertical trajectories connecting poles). As such, one might want to interpret the ribbon graphs of the two-fermion integral as being in correspondence with \textit{families} of non-critical horizontal trajectories. These trajectories never include any poles, nor zeroes of the differential. This, in turn, explains why these diagrams have three faces. The fermionic partial duals are, in some sense, more mysterious. From the diagrams shown in \ref{fig:onefermout}, we can see that their edges connect only one type of vertex. As far as we can tell, there are no natural candidate trajectories on the worldsheet connecting double poles of the same type. It would be good to return to this point in the future.

\subsection{Open-Closed-Open Triality in Target Space}\label{sec:OCOinTS}

In Sec. \ref{sec:embedding}, we showed how every Feynman diagram encoded the embedding map of the string into the target space. This map was found to be a holomorphic covering map of the Riemann sphere, $X(z)$, branched over exactly three points $X=w_1, X=w_2$ and $X=(w_1+w_2)/2$. $X(z)$ was completely specified by its three ramification profiles, which could be repackaged in terms of three permutations. Here, we explain how Open-Closed-Open triality manifests itself at the level of these maps. In other words, we find a target space picture for the equivalence of the six open descriptions we have found. 

As explained in Sec. \ref{sec:FDaspreimage}, one of the easiest ways to understand the relation between the $K,M$ model Feynman diagram and the map $X(z)$ is to view the graph as the pre-image of a line segment in target space, connecting $X=w_1$ and $X=w_2$. See Fig. \ref{fig:FDasPreim}.  Schematically, we might write this as 

\begin{equation}
    K,M-\text{Model Feynman Diagram} \leftrightarrow X^{-1} \left( [w_1,w_2] \right) 
\end{equation}

The two endpoints of the line segment are the two types of vertices in the $K,M$-model ribbon graph. However, we could have just as well considered a different interval in target space, connecting two other branchpoints. Its pre-image, under the \textit{same} Belyi map, would have given us a different graph, whose vertices correspond to the endpoints of this newly chosen interval in target space. This new graph, however, encodes the same information.

For example, we could look at the pre-image of the target space interval $[w_1,(w_1+w_2)/2]$. The vertices of this graph would now be the white crossed vertices ($X=w_1$) and the black dots (corresponding to $X=(w_1+w_2)/2$), just as in the Feynman diagrams of the $C,D$ model. This is no coincidence. The pre-image of this interval, under the Belyi map derived from a given $K,M$ model Feynman diagram, gives us precisely the equivalent $C,D$-model diagram we had found in Sec. \ref{sec:OCOpartials}. As shown below in Fig. \ref{fig:sameperms}, we see these two diagrams indeed specify the same three permutations, and hence the same Belyi map.

\begin{figure}[H]
    \centering
\includegraphics[scale=0.70]{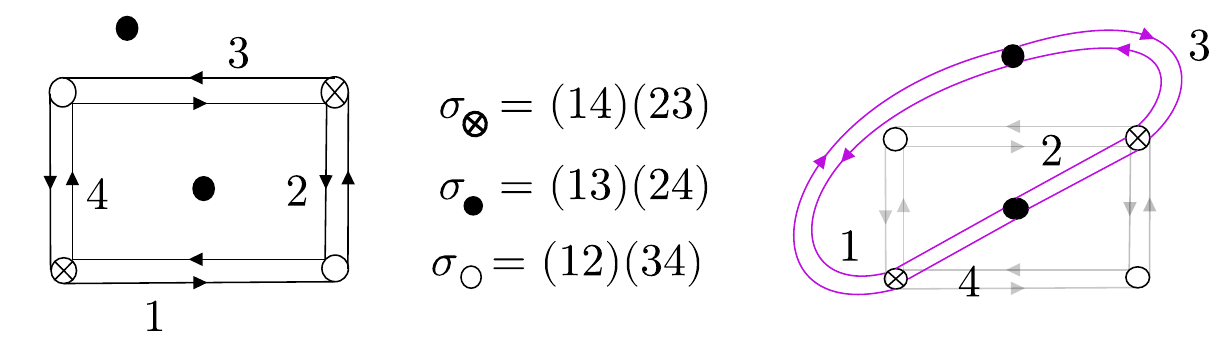} 

    \caption{\textbf{Graphs related via (partial) graph duality specify the same Belyi Maps } The pre-image, under the same Belyi map, of two different intervals in target space connecting two branchpoints, gives rise to partial dual graphs. Viewed in reverse, we can see explicitly how the original $K,M$ model Feynman diagram and the corresponding one of the $C,D$ model encode the same permutation triple, and hence the same Belyi map. } 
    \label{fig:sameperms}
\end{figure}

\begin{figure}[ht!]
    \centering
\includegraphics[scale=0.65]{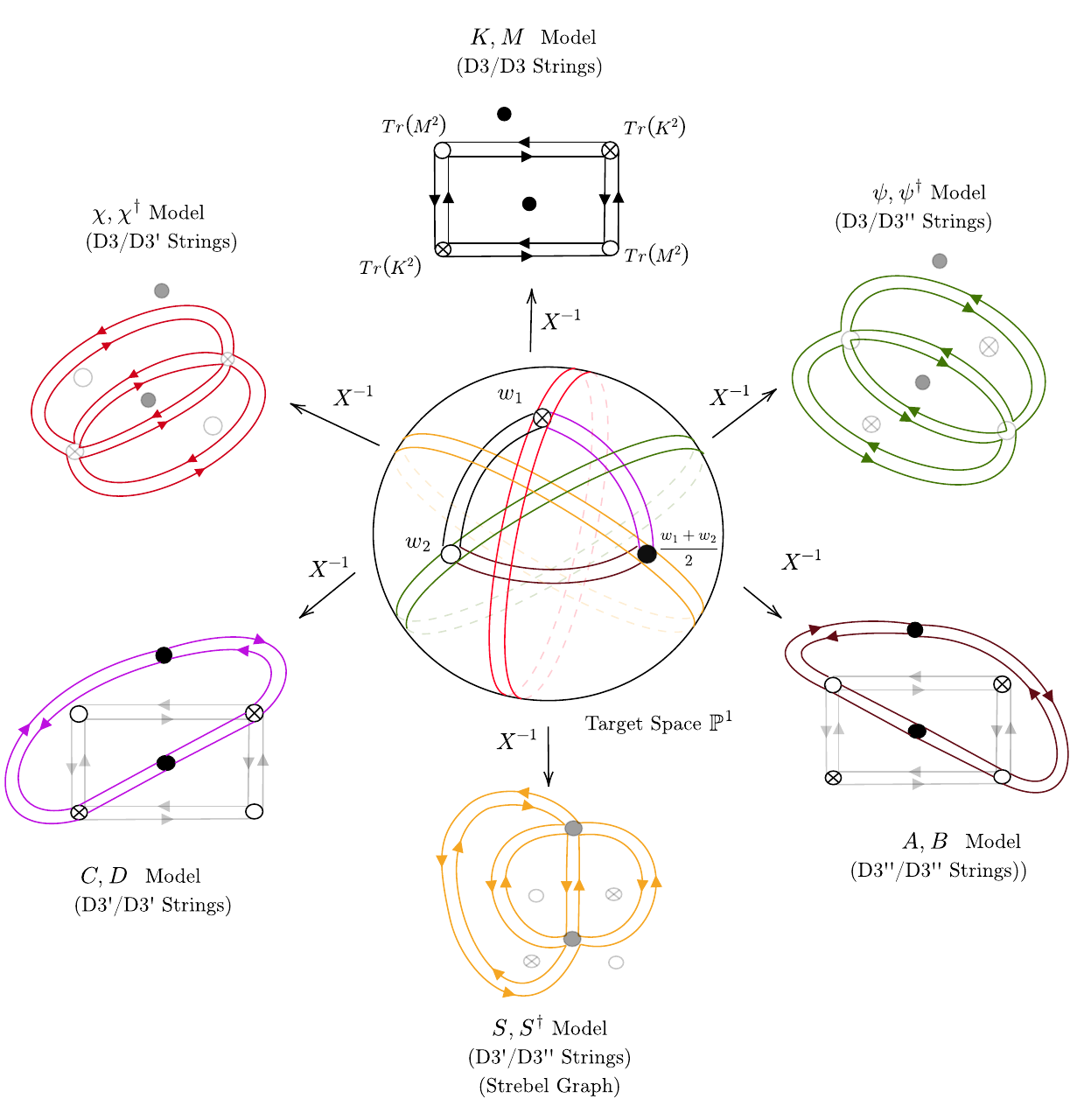} 

    \caption{\textbf{Open-Closed-Open Triality in Target Space } The Feynman diagrams of the six open string descriptions, related to one another via the integrating in/out procedure of Secs. \ref{sec:intinout} and \ref{sec:OCOgraphduality}, can all be viewed as pre-images of various target space intervals or loops under the same Belyi map $X(z)$. This explains how they all encode the same information about the closed string's embedding in the bulk. The Feynman graphs generated by one-matrix integrals are pre-images of closed loops, anchored at one of the three branchpoints. The two-matrix integrals give rise to Feynman diagrams, whose edges all correspond to intervals on the $\mathbb{CP}^{1}$, connecting two branchpoints. The two types of vertices of those diagrams are simply the two endpoints of the interval. The graph duality relating the various diagrams becomes manifest from this target space perspective.} 
    \label{fig:OCOfromTS}
\end{figure}

In the same vein, we can view the corresponding Feynman diagram of the $A,B$ model as the pre-image under the Belyi map of a line on the target Riemann sphere connecting instead $X=(w_1+w_2)/2$ to $X=w_2$. This story is quite general. All the Feynman diagrams of the six open descriptions related to each other via the integrating in/out procedure of Secs. \ref{sec:intinout} and \ref{sec:OCOgraphduality} give rise to the same Belyi map. Said a different way, they can all be viewed as pre-images, under the same Belyi map, of various target space intervals or loops. See Fig. \ref{fig:OCOfromTS}. 

The one-matrix models, which include the two fermionic descriptions and the $S,S^{\dagger}$ model, generate Feynman diagrams with a single type of vertex but two types of faces. Clearly, they cannot arise as pre-images of an interval with distinct end-points. Instead, the edges of their ribbon graphs map onto a closed loop in target space. This loop wraps around the $\mathbb{CP}^{1}$, from one branchpoint back to itself. It encircles precisely one other branchpoint. This way, the two faces correspond to the two other branchpoint, one on each side of the loop \footnote{Another way to phrase this is that we thicken this loop into a ribbon. }. These are the orange, green, and red circles depicted in Fig. \ref{fig:OCOfromTS}.   

This new perspective on open-closed-open triality provides a target space explanation for the number of open string descriptions: there are 3 closed loops, one for each branch point, and 3 intervals connecting two branchpoints at a time. These are the 3 one-matrix and 3 two-matrix integrals we found. The graph duality relating the various Feynman diagrams is also made manifest from this point of view. For example, as shown in Fig. \ref{fig:OCOfromTS}, the Feynman diagrams of the $S, S^{\dagger}$ model are the pre-images of a loop anchored at $X=(w_1+w_2)/2$. Such a loop crosses the interval connecting $X= w_1$ to $X=w_2$ precisely once. Since the pre-images of that interval are none other than the edges of the $K,M$ model diagrams, we see that the two diagrams must be graph dual to each other, as expected. 

As a final comment, we note that such a target space perspective on open-closed-open triality had to exist. The logic goes as follows. We found in Sec. \ref{sec:OCOonWS} that the related Feynman diagrams of the six open string descriptions all reconstructed the same worldsheet. In particular, they give rise to the same Strebel differential, since that differential is unique to each Riemann surface. In Sec. \ref{sec:StrebelAsPullback}, we showed how the Strebel differential was the pullback of a target space Kähler form on to the worldsheet by embedding map $X(z)$. We therefore conclude that if all related Feynman diagrams give rise to the same Strebel differential, they must necessarily encode the same Belyi map.

\pagebreak

\section{Conclusion}

\subsection{Main Takeaways So Far} \label{sec:Takeaways}
In the context of a very simple model of gauge/string duality, we have laid out a concrete mechanism to translate between gauge theory Feynman diagrams and closed strings. After the rather lengthy discussion in the body of the paper, we distill here what we believe to be the most salient points of this work. We have also included links to the figures which best illustrate them.

\begin{itemize}
    \item We recast the Feynman diagram expansion of certain protected correlators in $\mathcal{N}=4$ SYM as a weighted sum over closed string worldsheets, to all orders in the $1/N$ expansion. $1/N$ plays the role of the string coupling, as expected. See Fig. \ref{fig:introsummary}.
    \item There exists an A-model topological string that generates this dual sum over worldsheets. It will be discussed at length in \cite{DSDIII}. A quick summary can be found in Sec. \ref{sec:StringLocalization}. This A-model string is equivalent to the $c=1$ string at self-dual radius \cite{MukhiVafa,ashok2006topological}. Single trace operators map to specific vertex operators in these dual string descriptions. This gives a one-to-one map between matrix and string correlators.
    
     \item The protected gauge theory correlators studied in this paper are multiple ($n$) single trace operators inserted at one of two spacetime points. They can also be computed from a certain integral over two Hermitian matrices. We study this matrix integral in the regular 't Hooft limit, rather than resorting to the double-scaling limit often thought to be necessary for the existence of a string dual.
     
    \item Each Feynman diagram is assigned to a unique point on the moduli space of punctured Riemann surfaces via the Strebel parametrization of $\mathcal{M}_{g,n}$ (cf. Fig. \ref{fig:StrebelDict}), following the program laid out in \cite{freefieldsadsI,freefieldsadsII,freefieldsIII}.
    
    \item  The Feynman diagram is not viewed as a discretization of the worldsheet. Instead, the sum over a discrete set of Feynman diagrams translates to an integral over moduli space delta-function localized at a discrete set of points. These points are labeled by integer-length Strebel graphs, as proposed by Razamat \cite{razamatGauss}, and lie on a lattice on $\mathcal{M}_{g,n}$, illustrated in Fig. \ref{fig:latticeMgn}.
    
    \item Every Riemann surface can be dissected into a collection of strips (Fig. \ref{fig:StrebelAsStrips}). Each edge of the Feynman diagram corresponds to one such strip (Fig. \ref{fig:FDtoStripstoWS}). How the various edges of each diagram meet at the vertices and surround the faces of the graph dictates the precise way in which the dual strips are glued together into a closed string worldsheet (cf. Figs. \ref{fig:GluingAtEdges}, \ref{fig:GlueingAtVertices}, \ref{fig:GlueingAtFaces}). 
    
    \item Each Feynman diagram equally encodes a holomorphic covering map of a target Riemann sphere. Physically, the worldsheet wraps the compactified plane ($\mathbb{CP}^{1}$) on the boundary of $AdS$, where the gauge theory operators are inserted (shown in Fig \ref{fig:EmbeddingMap}). The vertices of the Feynman diagram map onto the insertion points.
    
    \item The covering map is branched over three points of the target space. Two of these points correspond to the insertion points of the operators on the boundary, we have chosen the third to lie midway between them. How the various edges of each diagram meet at the vertices and surround the faces of the graph determines the ramification profile of the map over these three points. This uniquely specifies the covering map. See Figs \ref{fig:FDtoPerms} \& \ref{fig:PermstoMaps}.
    \item The Feynman diagram can be thought of as the pre-image of a line drawn on the target space Riemann sphere. This line connects the two operator insertion points, as shown in Fig. \ref{fig:FDasPreim}.
    
    \item The weight of each Feynman diagram maps onto the exponential of the Nambu-Goto action, weighting each dual string configuration in the closed string path integral. The (regulated) area of the worldsheet (computed in Strebel gauge) reproduces the position space Feynman propagator of the gauge theory fields appearing in the correlator (cf. Fig. \ref{fig:WeightFD}).
    
    \item By considering determinant insertions in the 2-matrix model, we find up to six different open string descriptions of the same dual closed string theory (summarized in Fig. \ref{fig:OCOfromBranes}). This has been dubbed Open-Closed-Open triality. 
    
    \item The Feynman diagrams generated by these various matrix models are all related via (partial) graph duality. As such, they all map to the same point on $\mathcal{M}_{g,n}$ under the Strebel construction. The edges of their ribbon graphs coincide with different unique curves traced out on each worldsheet by its Strebel differential (cf. Fig. \ref{fig:OCOfromWS}).
    
    \item One of these open strings descriptions plays a role similar to the Kontsevich model for the 1/2 SUSY sector of $\mathcal{N}=4$ SYM (cf. Fig. \ref{fig:OCON=4}). It arises from open strings on giant graviton branes (depicted in Fig. \ref{fig:GiantGravitons}). 
    
    \item This Open-Closed-Open triality allows us to establish that certain extremal correlators in $\mathcal{N}=4$ SYM are equal, to all orders in $1/N$, to tachyon momentum correlators in the $c=1$ string at self-dual radius, as conjectured in \cite{Jevicki:2006tr,halfBPSc=1followup} and predicted by \cite{AlexandrovNormal,brown2011complex}. 
\end{itemize}

\subsection{Looking Forward} \label{sec:lookingforward}
 We are hopeful that some of the lessons learned from this toy model of the AdS/CFT correspondence might generalize to more interesting examples of holography, paving a way towards a future derivation. We are of course still very far from that dream. Below, we highlight some possible small steps in the right direction along with important relations to other works.

\begin{enumerate}
\item \textbf{Turning on interactions: } The correlators we considered were free-field correlators, governed by simple Wick contractions. If we add an interaction term to the matrix model action, say for example $\Tr K^4$, we can treat it in perturbation theory around the free point. Observables in the interacting theory are therefore (perturbatively) matched to correlators in the free theory with additional insertions of the perturbation  $\Tr K^4$. 
Note that perturbation theory in the 't Hooft coupling is convergent (with finite radius of convergence), at each order in $1/N$.\footnote{What we are calling `t'Hooft coupling' is distinct from the 't Hooft coupling appearing from the reduction of $\mathcal{N}=4$ to the matrix model. Rather, it is a particular combination of $N$ and the coupling constant multiplying the perturbation $\Tr K^4$ in this example.} 

Since we know how to write all expectation values in the free matrix model in terms of stringy correlators, we can in principle give a string description of the interacting two-matrix integrals as well (again, to all orders in perturbation theory). Since single trace operators map on to vertex operators in the A-model topological string (or its $c=1$ reformulation), a generic two-matrix model action of the form 
\begin{equation} \label{eq:defmatrixaction}
    \Tr(KM) +\sum_{n} \frac{c_{n}}{n} \Tr(K^{n}) + \frac{\tilde{c}_{n}}{n} \Tr(M^{n})
\end{equation}
would deform the dual string by adding integrated vertex operators to the tensionless string action. Schematically, we would write
\begin{equation}
    S_{0}^{WS} +\sum_{n} c_{n} \int d^2z \  \mathcal{V}_{n} + \sum_{n}  \tilde{c}_{n} \int d^2 z \ \tilde{\mathcal{V}}_{n} \label{eq:defstringaction}
\end{equation}

The additional perturbation generates \textit{internal} vertices to the Feynman diagrams. On the worldsheet, these internal vertices correspond to the additional insertions coming from the integrated vertex operators in Eq.(\ref{eq:defstringaction}). 

It would be very interesting to understand how the picture of discrete lattice points on moduli space gets modified as we turn on interactions. As suggested in \cite{gaberdiel2021symmetric}, there is an expectation that the delta-functions in the moduli space integrand supported on these discrete points get smoothed out, with a characteristic width set by the 't Hooft coupling constant.\footnote{EAM thanks Beat Nairz for discussions of this mechanism in the context of the string dual to the symmetric orbifold of $\mathbb{T}^{4}$.} 

\item{\textbf{Winding Modes in the $c=1$ String:} In Sec. \ref{sec:IMmodel}, we found a dictionary between traces built solely from $K$ ($M$) and positive (negative) momentum tachyon vertex operators in the $c=1$ string at self-dual radius (see Eq.(\ref{eq:tracesdict})). The $c=1$ string has another set of vertex operators which excite winding \cite{grossKlc=1}. We do not know how to describe them from the matrix model point of view so far. We have also not found any natural candidates for them in the A-model topological string. It is possible that the matrix model only captures a subsector of the $c=1$ string theory. Another possibility is that matrix observables other than single trace operators are dual to the winding modes. One idea is that they might be described by the faces of matrix model Feynman diagrams. One would in principle need to engineer a matrix integral generating Feynman diagrams with two types of faces and two types of vertices. Similar matrix model ideas aiming to incorporate winding modes were explored in \cite{Dijkgraaf:2003xk, AlexandrovNormal, Alexandrov:2002fh, KostovWindingMM}. Towards that end, we have begun to explore a two complex matrix model with rectangular matrices. If true, it would be the single trace operators of the F-type description which would be dual to the winding modes: graph duality would implement T-duality on the worldsheet. }

\item{\textbf{Twisted Holography:}} What constitutes a tractable field theory generalization of the two-matrix model studied here, which also embeds neatly in the broader context of AdS/CFT? Twisted holography makes for a promising candidate  \cite{CostelloGaiottoTH, Budzik:2021fyh, Costello:2020jbh}. On the open string side, the actions look quite similar:
\begin{equation}
    \Tr_{N}(KM) \rightarrow \int d^{2}z \Tr_{N}(\beta \bar{\partial} \gamma) 
\end{equation}
The full gauge theory action consists of this bosonic gauged $\beta, \gamma$ system along with a fermionic counterpart. This action was in fact derived from the open string field theory on $N$ branes in the B-model on $\mathbb{C}^{3}$ \cite{CostelloGaiottoTH}, which itself is known to reduce to holomorphic Chern-Simons thanks to the work of \cite{WittenCSasOSFT}.

There is a subset of the so-called $A_{n}$-tower of physical operators that play the field theory counterpart to the simple matrix traces considered here. Their correlators generate the same ribbon graphs in their Feynman expansion. However, by inserting operators at different points, each edge of the Feynman diagram would now carry a position space propagator dictated by the OPE

\begin{equation}
   \beta_{ab}(z_1) \gamma_{cd}(z_{2})  \sim \frac{1}{z_{1}-z_{2}} \delta_{ad}\delta_{bc}
\end{equation}

This simple gauge theory encodes the chiral algebra of $\mathcal{N}=4$ SYM. It captures a subset of $\mathcal{N}=4$ SYM operators living on a two-dimensional plane contained in the four-dimensional boundary of $AdS_{5}$, generalizing our own setup shown in Fig. \ref{fig:IntroEmbedding}. There is a known closed topological B-model string dual to this chiral algebra. One might speculate there exists an A-model string theory, similar to the one presented here, which localizes to covering maps of the (compactified) boundary subplane. These maps would presumably be branched over more than three points, namely over the arbitrary insertion points of operators in the chiral algebra.  The naïve guess is that the branching profile would be similarly dictated by the R-charge of the operator. This would presumably also have a relation to the branched coverings in the tensionless limit of the $AdS_3\times S^3\times T^4$ theory. This is currently under early investigation. 

\item{\textbf{Relation to Double-Scaled Matrix Models:} How does the two-matrix integral/string duality presented here compare to their well-known double-scaled counterparts? We begin by noting that all $(p,q)$ minimal strings can be captured by various scaling limits of the same two-matrix chain we have studied here \cite{2matrixMinMods} \footnote{The recently proposed `complex Liouville string' \cite{complexLiouString} also arises as the double-scaling limit of a two-matrix integral. At present, it appears harder, though certainly worthwhile, to make a direct connection to their work.}. From an integrability perspective, this reflects the fact that the 2d Toda hierarchy governing our two-matrix integral \cite{BonoraXiong2matrix,tophierarchy} has the familiar KdV hierarchies as various limits \cite{BonoraLimits}. On the worldsheet side, it suggests that deformations of the A-model Kazama-Suzuki, or its $c=1$ string reformulation, contain the $(p,q)$ minimal strings in some limit. It would be very interesting to understand precisely how one implements the double-scaling limit directly at the level of the worldsheet. 

One vague idea is suggested by the work of Hsu and Kutasov \cite{HsuKutasov} on the gravitational Sine-Gordon model\footnote{EAM thanks Bruno Balthazar for several discussions on this point.}. This model can be viewed as a perturbation of the $c=1$ string by a linear combination of momentum $2$ tachyon vertex operators\footnote{Similar constructions have recently been studied by \cite{RodriguezDefs,BalthazarDefs} in the search for solvable models of time-dependent string backgrounds.}. In the language of our $K,M$ model, this would correspond to adding quadratic $\Tr(K^2)$ and $\Tr(M^2)$ terms to the action (using the dictionary in Eq.(\ref{eq:tracesdict})). The authors of \cite{HsuKutasov} found that the deformation triggered an RG flow on the worldsheet, ending at the pure topological gravity fixed point. One is tempted to compare this to the double-scaling of the Gaussian matrix model giving rise to the matrix Airy function. This matrix Airy model is well-known to capture precisely the correlators of pure topological gravity on the worldsheet. A generalization of the field theoretic arguments given by A.B. Zamolodchikov in \cite{ZamolodchikovMinMods} to the string setting would suggest that more complicated deformations of the matrix model potential would trigger RG flows on the worldsheet, whose fixed points would be the other minimal strings.}
\end{enumerate}

\section*{Acknowledgements}

We have greatly benefited from many discussions with numerous people over the long course of this work. 

RG and EAM would particularly like to thank Matthias Gaberdiel and Wei Li for related collaboration on the worldsheet theory dual to the two-matrix model and the more general gauged $\beta,\gamma$ system. RG and EAM also thank Alessandro Giacchetto and Pronobesh Maity for collaboration on applications of these ideas to the DSSYK matrix model and discrete Weil-Petersson volumes.  

EAM would like to thank the theory groups at Berkeley, Caltech, CERN, KCL, LMU, McGill, MIT, NYU, SGCP \& UTorino, as well as the organizers of the KITP "What is String Theory?" programme, of IGST 2024, of the joint KIAS-Saclay String Workshop, of the PCTS Confinement workshop, and of the "Solving Holographic Theories" workshop for the opportunity to present this work, and for the useful feedback and questions which helped shape several sections of this paper. EAM and SK thank the SwissMAP Research Station in Les Diablerets for hosting a research stay, which helped to complete this work. EAM gratefully acknowledges the KITP in Santa Barbara for hospitality during Winter/Spring 2024. This research was supported in part by grant NSF PHY-2309135 to the Kavli Institute for Theoretical Physics (KITP). RK would like to thank the organizers of Les Houches School on Quantum Geometry, 2024, where part of this work was presented in the form of a poster.

EAM also greatly benefited from countless ideas and discussions with many people over the past two years. They have been crucial in sharpening and helping craft the presentation of the proposal put forward here. These include (and it is still by no means an exhaustive list!): Ofer Aharony, Ahmed Almheiri, Dionysios Anninos, Anthony Ashmore, Bruno Balthazar, Panagiotis Betzios, Antoine Bourget, Robert Brandenberger, Edouard Brezin, Johannes Brödel, Kasia Budzik, Minjae Cho, Scott Collier, Frank Coronado, Andrea Dei, Nadav Drukker, Sibylle Driezen, Atakan Firat, Barak Gabai, Alessandro Giacchetto, David Gross, Simeon Hellerman, Sean Hartnoll, Akikazu Hashimoto, Daniel Harlow, Veronika Hubeny, Anthony Houppe, Camillo Imbimbo, Clifford Johnson, Taro Kimura, Vladimir Kazakov, Manki Kim, Igor Klebanov, Bob Knighton, David Kolchmeyer, Maxim Kontsevich, Ivan Kostov, Zohar Komargodski, Ji Hoon Lee, Adam Levine, Yuri Lensky, Carlo Maccaferri, Juan Maldacena, Emil Martinec, Nathan McStay, Joe Minahan, Ruben Minasian, Sunil Mukhi, Sameer Murthy, Hirosi Ooguri, Beat Nairz, Kiarash Naderi, Olga Papadoulaki, Geoff Pennington, Eric Perlmutter, Boris Pioline, Silviu Pufu, Daniel Ranard, Leonardo Rastelli, Victor Rodriguez, Phil Saad, Max Schwick, Ashoke Sen, Steve Shenker, Zhengyan Darius Shi, Eva Silverstein, Julian Sonner, Vit Sriprachyakul, Bogdan Stefanski, Washington Taylor, Cumrun Vafa, Herman Verlinde, Erik Verlinde, Spenta Wadia, Xi Yin, Barton Zwiebach. EAM also thanks Anthony Ashmore and Daniel Brennan for feedback on the draft, as well as Daniel Ranard for a figure. DS would like to thank Pronobesh Maity, Xi Yin, Daniel Jafferis, Cumrun Vafa, and Kasia Budzik for important discussions at different stages of the project. RK would like to thank Ashoke Sen and Raghu Mahajan for some related discussions.

RG is supported by a J. C. Bose Fellowship of the SERB as well as more broadly through the framework of support for the basic sciences by the people of India. 
EAM is supported by a SwissMAP Research fellowship and the ITP of ETH Zurich.
DS is supported by a Purcell fellowship at Harvard University. RG and RK are also supported by the Department of
Atomic Energy, Government of India, under project no. RTI4001.

\pagebreak
\appendix
\section*{Appendix}
\addcontentsline{toc}{section}{Appendices}
\renewcommand{\thesubsection}{\Alph{subsection}}
\subsection{Permutations and Feynman diagrams: Explicit example}\label{sec:FDperm}
\begin{figure}[ht!]
    \centering
    \includegraphics[scale = 0.75]{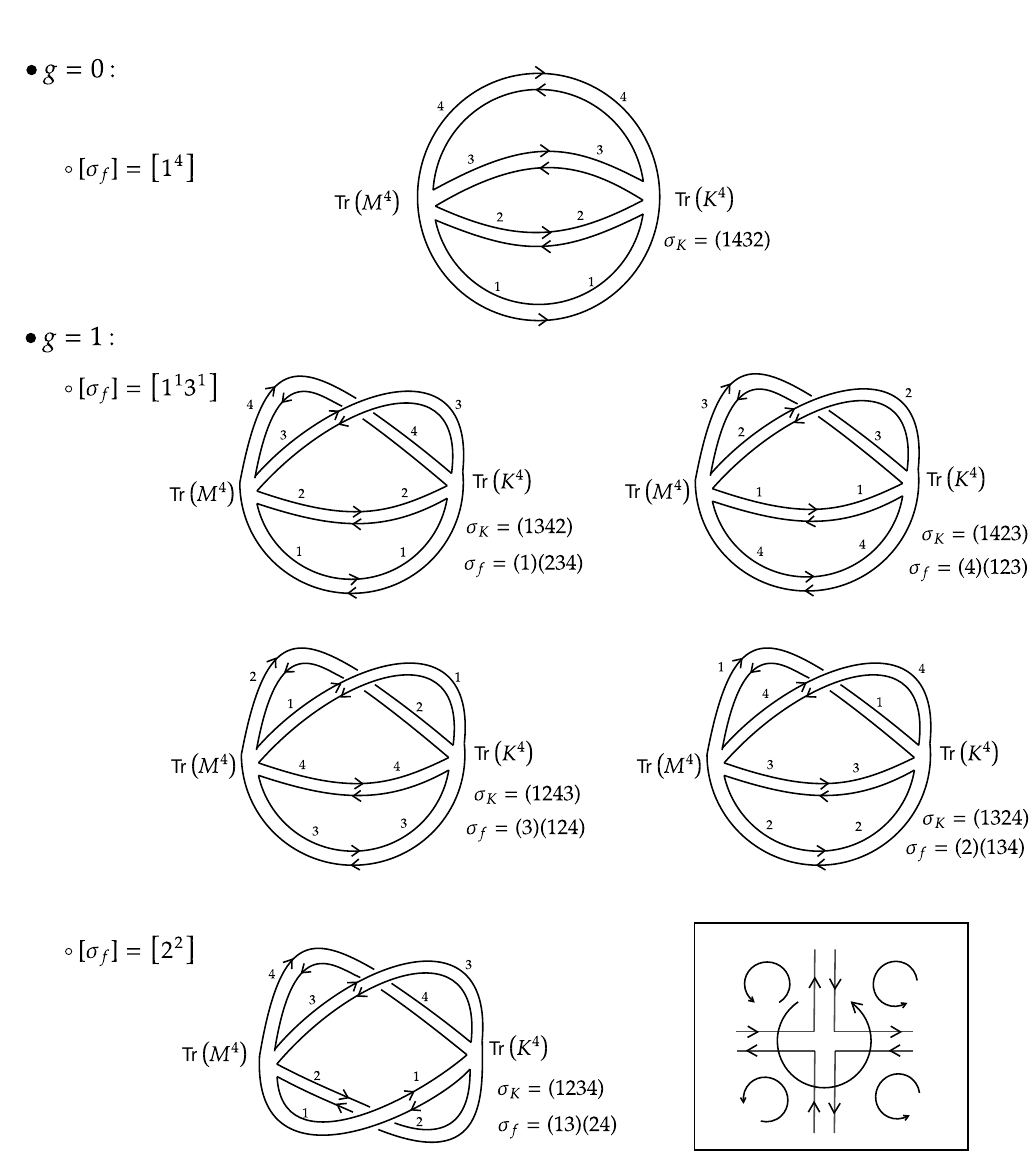}
    \caption{All the (labelled) Feynman diagrams contributing to $\langle \text{Tr}(M^4)\text{Tr}(K^4)\rangle$.}
    \label{fig: perm_example}
\end{figure}

We will demonstrate the role of all 3 permutations through a simple but non-trivial example of the correlator $\langle \text{Tr}(M^4)\text{Tr}(K^4)\rangle$. It is clear that the $\sigma_{M}$ and $\sigma_{K}$ are permutations with one cycle of length $4$. Hence, 
\begin{align}
    [\sigma_K]=[\sigma_{M}]=\{(1432),(1342),(1423),(1243),(1324),(1234)\}
\end{align}
First of all, in \eqref{eq:permresult_main}, we have fixed the $\sigma_{M}$. We pick a representative of $\sigma_{M}$, say $(1234)$ and now on RHS, we traverse over the $[\sigma_K]$. $\sigma_f$ will automatically be fixed to $(\sigma_{K}\sigma_M)^{-1}$. The $\sigma_{M}$ cycle structure leads to the different Feynman diagrams as shown in figure \ref{fig: perm_example}. We again label each edge of the diagram by labeling the right bank of the edge while going from the $K$ vertex to the $M$ vertex. With this labeling at hand, we can read off the different permutations from a diagram. $\text{Tr}(M^4)$ insertion reads $1\rightarrow 2\rightarrow 3\rightarrow 4$ in the anticlockwise direction in all the diagrams, and hence, the $\sigma_{M}$ is fixed to be $(1234)$. We have drawn the six diagrams in total each corresponding to one $\sigma_K\in [\sigma_K]$. Again, this can be read from the diagram by going in an anti-clockwise direction around the $K$ vertex. Since $\sigma_{f}=(\sigma_{K}\sigma_M)^{-1}$, the sum over $\sigma_f$ clicks six times, due to the delta function in Eq.(\ref{eq:permresult_main}). We have orientation at both $K$ and $M$ vertices(anticlockwise) and it induces an automatic orientation on the faces(as indicated in the box in figure \ref{fig: perm_example}). Hence, we can finally read the $\sigma_f$ permutation by following the arrows of the ribbon graph and noting down the permutation cycles that appear, and verify that this way of obtaining $\s_f$ from the faces agrees with the original definition of $\s_f = (\s_K\s_M)^{-1}$. This establishes the relation between Feynman diagrams and permutations.

We can now calculate the correlator by simply noting that there are three possible cycle structures for $\sigma_f$, $[1^4]$, $[1^13^1]$ and $[2^2]$. $[1^4]$ has four cycles, occurs once in the sum and hence, contributes $N^0$. $[1^13^1]$ has two number of cycles, occurs $4$ times in the sum and hence, contributes $4N^{-2}$. Finally, $[2^2]$  also has two number of cycles, occurs once in the sum and hence, contributes $N^{-2}$. We have used the fact the all the diagrams have $E=4$.
\begin{align}\label{eq: perm_example_1}
    \langle \Tr(M^4)\Tr(K^4)\rangle=\frac{4!}{6}g^4(1+4N^{-2}+N^{-2})=g^4(4+20N^{-2})
\end{align}
As a sanity check, the sum of coefficients is $24$ which is exactly the order of $S_4$. And we expect $24$ possible wick contractions when we look at the sum over $\gamma$ in Eq.(\ref{eq: sum_gamma}), each term contributing either $N^0$ or $N^{-2}$.

The prefactor $n!/|[\sigma_K]|$ carries a group theoretic meaning that it is exactly the size of the centralizer of some $\sigma_K\in [\sigma_K]$. Hence, this is always a positive integer. By comparing \eqref{eq: sum_gamma} to \eqref{eq:permresult_main}, we note the difference of this overall factor, which we have identified as the size of the centralizer of the $[\sigma_K]$. We can interpret this as an indication that not all wick contractions lead to inequivalent Feynman diagrams, rather an $n!/|[\sigma_K]|$ number of equivalent diagrams.

Finally, we could have further summed over $[\sigma_M]$ on both sides to get a more symmetric form of the expression, 
\begin{align}
     \left\langle \prod_{j=1}^{V_M}\text{Tr}(M^{l_j})\prod_{i=1}^{V_K}\text{Tr}(K^{k_i})\right\rangle&=\frac{E!}{|[\sigma_K]||[\sigma_M]|}\sum_{\substack{\sigma_K\in [\sigma_K],\\\sigma_{M}\in [\sigma_{M}],\sigma_f\in S_E}}g^EN^{C_{\sigma_f}-E}\delta(\sigma_M\sigma_f\sigma_K)
\end{align}
where $|[\sigma_{M}]|$ is the size of the conjugacy class of $[\sigma_{M}]$. The right-hand side is no longer representative $\sigma_{M}$ dependent, rather it depends on the conjugacy class $[\sigma_{M}]$. This sum can be related to the sum over the more familiar unlabelled Feynman diagram expansion in some normalization of the vertex operators, 
\begin{align}
\left\langle \prod_{j=1}^{V_M}\left(\inv{l_j} \text{Tr}(M^{l_j})\right)\prod_{i=1}^{V_K}\left(\inv{k_i}\text{Tr}(K^{k_i})\right)\right\rangle \sim \sum_{G\in \text{Feynman Diagrams}}\frac{g^E N^{F-E}}{|\text{Aut}(G)|}= \sum_{G}\frac{g^EN^{2-2h-(V_K + V_M)}}{|\text{Aut}(G)|},
\end{align}
where $h$ is the genus of the Feynman diagram, computed in the usual way. For the example at hand, we will have one Feynman diagram at genus $0$ and two Feynman diagrams at genus $1$ as shown in figure \ref{fig: perm_example}. The ratio of the |Aut($G$)|'s for the more symmetric genus 1 diagram to less symmetric genus 1 diagram (figure \ref{fig: perm_example}) will be 4 as reflected in \eqref{eq: perm_example_1}.

\subsection{The different dual descriptions} \label{sec: diff_duals}
In this section we will discuss in detail how to obtain different matrix model descriptions, whose Feynman diagrams are the (partial) graph duals of the Feynman diagrams of the original $K-M$ model. See \cite{brown2011complex} for very similar manipulations, and a beautiful discussion of the "complex" version of the Kontsevich model.
\subsubsection*{The F type dual} \label{sec:appFtype}
We start with the $K-M$ matrix model with determinant insertions,
\begin{align}
    Z(X,Y,V) =& \frac{\mathcal{K}}{Z_N}\int dK dM_{N \times N} e^{-\frac{N}{g}\text{Tr}_N(K\sqrt{Y}M\sqrt{Y})} \prod_{a=1}^Q \text{det}_{N}(x_a \mathbb{I}_{N}-M)\prod_{\mu=1}^R \text{det}_{N}(v_{\mu}\mathbb{I}_{N}-K)\\
    \text{where, }&\mathcal{K}:=\left( \frac{\det_{N}(Y)}{\det_{Q}(X) \det_{R}(V)} \right)^{N}
\end{align}
where $Z_N = \left(2\pi g/N\right)^{N^2}$. We will write the determinant as a fermionic integral and then integrate out $K$ followed by integrating out $M$ to get a completely fermionic integral($Z(X,Y,V)\equiv Z$),
\begin{align}
    Z=\frac{\mathcal{K}}{Z_N}\int dK dMd\psi d\psi^{\dagger}d\chi d\chi^{\dagger}\exp\bigg(&-\frac{N}{g}\text{Tr}_N(K\sqrt{Y}M\sqrt{Y})+\psi_{ia}^\dagger(X_{ab}\delta_{ij}-\delta_{ab}M_{ij})\psi_{jb}\nonumber\\
    &\hspace{140 pt}+\chi_{i\mu}^{\dagger}(V_{\mu\nu}\delta_{ij}-K_{ij}\delta_{\mu\nu})\chi_{j\nu}\bigg)\nonumber\\
    =\frac{\mathcal{K}}{Z_N}\int dK dM d\psi d\psi^\dagger d\chi d\chi^\dagger \exp\bigg(&-\frac{N}{g} K_{ij}((\sqrt{Y}M\sqrt{Y})_{ji}+g\chi_{i\mu}^{\dagger}\chi_{j\mu})+\psi_{ia}^\dagger(X_{ab}\delta_{ij}-\delta_{ab}M_{ij})\psi_{jb}\nonumber\\
    &\hspace{170 pt}+\chi_{i\mu}^{\dagger}V_{\mu\nu}\chi_{i\nu} \bigg)
\end{align}
Integrating out $K$ will give a $(2\pi g/N)^{N^2}\delta\left((\sqrt{Y}M\sqrt{Y})_{ji}+\frac{g}{N}\chi^{\dagger}_{i\mu}\chi_{j\mu}\right)$, which upon changing the integral variables $\tilde{M}=\sqrt{Y}M\sqrt{Y}$ sets $M_{ij}=-\frac{g}{N}(Y^{-\half})_{ik}\chi_{l\mu}^{\dagger}\chi_{k\mu}(Y^{-\half})_{lj}$, and also cancels the $Z_N$ factor. We end up with fully fermionic integral,
\begin{align}\label{fullyfermion}
    Z=\mathcal{K}'\int d\psi d\psi^{\dagger}d\chi d\chi^{\dagger}\exp\bigg(\psi^{\dagger}_{ia}X_{ab}\psi_{ib}+\chi^{\dagger}_{i\mu}V_{\mu\nu}\chi_{i\nu}+\frac{g}{N}\psi^{\dagger}_{ia}(Y^{-\half})_{ik}\chi_{l\mu}^{\dagger}\chi_{k\mu}(Y^{-\half})_{lj}\psi_{ja}\bigg)
\end{align}
where $\mathcal{K}'=(\text{det}_Q(X)\text{det}_R(V))^{-N}$. This is the equivalent description of our model in terms of the two kinds of fermions. \\
Next, we integrate in a gaussian integral of complex bosonic fields $S_{a\mu}$ with the shifts ensuring that the quartic term($\sim \psi\psi^{\dagger}\chi\chi^{\dagger}$ term) is exactly cancelled, 
\begin{align}\label{fermionic+bosonic}
Z=&\frac{\mathcal{K}'}{Z_{QR}}\int d\psi d\psi^{\dagger}d\chi d\chi^{\dagger}d S dS^{\dagger}\exp\bigg( \frac{N}{g}(S^{\dagger}_{\mu a}-\frac{g}{N}\psi_{ia}^{\dagger}(Y^{-\half})_{ik}\chi_{k\mu})(S_{a\mu}-\frac{g}{N}\chi^{\dagger}_{l\mu}(Y^{-\half})_{lj}\psi_{ja})  \nonumber\\
&\qquad\quad\qquad\qquad\qquad\qquad+\psi^{\dagger}_{ia}X_{ab}\psi^{\dagger}_{ib}+\chi^{\dagger}_{i\mu}V_{\mu\nu}\chi_{i\nu}-\frac{g}{N}\psi^{\dagger}_{ia}(Y^{-\half})_{ik}\chi_{k\mu}\chi_{l\mu}^{\dagger}(Y^{-\half})_{lj}\psi_{ja}\bigg)\nonumber\\
=&\frac{\mathcal{K}'}{Z_{QR}}\int d\psi d\psi^{\dagger}d\chi d\chi^{\dagger}d S dS^{\dagger}\exp\bigg( \frac{N}{g}S^{\dagger}_{\mu a}S_{a\mu}-S^{\dagger}_{\mu a}\chi^{\dagger}_{l\mu}(Y^{-\half})_{lj}\psi_{ja}-\psi_{ia}^{\dagger}(Y^{-\half})_{ik}\chi_{k\mu}S_{a\mu} \nonumber\\ &\qquad\qquad\qquad\qquad\qquad\qquad\qquad\qquad\qquad\qquad\quad+\psi^{\dagger}_{ia}X_{ab}\psi_{ib}+\chi^{\dagger}_{i\mu}V_{\mu\nu}\chi_{i\nu}\bigg)
\end{align}
where $Z_{QR}= (\frac{2\pi g}{N})^{QR}$. We can now integrate out the fermions to get a bosonic theory. Terms involving fermions ($\psi, \chi$) in the exponent can be written in the following matrix product form,
\begin{align}
    \begin{pmatrix}
\psi^{\dagger}_{ia}& \chi_{i\mu}^\dagger
\end{pmatrix}\begin{pmatrix}
    X_{ab}\delta_{ij} & -(Y^{-\half})_{ij}S_{a\nu}\\ 
    - S^{\dagger}_{\mu b}(Y^{-\half})_{ij}& V_{\mu\nu}\delta_{ij}
\end{pmatrix}\begin{pmatrix}
 \psi_{jb} \\ \chi_{j\nu}  
\end{pmatrix}=\begin{pmatrix}
\psi^{\dagger}_{ia}& \chi_{i\mu}^\dagger
\end{pmatrix}\begin{pmatrix}
    X_{ab}\psi_{ib}-(Y^{-\half})_{ij}S_{a\nu}\chi_{j\nu}\\
    - S^\dagger_{\mu b}(Y^{-\half})_{ij}\psi_{jb}+V_{\mu\nu}\chi_{i\nu}
\end{pmatrix}\nonumber\\
=\psi^{\dagger}_{ia}X_{ab}\psi_{ib}-\psi^{\dagger}_{ia}Y^{-\half}_{ij}S_{a\nu}\chi_{j\nu}-\chi_{i\mu}^{\dagger} S^\dagger_{\mu b}(Y^{-\half})_{ij}\psi_{jb}+\chi_{i\mu}^{\dagger}V_{\mu\nu}\chi_{i\nu}
\end{align}
which is the same as the terms we had in the \eqref{fermionic+bosonic}. Now, we can carry out the $\psi, \chi$ integral,
\begin{align}
Z=&\frac{\mathcal{K}'}{Z_{QR}}\int d S d S^{\dagger}e^{\frac{N}{g} S^{\dagger}_{\mu a}S_{a\mu}} \int d\psi d\psi^{\dagger}d\chi d\chi^{\dagger}\exp\bigg\{\begin{pmatrix}
\psi^{\dagger}_{ia}& \chi_{i\mu}^\dagger
\end{pmatrix}\begin{pmatrix}
    X_{ab}\delta_{ij} & -(Y^{-\half})_{ij}S_{a\nu}\\ 
    - S^{\dagger}_{\mu b}(Y^{-\half})_{ij}& V_{\mu\nu}\delta_{ij}
\end{pmatrix}\begin{pmatrix}
 \psi_{jb} \\ \chi_{j\nu}  
\end{pmatrix}\bigg\}
\end{align}
Now, upon integrating out fermions we are left with a determinant of the matrix (we have assumed $Y_{ij} = y_i \de_{ij}$ below for simplicity),
\begin{align}
    Z=\frac{\mathcal{K}'}{Z_{QR}}\int d S d S^{\dagger}\exp\bigg( \frac{N}{g} S^{\dagger}_{\mu a}S_{a\mu} \bigg)\prod_{i=1}^N\text{det}_{Q+R}\begin{pmatrix}
    X_{ab} & -(y_i)^{-\half}S_{a\nu}\\ 
    - S^{\dagger}_{\mu b}(y_{i})^{-\half}& V_{\mu\nu}
\end{pmatrix}
\end{align}
The determinant insertion can be manipulated and turned into a potential term in the action as 
a power series in $ S^\dagger  S$. Anyways, this is the bosonic theory we wanted to get. This matrix can be written as follows,
\begin{align}
    \text{det}_{Q+R}\begin{pmatrix}
    X & -(y_i)^{-\half} S\\ 
    - S^{\dagger}(y_{i})^{-\half}& V
\end{pmatrix}=\text{det}_{Q+R}\bigg(\begin{pmatrix}
    X& 0\\ 
    0& V
\end{pmatrix}\begin{pmatrix}
    1&-(y_i)^{-\half} X^{-1} S \\ -(y_i)^{-\half}V^{-1} S^{\dagger}&1
\end{pmatrix}\bigg)\nonumber\\
=\text{det}_{Q}(X)\text{det}_R(V)\text{det}_{Q+R}(\mathbb{I}_{Q+R}-\Lambda_i)\\
\text{where,  }
    \Lambda_i=\begin{pmatrix}
    0&(y_i)^{-\half} X^{-1} S \\ (y_i)^{-\half}V^{-1} S^{\dagger}&0
\end{pmatrix}\text{ and } \text{Tr}_{Q+R}(\Lambda_i^{2m})=2y_i^{-m}\text{Tr}_Q((X^{-1} S V^{-1} S^{\dagger})^m)
\end{align}
So, finally we end up with the partition function upon exponentiating the determinant as trace of the logarithm and then, expanding the logarithm.
\begin{align} 
Z = \frac{1}{Z_{QR}}  \int d S d S^{\dagger} \exp\bigg(\frac{N}{g} S^{\dagger}_{\mu a}S_{a\mu} - \sum_k \inv{k} (\sum_{i=1}^N y_i^{-k}) \text{Tr}_Q((X^{-1} S V^{-1} S^{\dagger})^k\bigg)
\end{align} 
Now we do the following transformations: $ S \to X^{-\half}  S V^{-\half},  S^{\dagger} \to  V^{-\half} S^{\dagger} X^{-\half} $. Doing these transformations and taking into account the Jacobians, we end up with the expressions
\begin{align} 
Z = \frac{(\text{det}_Q(X))^{R}(\text{det}_{R}(V))^{Q}}{Z_{QR}} \int d S d S^{\dagger} \exp(\frac{N}{g}Tr(V S^{\dagger} X S) - \sum_k  \frac{s_k}{k} Tr(( S S^{\dagger})^k)
\end{align}
Where $s_k =  Tr_{N}(Y^{-k})$. Now we can see that the source $Y$ is associated to vertices in this picture, and the sources $X$ and $V$ are associated to the `$Q$ faces' and `$R$ faces' of the Feynman diagram respectively.

\subsubsection*{Intermediate Single fermionic integrals}
We can also get single fermionic integrals as well from  \eqref{fullyfermion} by selectively integrating out $\psi$ or $\chi$,
\begin{align}
    Z=\bigg(\frac{\mathcal{K}}{det_N(Y)^N}\bigg)\int d\psi d\psi^{\dagger}d\chi d\chi^{\dagger}\exp\bigg(\psi^{\dagger}_{ia}X_{ab}\psi_{ib}+\chi^{\dagger}_{i\mu}V_{\mu\nu}\chi_{i\nu}+\frac{g}{N}\psi^{\dagger}_{ia}(Y^{-\half})_{ik}\chi_{l\mu}^{\dagger}\chi_{k\mu}(Y^{-\half})_{lj}\psi_{ja}\bigg)
\end{align}
Integrating out $\chi$ will give us a $\prod_{\mu=1}^R\det_{N}(v_{\mu}\delta_{ij}-\frac{g}{N}(Y^{-\half})_{il}(\psi\psi^{\dagger})_{lk}Y^{-\half}_{kj})=\det(V)^N\prod_{\mu=1}^N\det_N(I - \frac{g}{Nv_{\mu}}Y^{-1}\psi\psi^{\dagger})$ which can further be written in terms of exponents to give,
\begin{align}\label{fermion1}
    Z&=\mathcal{K}\bigg(\frac{\det(V)}{\det(Y)}\bigg)^N\int d\psi d\psi^{\dagger}\exp\bigg(\psi^{\dagger}_{ia}X_{ab}\psi_{ib} - \sum_{k=1}^{\infty}\frac{\Tr_{R}V^{-K}}{k}\Tr_N(\frac{g}{N}Y^{-1}\psi\psi^{\dagger})^k\bigg) \nonumber\\
    &=\mathcal{K}\bigg(\frac{\det(V)^N}{\det(Y)^{N-Q}}\bigg)\int d\psi d\psi^{\dagger}\exp\bigg(\psi^{\dagger}_{ia}X_{ab}Y_{ij}\psi_{jb} - \sum_{k=1}^{\infty}\frac{\Tr_{R}V^{-K}}{k}\Tr_N(\frac{g}{N}\psi\psi^{\dagger})^k\bigg)
\end{align}
Where we did $\psi \to Y^{-\half}\psi$ in the second line. If we had integrated out $\psi$, we would have ended up with,
\begin{align}\label{fermion2}
    Z&=\mathcal{K}\bigg(\frac{\det(X)}{\det(Y)}\bigg)^N\int d\chi d\chi^{\dagger}\exp\bigg(\chi^{\dagger}_{i\mu}V_{\mu\nu}\chi_{i\nu} -\sum_{k=1}^{\infty}\frac{\Tr_{Q}X^{-K}}{k}\Tr_N(\frac{g}{N}Y^{-1}\chi\chi^{\dagger})^k\bigg)\nonumber\\
    &=\mathcal{K}\bigg(\frac{\det(X)^N}{\det(Y)^{N-R}}\bigg)\int d\chi d\chi^{\dagger}\exp\bigg(\chi^{\dagger}_{i\mu}V_{\mu\nu}Y_{ij}\chi_{j\nu} - \sum_{k=1}^{\infty}\frac{\Tr_{Q}X^{-K}}{k}\Tr_N(\frac{g}{N}\chi\chi^{\dagger})^k\bigg)
\end{align}
where we did $\chi \to Y^{-\half}\chi$ in the second line. \eqref{fermion1} and \eqref{fermion2} are the fermionic integrasls which are obtained as an intemediate step in going from $KM$ to $AB$ and $KM$ to $CD$ model respectively.

\subsubsection*{The partial duals}
We can start with (\ref{fermion1}), and write the interaction terms in a `color-flavor transformed' form \cite{AltlandSonner2020, Altland2022,goel2021string}
\be 
\Tr_N(\psi\psid)^k = - \Tr_Q(\psid\psi)^k
\ee 
And introduce a delta function of the form
\be 
1 = \int dA \de (A_{ba} + \frac{g}{N}\psid_{ia}\psi_{ib})
\ee 
and write it as an exponential. So we end up with 
\be 
\begin{split}
    Z&=\mathcal{K}\bigg(\frac{\det(V)^N}{\det(Y)^{N-Q} Z_{Q}}\bigg)\int d\psi d\psi^{\dagger}\exp\bigg(\psi^{\dagger}_{ia}X_{ab}Y_{ij}\psi_{jb}+ \sum_{k=1}^{\infty}\frac{\Tr_{R}V^{-K}}{k}\Tr_Q(\frac{g}{N}\psid\psi)^k\bigg)\\
    &\hspace{30pt}\int dA dB \exp\bigg( -\frac{N}{g} B_{ab}(A_{ba} + \frac{g}{N}\psid_{ia}\psi_{ib}) \bigg)
\end{split}
\ee 
Using the delta function relation, we can rewrite the fermion interaction term as 
\be 
 \sum_{k=1}^{\infty}\frac{\Tr_{R}V^{-K}}{k}\Tr_Q(\frac{g}{N}\psid\psi)^k \to \sum_{k=1}^{\infty}(-1)^{k}\frac{\Tr_{R}V^{-K}}{k} \Tr_Q(A^k) 
\ee 
On the other hand, now we can integrate out the $\psi_{ia}$ fermions as they are quadratic, and end up with $\prod_{i}det(y_i X_{ab} - B_{ab})$. Doing both the steps, we end up with  
\begin{align}\label{AB model}
    Z=&\bigg( \frac{\text{det}_{Q}(X) \text{det}_{R}(V)}{\text{det}_N(Y)}\bigg)^{Q}\frac{1}{Z_Q}\int dA d B e^{-\frac{N}{g}\text{Tr}_Q(\sqrt{X}B\sqrt{X} A)} \frac{\text{det}_{N+Q}(I_{Q}\otimes Y -B\otimes I_{N})}{ \text{det}_{Q+R} (I_{Q}\otimes V + A\otimes I_{R})}\\
    =&\bigg( \frac{\text{det}_{Q}(X) \text{det}_{R}(V)}{\text{det}_N(Y)}\bigg)^{Q}\frac{1}{Z_Q}\int dA dB e^{-\frac{N}{g}\text{Tr}_Q(\sqrt{X}B\sqrt{X} A)} \prod_{i=1}^N\text{det}_Q(y_i-B)\prod_{\mu=1}^R(\text{det}_Q(v_{\mu}+A))^{-1}
\end{align}
where we have rescaled $B \to X^{-\half}BX^{-\half}$ in the intermediate step. This action consists of $Q\times Q$ matrices $A, B$, and the sources $Y$ and $V$ are associated to $B$ and $A$ vertices respectively, and the source $X$ is associated to the faces of its Feynman diagram. So the Feynman diagrams of this model are partial duals of the $KM$ model. \\
We can alternatively start with (\ref{fermion2}), and do the following `color flavor transformation'
\be 
\Tr_N(\chi\chid)^k = - \Tr_R(\chid\chi)^k
\ee 
and introduce the delta function
\be 
1 = \int dD \de\bigg( D_{\nu\mu} + \frac{g}{N}\chid_{i\mu}\chi_{i\nu} \bigg) = \inv{Z_R} \int dC dD \exp\bigg(-\frac{N}{g} C_{\mu\nu}(D_{\nu\mu} + \frac{g}{N}\chid_{i\mu}\chi_{i\nu}) \bigg)
\ee 
which will allow us to integrate out the $\chi$ fermions, similar to the last derivation. We will end up, after certain rescalings, with 
\be 
\begin{split}
    Z&= \bigg(\frac{\text{det}_Q(X)}{\text{det}_N(Y)}\bigg)^R\frac{1}{Z_R} \int dCdD e^{-\frac{N}{g}\text{Tr}_R(CD)}\frac{\text{det}_{N+R}(I_R\otimes Y-V^{-1}C\otimes I_N )}{\text{det}_{Q+R}(I_R\otimes X + D\otimes I_Q)}\\
    &\equiv \bigg(\frac{\text{det}_R(V)\text{det}_Q(X)}{\text{det}_N(Y)}\bigg)^R\frac{1}{Z_R} \int d{C}dD e^{-\frac{N}{g}\text{Tr}_R(\sqrt{V}{C}\sqrt{V}D)}\prod_{i=1}^N \text{det}_R(y_i - { C}) \prod_{a=1}^{Q} (\text{det}_R(x_a + D))^{-1}
\end{split}
\ee 
This is the other `partial dual' model, where the sources $Y$ and $X$ are associated to $ C$ and $D$ vertices respectively, and the source $V$ is associated to the face of a Feynman diagram.

\bibliographystyle{utphys}
\bibliography{refs}

\end{document}